\documentclass[fleqn,usenatbib]{mnras}

\usepackage{newtxtext,newtxmath}

\usepackage[T1]{fontenc}
\usepackage{ae,aecompl}

\usepackage{arydshln}

\usepackage[flushleft]{threeparttable}

\usepackage{etoolbox}

\usepackage{graphicx}	
\usepackage{amsmath}	
\usepackage{float}
\usepackage{gensymb}

\title[X-SHOOTER Lyman-$\alpha$ Survey at $z=2$: Overview]{The X-SHOOTER Lyman-$\alpha$ survey at $z=2$ (XLS-$z2$) I: What makes a galaxy a Lyman-$\alpha$ emitter?}
\author[Matthee et al.]{Jorryt Matthee$^{1}$\thanks{Zwicky Fellow. E-mail: mattheej@phys.ethz.ch}, David Sobral$^{2}$, Matthew Hayes$^{3}$, Gabriele Pezzulli$^{4}$, 
\newauthor Max Gronke$^{5}$, Daniel Schaerer$^{6,7}$, Rohan P. Naidu$^{8}$, Huub R\"ottgering$^{9}$, 
\newauthor Jo\~ao Calhau$^{10,11}$, Ana Paulino-Afonso$^{12}$, S\'ergio Santos$^{2}$ and Ricardo Amor\'in$^{13,14}$ \\
$^{1}$ Department of Physics, ETH Z\"urich, Wolfgang-Pauli-Strasse 27, 8093 Z\"urich, Switzerland \\
$^{2}$ Department of Physics, Lancaster University, Lancaster, LA1 4YB, UK \\
$^{3}$ Stockholm University, Department of Astronomy and Oskar Klein Centre for Cosmoparticle Physics,\\ AlbaNova University Centre, SE-10691, Stockholm, Sweden \\
$^{4}$ Kapteyn Astronomical Institute, University of Groningen, Landleven 12, 9747 AD Groningen, The Netherlands \\
$^{5}$ Department of Physics \& Astronomy, Johns Hopkins University, Bloomberg Center, 3400 N. Charles St., Baltimore, MD 21218, USA\\
$^{6}$ Observatoire de Gen\`eve, Universit\' e Gen\`eve, 51 Ch. des Maillettes, 1290 Versoix, Switzerland \\
$^{7}$ Universit\'e de Toulouse; UPS-OMP; IRAP; Toulouse, France \\
$^{8}$  Center for Astrophysics $|$ Harvard \& Smithsonian, 60 Garden Street, Cambridge, MA 02138, USA\\
$^{9}$ Leiden Observatory, Leiden University, PO\ Box 9513, NL-2300 RA Leiden, The Netherlands\\
$^{10}$ Instituto de Astrof\'isica de Canarias, E-38200 La Laguna, Tenerife, Spain \\
$^{11}$ Departamento de Astrof\'isica, Universidad de La Laguna, E-38205 La Laguna, Tenerife, Spain \\ 
$^{12}$ Centro de Astrof\'{\i}sica e Gravita\c c\~ao  - CENTRA, Departamento de F\'{\i}sica, \\ Instituto Superior T\'ecnico - IST, Universidade de Lisboa - UL, Av. Rovisco Pais 1, 1049-001 Lisboa, Portugal\\
$^{13}$ Instituto de Investigaci\'on Multidisciplinar en Ciencia y Tecnolog\'ia, Universidad de La Serena, Ra\'ul Bitr\'an 1305, La Serena, Chile\\
$^{14}$ Departamento de Astronom\'ia, Universidad de La Serena, Av. Juan Cisternas 1200 Norte,  La Serena, Chile}

\pubyear{2021}

\begin{document}
\label{firstpage}
\pagerange{\pageref{firstpage}--\pageref{lastpage}}
\maketitle

\begin{abstract}
We present the first results from the X-SHOOTER Lyman-$\alpha$ survey at $z=2$ (XLS-$z2$). XLS-$z2$ is a deep spectroscopic survey of 35 Lyman-$\alpha$ emitters (LAEs) utilising $\approx90$ hours of exposure time with VLT/X-SHOOTER and covers rest-frame Ly$\alpha$ to H$\alpha$ emission with R$\approx4000$. We present the sample selection, the observations and the data reduction. Systemic redshifts are measured from rest-frame optical lines for 33/35 sources. In the stacked spectrum, our LAEs are characterised by an interstellar medium with little dust, a low metallicity and a high ionisation state. The ionising sources are young hot stars that power strong emission-lines in the optical and high ionisation lines in the UV. The LAEs exhibit clumpy UV morphologies and have outflowing kinematics with blue-shifted Si{\sc ii} absorption, a broad [O{\sc iii}] component and a red-skewed Ly$\alpha$ line. Typically 30 \% of the Ly$\alpha$ photons escape, of which one quarter on the blue side of the systemic velocity. A fraction of Ly$\alpha$ photons escapes directly at the systemic suggesting clear channels enabling a $\approx10$ \% escape of ionising photons, consistent with an inference based on Mg{\sc ii}. A combination of a low effective H{\sc i} column density, a low dust content and young starburst determine whether a star forming galaxy is observed as a LAE. The first is possibly related to outflows and/or a fortunate viewing angle, while we find that the latter two in LAEs are typical for their stellar mass of 10$^9$ M$_{\odot}$.
\end{abstract}

\begin{keywords}
galaxies: formation -- galaxies: high-redshift -- cosmology: dark ages, reionization, first stars --  galaxies: starburst -- galaxies: ISM
\end{keywords}



\section{Introduction}
In the last two decades, the Lyman-$\alpha$ emission line (Ly$\alpha$; $\lambda_0=1215.67$ {\AA}) has fulfilled its longstanding promise \citep{PartridgePeebles1967} of being a powerful tool to study galaxies in the early Universe \citep[e.g.][]{Rhoads2000,Gawiser2007,Ouchi2008,Hayes2010,Kashikawa2011,Matthee2015,Konno2016,Drake2017,Zheng2017,Taylor2020}. Although perhaps not as bright as intrinsically expected \citep[e.g.][]{CharlotFall1993,Hayes2015}, its high equivalent width, its rest-frame UV wavelength, the adjacent continuum breaks in the spectrum and the peculiar line-shape have made Ly$\alpha$ an extremely useful emission line to find and spectroscopically identify the redshifts of galaxies out to the most distant Universe \citep[e.g.][]{Finkelstein2013,Oesch2015,Zitrin2015}.

A key uncertainty in the study of Ly$\alpha$ emission from galaxies is the Ly$\alpha$ escape fraction, $f_{\rm esc, Ly\alpha}$, and how this depends on properties of the interstellar medium (ISM). The Ly$\alpha$ transition has a high scattering cross section and thus is resonant (see \citealt{DijkstraReview} for a review). This means that only small amounts of neutral hydrogen in the ISM are needed to cause Ly$\alpha$ photons to scatter significantly. Scattering increases the effective path-length and the likelihood of absorption by dust while also leading to a diffusion in frequency and space \citep{Neufeld1990,MasRibas2017}. How this exactly happens depends on gas turbulence, the column density distribution and clumpiness, the velocity field and the dust content in a complex way \citep[e.g.][]{Verhamme2006,GronkeDijkstra2016,Gronke2017b}. 

The same characteristics that make Ly$\alpha$ observations so attractive at high-redshift also make it challenging to determine the physical properties of Ly$\alpha$ emitters (LAEs) in great detail. First, the high equivalent widths (EW) are often accompanied by a faint UV continuum and some LAEs are not even detected in the deepest imaging that exists \citep[e.g.][]{Maseda2018}. Second, the well-understood strong rest-frame optical emission lines such as [O{\sc iii}]$_{5008}$ and H$\alpha$ are currently difficult or even impossible to observe at $z>3$ due to the atmospheric emission and background in the infrared. This redshift coincides with the redshift where the majority of LAEs are found with Ly$\alpha$ being redshifted into the optical. Therefore, many open questions remain. What is the typical $f_{\rm esc, Ly\alpha}$? Why are not all distant star-forming galaxies observed as LAEs \citep[e.g.][]{Hayes2010,Hagen2016,Matthee2016}? What makes a galaxy a LAE?

Pioneering studies have shown that $f_{\rm esc, Ly\alpha}$ is $\approx 30$ \% in LAEs \citep{Nakajima2012,Blanc2011,Song2014,Trainor2016,Sobral2016}, but where these photons escape in the spectral and spatial domain has been poorly explored. Samples of star-forming galaxies at $z\approx2$ that are selected irrespective of their Ly$\alpha$ luminosity have much lower $f_{\rm esc, Ly\alpha}$ in the range $1-5$ \% \citep{Hayes2010,Matthee2016}. Furthermore, $f_{\rm esc, Ly\alpha}$ has been found to correlate with nebular dust attenuation \citep{Atek2008,Blanc2011,Yang2017}, but additional independent correlations with other properties such as the H{\sc i} column density, gas-phase metallicity and possibly viewing angle have been reported as well \citep{Shibuya2014,Henry2015,Trainor2016,Yang2017}. It is likely that several processes impact $f_{\rm esc, Ly\alpha}$, but it is yet to be explored in detail \citep[e.g.][]{Runnholm2020} how these vary with mass and redshift.

Besides $f_{\rm esc, Ly\alpha}$, the Ly$\alpha$ output of a galaxy is also determined by the production rate of Ly$\alpha$ photons, which is tightly linked to the production rate of ionising photons. The intrinsic Ly$\alpha$ EW is related to the production rate of ionising photons relative to the UV continuum and therefore sensitive to the spectrum of the ionising sources \citep[e.g.][]{CharlotFall1993,Raiter2010}. In particular, an intrinsically high Ly$\alpha$ EW is an indicator of galaxies with very young and extremely low-metallicity stars \citep[e.g.][]{CharlotFall1993,Raiter2010,Sobral2015,Maseda2020}. A high EW could also be indicative of additional sources of Ly$\alpha$ emission such as fluorescence in the proximity of a luminous ionising source (such as a quasar; e.g. \citealt{Cantalupo2012,Marino2018}) or collisional excitation from gravitational collapse \citep[e.g.][]{Dijkstra2006}. In this work we assume that the main production mechanism of Ly$\alpha$ photons is recombination associated to star formation within galaxies.

LAEs at intermediate redshift are of interest as they may be good and practically useful analogues to the galaxies responsible for reionisation. This is because the galaxies that are known to have the highest Lyman Continuum (LyC; $\lambda_0<912$ {\AA}) escape fractions are strong LAEs \citep[e.g.][]{Izotov2018} and because the Ly$\alpha$ EW is observed to correlate with the LyC escape fraction \citep{Marchi2018,Steidel2018}. It is thus plausible that galaxies that are significantly leaking LyC photons can be found more easily in samples of LAEs compared to general galaxy samples. The LyC escape fraction is a key parameter for understanding the sources of cosmic reionisation \citep{Robertson2013,Naidu2019}. It is however very challenging to measure for individual systems at $z>3$ due to the stochastic opacity of the intergalactic medium (IGM; \citealt{Madau1995,Inoue2014,Vanzella2018}). Such measurements are possible in low-redshift analogues of distant galaxies \citep[e.g.][]{Izotov2018,Jaskot2019}, but this requires challenging UV spectroscopy and selection functions are typically complicated. It is furthermore unclear whether the star formation histories (SFHs) of local analogues truly resemble galaxies in the early Universe \citep{Amorin2012a}. The Ly$\alpha$ escape fraction and line-shape that emerges from the ISM are correlated with the escape fraction of LyC photons \citep{Verhamme2015,Dijkstra2016,Izotov2018,Gazagnes2020}, but are much simpler to measure for larger samples and over a range in cosmic times. Measurements of the Ly$\alpha$ profile over a range of galaxy properties and cosmic times are therefore a promising avenue for mapping the contribution of various galaxies to the epoch of reionisation \citep{Matthee2018} and currently $z\approx2$ is the highest redshift where it is possible to control for the intrinsic Ly$\alpha$ production from the ground.

Due to its sensitivity to intervening neutral hydrogen, the evolution of the Ly$\alpha$ luminosity function \citep[e.g.][]{Konno2018} and the observed distributions of Ly$\alpha$ EWs among galaxies are used as a tracer of the evolution of the neutral fraction into the epoch of reionisation \citep[e.g.][]{Stark2010,Jung2018,Mason2018b}. The impact of the (neutral) IGM however depends on the specific velocity at which Ly$\alpha$ photons escape the ISM \citep[e.g.][]{Dijkstra2014}, and the EW may also vary due to evolution in the intrinsic Ly$\alpha$ luminosity. Therefore, to fully interpret the results from such surveys in the context of an evolving IGM, we require a complete understanding of the variation of the Ly$\alpha$ line profile and the production and escape of Ly$\alpha$ photons among galaxies. In particular, it is crucial to map out how these variations are dependent on galaxy properties that are not affected by the evolution of the IGM, such as rest-frame optical line strengths.

To make progress on these aspects, we have undertaken a large narrow-band survey of LAEs at $z\approx2$ \citep{Sobral2016}, which is currently the only redshift where it is possible to measure all the important lines in the wavelength range between Ly$\alpha$ and H$\alpha$ with ground-based facilities. Here we present the first results of the X-SHOOTER Lyman-$\alpha$ Survey at redshift $z=2$ (XLS-$z2$), which constitutes the spectroscopic component of this program. The sample is composed of 35 objects, of which 20 newly observed and 15 with archival data. This survey optimally uses the large wavelength coverage ($\lambda=0.3-2.5 \mu$m) of the X-SHOOTER instrument \citep{Vernet2011} on the Very Large Telescope (VLT), meaning that we observe all emission lines from Ly$\alpha$ to H$\alpha$ simultaneously. Moreover, the spectral resolution of the Ly$\alpha$ observations of our set-up ($R\sim4000$) is significantly higher than most Ly$\alpha$ studies at $z>2$ \citep[e.g.][]{Kulas2012,Trainor2015,Verhamme2018,Hayes2020} and the non-resonant lines in the rest-frame optical allow stringent estimates of the systemic redshift that most high-redshift studies lack. These data allow us to connect faint spectroscopic features in the rest-frame UV to the well-known optical lines. With these data we thus simultaneously measure the intrinsic Ly$\alpha$ budget, the attenuation and various escape mechanisms as the lowest HI column density paths and related ionisation parameter, and the presence of outflows and their velocities.

In this paper we present the selection of the sample and discuss how representative this sample is at $z\approx2$ (\S $\ref{sec:sample}$). The observations and data reduction are detailed in \S $\ref{sec:data}$. We present our methods for extracting aperture-matched 1D spectra from individual objects and stacks in \S $\ref{method:1D}$ and how we measure systemic redshifts for the majority of the sample (\S $\ref{sec:zsys}$). In order to address which properties make galaxies LAEs, we focus on a stack of LAEs that are representative for the population of LAEs at $z=2.2$. In \S $\ref{sec:stack_measurements}$ we present measurements of various absorption and emission-lines from the UV to the rest-frame optical. These are used to determine the nature of the ionising sources, the Ly$\alpha$ escape fraction, the star formation rate and various properties of the ISM (\S $\ref{sec:results}$). In \S $\ref{sec:discussion}$, we discuss what these results imply for the nature of LAEs, in particular what determines whether galaxies are observed as LAEs and what these results imply for galaxies in the epoch of reionisation. \S $\ref{sec:summary}$ summarises our results.

We use a flat $\Lambda$CDM cosmology with $\Omega_{\rm M}=0.3$ and $H_0=70$ km s$^{-1}$ Mpc$^{-1}$ and a \cite{Chabrier2003} initial mass function (IMF). For solar abundances we use the reference values $Z_{\star}=0.0142$ and 12+log$_{10}$(O/H)=8.7. Emission-line wavelengths are presented as vacuum wavelengths. All equivalent widths are in the rest-frame. Magnitudes are in the AB system.

\section{Sample} \label{sec:sample}
\subsection{Selection criteria} \label{sec:criteria}
Our full sample consists of 35 Ly$\alpha$ flux-limited selected galaxies at redshifts $z=2.00-2.47$ (with a median redshift $z=2.22$) that have mostly been pre-selected with narrow-band imaging and that have been observed by the X-SHOOTER spectrograph. 

The main selection criterion for our sample is that targets are known Ly$\alpha$ emitters at $z\approx2$ with Ly$\alpha$ EW$_0>25$ {\AA}. This redshift is high enough for observed Ly$\alpha$ photons to be shifted beyond the atmospheric transmission cut-off in the UV, but low enough for H$\alpha$ to avoid the high thermal background.
At this particular redshift the strong rest-frame optical H$\alpha$, [O{\sc iii}], H$\beta$ and [O{\sc ii}] lines all lie in regions with high atmospheric transmission \citep[e.g.][]{Nakajima2012,Sobral2013,Khostovan2015}.  

\begin{figure}
\includegraphics[width=8.5cm]{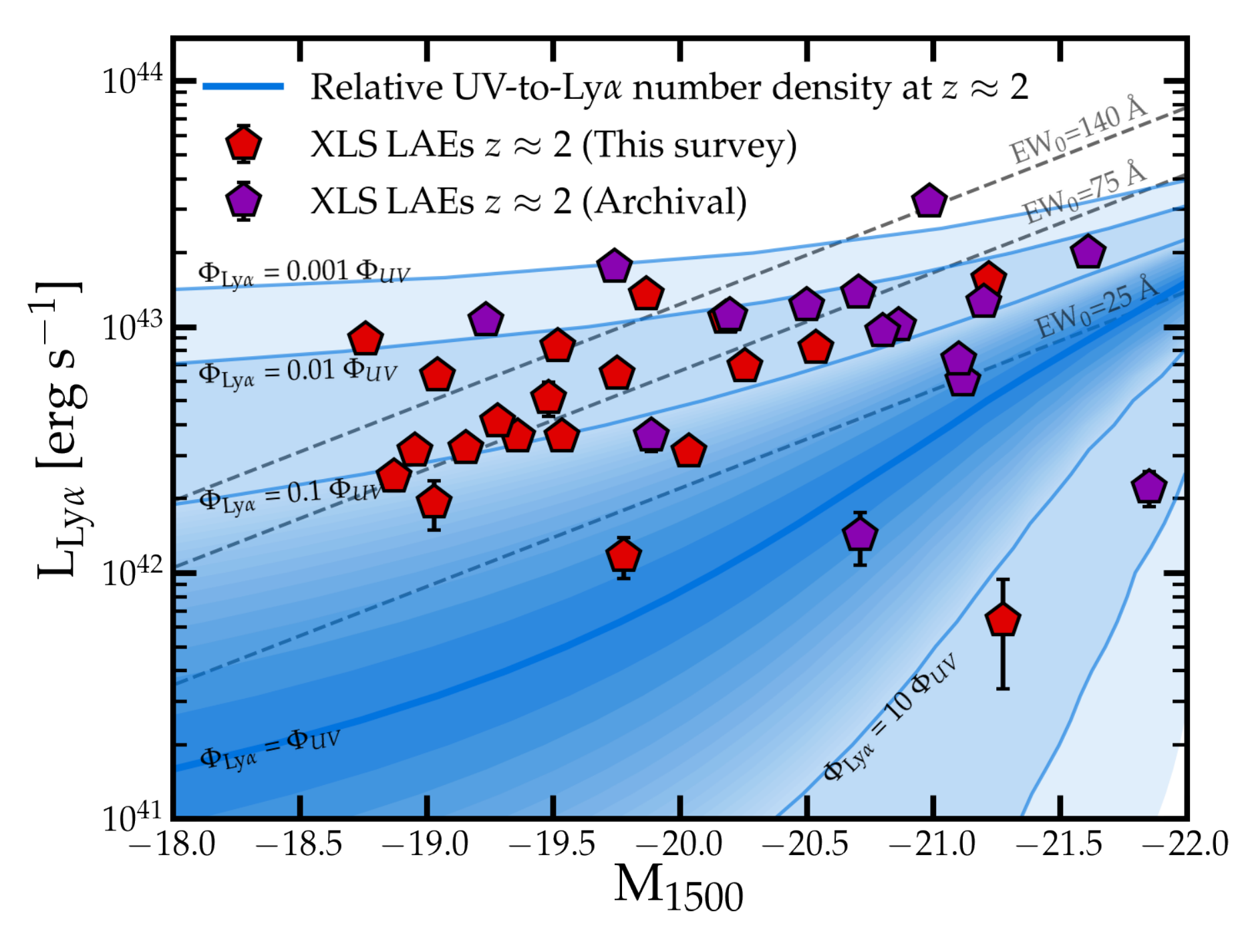}
\caption{The UV and Ly$\alpha$ luminosities of the XLS galaxies. Targets that have been observed as part of our own survey are shown in red and archival targets are shown in purple. Grey dashed lines illustrate lines at constant Ly$\alpha$ equivalent width assuming a fixed UV slope $\beta=-2$. The blue line illustrates the L$_{\rm Ly\alpha}$ - M$_{1500}$ relation that is the intercept along which galaxies have the same number densities per log luminosity interval according to their UV (\citealt{Parsa2016}) and Ly$\alpha$ luminosity (\citealt{Sobral2016} ) at $z\approx2$. We also show the relations required for various relative number densities. While XLS LAEs are representative for the Ly$\alpha$-selected galaxy population, they are a rare (typically 1:15) subsample of the general UV-selected galaxy population at $z\approx2$.}
\label{fig:sample}
\end{figure}

The majority (25/35) of our targets are directly selected based on strong Ly$\alpha$ emission identified in narrow-band imaging spanning a combined volume of $\approx5\times10^6$ cMpc$^{3}$ in well-studied extragalactic fields as COSMOS, UDS and CFHTLS-W4. This selection technique effectively implies a Ly$\alpha$ flux and EW limit. The combination of typical narrow-band filter widths and a 3$\sigma$ excess significance typically means imposing an EW limit of $>25$ {\AA} \citep[e.g.][]{Gronwall2007,Ouchi2008,Sobral2016,Matthee2017Bootes}. However, by prioritising the spectroscopic follow-up to objects with a somewhat higher Ly$\alpha$ flux ($\gtrsim3\times10^{-17}$ erg s$^{-1}$ cm$^{-2}$; on average $2\times10^{-16}$ erg s$^{-1}$ cm$^{-2}$) we are effectively slightly skewed towards higher EWs at fainter UV luminosities (Fig. $\ref{fig:sample}$). A few targets have Ly$\alpha$ EWs $\lesssim10$ {\AA} as those were identified with a narrower filter \citep{Matthee2016} or their initial EW measurement was overestimated. Where possible, we initially removed objects from the parent sample that are identified as being powered by AGN through their radio or X-Ray emission \citep{Calhau2020}, or broad (FWHM$>1000$ km s$^{-1}$) emission-lines \citep{Sobral2018}. We note that we only performed shallower spectroscopy for a small subsample prior to this program meaning that the broad-line rejection could not be performed homogeneously and that due to the limiting sensitivity of the X-Ray data faint AGNs could have been missed.

The sample comprises 20 targets from our own program (ESO program ID 102.A-0652; PI Matthee) and 15 targets for which archival X-SHOOTER data were publicly available. The coordinates, program IDs and observation details of the targets are listed in Table $\ref{tab:observationlog}$. Targets were assigned an XLS-ID as follows: XLS-1 to XLS-20 originate from our main survey, while XLS-21 to 35 are archival objects. Within these two groups the IDs are ranked by right ascension. A comparison of the XLS-ID to other IDs of these galaxies is listed in Table $\ref{tab:IDs}$.  

The main 20 targets are selected from a wide-field narrow-band survey \citep{Sobral2016}. We preferentially selected galaxies in regions with best ancillary data (e.g. {\it HST} coverage from the CANDELS program). The archival objects are selected in various ways, listed in Table $\ref{tab:observationlog}$ and described in more detail here:
\begin{itemize}
\item XLS-21, 22 and 23 were initially selected to have detections in the rest-frame UV emission lines C{\sc iii}] and O{\sc iii}], in addition to strong Ly$\alpha$ \citep{Amorin2017}. It is unclear whether the additional selection criterion of a C{\sc iii}] and O{\sc iii}] detection necessarily implies that these objects are not representative of LAEs with similar Ly$\alpha$ luminosity and EW. We note that these objects are all also detected by an independent Ly$\alpha$ narrow-band survey \citep{SC4K}, where they were found to be part of a larger sample with similar luminosity and EW. 
\item XLS-24, 25, 26, 27 and 28 are identified as Ly$\alpha$ emitters in narrow-band surveys by \cite{Nilsson2009}, \cite{Nakajima2012} and \cite{Hayes2010}, and were chosen for spectroscopic follow-up observations based on their high luminosity compared to the other LAEs in these respective surveys. The luminosity and EW ranges of these objects are comparable to the main 20 targets, indicating these are representative LAEs. XLS-24 is also part of our own parent catalog and has been studied in \cite{Rhoads2014}. We note that XLS-26 and 27 have additionally been observed by a program that selected them as a candidate Lyman-Continuum leaker \citep{Naidu2017,Oesch2018}.
\item XLS-29, 30 and 31 are selected through their high H$\alpha$ EW ($>300$ {\AA}; \citealt{Terlevich2015}) with respect to other objects in a catalogue of UV-selected galaxies at $z\approx2$ \citep{Erb2006}. While the selection of these objects is to first order independent of the Ly$\alpha$ emission properties, the fact that these objects have detectable Ly$\alpha$ emission is perhaps not surprising (see also \citealt{Erb2010}). The H$\alpha$ EW criterion implies that these galaxies produce significant amounts of Ly$\alpha$ photons and the fact that the parent sample is UV-selected implies that there is little dust attenuation, which is associated to a higher escape fraction of Ly$\alpha$ photons \citep[e.g.][]{Atek2008}. 
\item XLS-32, 33, 34 and 35 are selected based on having strong and symmetric Ly$\alpha$ lines in low-resolution rest-frame UV spectra \citep{Erb2016} and initially originate from a sample of UV-selected galaxies. 
\end{itemize}
Fig. $\ref{fig:sample}$ shows that the archival objects are typically more luminous. Several archival objects that were not directly selected as LAEs show Ly$\alpha$ emission lines with relatively low EW. We include these in order to expand the dynamic range of EWs and Ly$\alpha$ escape fractions. 

\subsection{How representative is the sample of the star-forming population at $z\approx2$?} \label{sec:rep}
When defining a survey it is crucial to first investigate how representative the galaxy sample is compared to the full galaxy sample at a similar cosmic time. Here we address this by comparing the number densities of LAEs to the number densities of all UV-selected star-forming galaxies at $z\approx2$.
 
As shown in Fig. $\ref{fig:sample}$, the Ly$\alpha$ luminosities span the range $0.7 - 40 \times10^{42}$ erg s$^{-1}$ ($\approx0.2-10 \times L^{\star}$; \citealt{SC4K}) and the UV magnitudes range from $-18.5$ to $-22$ ($\approx0.2-6 \times L^{\star}$; \citealt{Parsa2016}). Diagonal lines of fixed Ly$\alpha$ EW (assuming a constant UV slope $\beta=-2$) illustrate that the typical EW of our sample is $\approx75$ {\AA}. When we estimate the number density associated with the Ly$\alpha$ luminosity of each target, we find that the median number density of our targets is $\approx1\times10^{-4}$ comoving Mpc$^{-3}$ per log luminosity interval. This is a factor 5 lower than the median number density of UV-selected galaxies associated with their UV luminosities. Fig. $\ref{fig:sample}$ also directly illustrates how representative our targets are in terms of their relative UV and Ly$\alpha$ luminosities compared to the UV-selected galaxy population at $z\approx2$ based on their relative abundances. We derive the relation between Ly$\alpha$ and UV luminosity that is required to match the number densities from the UV luminosity function from \cite{Parsa2016} to the Ly$\alpha$ luminosity function from \cite{Sobral2016} at $z\approx2.2$. This method assumes a linear relation between the UV and Ly$\alpha$ luminosity. The correction factor for a sub-linear relation would be in the range 1-2 for a (conservative) slope range 0.5-1.0. Even a factor of 2 would be a small correction in this context and leave our conclusions basically unchanged. The number densities are similar for objects with EWs$\approx10-15$ {\AA}. For the EWs of the objects in our sample the number density is on average 15 times lower than the UV-selected population. This implies that while our sample is by selection broadly representative of the population of LAEs at $z\approx2$, this sample is a rare sub-sample of the UV-selected galaxy population at $z\approx2$. This is in agreement with typical population-averaged \citep{Hayes2011,SC4K} and mass-averaged \citep{Matthee2016} Ly$\alpha$ escape fractions of $\approx2$ \% at $z\approx2$, while Ly$\alpha$ escape fractions measured in LAEs are typically $\approx30$ \% (e.g. \citealt{Hayes2010,Song2014,Trainor2016,Sobral2016} and \S $\ref{sec:derived}$).

\subsection{Fields, photometry and ancillary data}
In this subsection we summarise the available photometry and ancillary data of the targets. The photometry is used to derive UV luminosities, colours and stellar masses as described in \S $\ref{sec:SEDmodel}$.
\begin{itemize}
\item XLS-1 to 14, and XLS-21 to 24 are located in the COSMOS field and are therefore covered by $>30$ bands of near-UV to mid-IR photometry \citep[e.g.][]{Ilbert2009,Laigle2016}. We use the aperture-corrected photometry that is described in \cite{Santos2020}, which includes the latest data-release (DR4) of the UltraVISTA survey in the near-infrared \citep{McCracken2012}. High-resolution {\it HST}/ACS imaging is available for all targets in the F814W filter \citep{Koekemoer2007}. We compile size measurements of the COSMOS targets based on these data from \cite{PaulinoAfonso2018}.

\item XLS-15 to 19 are in the UDS/SXDS field, which is covered by very deep ground-based imaging in the optical \citep{Furusawa2008}, near-infrared \citep{Lawrence2007,Jarvis2013} and mid-infrared \citep{Mehta2018}. We use the multi-wavelength aperture-corrected photometry from \cite{Mehta2018}. XLS-16 and XLS-19 are covered by the {\it HST} CANDELS survey, from which we compile their size measurements \citep{vdWel2012}.

\item XLS-20 is located in the SA22/CFHTLS-W4 field and is covered by the deep part of the Subaru HyperSuprimeCam survey \citep{HSC_SURVEYPAPER}. These data are about 2 magnitudes shallower than the data in the COSMOS field. We use our own aperture-corrected photometry in the $g, r, i, z, y$ filters with the same technique as described in detail in \cite{Santos2020}. 

\item XLS-25 to 28 targets are in the Extended Chandra Deep Field-South field. XLS-25 and 28 have high-resolution {\it HST}/ACS imaging from the GEMS survey \citep{Rix2004}. We compile size measurements from \cite{Haussler2007} and we use the multi-wavelength ground-based photometry in $>30$ filters from the MUSYC survey \citep{Cardamone2010}. XLS-26 and 27 are located in the {\it HST} extreme deep field with multi-wavelength photometry from the CANDELS survey \citep{Grogin2011,Guo2013} and we use size measurements from \cite{vdWel2012}. 

\item XLS-29 to 35 are located in extra-galactic fields that were selected to have a bright background QSO \citep{Steidel2004}. For XLS-29 to 31 we collected Palomar photometry in the $U, G, R, J, K_s$ filters from \cite{Erb2006_Mass}. We have no photometry for XLS-32 to 35, but we instead use the X-SHOOTER spectra directly to measure the UV continuum luminosity. We collected and reduced archival high-resolution {\it HST} imaging data for XLS-29 to 32 and XLS-35 from programs with IDs 9133 (PI Falco), 9367 (PI Hazard), 11694 (PI Law) and 12471 (PI Erb) using the {\it HST} Legacy Archive.  
\end{itemize}

\subsection{SED modeling} \label{sec:SEDmodel} 
In order to obtain the rest-frame UV luminosity (M$_{1500}$) and the stellar mass (M$_{\rm star}$), we model the spectral energy distributions (SEDs) of the galaxies using the {\sc Magphys} code \citep{daCunha2008}. We use aperture-corrected photometry obtained with the same methodology as described in detail in \cite{Santos2020}. For most sources, the photometric information is quite homogeneous in terms of wavelength coverage and depth (0.3-5.0 $\mu$m, $\approx28-25$ AB magnitude, with higher sensitivity in bluer bands). The COSMOS and ECDFS objects additionally benefit from several filters with medium-bandwidth. For XLS-20 and XLS-29 to 31 the coverage is shallower and limited to the optical ($0.3-1.0 \mu$m, $\approx26$ AB).

In short, {\sc Magphys} uses dust attenuation models from \cite{CharlotFall2000} and stellar populations using \cite{BruzualCharlot2003} models where a \cite{Chabrier2003} initial mass function with mass range $0.1-100$ M$_{\odot}$ is assumed. The star formation histories (SFHs) are a combination of a continuous exponentially decaying history that follows an initial rise and an additional instantaneous burst (with a duration between 30-300 Myr and a mass-fraction of 0.1-100 of the integrated mass from the continuous SFH). As {\sc Magphys} does not model nebular emission, we exclude the medium and broad-band filters that are contaminated by strong Ly$\alpha$, H$\beta$+[O{\sc iii}] and H$\alpha$ emission from the fitting procedure as they may lead to over-estimated stellar masses \citep[e.g.][]{SchaererBarros2009}. Except for XLS-20, 29, 30 and 31, all objects are covered by deep {\it Spitzer}/IRAC data, which is particularly useful for constraining the stellar masses. For XLS-32 to 35 we use the stellar masses derived by \cite{Erb2016} and we measure the UV continuum luminosity directly from the X-SHOOTER spectrum. Independently from the SED fitting, the UV slope $\beta$ is measured by fitting a power-law of the form $f_\lambda \propto \lambda^{\beta}$ to all photometric bands that cover rest-frame wavelengths 1300 to 2100 {\AA}. The measurements are listed in Table $\ref{tab:flux_measurements}$.

 \begin{table*}
\caption{Observation log. Exposure times are in ks. The spectral resolution around the Ly$\alpha$ wavelength is based in the nominal resolution of R=6700, 5400 and 4100 for slit widths 0.8$''$, 1.0$''$, 1.3$''$, respectively. The listed seeing is the median seeing in arcsec. Selection methods: 1) Ly$\alpha$ flux, 2) strong rest-frame UV emission line galaxy, 3) Lyman-continuum leaker candidate, 4) BX-galaxy with H$\alpha$ EW $>300$ {\AA}, 5) high [O{\sc iii}]/H$\beta$ and symmetric Ly$\alpha$ emission at low resolution. }
\begin{tabular}{lrrrrrrrp{2cm}r}
ID & R.A. (J2000) & Dec. (J2000) & R$_{\lambda \rm = Ly\alpha}$ & t$_{\rm exp, UVB}$ & t$_{\rm exp, VIS}$ & t$_{\rm exp, NIR}$& Seeing & Program ID & Selection \\ \hline
XLS-1 & 09:57:59.73 & +02:18:04.86 & 4100 & 6.1 & 5.4 & 6.4 & 0.8 & 102.A-0652 & 1\\
XLS-2 & 10:00:13.91 & +01:39:24.30 & 4100 & 16.8 & 12.8 & 15.0 & 0.7 & 098.A-0819, 099.A-0254, 102.A-0652 & 1\\
XLS-3 & 10:00:24.61 & +02:27:01.07 & 4100 & 10.7 & 9.3 & 11.2 & 0.5 & 102.A-0652& 1\\
XLS-4 & 10:00:26.65 & +02:17:14.42 & 4100 & 10.7 & 9.3 & 11.2 & 0.6 & 102.A-0652& 1\\
XLS-5 & 10:00:33.97 & +02:13:15.92 & 4100 & 6.1 & 5.4 & 6.4 & 0.5 & 102.A-0652& 1\\
XLS-6 & 10:00:35.73 & +02:15:06.66 & 4100 & 10.7 & 9.3 & 11.2 & 0.7 & 102.A-0652& 1\\
XLS-7 & 10:00:38.66 & +02:09:20.72 & 4100 & 13.4 & 11.6 & 14.0 & 0.5 & 102.A-0652& 1\\
XLS-8 & 10:00:42.21 & +02:08:09.62 & 4100 & 8.0 & 7.0 & 8.4 & 0.5 & 102.A-0652& 1\\
XLS-9 & 10:00:50.66 & +02:07:42.06 & 4100 & 10.7 & 9.3 & 11.2 & 0.6 & 102.A-0652& 1\\
XLS-10 & 10:00:50.87 & +02:06:31.24 & 4100 & 10.7 & 9.3 & 11.2 & 0.7 & 102.A-0652& 1\\ 
XLS-11 & 10:01:06.55 & +01:45:45.47 & 4100 & 14.5 & 11.1 & 15.4 & 0.7 & 099.A-0254, 102.A-0652& 1\\
XLS-12 & 10:01:36.21 & +02:15:16.80 & 4100 & 11.9 & 10.1 & 12.5 & 0.8 & 098.A-0819, 102.A-0652& 1\\ 
XLS-13 & 10:02:16.16 & +02:32:18.87 & 4100 & 6.1 & 5.4 & 6.4 & 0.6 & 102.A-0652& 1\\
XLS-14 & 10:02:35.32 & +02:12:13.42 & 4100 & 13.6 & 13.2 & 15.1 & 0.7 & 0100.A-0213, 102.A-0652& 1 \\ 
XLS-15 & 02:17:15.52 & -05:07:14.97 & 4100 & 10.7 & 9.3 & 11.2 & 0.7 & 102.A-0652& 1\\
XLS-16 & 02:17:26.42 & -05:13:40.95 & 4100 & 10.7 & 9.3 & 11.2 & 0.8 & 102.A-0652& 1\\ 
XLS-17 & 02:17:41.39 & -05:06:49.61 & 4100 & 10.7 & 9.3 & 11.2 & 0.5 & 102.A-0652& 1\\ 
XLS-18 & 02:17:46.13 & -05:02:55.51 & 4100 & 21.0 & 15.2 & 25.0 & 0.8 & 098.A-0819, 099.A-0254, 102.A-0652& 1\\ 
XLS-19 & 02:17:55.77 & -05:12:41.00 & 4100 & 10.7 & 9.3 & 11.2 & 0.8 & 102.A-0652& 1\\ 
XLS-20 & 22:15:48.23 & +00:23:57.46 & 4100 & 9.4 & 8.1 & 9.8 & 0.9 & 102.A-0652& 1\\  \hdashline[2pt/2pt]
XLS-21 & 10:00:10.95 & +01:51:46.66 & 5400 & 10.6 & 11.1 & 10.8 & 0.6 & 0101.B-0779 & 2\\ 
XLS-22 & 10:00:39.56 & +02:15:38.44 & 5400 & 10.6 & 11.1 & 10.8 & 0.6 & 0101.B-0779 & 2\\ 
XLS-23 & 10:01:20.81 & +02:36:19.27 & 5400 & 7.3 & 7.2  & 7.6 & 0.6 & 0101.B-0779 &2\\ 
XLS-24 & 10:00:49.22 & +02:01:21.30 & 4100 & 3.6 & 3.6 & 3.6 & 0.9 & 084.A-0303 & 1 \\  
XLS-25 & 03:32:32.31 & -28:00:52.20 & 5400 & 2.4 & 4.4 & 4.8 & 0.9 & 088.A-0672 & 1 \\ 
XLS-26 & 03:32:35.48 & -27:46:16.91 & 5400 & 17.3 & 13.3 & 14.4 & 0.7 & 092.A-0774, 099.A-0758 & 1, 3\\ 
XLS-27 & 03:32:46.46 & -27:50:36.63 & 5400 & 10.6 & 10.0 & 10.8 & 0.8 & 099.A-0758 & 3\\ 
XLS-28 & 03:32:49.34 & -27:59:52.35 & 5400 & 4.8 & 8.8 & 9.6 & 1.3 & 088.A-0672 & 1\\ 
XLS-29 & 23:46:09.06 & 12:47:56.00 & 6700 & 4.4 & 4.0 & 4.6 & 0.7 & 091.A-0413 & 4 \\ 
XLS-30 & 23:46:18.57 & 12:47:47.38 & 6700 & 4.4 & 4.0 & 4.6 & 0.8 & 091.A-0413 & 4\\ 
XLS-31 & 23:46:29.43 & 12:49:45.54 & 6700 & 8.8 & 8.0 & 9.3 & 0.6 & 091.A-0413 & 4\\ 
XLS-32 & 02:09:49.21 & -00:05:31.67 & 5400 & 2.7 & 2.8 & 3.6 & 0.7 & 097.A-0153 & 5 \\ 
XLS-33 & 02:09:44.23 & -00:04:13.51 & 5400 & 7.2 & 7.4 & 7.2 & 0.8 & 097.A-0153 & 5\\  
XLS-34 & 02:09:43.15 & -00:05:50.21 & 5400 & 6.3 & 6.5 & 6.3 & 0.9 & 097.A-0153 & 5\\ 
XLS-35 & 01:45:16.87 & -09:46:03.47 & 5400 & 7.2 & 7.4 & 7.2 & 0.7 & 097.A-0153 & 5\\ 
\end{tabular}
\label{tab:observationlog}
\end{table*}

\subsection{Rest-frame UV morphology} \label{sec:morph}
High-resolution {\it HST} data is available for 31 objects and we show cut-out images centred on the objects in Figures $\ref{fig:thumbnails}$ and $\ref{fig:thumbnails2}$. For the majority of targets (XLS-1 to 14 and 21 to 24) ACS/F814W data consist of a single orbit, but XLS-16, 19, 26, 27, 29 and 30 have deeper data (sources of these data are listed above). XLS-25 and 28 have data in a similar filter (F850LP) at moderate depth. For XLS-31 and 32 the only available {\it HST} data has been taken with WFC3 in the F140W and F160W NIR filters. The {\it HST} data of XLS-35 consists of F814W imaging with WFPC2. 

The rest-frame UV morphologies of several LAEs show multiple clumps on $\approx1$ kpc scales. From simple visual inspection (see also \citealt{PaulinoAfonso2018}), we find that 12 out of the 31 objects with high-resolution {\it HST} data appear as multiple clumps, see Table $\ref{tab:measurements}$.

\section{VLT/X-SHOOTER spectroscopy} \label{sec:data}
In this section we describe the observations and data reduction of the X-SHOOTER data.

X-SHOOTER \citep{Vernet2011} is a wide-band (0.3-2.5 $\mu$m) echelle spectrograph on the VLT. Two dichroics split the light into three arms, each with independent shutter and slit mask and with simultaneous exposures. These UVB, VIS and NIR arms are each optimised for their respective wavelength coverages of 300-560 nm (UVB), 560-1024 nm (VIS) and 1025-2480 nm (NIR). The exposures in the UVB and VIS arm are read out sequentially, meaning that in practice exposure times are longer in one of these arms compared to the other, depending on the observing strategy.

In the following sections we describe the observing strategy, characteristics and the data reduction. The data consist of typically 3 hours per source, for a total of 89.5 hours of on-source integration time (of which 62 hours is from our own program) in the UVB arm and similar times in the other arms. The typical spectral resolution around Ly$\alpha$ is R=4100-5400. The majority of the data reduction was performed with the standard ESO pipeline complemented with the {\sc Molecfit} tool \citep{molecfit1} to account for atmospheric transmission. We used our own {\sc Python}-based algorithms for optimally combining 2D spectra observed over multiple dates onto a common barycentric velocity grid centred on the spatial peak of the Ly$\alpha$ line.

 \subsection{Observations}
Observations of our own program were performed in service mode between 2 October 2018 and 28 February 2020. Archival data were taken between March 2010 and March 2019. In general observations were performed in dark conditions and with $V$-band seeing $\approx0.8''$. The nominal spectral resolution based on the slit-width at the redshifted Ly$\alpha$ wavelength, the total integration times in the various arms and the typical seeing of all observations are listed in Table $\ref{tab:observationlog}$. All observations use a blind offset from an acquisition star and are nodding between two positions along the slit in an ABBA sequence. 

For our own program, we identified acquisition stars with $R<17$ by matching the parent catalogues to the {\it Gaia} DR2 catalogue \citep{GAIADR2}. We then selected the star within 120$''$ from the target with the lowest proper motion and not blended with another object in projection. Offsets were calculated based on the relative position of the star and our targets in the narrow-band data, taking the small proper motion between the time of the narrow-band observation and the ESO semester into account. A distance of 3$''$ between the two nodding positions was used. The slits are placed at the parallactic angle at the start of the first exposure. 

We used 1.3$''$, 1.2$''$ and 0.9$''$ slits in the UVB, VIS and NIR arms corresponding to resolutions R=4100, 6500, 5600, respectively. Individual exposure times were 670s (UVB), 580s (VIS) and 4x175s (NIR; using four integrations at each nodding position), which were repeated in cycles of 4 per observing block that lasted roughly 1 hour. Typically targets were observed in three independent one-hour observing blocks (exact number of observing blocks ranging from 2 to 5). Most archival programs used a similar observing strategy with slight variations in exposure times and slit-widths. XLS-29, 30 and 31 were observed with a $K$-band blocking filter such that there is no coverage of H$\alpha$ emission. 

When the program started, the majority of our narrow-band selected targets were not yet spectroscopically confirmed, leading to the non-zero risk that they were interlopers or spurious sources. Therefore, while our observations were performed remotely in service mode, we specifically designed the execution strategy such that each target would only be observed for a maximum of one hour during the first attempt. Remaining observing blocks were scheduled with the time constraint that they would follow at least three days later. This allowed us to reduce and analyse the data and communicate any target change in case that would be necessary. In practice, we changed target only twice. XLS-20 is a replacement (and therefore has less total exposure time) for a target that turned out to be a star with colours similar to a blue galaxy at $z\sim2$ and where variability boosted the narrow-band mimicking an emission-line. We also decided to move one OB from XLS-8, which turned out to have a very low Ly$\alpha$ EW, to XLS-7. 

For both our own program and the archival data we visually inspected each raw exposure in the UVB arm for any issues with the data. We removed a handful of exposures with poor acquisition, bad seeing, contamination by spurious light from within the telescope or the laser from the AO-system on UT4. The exposure times listed in Table $\ref{tab:observationlog}$ only include the data that have eventually been used.

\subsection{Data reduction} \label{sec:reduction}
The X-SHOOTER data are reduced as follows. For each observing block (OB) of $\sim1$ hour, we use the X-SHOOTER pipeline version 3.2.0 \citep{Modigliani2010} implemented in {\sc EsoRex} to apply the standard reduction steps: bias (UVB and VIS) or dark (NIR) subtraction, flat-fielding, flexure correction and 2D mapping, wavelength calibration and flux calibration with standard stars. The same reduction steps are applied to telluric stars that are observed in the same nights. The {\sc Molecfit} tool \citep{molecfit1,molecfit2} is used to apply telluric corrections to the science observations.

Individual OBs are co-added as follows. We first resample the 2D spectra to a new grid where we converted the wavelength calibration of the 2D spectra to vacuum wavelengths using the IAU standard and shifted each spectrum to the barycentric reference frame. To improve the accuracy of the re-sampling, the new 2D spectrum is over-sampled by a factor two using a nearest interpolation. We then shift the 2D spectra in all arms such that the spatial centre coincides with the spatial peak position of the Ly$\alpha$ line. The peak position is identified by fitting a 2D Gaussian model to the 2D Ly$\alpha$ spectrum after this has been convolved with a 2D Gaussian ($\sigma_{\rm spatial}=0.32''$, $\sigma_{\lambda}=2$ {\AA}) in order to improve the S/N and wash out the detailed spectral structure of the Ly$\alpha$ profile. In most cases, the spatial peak of the Ly$\alpha$ and UV continuum emission are found to be slightly off-center due to slight inaccuracies in the acquisition and pointing of the telescope on the order of $0.2''$ (mean absolute deviation). In the case of XLS IDs 16, 25 and 27, we find that the Ly$\alpha$ line is further offset by $\approx0.2-0.3''$ from the UV continuum and nebular lines. For XLS-27 we find an offset of 1.1$''$ between Ly$\alpha$ and the UV. Finally, we combine the OBs with an inverse-variance weighted average where the variance is determined over the 400-500nm (UVB), 600-800 nm (VIS) and 1500-1600 nm (NIR) wavelength ranges.

\begin{figure}
\includegraphics[width=8.6cm]{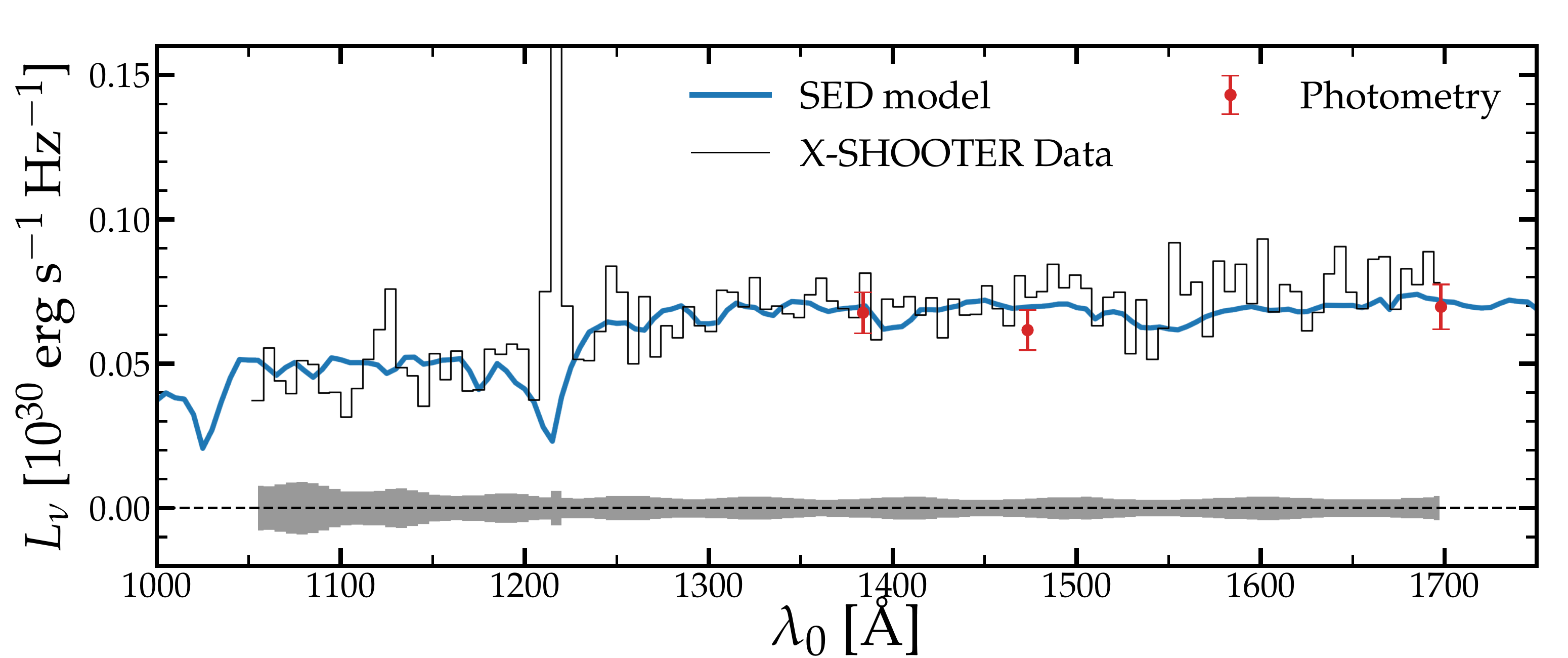}\\
\includegraphics[width=8.6cm]{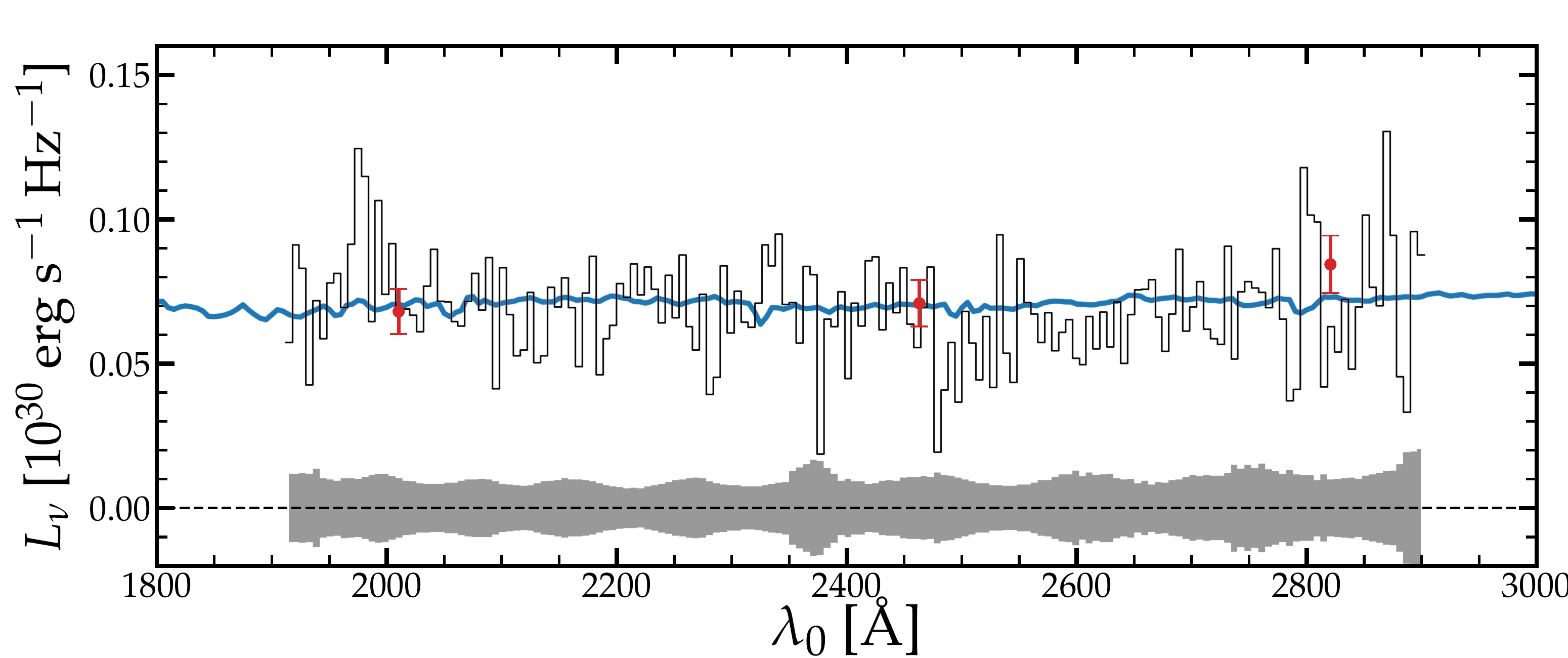}
\caption{Verification of the flux calibration of the spectra. The black line shows the median stacked spectrum of XLS-1 to 28 (except 14, 20, 22 and 27). The red points show the stacked photometry in the $B, g, V, R, I, z$ bands. The blue line shows the median stacked SED model from the same objects. The SED model does not include nebular lines. The UVB spectrum is binned by a factor 12  and the VIS spectrum by a factor 25 ($\Delta\lambda=2.5$ {\AA} and $\Delta\lambda=5$ {\AA}, respectively). The top panel shows the spectrum in the UVB arm and the bottom panel shows the spectrum in the VIS arm. We only show the stack at rest-frame wavelengths where all objects are covered in the same arm of the X-SHOOTER spectrograph. The axis ranges are chosen to highlight the continuum emission and we note the Ly$\alpha$ line is cut-off. No significant continuum is detected in the NIR arm. The grey shaded region shows the 1$\sigma$ rms of different bootstrap realisations of the stacked spectrum (\S $\ref{sec:stack}$). This is binned by the same factor as the spectrum. The normalisation of the spectra is matched to the average UV continuum of the SED model over $\lambda_0=1280 - 1500$ {\AA}.  } \label{fig:fluxcal}
\end{figure}

\section{Methods}
\subsection{Extraction of 1D spectra} \label{method:1D}
Here we describe how 1D spectra were extracted from the 2D data. We first motivate the choice for the centre and width of the extraction and then describe the way we optimise the spectrophotometric calibration and how we measure the noise level of the data.

\subsubsection{Centroid and aperture}
The centre of the extraction is based on the peak position of the Ly$\alpha$ line that we identified in the co-addition step in the data reduction (\S $\ref{sec:reduction}$). We extract the 1D spectra using an optimal extraction \citep{Horne1986} assuming a Gaussian-profile with a width that is optimised for each object individually. We collapse each 2D spectrum over a velocity range of $-500$ to $+500$ km s$^{-1}$ from the peak of the Ly$\alpha$ line and we measure the full-width half maximum (FWHM) of the Ly$\alpha$-light distribution. We repeat this process for a collapse of rest-frame wavelengths $\lambda_{0}=1260-1500$ {\AA} to identify the FWHM of the UV continuum-light distribution. In this paper, we choose to use the continuum-based FWHM to extract the 1D profiles. These FWHM are much larger than the typical offsets between the Ly$\alpha$ and the UV. For three objects (XLS-9, 14 and 22) we use a Ly$\alpha$ based aperture as the continuum is not detected with sufficient S/N. Typical FWHM of the continuum are $0.6''$, ranging from $0.4-0.9''$. The typical FWHM of the Ly$\alpha$ line is a factor 1.1 larger than that of the UV continuum. As described in detail below, the extraction aperture varies with wavelength in order to fix the encapsulated fraction of the flux.

For the majority of objects there are no large shifts between the spatial peak of the UV continuum and Ly$\alpha$. Most of the objects with offsets appear as multiple component systems in either the UV continuum imaging or through multiple narrow components in the [O{\sc iii}] emission-line, e.g. XLS-16, 25 and 35. The offsets between Ly$\alpha$ and the UV are sufficiently small that the extraction windows centred on the Ly$\alpha$ peak capture the large majority ($>80$ \%) of the flux. XLS-27 is a special case where Ly$\alpha$ is offset by $\approx1.1''$ ($\approx9$ kpc) from the UV continuum (and the rest-frame optical lines). For this object we therefore extract the Ly$\alpha$ spectrum on the Ly$\alpha$ position and the UV continuum and rest-frame optical spectrum on the position of the UV continuum. We note that we identify a spatial drift of the UV continuum across the slit in the UVB and VIS arms in the observations of XLS-29 to XLS-35. This is accounted for by increasing the extraction aperture by a factor 3 at the expense of adding some noise.

As the seeing is wavelength dependent, using the same extraction size over the full UVB to NIR wavelength range would result in a higher enclosed flux in redder wavelengths compared to bluer wavelengths. We use the standard stars that have been observed with a very wide 5$''$ slit and seeing conditions in the range of the observations to empirically obtain spectroscopy that encapsulated the same fraction of total flux over the full wavelength range. We measure the FWHM of the light distribution in the 2D spectra of the standard stars and store these in various wavelengths. Then, for each science object, we match the FWHM in the UVB arm to the closest standard star in terms of FWHM. We then match the extraction FWHM in the redder part of the spectrum to encapsulate the same fraction of the total flux and use this wavelength-dependent FWHM for our optimal Gaussian extraction.

\subsubsection{Spectrophotometric calibration} \label{sec:photcor} 
After the extraction, we optimise the spectrophotometric calibration by applying an achromatic normalisation correction. The average flux in the wavelength range $\lambda_0=1280 - 1500$ {\AA} that is measured in the extracted 1D spectra is thus matched to the average flux over the same wavelength region in the SED model that is best-fitted to the aperture-corrected multi-wavelength photometry (\S $\ref{sec:SEDmodel}$). This final calibration step simultaneously accounts for slit losses and uncertainties in the flux calibration. The correction derived in the UV continuum is applied to the full wavelength range from UVB to NIR. The SED model fits the various photometric bands in the rest-frame UV wavelength range (observed $B$ to $z$ band) very well. Propagated uncertainties in the flux calibration of our spectrum that would originate from the validity of the SED model would be more important in the rest-frame optical (observed NIR). In the observed NIR the majority of the photometric bands is contaminated by emission lines, such that the model is more dependent on choices regarding for example the star formation history. However, as we do not use the NIR data to optimise the flux calibration, these concerns are not relevant here as long as the uncertainties in the flux calibration are achromatic.

On average, we find that the flux normalisation of the spectrum is a factor $1.2\pm0.4$ higher than the photometry (where the error is the standard deviation and the extremes are 0.5 and 2.5). This suggests that uncertainties in the flux calibration dominate over slit losses. Indeed, because the sources are very compact and the seeing is typically good, we estimate slit losses $<10$ \% by simulating fake sources with the FWHM of the UV-continuum and by measuring the fraction of the flux that is retrieved in the slit. The wavelength-collapsed UV continuum is detected with S/N$>5$ in the spectra of all objects except XLS-14 and 22. We do not have aperture-corrected photometry for XLS-32 to 35. For these objects we do not apply a correction to the flux calibration of the spectrum.
 
By comparing how the average fluxes vary between single observing blocks we can empirically estimate the uncertainties associated with the acquisition and flux calibration. For a few sources, we are able to measure the continuum flux levels in the UVB (here we collapse $\lambda=400-530$ nm) and in the VIS arms (collapsing $\lambda=600-830$ nm) with a S/N$>10$ in single observing blocks. For these sources, we find a standard deviation of 9 \% in the flux levels in the UVB arm and 13 \% in the continuum level in the VIS arm. For sources where we detect the continua in individual observing blocks with a S/N ranging from 5-10 we find a typical standard deviation of 25 \%, but it is plausible that this additional variation can be explained by measurement errors. This suggests that the uncertainties on the fluxes are about $\approx10$ \%. We also find that the deviations from the mean typically occur coherently for the UVB and VIS arm, which suggests that the uncertainties are achromatic. 

In Fig. $\ref{fig:fluxcal}$ we show a comparison of the stacked spectrum of XLS-1 to 28 (except 14, 20, 22 and 27 due to their non-detection of the UV continuum, inhomogeneous photometry or large $>0.3''$ offset between Ly$\alpha$ and the continuum) to the stacked SED of the same objects. The spectrum is achieved in the same way as described in \S $\ref{sec:stack}$ and is significantly binned in the wavelength direction to highlight the continuum level. While the normalisations of the stacked spectrum and the SED are matched at $\lambda_0=1280-1500$ {\AA}, we show that the same corrections also lead to consistent continuum levels at $\lambda=2000-2800$ {\AA} (i.e. the VIS arm of X-SHOOTER). We also show the stacked photometric data points that were used to derive the SED fits, demonstrating that the average fit is a good fit. We cannot test how well the continuum is matched in redder wavelengths as we do not detect continuum in the NIR arm. For the 18 objects with a continuum detection in the VIS arm (collapsing $\lambda_0=2000 - 2800$ {\AA}) with a S/N$>5$, we retrieve fully consistent corrections on a source-by-source level. Both these results validate our wavelength-dependent extraction window described above.

\begin{figure}
\includegraphics[width=8.6cm]{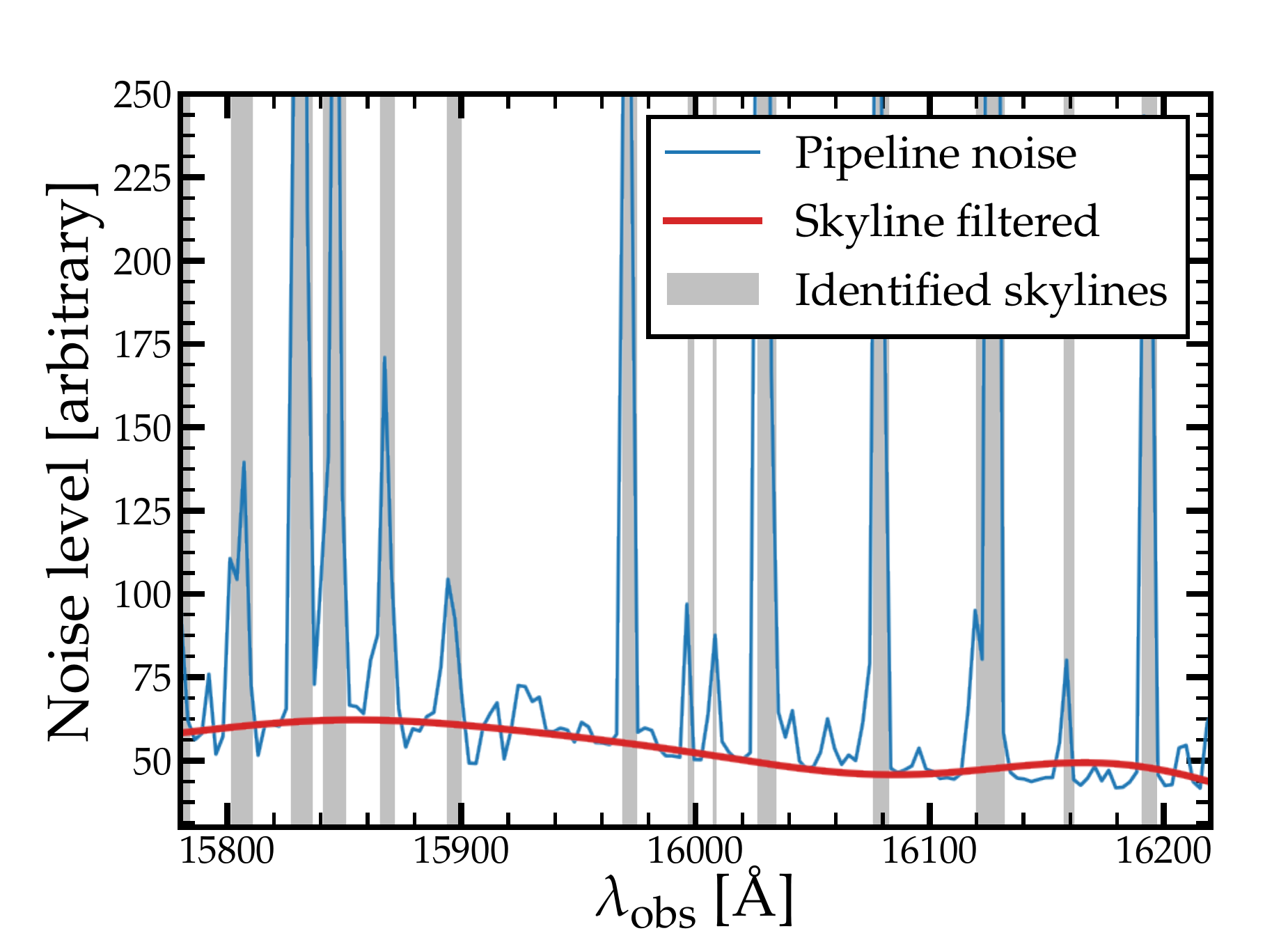}\\
\caption{Illustration of our skyline identification method. The blue line shows the propagated noise-level from the X-SHOOTER pipeline. The red line shows the reconstructed underlying wavelength dependency of the noise-level after identifying and removing skylines. The identified skylines are marked as grey shaded regions.} \label{fig:skylines}
\end{figure}

\subsubsection{Noise level}
We estimate the noise-level of the spectra by rescaling the wavelength-dependent noise level that is propagated from the pipeline with the actual noise level measured in the 2D spectra. As we only use wavelength ranges that are free from skyline emission, it is important to first identify skylines automatically, which we do in a two-step process. First, the strongest skylines are identified as inflection points in the propagated pipeline-noise model. Then, after masking these strong skylines, we use a Fourier filtering technique to reconstruct the part of the wavelength-dependence of the noise that is related to instrumental throughput and thermal background. The remaining fainter skylines are identified as modes with small scale power and can thus be removed. We illustrate this in Fig. $\ref{fig:skylines}$, where we show an example wavelength range around the redshifted H$\beta$ and [O{\sc iii}] of our target sample. 

\begin{figure}
\includegraphics[width=8.6cm]{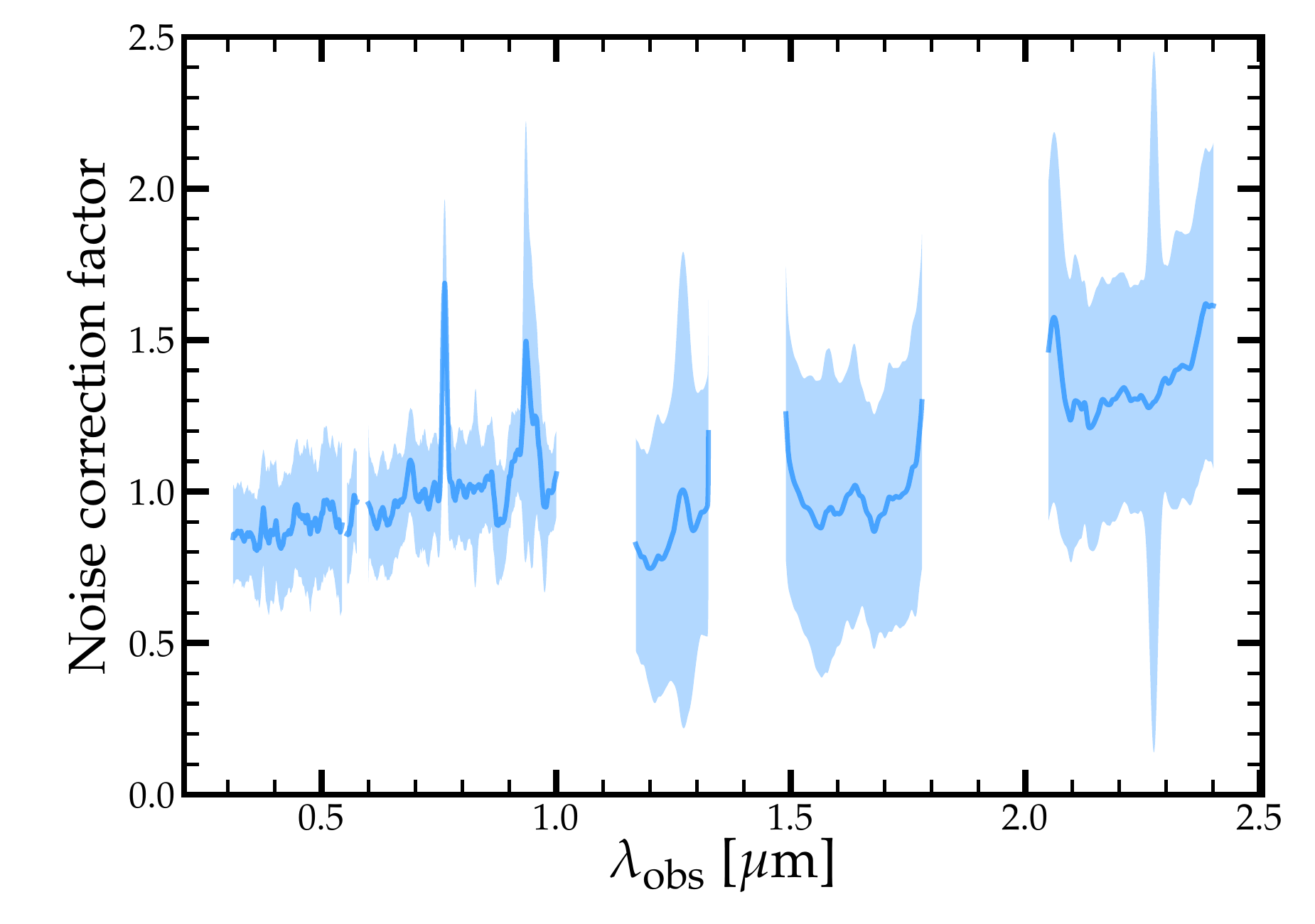}\\
\caption{Final correction factor applied to the pipeline-propagated noise level in order to match the noise level measured directly on the empty sky in the 2D spectra. The blue line shows the median correction factor of all objects and the shaded region shows the 1$\sigma$ range.} \label{fig:noise}
\end{figure}

The challenge in measuring the noise level on the data itself is that there are only limited number of empty sky-pixels in the 2D spectra available. It is possible to extract the 1D spectrum of the empty sky with the same optimal extraction aperture in 6-8 apertures that are independent (depending on the width of the extraction-profile) and away from the source itself or the negatives due to the nodding strategy. Then, for each wavelength-interval we could estimate the noise level from the standard deviation of these various 1D noise-spectra. However, due to the low number of independent apertures this measurement is noisy. Away from skylines, where the wavelength-dependence of the noise is weak and relatively smooth, we can circumvent this issue by calculating the standard deviation in a running tophat-kernel of width 20 {\AA} using the {\sc Pandas} package. After measuring the noise in the sky regions this way, we calculate the noise-correction factor as a function of wavelength and convolve this correction factor with a Gaussian with $\sigma=30$ {\AA}. In Fig. $\ref{fig:noise}$ we show that the noise-correction factors range within $\approx0.5-2.0$ and are mostly important in the $K_s$ band.

\subsection{Systemic redshift} \label{sec:zsys} 
It is well known that, due to resonant scattering, the peak redshift of the Ly$\alpha$ emission does not coincide with the systemic redshift  in the majority of galaxies \citep[e.g.][]{Steidel2010,Hashimoto2015,Verhamme2018,Muzahid2020}. 

In our data, the systemic redshift is best measured with the [O{\sc iii}]$_{4960,5008}$ doublet. Unlike Ly$\alpha$ and e.g. C{\sc iv}, the [O{\sc iii}] doublet is not a resonant transition and it is relatively unaffected by attenuation. There are also two practical reasons why [O{\sc iii}] is particularly helpful. First, after the Ly$\alpha$ line, it is the emission-line that is typically detected with highest signal-to-noise ratio. Besides, H$\alpha$ is redshifted into the $K_s$ band with a higher sky background compared to the observed wavelength of [O{\sc iii}].  
Second, as the [O{\sc iii}]$_{4960,5008}$ doublet has a fixed flux-ratio of 1:2.98, it is very useful to jointly fit both lines in the presence of skylines. For a single emission-line it often occurs that part of the line is affected by skyline residuals, challenging the measurements of the width and the peak flux in particular if the line-profile is not described by a single Gaussian profile. In most cases, this limitation can be overcome by jointly fitting the [O{\sc iii}] doublet, because the chance that both lines are affected by skyline residuals at the same rest-frame velocity is low.

\begin{figure*}
\includegraphics[width=14.6cm]{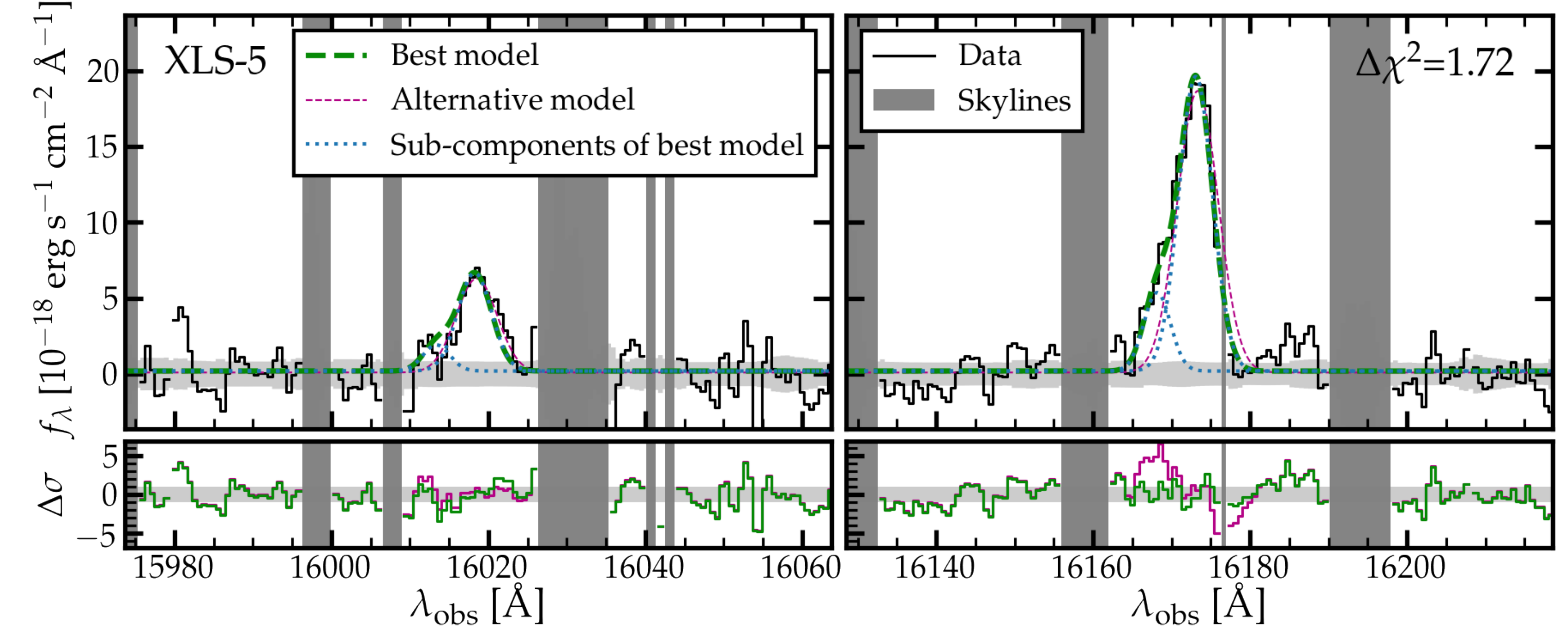} \\
\includegraphics[width=14.6cm]{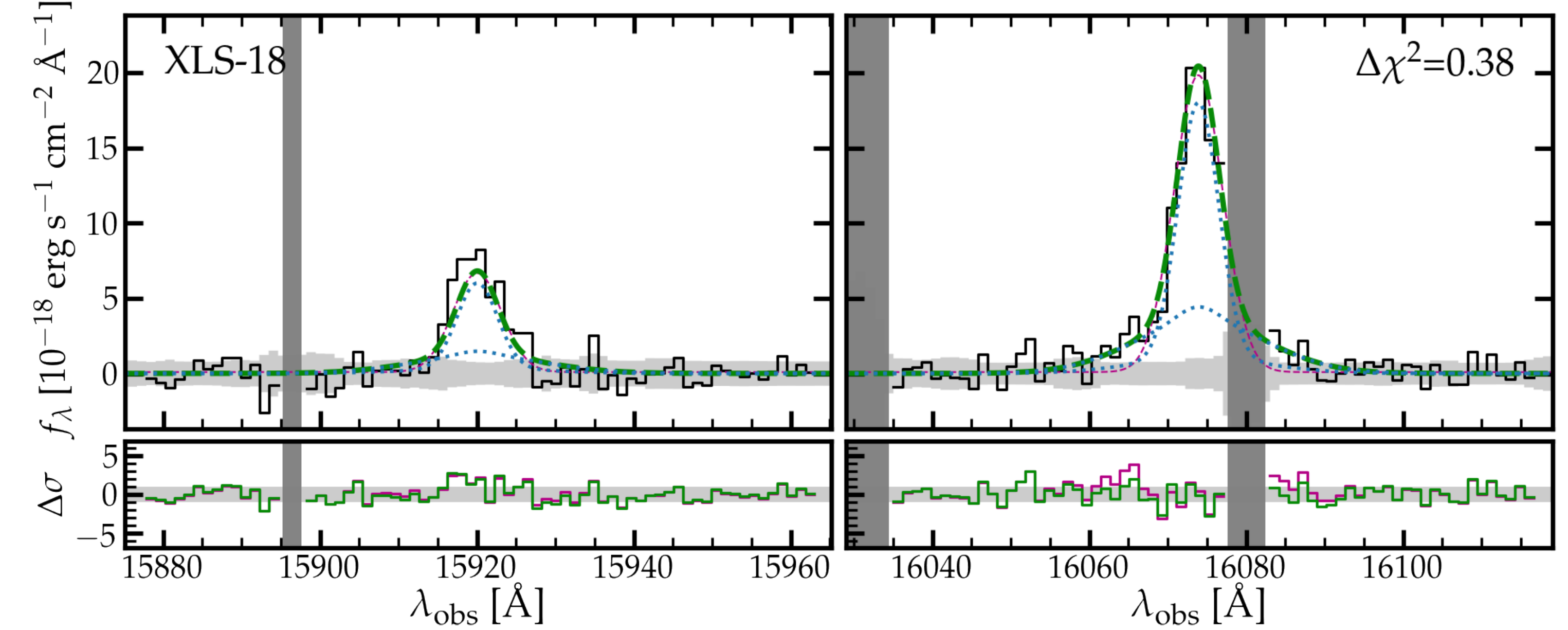} \\
\includegraphics[width=14.6cm]{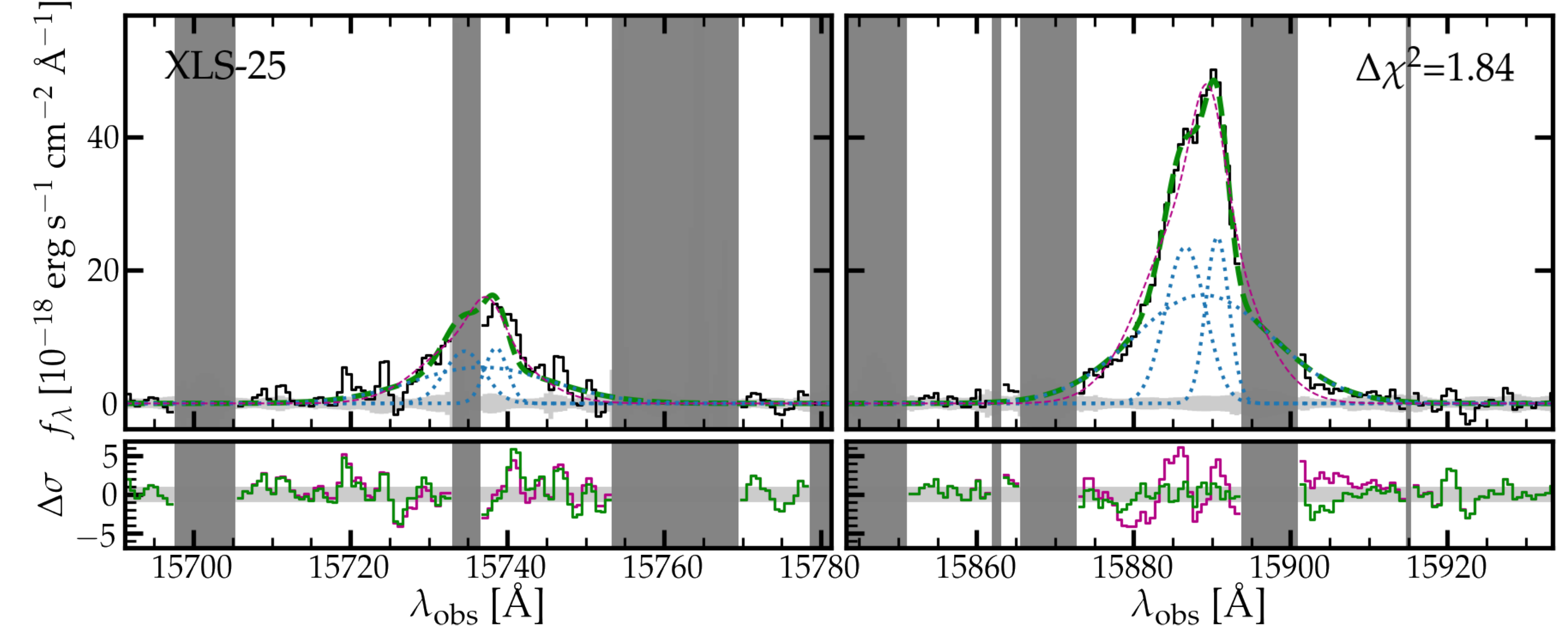} \\
\caption{Example line-profile fits of the [O{\sc iii}]$_{4960, 5008}$ doublet. In each row, the top panel shows the data and the bottom panel shows the residuals of the fits. The 1D spectrum is shown in black. The horizontal grey shaded region shows the 1$\sigma$ errors. The vertical shaded regions show the location skyline residuals that are masked in the fitting. The dashed green line shows the best-fit (multi-component) model, while the purple line shows the alternative model with less components. Dotted blue lines show the sub-components. $\Delta\chi^2$ is the difference between the reduced $\chi^2$ of the two models that are compared. Top row: XLS-5 is best-fitted by a combination of two narrow components. Middle row: XLS-18 is best-fitted by a narrow and a broad component. Bottom row: XLS-25 is best-fitted by a complicated [O{\sc iii}] profile that consists of two narrow and one broad component.}
\label{fig:O3_examples}
\end{figure*}

For each galaxy we first manually obtain a first-guess redshift using one of the [O{\sc iii}] lines or H$\alpha$ (in the rare case that the peaks of both [O{\sc iii}] lines are affected by skyline residuals). Visual inspection of the 2D spectra reveals that a significant fraction of objects consists of two (narrow) line-emitting components, in line with the high fraction of multiple component systems in the UV imaging (\S $\ref{sec:morph}$). We also notice that several bright objects show an [O{\sc iii}] profile with relatively broad wings, which has also been observed in low-redshift analogues of LAEs \citep[e.g.][]{Amorin2012,Henry2015,Hogarth2020}. These broad wings are discussed further in \S $\ref{sec:O3discuss}$. Two galaxies (XLS-25 and 35) visually show two narrow-components in addition to a broad component, see Fig. $\ref{fig:O3_examples}$. We identify [O{\sc iii}] and/or H$\alpha$ emission for 33/35 objects. These are shown in Fig. $\ref{fig:O3spec}$. 

For XLS-9 and XLS-13 we were not able to detect any emission-line within the full spatial area covered by the slit besides Ly$\alpha$ in rest-frame wavelengths $\lambda=1000-7000$ {\AA}. We note that a faint blue Ly$\alpha$ peak may be seen in XLS-13 at $\approx-300$ km s$^{-1}$ from the red Ly$\alpha$ line, suggesting the systemic redshift is at $\approx-150$ km s$^{-1}$ from the peak of the red Ly$\alpha$ line (Fig. $\ref{fig:Lyaspec}$). In order to understand why no rest-frame optical lines are detected, we determine upper limits of the strongest lines H$\alpha$ and [O{\sc iii}] assuming a single gaussian with line-width FWHM of 150 km s$^{-1}$ and velocity shifts of -300 to -100 km s$^{-1}$ with respect to Ly$\alpha$. The limiting flux of [O{\sc iii}] strongly depends on this shift while the expected wavelength for H$\alpha$ is less affected by skylines. The limiting H$\alpha$ fluxes can be translated into a lower limit of the Ly$\alpha$ escape fraction (assuming zero attenuation). For XLS-9 and XLS-13 we find $2\sigma$ limiting escape fractions of $>59$ \% and $>131$ \%. This implies that the non-detection of H$\alpha$ is not particularly unexpected for XLS-9, but we would expect an H$\alpha$ detection for XLS-13 with a S/N$>2.7$. This slight discrepancy can be a statistical noise effect or indicate that the H$\alpha$ line-width of XLS-13 is broader (resulting in a higher flux limit). Regarding the (typically) brighter [O{\sc iii}] lines we note that for the typical velocity shift of $-150$ km s$^{-1}$ with respect to Ly$\alpha$, the location of the redshifted [O{\sc iii}] lines of XLS-9 and XLS-13 are both heavily affected by skylines, plausibly explaining their non-detection.

For each object with a detected [O{\sc iii}] line, we use the {\sc lmfit} module for {\sc Python} to fit the [O{\sc iii}] doublet both using a single Gaussian component and as a combination of two Gaussian components. We assess which fit is preferred based on the reduced $\chi^2$.  The spectral resolution in the NIR data is $\approx50$ km s$^{-1}$ and we do not include the instrumental dispersion in the fitting procedure. We fit three components for XLS-25 and 35 (two narrow components and a broad one). For a single Gaussian fit, we allowed the initial redshift estimate to vary by $\pm500$ km s$^{-1}$. Both the [O{\sc iii}] lines have the same line-width (in km s$^{-1}$) and a fixed relative flux of 1/2.98. We allow the line-width FWHM to vary from 50 to 1000 km s$^{-1}$with initial guess at 150 km s$^{-1}$. For a two-component Gaussian fit we set the initial redshifts of the two components to $\Delta z=\pm0.001$, respectively, and the line-widths 100 and 400 km s$^{-1}$. The redshifts of these components are allowed to vary by 50 km s$^{-1}$ and the widths can vary freely between 50 and 1000 km s$^{-1}$ as long as the broad component is broader than the narrow component. For XLS-5, 10, 11, 12, 16, 26 and 27, where the shape of the [O{\sc iii}] line suggests two narrow-components or where two clumps are seen in the imaging data, we fit two narrow components both with initial width of 100 km s$^{-1}$ and maximum allowed separation of 200 km s$^{-1}$. In these objects the S/N is not sufficient to allow the detection of any additional broad component. The fits are highly sensitive to the presence of skyline residuals, which we therefore mask. 

Three example [O{\sc iii}] fits are shown in Fig. $\ref{fig:O3_examples}$. In these three example cases a two-component fit is preferred over a single component, as can clearly be seen from the residuals in the bottom panels. XLS-25 is a good example illustrating the use of simultaneously fitting both [O{\sc iii}] lines due to skyline contamination. Out of the 33 objects with [O{\sc iii}] detections, 11 objects are fitted with a single component, 5 with two narrow components, 13 with a narrow and a broad component and two objects with two narrow and one broad component. The narrow components have line-widths FWHM ranging from 60-160 km s$^{-1}$, typically 110 km s$^{-1}$. This means they typically are marginally resolved. The broad components have FWHM ranging from 200 to 700 km s$^{-1}$, typically 280 km s$^{-1}$. We define the redshift of the narrow component to be the systemic redshift. In case we fit two narrow components (see Table $\ref{tab:measurements}$) we define the systemic redshift to be at the redshift of the narrow component that is closest to Ly$\alpha$ along the spatial direction as this is likely the component or H{\sc ii} region that is the origin of the Ly$\alpha$ emission. For half of these systems, the systemic redshift corresponds to a fainter component of the galaxy. However, the 2D spectrum in these cases clearly shows that the brighter components are spatially offset. If we would use the luminosity-weighted average redshift of the two components, the systemic redshift would on average change by $+30$ km s$^{-1}$ (ranging from -19 to +100 km s$^{-1}$).

\begin{figure}
\includegraphics[width=8.6cm]{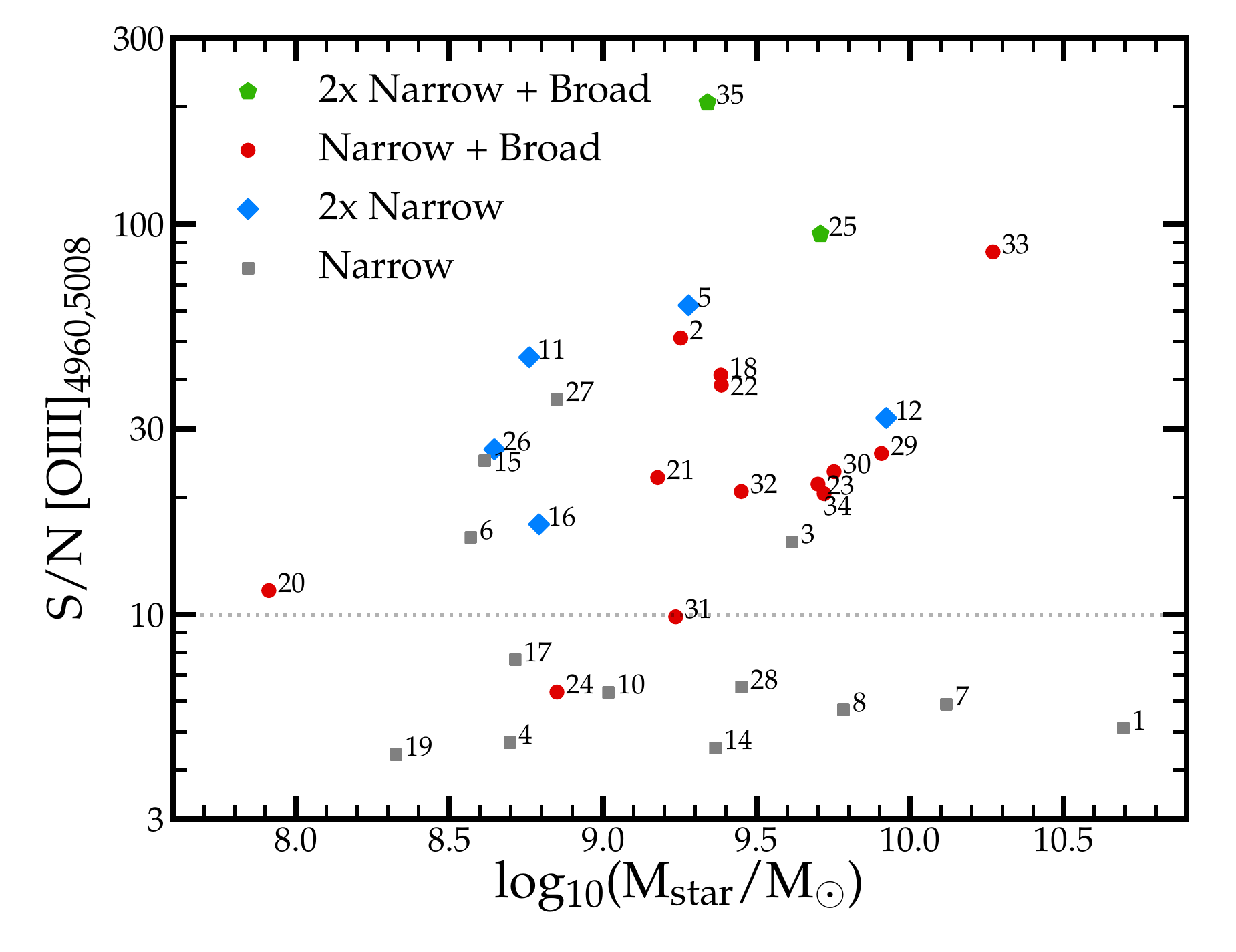} 
\caption{Integrated S/N ratio of the [O{\sc iii}]$_{4960, 5008}$ flux versus the stellar mass of the XLS targets. The IDs of the targets are labeled on each data-point. The objects that are best-fit by a single Gaussian are shown as grey squares. Red points show objects for which an additional broad component is required. Blue diamonds show objects that are best-fit by two narrow profiles and the green pentagons mark the objects for which two narrow and one broad component are required to accurately fit the [O{\sc iii}] profile. It is clear that a S/N of $\gtrsim10$ is typically required in order to identify complex features in the [O{\sc iii}] profile. }
\label{fig:O3_SN}
\end{figure}

As shown in Fig. $\ref{fig:O3_SN}$, the detection of more complex features in the [O{\sc iii}] spectrum depends on the integrated S/N. Below a S/N of 10 a single component is typically preferred. Above a S/N of 10 the galaxies with lower masses tend to be described by two narrow components. \footnote{We note that the [O{\sc iii}] line from XLS-1 is fitted by a single component with FWHM of 600 km s$^{-1}$. XLS-1 is likely an AGN as indicated from the very broad nebular lines, the detection of broad C{\sc iv} and Mg{\sc ii} emission and the red SED.} For all galaxies with [O{\sc iii}] detection, we measure the H$\beta$ and H$\alpha$ fluxes assuming the same line-profile as our best-fit [O{\sc iii}] profiles, but we note that these have relatively low S/N $\approx3$ in most individual objects. For comparison to other studies at $z\approx2$, we list the H$\alpha$-based SFR (see \S $\ref{sec:sfr}$, removing the contribution from broad emission to the H$\alpha$ flux in case a broad component is detected assuming it is not produced by recombination radiation associated to young stars) in Table $\ref{tab:flux_measurements}$.\footnote{It is possible to revert this correction based on the relative fraction of the line-flux that is in the broad component listed in Table $\ref{tab:measurements}$.} For comparison to studies at $z>6$, we also list the combined EW of H$\beta$ and [O{\sc iii}] which has been measured using the continuum estimate from the SED model.

We have verified the systemic redshift in the majority of sources using other emission lines, particularly the H$\alpha$ and H$\beta$ lines. The typical S/N in the H$\alpha$ line is 3 times lower than [O{\sc iii}], while the S/N in the H$\beta$ line is 5 times lower than [O{\sc iii}]. For a few objects we also verified the systemic redshift with detections of faint He{\sc ii}, O{\sc iii}]$_{1661,1666}$ and/or C{\sc iii}] line-emission in the rest-frame UV, which is a useful consistency check as this ensures a stable wavelength calibration over the UVB to NIR arms.

\begin{table*} 
\caption{General properties of the galaxies in the XLS sample. The half-light radius is determined from rest-frame UV imaging with {\it HST}/ACS when available.} 
\begin{tabular}{lrrrrrrrrrr}

ID & $z_{\rm sys}$ & M$_{1500}$ & $\beta$ & log$_{10}$(M$_{\rm star}$/M$_{\odot}$) & SFR$_{\rm H\alpha}$/M$_{\odot}$yr$^{-1}$ & r$_{1/2}$/kpc & L$_{\rm Ly\alpha}$/$10^{42}$erg s$^{-1}$ & EW$_{\rm Ly\alpha}$/{\AA} & EW$_{\rm H\beta + [OIII]}$/{\AA} \\ \hline
XLS-1$\dagger$ & 2.1961 & $-20.3$ & $0.3$ & $10.7\pm0.1$ & $9.3^{+85.6}_{-7.5}$ & $0.7$ & $6.9\pm0.7$ & $98^{+10}_{-10}$ & $155^{+81}_{-76}$\\
XLS-2 & 2.2296 & $-19.9$ & $-2.1$ & $9.3\pm0.2$ & $7.0^{+13.0}_{-2.2}$ & $0.8$ & $13.5\pm0.7$ & $182^{+13}_{-13}$ & $967^{+104}_{-103}$\\
XLS-3 & 2.2225 & $-19.0$ & $-2.1$ & $9.6\pm0.3$ & $1.0^{+1.5}_{-0.3}$ & $0.5$ & $3.1\pm0.2$ & $96^{+10}_{-9}$ & $184^{+13}_{-13}$\\
XLS-4 & 2.2279 & $-19.4$ & $-1.9$ & $8.7\pm0.1$ & $2.0^{+7.4}_{-1.3}$ & $1.2$ & $3.6\pm0.4$ & $76^{+12}_{-11}$ & $688^{+531}_{-480}$\\
XLS-5 & 2.2293 & $-20.5$ & $-2.0$ & $9.3\pm0.2$ & $10.3^{+30.2}_{-5.8}$ & $1.2$ & $8.2\pm0.5$ & $63^{+4}_{-4}$ & $422^{+33}_{-33}$\\
XLS-6 & 2.2218 & $-19.0$ & $-2.5$ & $8.6\pm0.3$ & $2.4^{+1.3}_{-0.8}$ & $0.6$ & $6.3\pm0.5$ & $179^{+16}_{-15}$ & $1391^{+164}_{-167}$\\
XLS-7 & 2.2229 & $-19.8$ & $-0.9$ & $10.1\pm0.1$ & $6.0^{+15.8}_{-4.1}$ & $1.4$ & $1.2\pm0.2$ & $16^{+4}_{-3}$ & $43^{+10}_{-9}$\\
XLS-8 & 2.0670 & $-21.3$ & $-1.3$ & $9.8\pm0.1$ & $4.8^{+5.3}_{-1.6}$ & $1.3$ & $0.1\pm0.3$ & $1^{+1}_{-1}$ & -\\
XLS-9 & 2.212* & $-19.5$ & $-1.8$ & $8.9\pm0.3$ & - & $0.8$ & $5.1\pm0.8$ & $95^{+15}_{-16}$ & -\\
XLS-10 & 2.2158 & $-19.3$ & $-1.9$ & $9.0\pm0.3$ & $29.6^{+379.6}_{-27.9}$ & $1.0$ & $4.1\pm0.5$ & $96^{+12}_{-13}$ & $555^{+88}_{-91}$\\
XLS-11 & 2.2172 & $-19.5$ & $-2.2$ & $8.8\pm0.3$ & $3.9^{+7.7}_{-1.5}$ & $1.0$ & $8.3\pm0.3$ & $152^{+11}_{-9}$ & $1554^{+144}_{-143}$\\
XLS-12 & 2.2064 & $-19.8$ & $-1.6$ & $9.9\pm0.1$ & $26.4^{+18.7}_{-10.7}$ & $0.9$ & $6.4\pm0.5$ & $99^{+11}_{-9}$ & $733^{+38}_{-37}$\\
XLS-13 & 2.234* & $-19.5$ & $-2.3$ & $8.8\pm0.1$ & - & $0.6$ & $3.6\pm0.3$ & $64^{+7}_{-6}$ & -\\
XLS-14 & 2.1418 & $-19.0$ & $-0.2$ & $9.4\pm0.3$ & $1.2^{+1.8}_{-0.5}$ & - & $1.9\pm0.4$ & $64^{+15}_{-15}$ & $180^{+129}_{-125}$\\
XLS-15 & 2.2302 & $-19.2$ & $-2.8$ & $8.6\pm0.1$ & $0.9^{+0.5}_{-0.4}$ & - & $3.2\pm0.3$ & $82^{+8}_{-9}$ & $1100^{+66}_{-64}$\\
XLS-16 & 2.2098 & $-20.0$ & $-1.9$ & $8.8\pm0.1$ & $3.0^{+1.2}_{-0.5}$ & $1.5$ & $3.1\pm0.4$ & $39^{+5}_{-5}$ & $461^{+35}_{-33}$\\
XLS-17 & 2.2015 & $-20.2$ & $-2.3$ & $8.7\pm0.2$ & $2.6^{+0.3}_{-0.3}$ & - & $10.9\pm0.4$ & $100^{+3}_{-3}$ & $466^{+59}_{-62}$\\
XLS-18 & 2.2095 & $-21.2$ & $-2.2$ & $9.4\pm0.1$ & $25.2^{+18.7}_{-9.5}$ & - & $15.6\pm0.5$ & $62^{+2}_{-2}$ & $862^{+28}_{-28}$\\
XLS-19 & 2.2186 & $-18.9$ & $-2.3$ & $8.3\pm0.1$ & $2.0^{+4.4}_{-1.0}$ & $1.1$ & $2.5\pm0.3$ & $89^{+11}_{-11}$ & $665^{+454}_{-440}$\\
XLS-20 & 2.2210 & $-18.8$ & $-1.2$ & $7.9\pm1.2$ & $4.6^{+42.4}_{-2.7}$ & - & $8.9\pm0.5$ & $294^{+104}_{-60}$ & $3212^{+3469}_{-1426}$\\
XLS-21 & 2.4197 & $-20.7$ & $-1.9$ & $9.2\pm0.5$ & $8.6^{+5.7}_{-1.9}$ & $1.0$ & $13.8\pm0.4$ & $88^{+6}_{-6}$ & $2455^{+262}_{-221}$\\
XLS-22 & 2.4518 & $-20.2$ & $-2.0$ & $9.4\pm0.1$ & $1.8^{+0.8}_{-0.4}$ & $0.6$ & $11.2\pm0.3$ & $111^{+4}_{-4}$ & $1594^{+61}_{-62}$\\
XLS-23 & 2.4706 & $-21.0$ & $-2.1$ & $9.7\pm0.1$ & $16.2^{+13.6}_{-5.1}$ & $1.0$ & $32.3\pm0.4$ & $156^{+3}_{-3}$ & $1440^{+70}_{-75}$\\
XLS-24 & 2.2463 & $-19.7$ & $-2.6$ & $8.8\pm0.1$ &$5.7^{+24.8}_{-2.2}$ & $0.9$ & $17.6\pm0.3$ & $261^{+7}_{-7}$ & $994^{+193}_{-202}$\\
XLS-25 & 2.1721 & $-21.6$ & $-2.1$ & $9.7\pm0.1$ & $10.3^{+4.3}_{-3.5}$ & $0.8$ & $20.0\pm0.5$ & $57^{+1}_{-1}$ & $745^{+15}_{-17}$\\
XLS-26 & 2.1723 & $-19.2$ & $-2.6$ & $8.6\pm0.1$ & $11.5^{+12.3}_{-4.2}$ & $0.8$ & $10.6\pm0.3$ & $209^{+7}_{-7}$ & $3420^{+232}_{-204}$\\
XLS-27 & 1.9981 & $-19.9$ & $-1.6$ & $8.8\pm0.1$ & $9.7^{+1.1}_{-1.2}$ & $1.4$ & $3.6\pm0.5$ & $42^{+6}_{-5} \ddagger$ & $2108^{+170}_{-169}$\\
XLS-28 & 2.2051 & $-20.9$ & $-1.7$ & $9.5\pm0.1$ & $7.4^{+4.7}_{-1.1}$ & $4.1$ & $10.2\pm0.7$ & $51^{+4}_{-4}$ & $462^{+151}_{-124}$\\
XLS-29 & 2.3282 & $-21.9$ & $-1.6$ & $9.9\pm0.3$ & - & - & $1.7\pm0.4$ & $4^{+1}_{-1}$ & $49^{+3}_{-3}$\\
XLS-30 & 2.3051 & $-21.1$ & $-2.2$ & $9.8\pm0.2$ & - & - & $6.1\pm0.3$ & $26^{+1}_{-1}\mathsection$ & $392^{+35}_{-23}$\\
XLS-31 & 2.1737 & $-20.7$ & $-2.3$ & $9.2\pm0.2$ & - & - & $1.4\pm0.3$ & $9^{+2}_{-2}\mathsection$ & $392^{+98}_{-88}$\\
XLS-32 & 2.1682 & $-21.1$ & - & $9.4$ & $36.6^{+52.0}_{-19.8}$  & - & $7.3\pm0.8$ & $32^{+4}_{-4}$ & -\\
XLS-33 & 2.1922 & $-20.8$ & - & $10.3$ & $14.9^{+6.4}_{-3.5}$ & - & $9.6\pm0.4$ & $60^{+5}_{-4}$ & -\\
XLS-34 & 2.1885 & $-20.5$ & - & $9.7$ & $3.6^{+4.4}_{-1.5}$ & - & $12.2\pm0.6$ & $127^{+20}_{-15}$ & -\\
XLS-35 & 2.3567 & $-21.2$ & - & $9.3$ &$20.4^{+10.3}_{-3.5}$ & - & $12.6\pm0.5$ & $40^{+2}_{-2}$ & -\\
\end{tabular}
\vspace{1ex}

{\raggedright $\dagger$ XLS-1 is identified as an AGN. \par}
{\raggedright * For XLS-9 and XLS-13 we list the redshifts of the red peak of the Ly$\alpha$ line as we do not detect a non-resonant emission-line in their spectra. \par}
{\raggedright  $\ddagger$ We caution the interpretation of the Ly$\alpha$ EW of XLS-27 as its Ly$\alpha$ line is spatially offset from the continuum that has been used to estimate the EW by $\approx9$ kpc. \par}
{\raggedright  $\mathsection$ A comparison to the measurements from \cite{Erb2016} suggests that it is plausible that a significant fraction of the Ly$\alpha$ flux for XLS-30 and 31 is missing (i.e. a factor 2-4, respectively) due to the use of a very narrow slit and a possible mis-alignment. \par}

\label{tab:flux_measurements}
\end{table*}

\subsection{Lyman-$\alpha$ flux measurements} \label{sec:lyaflux} 
\begin{figure*}
\includegraphics[width=17.8cm]{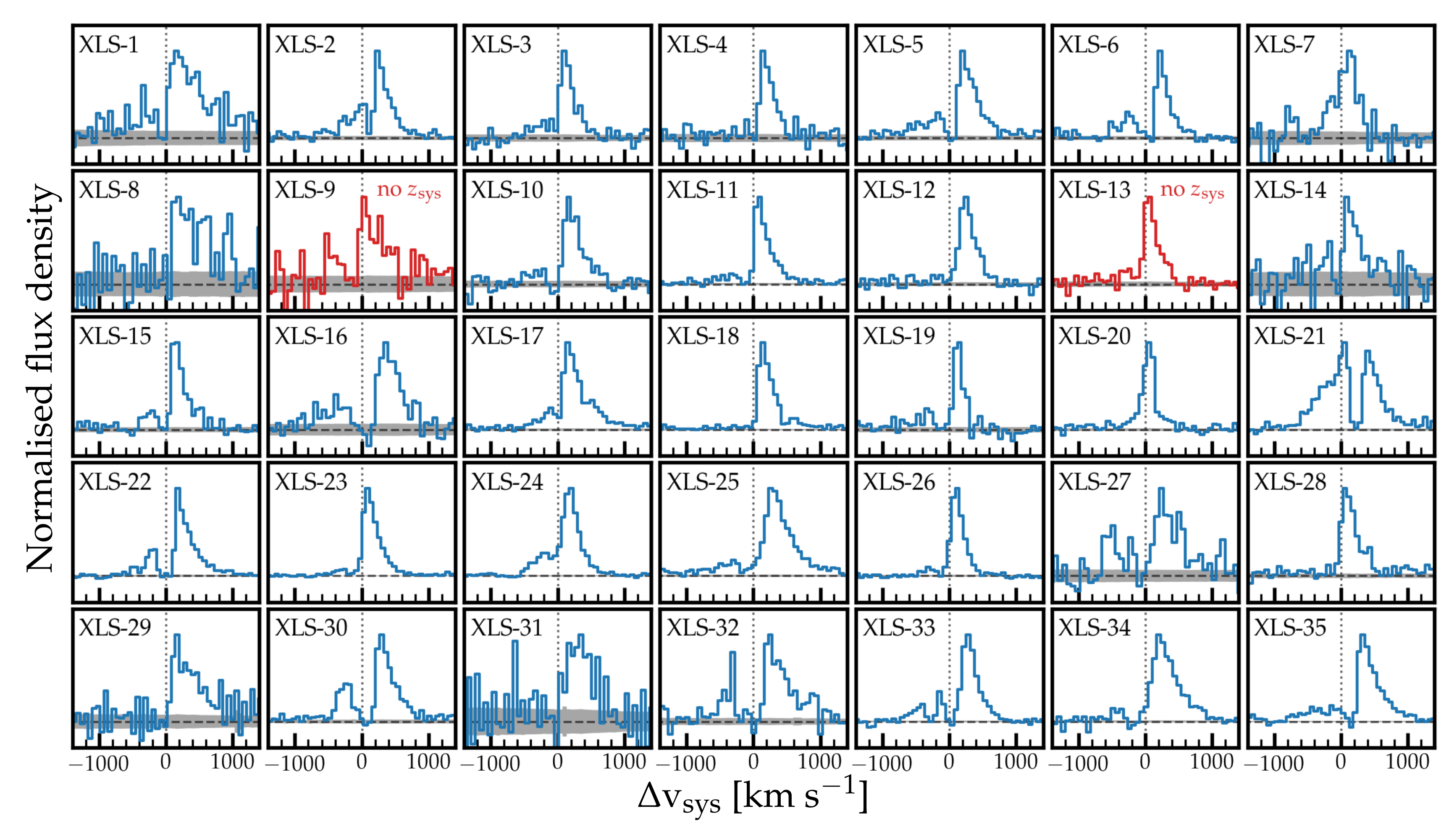} \vspace{-0.8cm}
\caption{Lyman-$\alpha$ profiles of the XLS sample. The spectra are binned in the velocity-direction by a factor two for visualisation purposes. Spectra are normalised to the peak Ly$\alpha$ flux density. The velocities are centred on the systemic redshift determined by [O{\sc iii}], except for XLS-9 and XLS-13 that are centred on the peak of the Ly$\alpha$ line as no non-resonant emission-line is detected. The grey shaded region shows the noise level. We note that we will explore these individual line-profiles in detail in an upcoming paper. }
\label{fig:Lyaspec}
\end{figure*}

The Ly$\alpha$ emission-lines of the XLS LAEs are shown in Fig. $\ref{fig:Lyaspec}$. The line-profiles are typically double-peaked\footnote{We will present a detailed investigation of the double peak fraction in a follow-up work, but note that 25 out of the 33 LAEs (i.e. 75 \%) with a systemic redshift show flux on the blue side of the systemic with a signal-to-noise above 5.} with the redder line being the strongest and being significantly skewed. We measure the Ly$\alpha$ flux non-parametrically by integrating the flux between $\pm1000$ km s$^{-1}$ from the systemic redshift. This velocity window captures the total Ly$\alpha$ flux for all LAEs. For the majority of LAEs with narrower lines the wide window will lead to very conservative uncertainties. The errors are obtained by re-measuring the flux on 1000 perturbations of the spectrum. The subtracted continuum level is estimated as the average continuum measured over the 1270-1300 {\AA} interval as motivated below. The rest-frame Ly$\alpha$ EW is computed as the ratio of the Ly$\alpha$ luminosity and the average continuum luminosity density over this interval.

The continuum level around the Ly$\alpha$ line can be fairly complicated to estimate accurately because of H{\sc i} absorption in the ISM, CGM or IGM \citep[e.g.][]{Laursen2011,McKinney2019}, possible absorption in A or B stars \citep{PenaGuerrero2013},  the nearby NV P Cygni profile at $\lambda_0\approx1245$ {\AA} \citep[e.g.][]{Chisholm2019} and the strong Si{\sc ii} interstellar absorption line at $\lambda_0=1260$ {\AA} \citep{Reddy2016b}. We therefore measure the continuum level throughout over the $\lambda_0=1270-1300$ {\AA} interval where the continuum is relatively featureless. For objects with low S/N in the continuum we estimate the continuum using SED fitting. We have verified that this results in similar Ly$\alpha$ EWs for the objects for which we could measure the continuum as well as in the stack.

For individual objects we do not apply an average CGM or IGM correction as it may be possible that the observed LAEs are on biased sight-lines that favour higher Ly$\alpha$ transmission. The average correction at $z=2.2$ only affects the blue side of the Ly$\alpha$ line and is rather moderate compared to high redshifts \citep{Inoue2014,Byrohl2020}. The Ly$\alpha$ luminosities and EWs are listed in Table $\ref{tab:flux_measurements}$.

\begin{table*}
\caption{The line-profile measurements of the fits to the [O{\sc iii}]$_{4960, 5008}$ doublet. Lines are fixed to the intrinsic 1:2.98 flux ratio. The FWHMs are listed in km s$^{-1}$ and not corrected for instrumental dispersion of 50 km s$^{-1}$. In case two narrow components are identified, $z_{\rm sys, 1}$ is the redshift of the [O{\sc iii}] line that is spatially most closely associated to the peak of the Ly$\alpha$ line. In the last column we list the results of visually inspecting the {\it HST} images for the presence of a clumpy structure.}
\begin{tabular}{lrrrrrrrrr}
ID & $z_{\rm sys, 1}$ & $z_{\rm sys, 2}$ & $z_{\rm broad}$ & FWHM$_{\rm sys, 1}$ & FWHM$_{\rm sys, 2}$ & FWHM$_{\rm broad}$ & f$_{\rm narrow, 2}$/f$_{\rm tot}$ & f$_{\rm broad}$/f$_{\rm tot}$ & Multiple clump {\it HST} \\ \hline
XLS-1 & 2.1961 & - & - & 596 & - & - & - & - & N\\
XLS-2 & 2.2296 & - & 2.2295 & 69 & - & 227 & - & 0.31 & N\\
XLS-3 & 2.2225 & - & - & 192 & - & - & - & -& N\\
XLS-4 & 2.2279 & - & - & 207 & - & - & - & -& N \\
XLS-5 & 2.2293 & 2.2283 & - & 94 & 75 & - & 0.17 & - & Y\\
XLS-6 & 2.2218 & - & - & 98 & - & - & - & - & N\\
XLS-7 & 2.2229 & - & - & 105 & - & - & - & - & Y\\
XLS-8 & 2.0670 & - & - & 161 & - & - & - & - & Y\\
XLS-9 & - & - & - & - & - & - & - & - & N\\
XLS-10 & 2.2158 & - & - & 124 & - & - & - & - & N\\
XLS-11 & 2.2172 & 2.2169 & - & 58 & 154 & - & 0.66 & - & Y\\
XLS-12 & 2.2064 & 2.2078 & - & 137 & 106 & - & 0.54 & - & Y\\
XLS-13 & - & - & - & - & - & - & - & - & N\\
XLS-14 & 2.1418 & - & - & 70 & - & - & - & - & N\\
XLS-15 & 2.2302 & - & - & 80 & - & - & - & - & N\\
XLS-16 & 2.2098 & 2.2117 & - & 101 & 94 & - & 0.57 & - & N\\
XLS-17 & 2.2015 & - & - & 169 & - & - & - & - & -\\
XLS-18 & 2.2095 & - & 2.2095 & 113 & - & 374 & - & 0.40 & -\\
XLS-19 & 2.2186 & - & - & 99 & - & - & - & - & N\\
XLS-20 & 2.2210 & - & 2.2210 & 94 & - & 196 & - & 0.43 & -\\
XLS-21 & 2.4197 & - & 2.4199 & 161 & - & 377 & - & 0.61 & N\\
XLS-22 & 2.4518 & - & 2.4516 & 97 & - & 246 & - & 0.80 & N\\
XLS-23 & 2.4706 & - & 2.4704 & 109 & - & 246 & - & 0.22 & N\\
XLS-24 & 2.2463 & - & 2.2465 & 84 & - & 422 & - & 0.20 & N\\
XLS-25 & 2.1721 & 2.1729 & 2.1725 & 113 & 66 & 397 & 0.15 & 0.60 & Y\\
XLS-26 & 2.1723 & 2.1717 & - & 103 & 125 & - & 0.34 & - & Y\\
XLS-27 & 1.9981 & - & - & 140 & - & - & - & - & Y\\
XLS-28 & 2.2051 & - & - & 128 & - & - & - & - & N\\
XLS-29 & 2.3282 & - & 2.3282 & 158 & - & 717 & - & 0.51 & Y\\
XLS-30 & 2.3051 & - & 2.3052 & 102 & - & 309 & - & 0.47 & Y\\
XLS-31 & 2.1737 & - & 2.1737 & 95 & - & 246 & - & 0.50 & Y\\
XLS-32 & 2.1682 & - & 2.1681 & 106 & - & 293 & - & 0.28 & N\\
XLS-33 & 2.1922 & - & 2.1923 & 105 & - & 235 & - & 0.48 & -\\
XLS-34 & 2.1885 & - & 2.1886 & 175 & - & 254 & - & 0.84 & -\\
XLS-35 & 2.3567 & 2.3578 & 2.3568 & 91 & 190 & 422 & 0.59 & 0.18 & Y\\

\end{tabular}
\label{tab:measurements}
\end{table*}

\subsection{Stacking} \label{sec:stack}
We use stacking to obtain the averaged spectrum of the LAEs. This stack is useful for identifying fainter features at the expense of losing information on the dispersion within the subset and complicating the analysis of line-profiles. 

We stack spectra in 2D because this allows us to investigate differences in the spatial extent of various wavelength regions. By stacking in 2D we are less sensitive to positional offsets between Ly$\alpha$ and the continuum \citep[e.g.][]{Hoag2019,Ribeiro2020} and uncertainties in the extraction apertures used for the 1D extraction in individual objects. 

First, individual 2D spectra are mapped to two common grids in rest-frame wavelength using a linear interpolation: one grid covers the UVB and VIS arms over $\lambda_0=980-3600$ {\AA} with $\Delta \lambda_0 = $0.06 {\AA} and the other one covers the NIR arm over $\lambda_0=3650-7250$ {\AA} with $\Delta \lambda_0 = $0.18 {\AA}. We scale the flux density of each object to the rest-frame luminosity density and only include objects with a measured systemic redshift. We also apply the same normalisation correction as described in \S $\ref{sec:photcor}$. Observed wavelengths between 545-560 nm and 1000-1110 nm are masked because of bad sensitivity in the highest orders of the VIS and NIR spectrographs. 

Second, we median combine the registered 2D spectra to obtain a typical spectrum of the specific subset. Errors are obtained through bootstrapping. We randomly resample the stacked subset 1000 times and repeat the stacking procedure for each resample to obtain the errors on the stacked spectrum.

Finally, we perform an optimal aperture-matched 1D Gaussian-extraction in a slightly modified way compared to individual spectra. We measure the FWHM along the spatial direction at wavelength intervals [$1290\pm50, 1390\pm50, 1490\pm50, 2200\pm125, 2500\pm125, 5008\pm2, 6564\pm2$] {\AA}. This means that our extraction is based on the size of the continuum, where we assume that the spatial extent of the nebular lines is similar to the rest-frame UV emission as we do not detect continuum emission in the NIR directly.\footnote{From an inspection of various 2D stacks, we note that the Ly$\alpha$ line is slightly more extended than the UV continuum, with a FWHM that we measure to be typically $\approx10$ \% higher. We find no difference in the spatial extent of the blue part of the Ly$\alpha$ line compared to the red part of the Ly$\alpha$ line.} The FWHM decreases slightly with wavelength from 0.96$''$ to 0.84$''$. Assuming that the wavelength dependence of the FWHM is smooth, we then fit a second-order polynomial and use that to derive the aperture-matched extraction size as a function of wavelength. The best-fit polynomial is slightly different for stacks of different subsets as each stack consists of a different combination of atmospheric conditions. For the {\it representative} stack (described in \S $\ref{sec:stack_measurements}$), we find FWHM $= 1.2 - 1.15\times10^{-4}  \lambda +0.89\times 10^{-8} \lambda^2$ where FWHM is in arcsec and $\lambda$ is in {\AA}. 

We have verified that the 1D extractions from 2D stacks described here are consistent with the stacks from the 1D extracted spectra of individual sources. Similarly we also derive the median fitted spectral energy distribution from the photometry and its uncertainty and find good agreement with the continuum levels between $1220-3000$ {\AA}.

\section{Stack of representative LAEs} \label{sec:stack_measurements}
\begin{figure*}
\includegraphics[width=18.1cm]{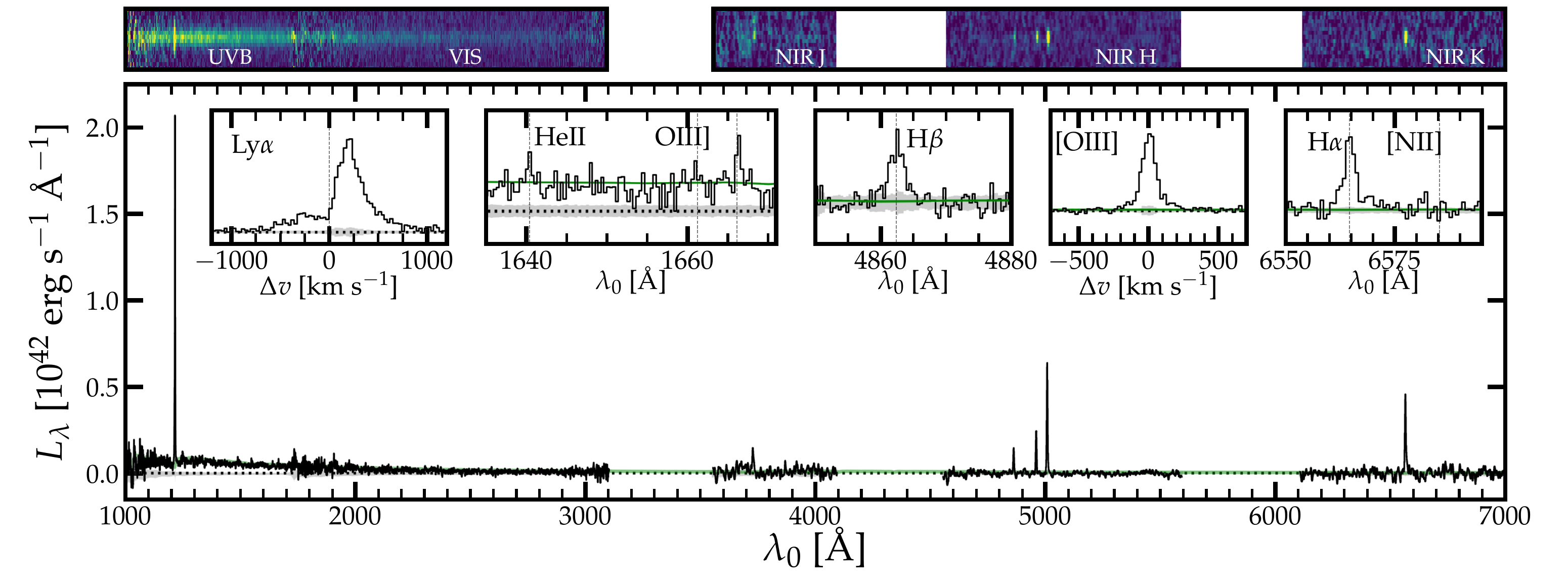}
\caption{Median stacked spectrum of our sample of LAEs at $z=2.2$.  In the top part of the Figure we show the 2D stack. In the main panel we show the 1D extraction of this stack. The black line shows the spectrum binned by a factor 10 (in the main panel). The grey region illustrates the uncertainty level. We show the spectrum zoomed-in on specific features: the Ly$\alpha$ (unbinned) line, the regions around the He{\sc ii} and O{\sc iii}] lines (binned by factor 2), the H$\beta$ line, the [O{\sc iii}]$_{5008}$ line and the H$\alpha$ line (the Balmer lines binned by factor 4, [O{\sc iii}] by a factor 2). The green line shows the stacked SED model derived from the photometry. }
\label{fig:stack}
\end{figure*}

Here we present a stack of LAEs that are {\it representative} for LAEs at redshift $z\approx2$. With this we mean specifically that we remove objects with Ly$\alpha$ EW$_0<10$ {\AA} (XLS-8) and objects that have been observed spectroscopically because of additional selection criteria (XLS-21 to 23, XLS-27 and XLS-29 to 31, see \S $\ref{sec:criteria}$). We also remove XLS-1 as it is an AGN. We further remove objects for which the data is not uniform: XLS-9 and XLS-13 as we did not measure a systemic redshift and XLS-29 to 31 because their H$\alpha$ line is not covered. The {\it representative} subset therefore includes 20 LAEs. We show the stacked spectrum of this sample in Fig. $\ref{fig:stack}$. The properties of the stack are presented in \S $\ref{sec:results}$.

\subsection{Line luminosity measurements}
\begin{figure}
\includegraphics[width=8.6cm]{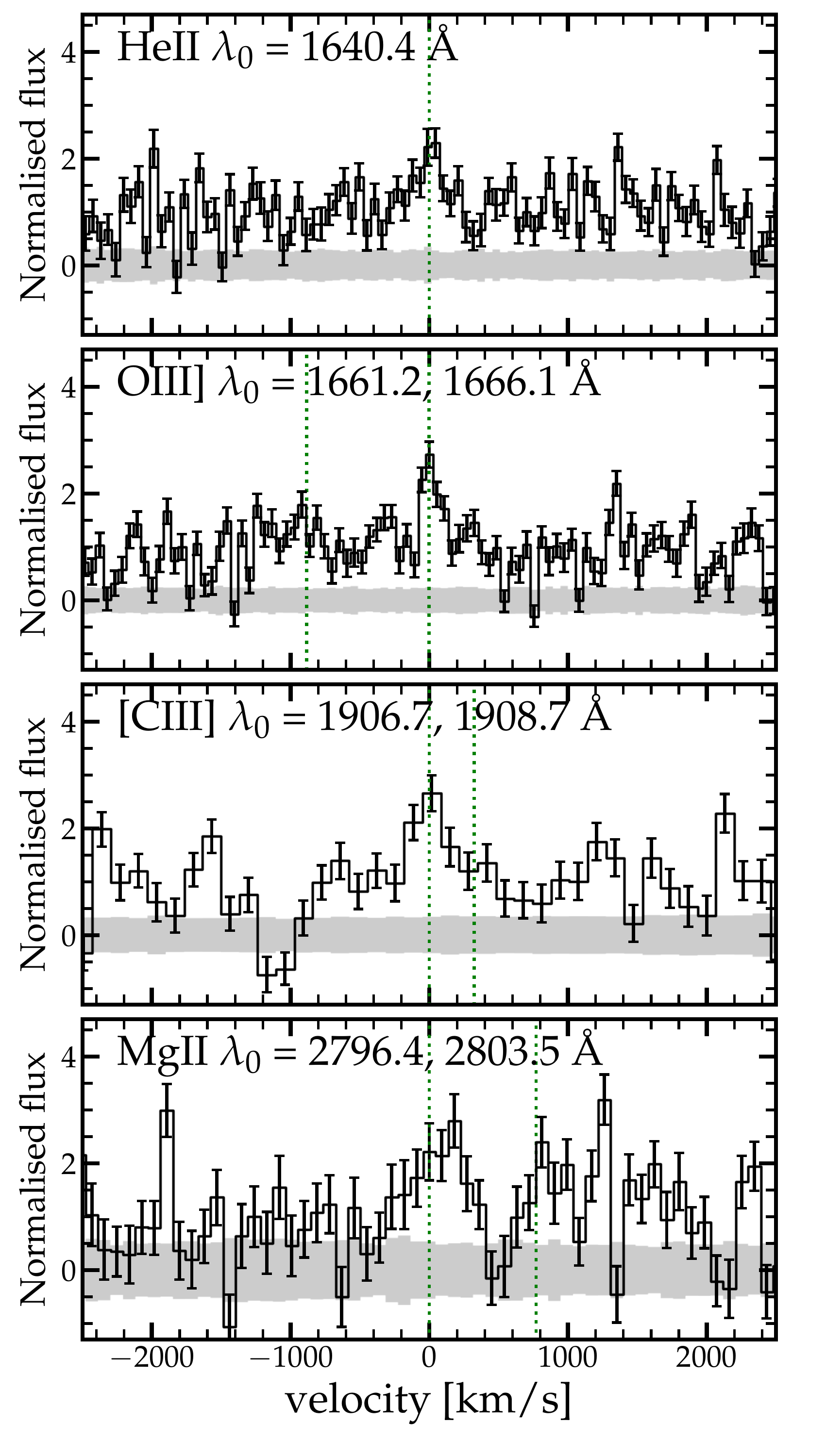} 
\caption{UV emission lines in the stacked XLS spectrum. Data were binned for visualisation. Grey shaded regions show the noise level. Green dotted lines mark the expected positions of emission-lines.}
\label{fig:UVemission}
\end{figure}

We visually inspect the stacked spectrum and find emission-line detections of Ly$\alpha$, C{\sc iv}$_{1548,1551}$, He{\sc ii}$_{1640}$, O{\sc iii}]$_{1661,1666}$, [C{\sc iii}]$_{1907}$, C{\sc iii}]$_{1909}$ and Mg{\sc ii}$_{2796,2803}$ in the rest-frame UV (Figures $\ref{fig:stack}$, $\ref{fig:CIV}$ and $\ref{fig:UVemission}$) and [O{\sc ii}]$_{3727,3729}$, [Ne{\sc iii}]$_{3870}$, H$\beta$, [O{\sc iii}]$_{4960,5008}$ and H$\alpha$ in the rest-frame optical (Fig. $\ref{fig:stack}$). 

\subsubsection{Rest-frame optical lines}
We notice that not all emission-line profiles are well described by a Gaussian profile, see the inset panels in Fig. $\ref{fig:stack}$. While this is not unexpected for Ly$\alpha$, we also notice more complex line-profiles in the case of H$\beta$, [O{\sc iii}] and H$\alpha$ that cannot be well described by the combination of a narrow and a broad Gaussian component. This complexity is likely explained by the fact that six representative LAEs show two narrow, closely separated [O{\sc iii}] lines and because five LAEs have strong optical lines that also include a broad component. Indeed, we have verified that removing the identified mergers leads to slightly more symmetric lines, but we note these still do not appear Gaussian. Therefore, we measure the luminosity in these lines non-parametrically by simply integrating the luminosity density within $\pm1000$ km s$^{-1}$ for Ly$\alpha$, $\pm280$ km s$^{-1}$ for [O{\sc iii}] and $\pm180$ km s$^{-1}$ for H$\alpha$ and H$\beta$. These boundaries were determined iteratively using a curve-of-growth approach. A larger window for the Balmer lines does not change the observed [O{\sc iii}]/H$\beta$ ratio. We note that the wings of the [O{\sc iii}] line have an FWHM$\approx280$ km s$^{-1}$. These wings could be present in the Balmer lines as well, but we do not detect them with the current sensitivity.

As we do not detect continuum in the NIR, we subtract the continuum measured in the median stack of the best-fitted spectral energy distribution models (this is shown in Fig. $\ref{fig:stack}$ as a green curve). This has a minimal impact on the luminosities of the emission-lines in the rest-frame optical. These continuum measurements are also used when we derive the EWs of the rest-frame optical lines. The 16-84th confidence percentiles of the line-luminosities and EWs are estimated by perturbing the spectrum and continuum levels with the propagated noise 1000 times. The measured luminosities and EWs are listed in Table $\ref{tab:measurements_stack}$.

The fainter emission lines that we detect can be well-fit by a single Gaussian, but this is probably a consequence of their lower S/N. The [O{\sc ii}] doublet is fit simultaneously, fixing the two lines to have the same line-width and fixing the continuum level to the level of the median SED. Similar to before, confidence intervals are estimated by perturbing the spectrum and the continuum level with their respective uncertainties. We measure similar luminosity for the two [O{\sc ii}] lines and a line-width FWHM of 120 km s$^{-1}$. This line-width is used to derive upper limits for the [N{\sc ii}]$_{6585}$ and [S{\sc ii}] lines and when fitting the [Ne{\sc iii}]$_{3870}$ line. 

\subsubsection{Rest-frame UV emission lines} \label{sec:measure_UV}
In the rest-frame UV, we find that the non-resonant emission-lines have FWHM around 120 km s$^{-1}$. The resonant Ly$\alpha$ and Mg{\sc ii} lines are broader. 

\begin{figure}
\includegraphics[width=8.8cm]{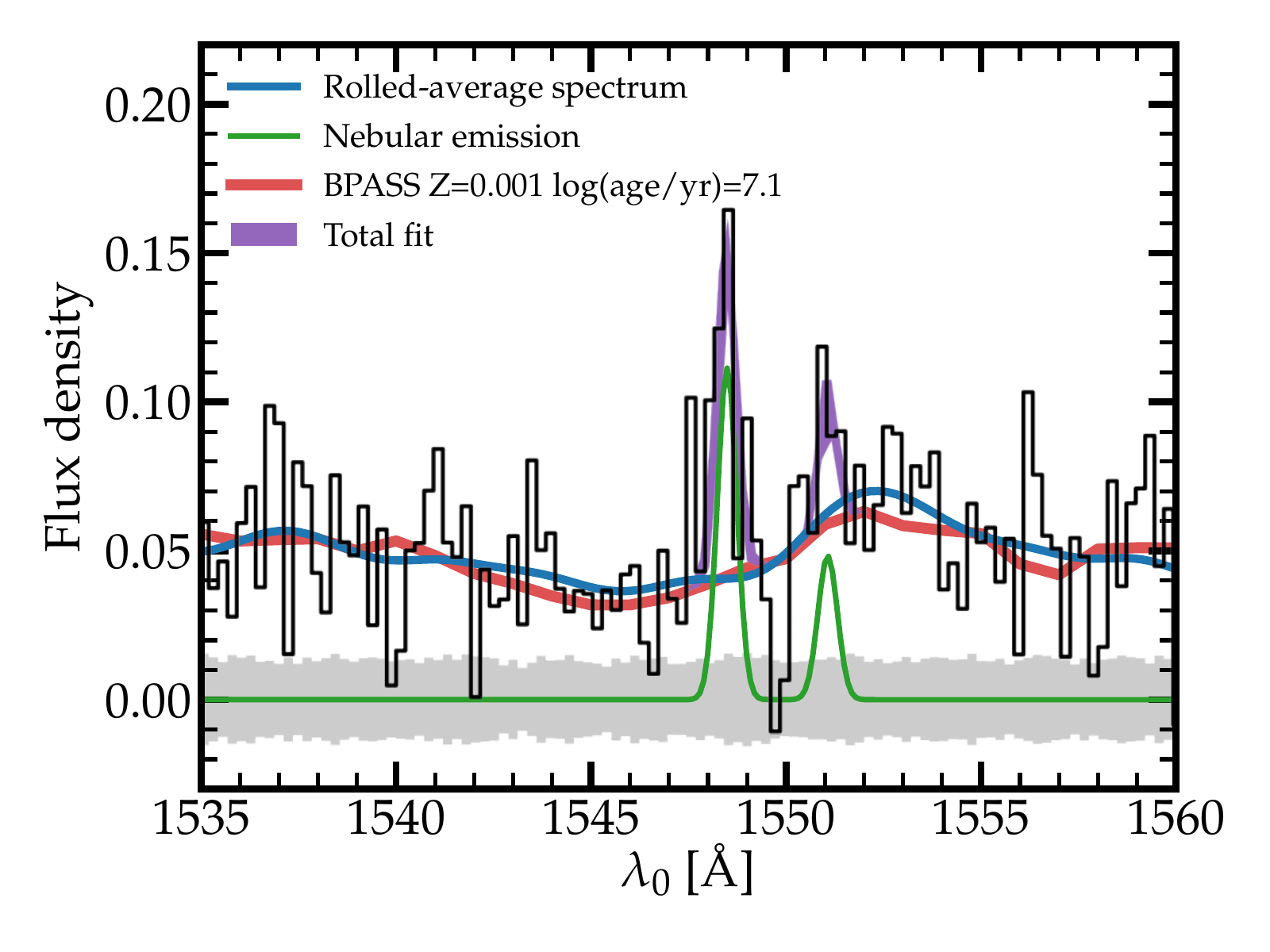} 
\caption{Zoom-in on the wavelength region around the C{\sc iv} feature in the stack of representative LAEs at $z\approx2$. The blue line shows rolled average spectrum obtained by masking $\pm130$ km s$^{-1}$ around the peaks of the nebular lines. The red line shows the best-fit stellar continuum model that agrees well with the rolled-average. The green lines show the best-fitted emission lines and the purple regions show the 1$\sigma$ uncertainties of the fit of the emission-lines on top of the continuum.}
\label{fig:CIV}
\end{figure}

For Ly$\alpha$ we estimate the continuum level between 1268-1300 {\AA} as described in \S $\ref{sec:lyaflux}$. The continuum around He{\sc ii}, O{\sc iii}] and C{\sc iii}] is well behaved and is fit simultaneously with the emission-lines. The uncertainty of the continuum-level is propagated while measuring uncertainties on the line-luminosity and EW. The continuum around the C{\sc iv} doublet is relatively complex due to the P-Cygni feature arising in the spectra of hot stars and possible interstellar absorption \citep[e.g.][]{VidalGarcia2017,Chisholm2019}. We therefore model the continuum by fitting a single-burst BPASS model over the wavelength ranges $\lambda_0 = [1530-1545, 1552-1570]$ {\AA} which are selected to mask the nebular line-emission and interstellar absorption. The best-fit model has an age 10$^{7.1}$ yr and a metallicity Z=0.001, see Fig. $\ref{fig:CIV}$. This model does not reproduce the full SED, but it serves its purpose for modelling the continuum around C{\sc iv}. We find that both the 1548.19, 1550.77 {\AA} lines are redshifted by $60\pm20$ km s$^{-1}$ \citep[indicating radiative transfer effects, e.g.][]{Berg2019}, have a FWHM $110\pm25$ km s$^{-1}$. The combined EW of the lines is $2.2\pm0.4$ {\AA}, where the 1548 line is $2.5\pm1$ times brighter than the 1551 line.

As the Mg{\sc ii} doublet is redshifted into a wavelength region with several skylines, the S/N ratio of the lines and continuum is very low (the S/N of the lines are 3.3 and 2.1, respectively). We assume a flat continuum around Mg{\sc ii} estimated by averaging over a 100 {\AA} wide window, masking the Mg{\sc ii} lines. Mg{\sc ii} lines are fitted with single Gaussians. We find that the peaks are redshifted by 50 km s$^{-1}$ with respect to the systemic. For the brighter Mg{\sc ii}$_{2796}$ line we measure FWHM$=380\pm50$ km s$^{-1}$ and we force the width of Mg{\sc ii}$_{2803}$ to be the same. The rest-frame EWs are EW$_{\rm MgII2796}$ = $6.1^{+3.1}_{-2.0}$ {\AA} and EW$_{\rm MgII2803}$ = $2.5^{+1.7}_{-1.3}$ {\AA}, respectively. These are a relatively typical EW given the UV luminosity of the stack \citep{Feltre2018}.

For the other UV lines, which all show no significant velocity offset compared to the systemic redshift, we subtract the continuum in a model-independent way by masking $\pm500$ km s$^{-1}$ around the line-centres and linearly interpolating the continuum level on both sides of this mask. For O{\sc iii}]$_{1661,1666}$, which has the highest S/N (O{\sc iii}]$_{1666}$ detection S/N=12.8), we measure a line-width $120\pm30$ km s$^{-1}$ and EWs $0.6\pm0.2$ {\AA} and $1.3\pm0.3$ {\AA}, respectively. The [C{\sc iii}]$_{1907}$ and C{\sc iii}]$_{1909}$ lines are in a noisy region of the stacked spectrum as they lie in the bluest part of the VIS arm of X-SHOOTER which has a lower sensitivity than the redder parts of the UVB arm. We therefore constrain the widths of these lines to the width of the O{\sc iii}] lines and measure EWs $2.8^{+0.7}_{-1.7}$ {\AA} and $1.5^{+0.5}_{-0.4}$ {\AA}, respectively. We find an indication that the He{\sc ii} line is somewhat broader (FWHM $210\pm90$ km s$^{-1}$) than the other lines, indicating possible contribution from broad stellar He{\sc ii} emission \citep{Brinchmann2008}. As we are interested in the nebular He{\sc ii} component but do not have the sufficient S/N to perform a two-component fit (He{\sc ii} detection S/N=8.6), we force the width to the range of widths of the O{\sc iii}] line and find an EW of $1.2^{+0.4}_{-0.4}$ {\AA}. Allowing the line-width to be larger, we would measure a line-flux that is a factor 1.3 higher.

\subsection{Si{\sc ii} absorption}
\begin{figure}
\includegraphics[width=8.6cm]{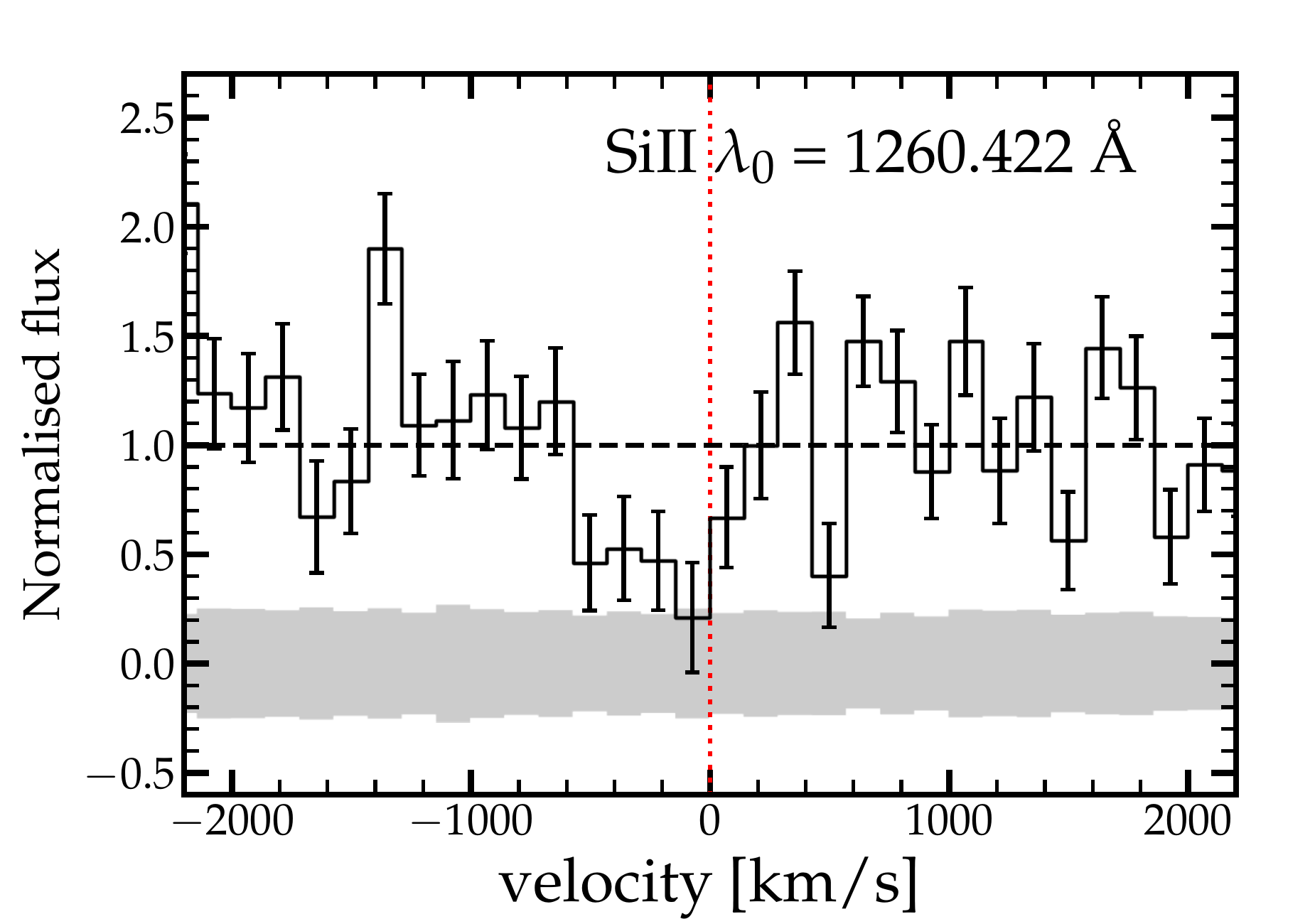} 
\caption{Detection of Si{\sc ii} in the stacked spectrum of representative LAEs. The velocity axis is with respect to the Si{\sc ii} line at the systemic redshift. Data were binned.   }
\label{fig:UVabsorption}
\end{figure}

We detect significant absorption from the low ionisation Si{\sc ii}$_{1260}$ line in our stacked spectrum, see Fig. $\ref{fig:UVabsorption}$. In this Figure the continuum level is estimated from the stacked SED model, which agrees well with the stacked spectrum. This absorption line has also been seen in several individual cases and stacks of LAEs \citep[e.g.][]{Shibuya2014,RiveraThorsen2015,Trainor2015}. We do not detect other absorption lines  at $>3\sigma$ significance. The absorption EW is measured by integrating between the maximum and minimum velocity at which absorption is detected. We perturb each spectrum 1000 times to estimate the uncertainties on the EW measured this way. For Si{\sc ii}, the EW is $-1.7^{+0.8}_{-0.3}$ {\AA} and the absorption-weighted average velocity is $-280^{+130}_{-70}$ km s$^{-1}$.

\begin{table}
\caption{Measurements of the stacked spectrum of representative LAEs at $z\approx2$. Upper limits are at the 3$\sigma$ level. Equivalent widths are all in the rest-frame. Line ratios are not corrected for attenuation, unless noted specifically.}
\begin{tabular}{lr}
Property & Measurement stack \\ \hline
M$_{1500}$ & $-19.9\pm0.2$ \\ 
$\beta$ & $-2.1\pm0.1$ \\ 
log$_{10}$(M$_{\rm star}$/M$_{\odot}$) & $9.0\pm0.1$ \\ 
SFR$_{\rm SED}$ & $4^{+5}_{-3}$ M$_{\odot}$ yr$^{-1}$ \\
age$_{\rm SED}$ & $250\pm200$ Myr \\
& \\
L$_{\rm Ly\alpha}$ & $(6.23^{+0.08}_{-0.08})\times10^{42}$ erg s$^{-1}$ \\
L$_{\rm [OII]3727,3729}$ & $(0.45^{+0.05}_{-0.06})\times10^{42}$ erg s$^{-1}$ \\
L$_{\rm [NeIII]3869}$ & $(0.18^{+0.03}_{-0.03})\times10^{42}$ erg s$^{-1}$ \\
L$_{\rm H\beta}$ & $(0.48^{+0.03}_{-0.02})\times10^{42}$ erg s$^{-1}$ \\
L$_{\rm [OIII]4960,5008}$ & $(2.97^{+0.07}_{-0.08})\times10^{42}$ erg s$^{-1}$ \\
L$_{\rm H\alpha}$ & $(1.67^{+0.05}_{-0.05})\times10^{42}$ erg s$^{-1}$ \\
L$_{\rm [NII]6585}$ & $<0.09\times10^{42}$ erg s$^{-1}$ \\
L$_{\rm [SII]6718,6733}$ & $<0.15\times10^{42}$ erg s$^{-1}$ \\

& \\ 
H$\alpha$/H$\beta$ & $3.5^{+0.2}_{-0.2}$\\
{[O{\sc ii}]$_{3729}$/[O{\sc ii}]$_{3727}$} & $1.0^{+0.2}_{-0.2}$ \\
O32 = ${\rm \frac{[OIII]_{5008}}{[OII]_{3727,3729}}}$ & $5.0^{+0.6}_{-0.5}$ \\
O3Hb = ${\rm \frac{[OIII]_{5008}}{H\beta}}$ & $4.7^{+0.3}_{-0.3}$ \\
R23 = ${\rm \frac{[OIII]_{4960, 5008} + [OII]_{3727,3729}}{H\beta}}$ & $7.1^{+0.4}_{-0.4}$ \\
Ne3O2 = ${\rm \frac{[NeIII]_{3870}}{[OII]_{3727,3729}}}$  & $0.4^{+0.1}_{-0.1}$ \\
log$_{10}$ N2Ha & $<-1.2$ \\
log$_{10}$ S2Ha & $<-1.0$ \\

& \\
E$(B-V)_{\rm gas}$ &  $0.22\pm0.06$ \\
SFR$_{\rm H\alpha}$ & $6^{+1}_{-1}$ M$_{\odot}$ yr$^{-1}$ \\
log$_{10}$($\rm \xi_{ion, UV dust}$/Hz erg$^{-1}$) & 25.3$^{+0.1}_{-0.1}$ \\
O32$_{\rm int}$ & $3.8^{+0.5}_{-0.5}$ \\
O3Hb$_{\rm int}$ & $4.5^{+0.3}_{-0.2}$ \\
R23$_{\rm int}$ & $7.1^{+0.5}_{-0.4}$ \\
Ne3O2$_{\rm int}$  & $0.4^{+0.1}_{-0.1}$ \\

log$_{10}(U)$ & $-2.44\pm0.03$ (stat.) $\pm0.10$ (sys.) \\
$n_{e, \rm [OII]}$ & $460^{+490}_{-280}$ cm$^{-3}$\\
$T_{e, {\rm O}^{2+}}$ & $14,800^{+700}_{-800}$ K\\
12+log(O/H)$_{\rm T_e}$ & $7.83^{+0.06}_{-0.05}$\\
12+log(O/H)$_{\rm R23}$ & $8.39^{+0.03}_{-0.04}$ \\
12+log(O/H)$_{\rm O32}$ & $8.12^{+0.04}_{-0.04}$ \\
12+log(O/H)$_{\rm O3Hb}$ & $8.29^{+0.04}_{-0.04}$ \\
12+log(O/H)$_{\rm Ne3O2}$ & $8.05^{+0.08}_{-0.06}$ \\
12+log(O/H)$_{\rm strong-line}$ & $8.21^{+0.03}_{-0.03}$ (stat.) $\pm0.14$ (sys.) \\
log$_{10}$(C/O) & $-0.8^{+0.2}_{-0.2}$ \\
log$_{10}$(N/O) & $<-0.8$ \\

& \\
EW$_{\rm Ly\alpha}$ & $73\pm4$ {\AA}  \\
EW$_{\rm H\alpha}$ & $531^{+131}_{-84}$ {\AA} \\ 
EW$_{\rm H\beta}$ & $70^{+20}_{-12}$ {\AA}\\ 
EW$_{\rm [OIII]_{4960,5008}}$ & $503^{+135}_{-87}$ {\AA}\\  
\end{tabular}
\label{tab:measurements_stack}
\end{table}

\begin{table}
\caption{Measurements of UV emission and absorption lines in the stack. Equivalent widths are all in the rest-frame.}
\centering
\begin{tabular}{lr}
Property & Measurement 2D stack \\ \hline
L$_{\rm CIV1548,1551}$ & $(9.9^{+1.8}_{-1.8})\times10^{40}$ erg s$^{-1}$ \\ 
L$_{\rm HeII 1640}$ & $(4.9^{+1.4}_{-1.6})\times10^{40}$ erg s$^{-1}$ \\
L$_{\rm OIII]1661}$ & $(2.6^{+1.0}_{-1.1})\times10^{40}$ erg s$^{-1}$ \\
L$_{\rm OIII]1666}$ & $(6.1^{+1.5}_{-1.4})\times10^{40}$ erg s$^{-1}$ \\
L$_{\rm [CIII]1907}$ & $(10^{+5}_{-5})\times10^{40}$ erg s$^{-1}$ \\
L$_{\rm CIII]1909}$ & $(6^{+3}_{-2})\times10^{40}$ erg s$^{-1}$ \\
L$_{\rm MgII2796}$ & $(6^{+2.2}_{-1.8})\times10^{40}$ erg s$^{-1}$ \\ 
L$_{\rm MgII2803}$ & $(3^{+1.3}_{-1.4})\times10^{40}$ erg s$^{-1}$ \\

& \\
EW$_{\rm CIV1548,1551}$ & $2.2^{+0.4}_{-0.4}$ {\AA}  \\ 
EW$_{\rm HeII1640}$ & $1.2^{+0.4}_{-0.4}$ {\AA}  \\
EW$_{\rm OIII]1661}$ & $0.6^{+0.2}_{-0.2}$ {\AA}  \\ 
EW$_{\rm OIII]1666}$ & $1.3^{+0.3}_{-0.3}$ {\AA}  \\
EW$_{\rm [CIII]1907}$ & $2.8^{+0.7}_{-1.7}$ {\AA}  \\
EW$_{\rm CIII]1909}$ & $1.5^{+0.5}_{-0.4}$ {\AA}  \\
EW$_{\rm MgII2796}$ & $6.1^{+3.1}_{-2.0}$ {\AA}  \\ 
EW$_{\rm MgII2803}$ & $2.5^{+1.7}_{-1.3}$ {\AA}  \\ 

& \\
Absorption lines & \\
EW$_{\rm SiII 1260}$ & $-1.7^{+0.8}_{-0.3}$ {\AA} \\  
$\overline{v}_{\rm SiII 1260}$ & $-260^{+160}_{-40}$ km s$^{-1}$\\  

& \\
Ly$\alpha$ profile & \\
$f_{\rm esc, Ly\alpha}$ & $0.27\pm0.04$ \\
Blue/Red & $0.26\pm 0.03$\\
Valley/Cont & $4.4\pm0.8$ \\
$\Delta v_{\rm red}$ & $+205\pm5$ km s$^{-1}$ \\
$\Delta v_{\rm blue}$ & $-294\pm5$ km s$^{-1}$ \\

\end{tabular}
\label{tab:measurements_stack_UV}
\end{table}

\subsection{Derived physical properties} \label{sec:derived}
We use the emission line measurements to derive the nebular attenuation, SFR$_{\rm H\alpha}$, the star formation rate surface density, Ly$\alpha$ escape fraction, the production efficiency of ionising photons, electron density, electron temperature and gas-phase oxygen abundance.

\subsubsection{Dust attenuation, SFR, ionising production efficiency and escape of Ly$\alpha$ photons} \label{sec:sfr}
The observed Balmer line-luminosities can be used to infer $Q_{\rm ion}$, the number of emitted ionising photons per second, following $Q_{\rm ion} = c_{\rm H\alpha} L_{\rm H\alpha}$,  where $c_{\rm H\alpha}=1.36\times10^{-12}$ erg$^{-1}$ \citep{Osterbrock1989}. The conversion is sensitive to the escape fraction of ionising photons and the dust content within HII regions, but these are assumed to be negligible here \citep[e.g.][]{Inoue2001,Dopita2006}. As the dominant source of ionising photons in the average LAE is star formation (see \S $\ref{sec:source}$), we can derive SFR$_{\rm H\alpha}$.

A main uncertainty in deriving $Q_{\rm ion}$ is the dust attenuation affecting the observed H$\alpha$ luminosity. We can estimate the nebular attenuation by using the observed Balmer decrement $\tau=\log(\frac{{\rm H}\alpha}{{\rm H}\beta} / 2.79)$, where the assumed intrinsic line-ratio of 2.79 depends slightly on electron density and temperature (we assume 100 cm$^{-3}$ and 15,000 K, see \S $\ref{sec:ISM}$), in combination with an assumed extinction law and dust geometry. We assume that dust is distributed as a uniform screen \citep[c.f.][]{Scarlata2009}. Following \cite{Reddy2015,Reddy2020} we assume that the nebular extinction curve $k(\lambda)$ follows the one from \cite{Cardelli1989}, such that E$(B-V)$ = 0.95$\tau$ and $k_{\rm H\alpha} =2.52$. For our stack we measure E$(B-V)_{\rm neb.} = 0.22\pm0.06$ which is then used to calculate the intrinsic H$\alpha$ luminosity following L$_{\rm H\alpha, intr}$ = L$_{\rm H\alpha, obs} \times 10^{0.4 E(B-V)_{\rm neb} k_{\rm H\alpha}}$.

The conversion of the intrinsic H$\alpha$ luminosity to SFR follows from the relations between $Q_{\rm ion}$ and H$\alpha$ luminosity, and $Q_{\rm ion}$ and SFR. The latter conversion depends on the SFH, the IMF, the properties of massive stars (e.g. binary fraction) and the stellar metallicity. We use the conversion SFR$_{\rm H\alpha} / $M$_{\odot}$ yr$^{-1} = 2.29\times10^{-42} \times \, {\rm L}_{\rm H\alpha, intrinsic}$/erg s$^{-1}$ derived by \cite{Theios2019} for BPASS v2.2 models with a \cite{Chabrier2003} IMF with a maximum stellar mass of 100 M$_{\odot}$, a constant star formation history with age $10^8$ yr and a metallicity 0.1 $Z_{\odot}$. Note that the conversion between SFR and intrinsic H$\alpha$ luminosity is a factor $\approx2$ smaller than the `standard' conversion with the same IMF but 10 times higher metallicity \citep{Murphy2011,KennicuttEvans2012}. This is due to the harder ionising spectra of low metallicity stars and the longer contribution to the ionising photon flux of binary stellar populations compared to populations of single stars \citep[e.g.][]{Gotberg2019}.

The production efficiency of ionising photons ($\xi_{\rm ion}$) is defined as $\xi_{\rm ion} = \frac{Q_{\rm ion}}{L_{\rm UV, intrinsic}}$ \citep{BouwensXION}. We note that $\xi_{\rm ion}$ is also related to the relative production of Ly$\alpha$ photons to the UV continuum and hence to the intrinsic Ly$\alpha$ EW \citep[e.g.][]{SM2019}. We estimate the intrinsic UV luminosity using $L_{\rm UV, intrinsic} = L_{\rm UV, observed} \times 10^{0.4 E(B-V)_{\star} k_{1500}}$. A crucial assumption is the relation between the stellar and nebular attenuation, the latter being estimated from the Balmer decrement. Several studies that investigated the relation between the stellar and nebular attenuation at $z\sim2-3$ have yielded conflicting results. Some indicate that the nebular and stellar extinction are similar, while others report a higher nebular attenuation \citep[e.g.][]{Kashino2013,Reddy2016,Steidel2016,Faisst2019,Theios2019}. Relevant for our sample is that \cite{Shivaei2020} report a higher nebular attenuation in systems with a gas-phase metallicity below 12+log(O/H)$<8.5$. Therefore, we assume the classical E$(B-V)_{\star} = 0.44 E(B-V)_{\rm gas}$ \citep{Calzetti2000} ratio here. For the UV attenuation, we use the \cite{Reddy2016} attenuation curve, which results in $k_{1500}=8.68$. The resulting $\xi_{\rm ion}$ is $10^{25.3\pm0.1}$ Hz erg$^{-1}$. This value is consistent with the value from the BPASS model that we used to convert H$\alpha$ luminosity to SFR. If we assume E$(B-V)_{\star} = E(B-V)_{\rm gas}$ we find a lower $\xi_{\rm ion} = 10^{24.9\pm0.1}$ Hz erg$^{-1}$. The values for E$(B-V)_{\star}$ from the SED fitting range from 0.0 to 0.17, with a mean of 0.03. This points towards an even lower stellar attenuation compared to the nebular attenuation.

Finally, having estimated the intrinsic H$\alpha$ luminosity we can calculate the Ly$\alpha$ escape fraction using $f_{\rm esc, Ly\alpha} = {\rm L}_{\rm Ly\alpha}/(8.7  {\rm L}_{\rm H\alpha, intrinsic})$. The intrinsic ratio between Ly$\alpha$ and H$\alpha$ depends slightly on gas temperature \citep[e.g.][]{Henry2015} but we use 8.7 for consistency with the literature. We measure a $f_{\rm esc, Ly\alpha} = 0.27\pm0.04$. This is consistent with the Ly$\alpha$ escape fraction measured using stacking of H$\alpha$ narrow-band imaging data on the parent sample of LAEs \citep{Sobral2016}, which suggests relative slit-losses are unimportant.

\subsubsection{Electron density, temperature, ionisation state, gas-phase abundances} \label{sec:ISM} 
The line ratios of the [O{\sc ii}]$_{3727,3729}$ doublet and the [C{\sc iii}]$_{1907}$/C{\sc iii}]$_{1909}$ lines are sensitive to the electron density \citep[e.g.][]{Keenan1992,Patricio2016}. Using Equation 7 from \cite{Sanders2016} we measure an electron density of $n_e=460^{+490}_{-280}$ cm$^{-3}$ based on the [O{\sc ii}] doublet consistent with \cite{Shirazi2014,Steidel2014}. The uncertainties on the [C{\sc iii}]/C{\sc iii}] ratio are too large to use it to obtain meaningful constraints on electron density.

The electron temperature is a key property of the ISM. We do not detect the temperature-sensitive [O{\sc iii}]$_{4363}$ line as this line is redshifted into a wavelength range with low atmospheric transmission. However, following the methodology from \cite{PerezMontero2017} we can estimate the temperature from the dust-corrected O{\sc iii}]$_{1661+1666}$/[O{\sc iii}]$_{5008}$ ratio. The intrinsic H$\alpha$/H$\beta$ ratio that is assumed to derive the attenuation depends on the electron temperature. Therefore we iteratively derive the attenuation and the electron temperature until convergence, which results in $T_{e, {\rm O}^{2+}} = 14,800^{+700}_{-800}$ K.

The relative strengths of the [O{\sc iii}], [O{\sc ii}] and H$\beta$ lines are sensitive to the electron temperature, gas-phase metallicity and ionisation state \citep[e.g.][]{NakajimaOuchi2014,Trainor2016}. In Table $\ref{tab:measurements_stack}$ we list the intrinsic ratios O32, O3Hb and R23 after correcting the emission line luminosities for attenuation based on the \cite{Cardelli1989} curve. We find a high O32, O3Hb and R23 of $3.8^{+0.5}_{-0.5}$, $4.5^{+0.3}_{-0.2}$ and $7.1^{+0.5}_{-0.4}$, respectively. These values point towards a low metallicity and a high ionisation parameter. Following the iterative methodology described in \cite{NakajimaOuchi2014} based on photoionisation models by \cite{KewleyDopita2002} we measure log$_{10}(U) = -2.38^{+0.04}_{-0.05}$.
The [Ne{\sc iii}] to [O{\sc ii}] ratio (Ne3O2) has also been proposed to trace the ionisation parameter \citep{Levesque2014}, similar to O32. While the lines originate from different species, the main benefit is that they are closer in wavelength and thus less susceptible to uncertainties in the attenuation correction. A caveat is that the ionisation energy from [Ne{\sc iii}] is 41 eV, which is somewhat higher than the ionisation energy of [O{\sc iii}] (35.1 eV). Therefore, changing the hardness of the ionisation field can lead to variations in the relations between $U$ and Ne3O2, and $U$ and O32. Applying the calibration based on photoionization modelling of LBGs at $z\sim2$ by \cite{Strom2018}, the observed Ne3O2 ratio implies log$_{10}(U) = -2.47\pm0.08$, which is consistent with the value derived using their calibration for O32  (log$_{10}(U) = -2.49\pm0.03$). Given the variation in $U$ for different methods, we retrieve the average $U$ and conservatively add $0.1$ dex systematic uncertainty, i.e. log$_{10}(U) = -2.44\pm0.03$ (stat.) $\pm0.10$ (sys.).

We can estimate the gas-phase metallicity directly using the electron temperature and the strength of the oxygen lines compared to H$\beta$. Following \cite{PerezMontero2014} we measure 12+log(O/H)=$7.83^{+0.06}_{-0.05}$. We use $n_e=500$ cm$^{-3}$, but the metallicity would only change by 0.01 dex within the constrained range of electron densities. Alternatively, we can estimate the gas-phase metallicity using empirical strong-line calibrations derived in low-redshift analogues of high redshift galaxies by \cite{Bian2018}. Besides R23, O32 and O3Hb, we also use the metallicity calibration based on the Ne3O2 ratio. Combining all our estimates (see Table $\ref{tab:measurements_stack}$) yields an average gas-phase metallicity of 12+log(O/H)$_{\rm strong-line} = 8.21^{+0.03}_{-0.03}$ (stat.) $\pm0.13$ (sys.), where the systematic error comes from the variation among the different line indices. This measurement is consistent with our upper limit on the [N{\sc ii}]/H$\alpha$ ratio that corresponds to 12+log(O/H)$<8.25$. The metallicity estimated from the strong-lines is on average 0.4 dex higher than our estimate inferred through the $T_e$ method, potentially indicating issues with using the UV lines with much higher critical density compared to the optical [O{\sc iii}] line (see also \citealt{Rigby2020}). Finally, following \cite{PerezMontero2017}, we use the dust-corrected C{\sc iv}, O{\sc iii}] and C{\sc iii}] lines to estimate a C/O abundance of log$_{10}$(C/O)$= -0.8\pm0.2$ and we use the upper limit on [N{\sc ii}] to derive log$_{10}$(N/O)$<-0.8$ (3$\sigma$).

\section{Results} \label{sec:results}
Here we address the average nature of LAEs at $z\approx2$ by predominantly synthesising the various measurements of our stack, but we also include results from the high-resolution rest-frame morphology of individual sources. The variation of the properties among the individual sources will be explored in an upcoming paper.
\begin{figure}
\includegraphics[width=8.6cm]{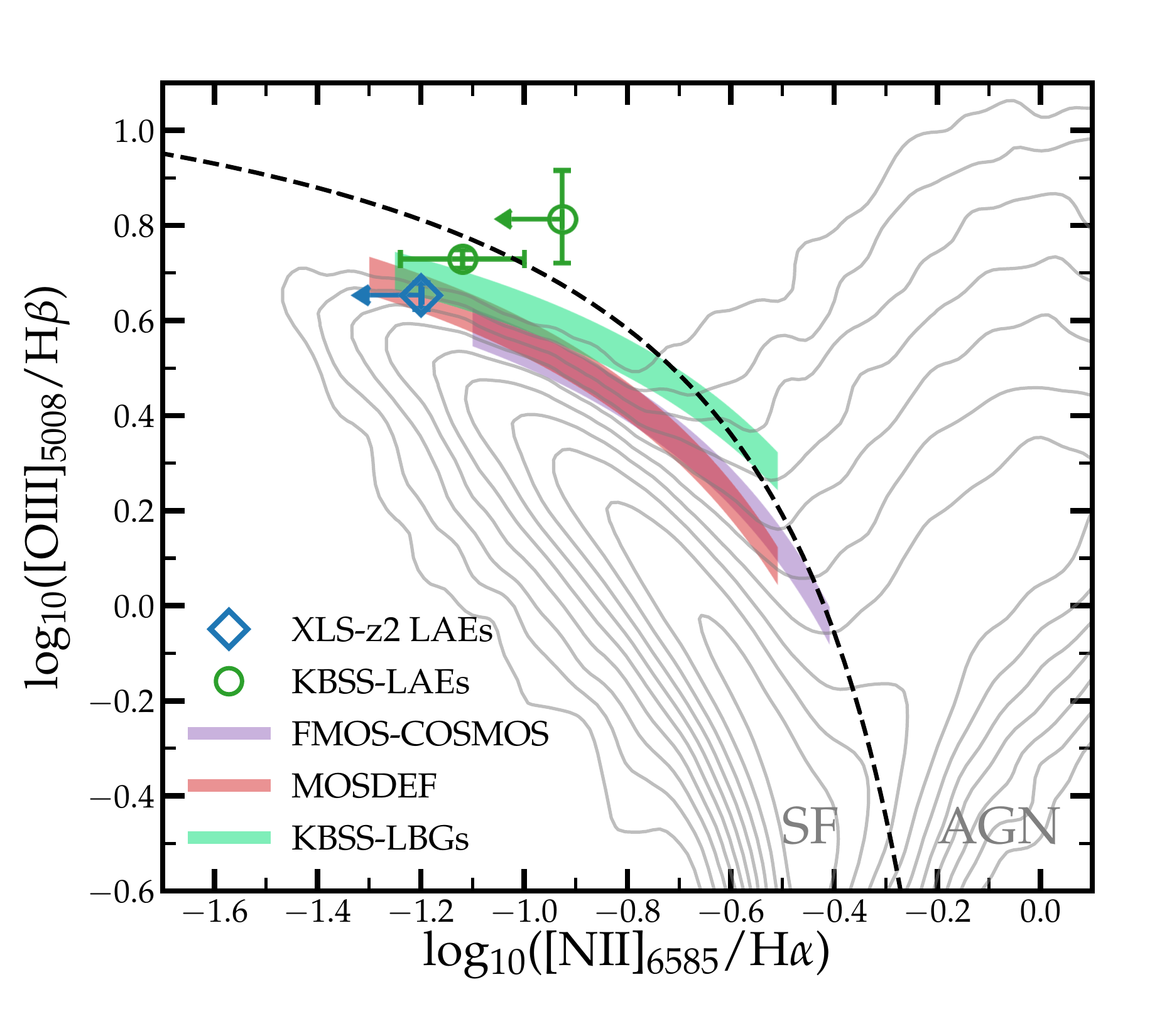}
\caption{Location of the stack of representative LAEs at $z\approx2.2$ on the `BPT' diagram (blue diamond). The grey contours mark the distribution of galaxies at $z=0.03-0.06$ in SDSS (\citealt{Brinchmann2004}). The dashed line is the demarcation line between ionisation due to young stars and AGN (\citealt{Kauffmann2003}). Green symbols show the average line-ratios of LAEs and LBGs with strong Ly$\alpha$ emission at $z\sim2$ measured by \citet{Trainor2016}, where LAEs have the highest O3Hb. These samples are intrinsically a factor two less and more luminous, respectively. The coloured shaded regions show the average locations of galaxies from various large surveys at $z\sim2$. Green shows UV-selected galaxies from KBSS (\citealt{Strom2017}), red shows $H$-band selected galaxies from MOSDEF (\citealt{Shapley2015,Sanders2020}) and purple shows the slightly more massive galaxies from FMOS-COSMOS (\citealt{Kashino2017}). The masses of our LAEs are generally lower than those in such continuum-selected galaxy samples. Upper limits are at the 3$\sigma$ level.}
\label{fig:BPT}
\end{figure}

\subsection{LAEs are powered by young metal poor stars} \label{sec:source}
Several emission line ratios that we observe in the stacked spectrum are sensitive to the source of ionisation. These include the well-known relative strengths of [O{\sc iii}]/H$\beta$ compared to [N{\sc ii}]/H$\alpha$ (i.e. the BPT diagram; \citealt{BPT}), but also the ratio of high-ionisation metal lines (such as C{\sc iv}, O{\sc iii}], C{\sc iii}]) to He{\sc ii} in the rest-frame UV \citep[e.g.][]{Feltre2016,Nakajima2018}.

\subsubsection{BPT diagram}
In Fig. $\ref{fig:BPT}$ we show the location of our stacked spectrum in the BPT diagram and compare this to galaxies at $z=0.03-0.06$ in the SDSS \citep{Brinchmann2004,Alam2015}. Similar to other samples of galaxies at $z\sim2$ \citep[e.g.][]{Shapley2015,Kashino2017,Strom2017}, we find elevated [O{\sc iii}]/H$\beta$ ratios at fixed [N{\sc ii}]/H$\alpha$ which are rare in the local Universe. The line ratios are however still well within the part of the diagram that implies photoionisation by young stars \citep{Kauffmann2003}. The observed ratios indicate a high ionisation state possibly due to a hard ionising spectrum from metal-poor stars \citep{Steidel2016,Topping2020}.

\subsubsection{Rest-frame UV ionisation diagram}
In addition to using the strong rest-frame optical lines, various line-ratios in the rest-frame UV have recently been proposed to be capable of identifying AGN activity \citep[e.g.][]{Feltre2016,Nakajima2018}. 

\begin{figure} 
\includegraphics[width=8.6cm]{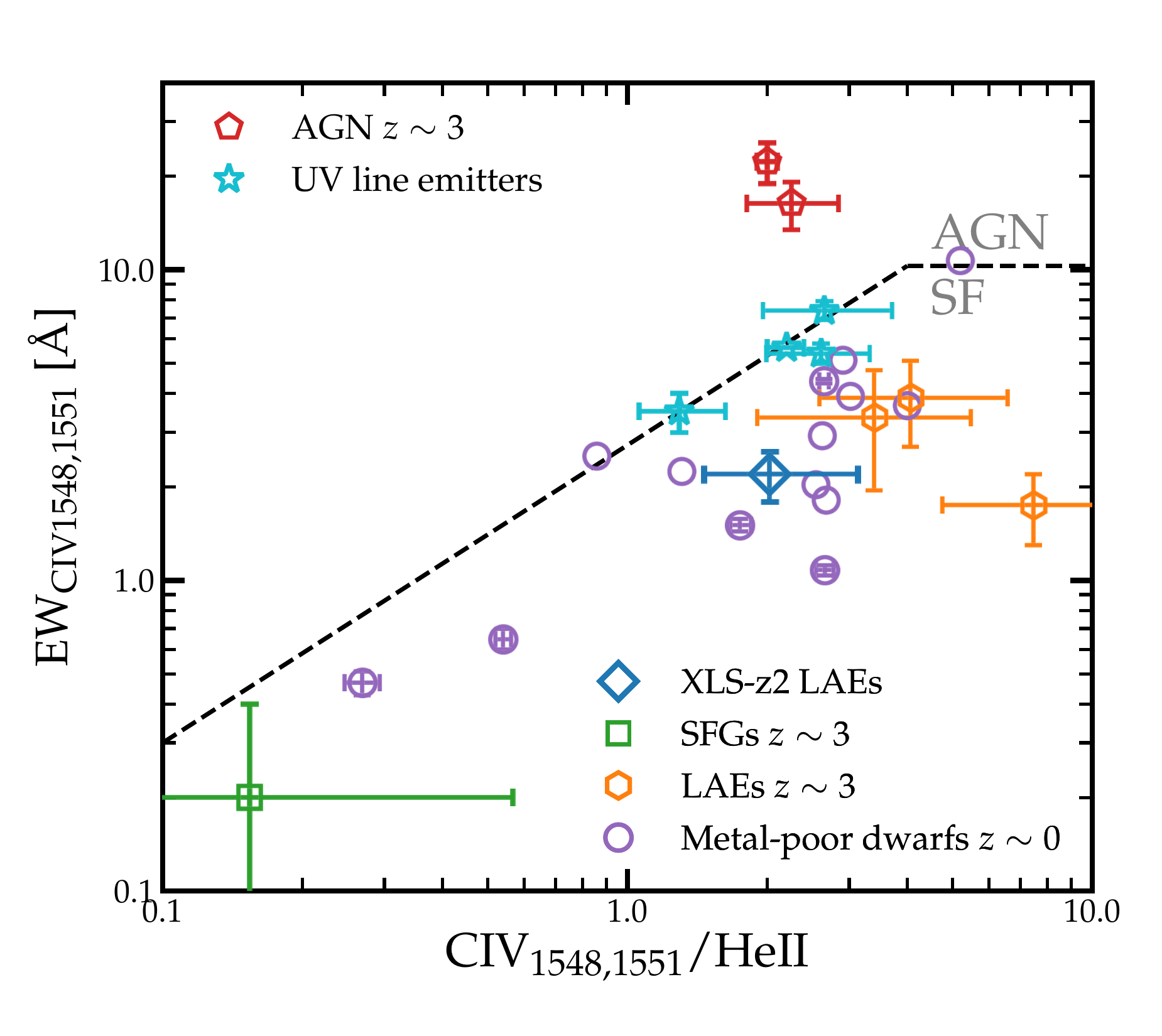} 
\caption{The relation versus the combined C{\sc iv}$_{1548+1551}$ EW and the C{\sc iv}/He{\sc ii} line-ratio. The dashed line shows the demarcator between ionisation by star formation or AGN (\citealt{Nakajima2018}).  The blue diamond shows the measurement of the typical LAEs at $z\approx2$, which is consistent with photoionisation by stellar atmospheres. The orange hexagons show stacks of LAEs at $z\approx3$ (\citealt{Nakajima2018b,Feltre2020}), the green square shows typical star-forming galaxies at $z\sim3$ from the VUDS survey (\citealt{Nakajima2018}), while the purple circles show metal-poor dwarfs at $z\sim0$ (\citealt{Senchyna2017,Senchyna2019,Berg2019b}). The cyan stars are galaxies at $z\sim2$ that have been selected on strong rest-frame UV line emission (\citealt{Amorin2017,Nakajima2018}). With red pentagons we show the average EW and line-ratio measured in stacks of UV-selected (\citealt{Hainline2011}) and Ly$\alpha$-selected AGN (\citealt{Sobral2018}) at $z\sim2-3$.  }
\label{fig:CIVAGN}
\end{figure}

Here, we focus on the C{\sc iv}/He{\sc ii} ratio. While we note that measurements of nebular C{\sc iv} emission may be complicated by stellar and interstellar absorption and emission as described in \S $\ref{sec:measure_UV}$, the attractive feature of C{\sc iv} is that its production by collisional excitation requires an energy of 47.9 eV. The main production mechanism for nebular He{\sc ii} is recombination, which requires an ionisation energy of 54.4 eV. We therefore expect that objects with a harder ionising spectrum have a relatively stronger He{\sc ii} line at fixed carbon abundance. In addition to the line ratio, we also focus on the C{\sc iv} equivalent width. There naturally is a maximum EW associated to a stellar population and given the blackbody-like shape of stellar spectra that exponentially drop around 50 eV for $T\approx50,000$ K, it is very challenging to reach EWs in excess of $\sim10$ {\AA}, unless the stellar atmospheres are extremely hot \citep{Nakajima2018}. Very high C{\sc iv} EWs thus indicate power-law type ionising sources such as AGN.
We show the C{\sc iv} EW and the C{\sc iv}/He{\sc ii} ratio of our stacked sample of LAEs in Fig. $\ref{fig:CIVAGN}$. The observed EW and line-ratio can be explained by photoionisation by stellar atmospheres and does not indicate significant AGN contribution, in agreement with the results from the BPT diagram. The C{\sc iv} EW of typical LAEs is significantly lower than the EWs measured in stacks of Ly$\alpha$ \citep{Sobral2018} and UV-selected AGN \citep{Hainline2011} at these redshifts, validating the use of this diagram to identify AGN activity. 

In addition to C{\sc iv}, we use the O{\sc iii}]/He{\sc ii} line ratio to identify AGN activitity \citep[e.g.][]{Amorin2017}. Similarly, with the O{\sc iii}]$_{1661+1666}$ flux being higher than the He{\sc ii} flux this line-ratio prefers photoionisation by young stars as well \citep{Feltre2016}.

\subsubsection{Young and metal-poor stellar populations}
The equivalent widths of the Balmer lines are sensitive to the age of stellar populations as they trace the amount of ionising photons originating from O stars relative to the stellar continuum from lower mass stars \citep[e.g.][]{Leitherer1999}. Likewise, the EW of the He{\sc ii}$_{1640}$ line is very sensitive to the hardness of the spectrum of a composite stellar population \citep[e.g.][]{Kehrig2015,Berg2018,Gotberg2019,Nanayakkara2019}, which traces the stellar iron abundance, the high-mass end of the IMF, the presence of high-mass binaries and other detailed properties of very massive stars  \citep[e.g.][]{Schaerer2003,GrafenerVink2015,Szecsi2015,Stanway2020}.

We explore the implications of the observed EWs of H$\alpha$ for the age of the stellar populations using the results from \cite{Xiao2018}, who performed CLOUDY modelling on single-burst BPASS models \citep{BPASS2018} with a range of metallicities ($Z$) and ionisation parameters ($U$). As in these models the nebular metallicity is fixed to the stellar metallicity and the relative metal abundances do not vary with metallicity \citep[c.f.][]{Steidel2016}, we do not use them to interpret line ratios involving metal lines. The H$\alpha$ EW of the stack is $531^{+131}_{-84}$ {\AA}. This implies a burst-age of $6^{+4}_{-2}$ Myr, which is only mildly sensitive to $Z$. This is in agreement with results from \cite{Leitherer1999}, where such H$\alpha$ EW corresponds to ages of 5 Myr. For a constant star formation history, we find that the observed H$\alpha$ EW corresponds to an age of $\approx40^{+20}_{-15}$ Myr. Independently of the H$\alpha$ EW, the light-weighted ages inferred from the SED fitting to the broad-band continuum data are on average $250\pm200$ Myr. Indeed, if these typical LAEs experienced a constant star formation rate of $\approx5$ M$_{\odot}$ yr$^{-1}$ they would have formed their total stellar mass in $\approx200$ Myr. We further note that the low C/O abundance suggests that there has not yet been time for carbon enrichment from evolved stars \citep[e.g.][]{Berg2019b}, consistent with young ages.

We note that the He{\sc ii} EW of $1.2\pm0.4$ {\AA} that we observe in typical LAEs at $z\approx2$ suggests very hot stars with a very low stellar metallicity. However, it is challenging to quantify this metallicity as current models are known to miss sources of He{\sc ii}-ionising photons \citep[e.g.][]{Berg2018,Saxena2020,Wofford2020}, such as X-ray binaries \citep[e.g.][]{Schaerer2019} or radiative shocks \citep[e.g.][]{Jaskot2013,Stasinska2015,Plat2019}. 
An indirect estimate of the stellar metallicity of the young stars can be obtained by assuming pure enrichment from core-collapse supernovae and that the gas-phase metallicity is the same as the stellar metallicity. Pure core-collapse enrichment yields O/Fe$\approx5$ O/Fe$_{\odot}$ \citep{Nomoto2006} and this is warranted by observations at $z\approx2$ \citep[e.g.][]{Steidel2016,Topping2020} and hydrodynamical simulations \citep{MattheeSchaye2018}. Under these assumptions the gas-phase metallicity 12+log(O/H)$\approx8.0$ corresponds to a stellar metallicity $Z\approx10^{-3}$, which is in agreement with the metallicity inferred through photoionisation modelling of galaxies at $z\sim2$ with similar locations in the BPT diagram \citep{Topping2020}.

\subsection{LAEs are clumpy and have turbulent and outflowing kinematics} \label{sec:outflows}
\subsubsection{Clumpy structures} \label{sec:clumpy}
As discussed in \S $\ref{sec:morph}$ and listed in Table $\ref{tab:measurements}$, the {\it HST} imaging reveals multiple clumps in 12/31 objects. We further identify seven galaxies with multiple narrow emission-line components that have a velocity shift with respect to each other (\S $\ref{sec:zsys}$). Except for XLS-16, these all appear clumpy in the imaging data as well. The remaining LAEs typically appear very compact, but we also note that we find a bias towards identifying multiple clumps in the more luminous systems. These results strongly suggest that the UV-light and the emission lines in these LAEs are dominated by one or a few massive complexes of star formation \citep[see also][]{Cornachione2018}, possibly associated to ongoing mergers.

\subsubsection{Outflowing ionised gas from stellar feedback}
From the stack, the clearest evidence of outflowing gas is blue-shifted interstellar absorption that we detect in the low-ionisation Si{\sc ii} 1260 {\AA} line (Fig. $\ref{fig:UVabsorption}$). The absorption-weighted velocity is shifted by $-260^{+160}_{-40}$ km s$^{-1}$ with respect to the systemic velocity. This is a slightly higher velocity than measured in continuum-selected galaxies at $z\sim3$ \citep{Shapley2003,Erb2015}. Such outflows could plausibly be driven by the feedback processes associated to the young starbursts in these LAEs \citep{Ma2016}. For the 17 representative LAEs that have rest-frame UV size measurements from {\it HST} data (Table $\ref{tab:flux_measurements}$), we measure a median (mean) $\Sigma_{\rm SFR_{\rm H\alpha}}=1.2 (1.8)$ M$_{\odot}$ yr$^{-1}$ kpc$^{-2}$ assuming $\Sigma_{\rm SFR_{\rm H\alpha}} = {\rm SFR}_{\rm H\alpha}/2 \pi r_{\rm eff, UV}^2$ \citep{Shibuya2019}, which is well above the typical threshold of $\Sigma_{\rm SFR}>0.1$ M$_{\odot}$ yr$^{-1}$ kpc$^{-2}$ for driving large scale galactic outflows \citep[e.g.][]{Heckman2001}.

\subsubsection{Complex [O{\sc iii}] profiles: outflows, turbulence, unresolved structures?} \label{sec:O3discuss}
The [O{\sc iii}] line-profiles in a significant fraction of the LAEs show evidence for a broader (FWHM$\approx280$ km s$^{-1}$) component. Fig. $\ref{fig:O3_SN}$ shows that these broad components can only be identified in the LAEs with higher S/N detections of [O{\sc iii}], and that the broader components are more often seen at $M_{\rm star}\approx3\times10^{9}$ M$_{\odot}$ than at lower masses. A broad [O{\sc iii}] emitting component with FWHM$\approx280$ km s$^{-1}$ is also seen in the stack of representative LAEs, which contains the majority of LAEs for which the individual S/N of [O{\sc iii}] was not sufficient to detect a broad component. Broad emission-lines have also been observed in local analogues of LAEs \citep[e.g.][]{Amorin2012,Bosch2019,Hogarth2020} and they have been proposed to indicate (shock)-ionised outflowing material \cite[e.g.][]{Heckman1990,Veilleux2005,Freeman2019}. However, these lines are typically broader than $>500$ km s$^{-1}$, unlike we observe. Indeed, using stacks of galaxies with M$_{\rm star} \gtrsim10^{10}$ M$_{\odot}$ at $z\sim2$, \cite{Davies2019} find that the width of the broad component increases with the star formation surface density. Following this observed correlation, the $\Sigma_{\rm SFR}$ of typical LAEs would imply a broader FWHM$\approx600$ km s$^{-1}$. As broader lines are more difficult to detect, deeper data are required to resolved this tension.
Alternatively, the observed broad components could be due the unresolved motions of several (potentially somewhat more evolved) HII regions within the galaxies \citep[e.g.][]{Ostlin2015} or line-emission originating from gas in turbulent mixing layers such that the width of the broad component reflects the turbulent velocity \citep[e.g.][]{Westmoquette2007}. In order to differentiate these scenarios we would need to measure line-ratios resolved for the narrow and broad components \citep[e.g.][]{Hogarth2020} which the current S/N does not allow for all lines.

\begin{figure}
\includegraphics[width=8.6cm]{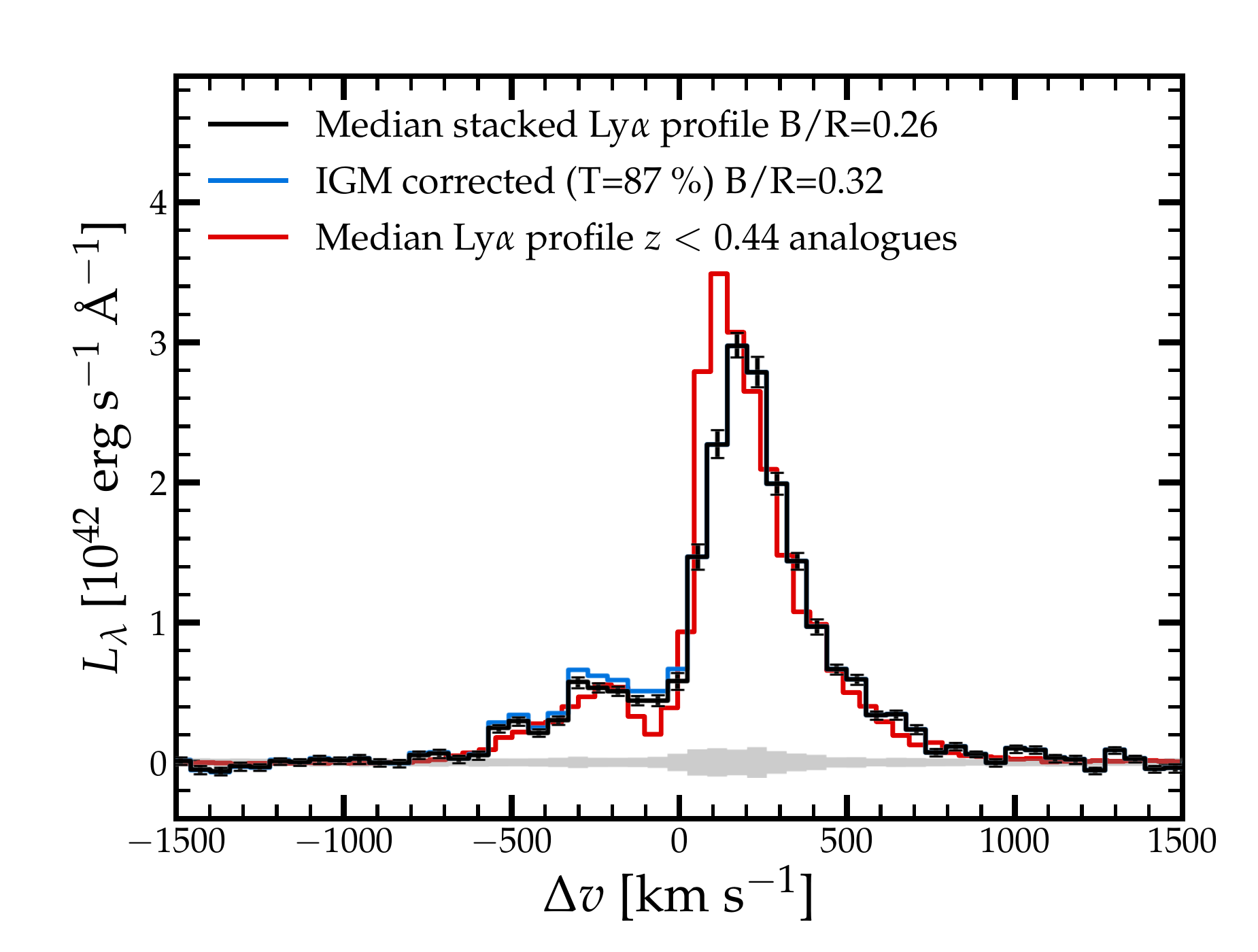}
\caption{Stacked Ly$\alpha$ profile of the representative sample of LAEs at $z\approx2$ (black). 26 \% of the flux emerges blue-wards of the systemic redshift. The blue line shows the Ly$\alpha$ profile with an inverse IGM correction from \citealt{Madau1995}. This slightly increases the flux on the blue-side of the systemic velocity. The red profile shows the median Ly$\alpha$ spectra from {\it HST}/COS observations of in low-redshift analogues of LAEs (\citealt{Hayes2020}). We renormalised this spectrum by a factor 1.5 to have the same integrated Ly$\alpha$ luminosity. The spectra are remarkably similar, particularly as there is significant variation among the spectra of individual objects (Fig. $\ref{fig:Lyaspec}$). The relative strength of the red peaks with respect to the blue peaks indicates that Ly$\alpha$ photons scatter through a mainly outflowing medium.}
\label{fig:Lyaprofile}
\end{figure}

\subsubsection{Outflows indicated from the Ly$\alpha$ profile}
The Ly$\alpha$ profiles of LAEs are another observable that is sensitive to the kinematics of the gas \citep[e.g.][]{Neufeld1990,Verhamme2006,KakiichiGronke2019}. While a detailed study of the Ly$\alpha$ line profiles of individual LAEs will be presented in Matthee et al. (in prep), we focus here on the stacked Ly$\alpha$ profile which is shown in Fig. $\ref{fig:Lyaprofile}$. The velocity axis is centred on the systemic redshift. It is clear that the median Ly$\alpha$ profile shows a prominent skewed red line that peaks at $\Delta v=+205\pm5$ km s$^{-1}$ and a flatter blue peak that peaks at $\Delta v = -294\pm5$ km s$^{-1}$. The observed blue-to-red flux ratio is $0.26\pm0.01$. As our stacked Ly$\alpha$ spectrum averages over many independent sight-lines, we can retrieve the intrinsic flux on the blue side of the line by correcting for the average IGM attenuation at $z=2.2$ following \cite{Madau1995}. This correction is small and yields an intrinsic blue-to-red flux ratio of $0.32\pm0.01$. This ratio is very comparable to the average blue-to-red flux ratio observed by \cite{Trainor2015} in slightly lower resolution spectra of LAEs in quasar fields at $z\sim2$. Fig. $\ref{fig:Lyaprofile}$ also shows the average Ly$\alpha$ profile of low redshift analogues of LAEs at $z<0.44$ observed with {\it HST}/COS \citep{Hayes2020}. It is remarkable that the average Ly$\alpha$ profiles at $z\approx2$ and the analogs are so similar, in particular as the individual Ly$\alpha$ profiles show a lot of scatter on an individual basis \citep[see Fig $\ref{fig:Lyaspec}$ and e.g.][]{Yang2017}. The red peak appears to be less sharp and the valley appears to by slightly shallower at $z\approx2$ compared to the average spectrum in the low redshift analogues. The dominant red peak is clear evidence that Ly$\alpha$ photons scatter through an outflowing medium \citep[e.g.][]{Barnes2011,GronkeDijkstra2016,Gurung2019}.

\begin{figure*}
\begin{tabular}{cc}
\includegraphics[width=8.7cm]{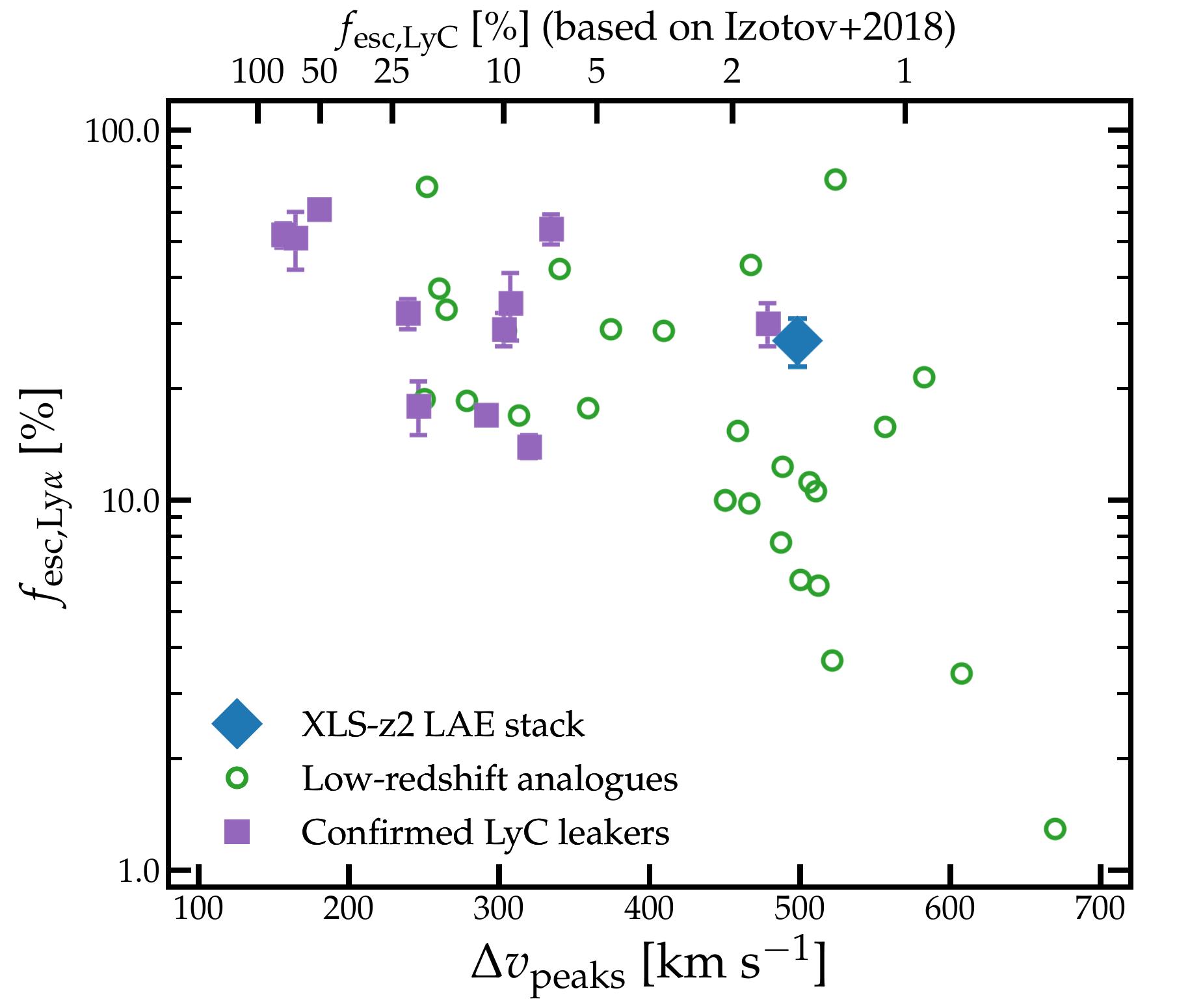} &\includegraphics[width=8.7cm]{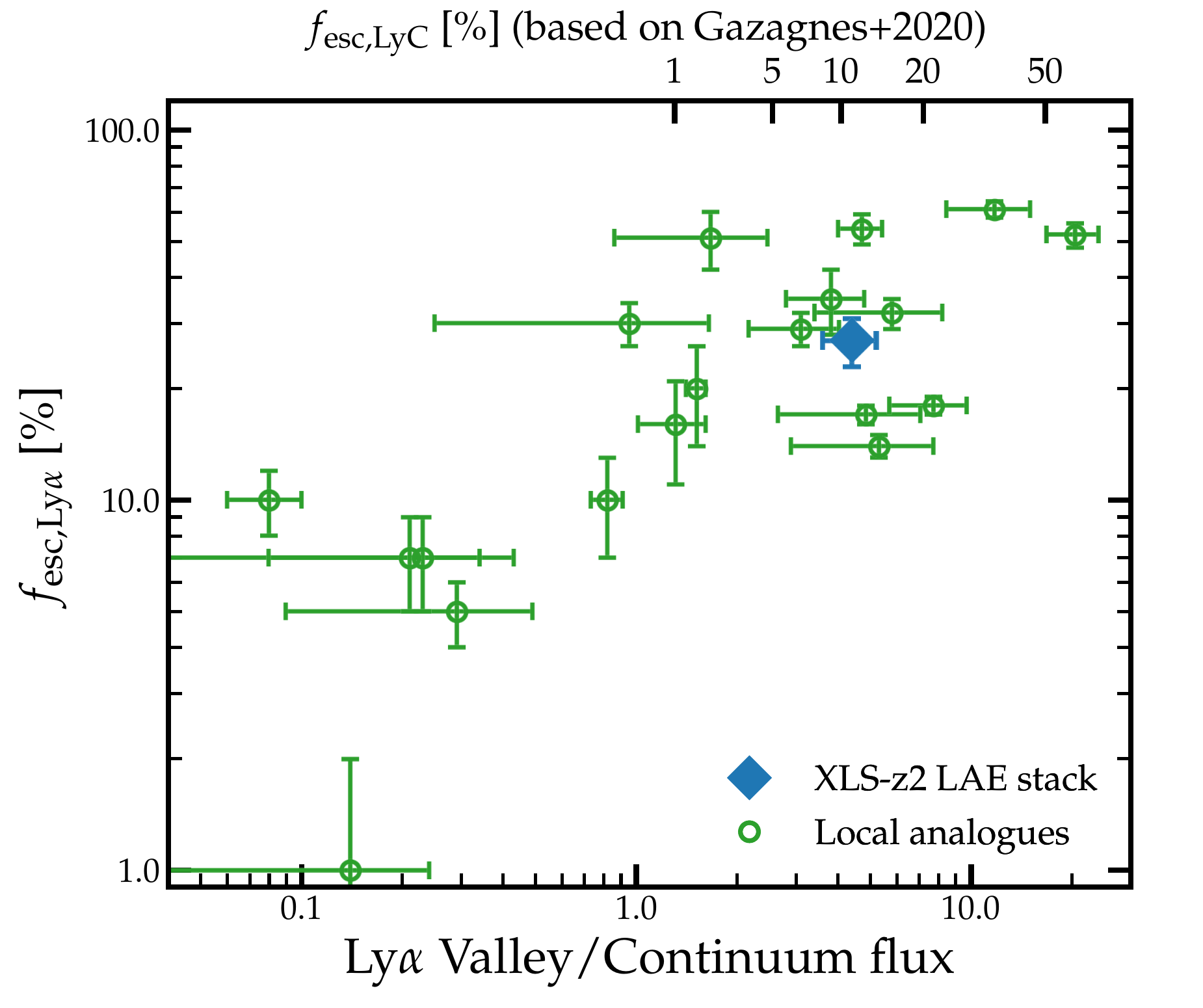} 
\end{tabular}
\caption{The dependence of the Ly$\alpha$ escape fraction on the line profile. In the left panel we show the separation of the red and blue peaks. For comparison we add data-points from low-redshift analogues (\citealt{Yang2017}) and a sub-set of those that have been confirmed as LyC leakers (\citealt{Izotov2018}). The Ly$\alpha$ escape fraction is anti-correlated with the peak separation, although there is significant scatter. Following the empirical relation between the peak separation and the LyC escape fraction we show the expected escape fraction on the top x-axis. The XLS-$z2$ stack of LAEs has a relatively high peak separation given the Ly$\alpha$ escape fraction, which is a stacking effect. In the right panel we show the flux that escapes at the valley (i.e. the local minimum between the two Ly$\alpha$ peaks close to the systemic redshift) relative to the continuum. Here we compare to low-redshift analogues as compiled by (\citealt{Gazagnes2020}). We also show the corresponding LyC escape fraction following the empirical relation from \citet{Gazagnes2020}. While the peak separation implies a fairly low LyC escape fraction, the relative valley-to-continuum flux implies $f_{\rm esc,LyC}\approx10$ \% in typical LAEs at $z\approx2$.}
\label{fig:lyaprofile}
\end{figure*}

\subsection{Typical LAEs are probably moderate LyC leakers}
Due to the resonant nature of the Ly$\alpha$ transition, a connection between the escape of LyC and Ly$\alpha$ photons is theoretically expected. These escape fractions may correlate with each other, as Ly$\alpha$ photons scatter through the low column density channel paths of least resistance \citep{Dijkstra2016} that allow for LyC escape. The lowest column density channels leave an imprint on the Ly$\alpha$ line profile \citep[e.g.][]{Verhamme2015,KakiichiGronke2019}. As LAEs are the class of galaxies with the highest Ly$\alpha$ escape fractions, it is plausible that they are also the types of galaxies with the highest LyC escape fraction \citep[indeed, observations of UV-selected galaxies indicate that galaxies with higher Ly$\alpha$ EW have a higher escape fraction][]{Marchi2018,Steidel2018}.\footnote{It is possible that the LyC escape fraction is so high that the production of Ly$\alpha$ photons is reduced. In {\it practice} however the Ly$\alpha$ luminosity of such systems may still be high as the mechanisms that will lead to LyC escape may also allow Ly$\alpha$ escape. Only extreme LyC escape fractions therefore realistically will lead to decreased Ly$\alpha$ output.} 

As discussed in \S $\ref{sec:ISM}$ the high value of O32 in the stack of typical LAEs indicates an ionisation parameter log$_{10}(U)=-2.4$. Combined with the gas-phase metallicity, this value indicates that the ISM, on average, is still mostly ionisation bounded \citep{NakajimaOuchi2014}, and it is not as high as the most extreme LyC leakers which have O32$>10$ \citep{Izotov2018}. This suggests that typical LAEs are not extreme LyC leakers. 

The Ly$\alpha$ profile is a direct tracer of the H{\sc i} column density. It is however tricky to interpret the stacked Ly$\alpha$ line profile shown in Fig. $\ref{fig:Lyaprofile}$. The peak separation of the Ly$\alpha$ lines is quite large with $498\pm5$ km s$^{-1}$, which suggests an escape fraction of $f_{\rm esc, LyC}\approx2$ \% based on comparison to the empirical trend identified in \cite{Izotov2018}, see the left panel of Fig. $\ref{fig:lyaprofile}$. However, the broadness of the blue part of the line compared to the broadness on the red part of the line may indicate that there is more variation in the velocity at which blue Ly$\alpha$ photons escape compared to red Ly$\alpha$ photons (see also \citealt{Henry2015}). Due to stacking, the observed `peak' of the blue line is thus the result of a complex interplay of various blue peak positions and strength, and therefore less directly connected to the effective H{\sc i} column density. Indeed, the left panel of Fig. $\ref{fig:lyaprofile}$ suggests that the observed peak separation is relatively high given the $f_{\rm esc, Ly\alpha}$.

On the other hand, the stacked Ly$\alpha$ spectrum also shows clear non-zero flux at the `valley' near the systemic velocity. The level of the flux in the valley traces the number of Ly$\alpha$ photons that have not experienced significant frequency redistribution  and is therefore also correlated strongly with $f_{\rm esc, LyC}$ \citep{RiveraThorsen2017,Vanzella2016,Gazagnes2020}. With a valley-flux that is $4.4\pm0.8$ times higher than the continuum we infer $f_{\rm esc, LyC}\approx11\pm6$ \% \citep[based on the empirical correlation in][]{Gazagnes2020}, see Fig. $\ref{fig:lyaprofile}$. This measurement may also be influenced by the stacking and the spectral resolution. As the majority of the flux on the red and blue parts of the line is at velocities that are $>200$ km s$^{-1}$ from the valley ($>2.5\times$ the resolution FWHM) the resolution effect is likely small. Therefore, while challenging to interpret, the stacked Ly$\alpha$ profile suggests a moderate average LyC escape fraction of $\approx10$ \%. This average may be driven by a few galaxies with very low H{\sc i} column density ($N_{\rm HI}<10^{17}$ cm$^{-2}$) channels through which a fraction of the LyC photons escape and Ly$\alpha$ photons experience negligible velocity shift\footnote{See for example XLS-2, XLS-20 and XLS-21 in Fig $\ref{fig:Lyaspec}$. We will investigate these Ly$\alpha$ spectra in detail in an upcoming paper.}. This average escape fraction is consistent with direct constraints from stacking LyC data of LAEs at $z\approx3$ \citep[e.g.][]{Micheva2017,Iwata2019}.

\bigskip

The line-ratio of the resonant Mg{\sc ii}$_{2796, 2803}$ doublet has been proposed as tracer of LyC escape \citep[e.g.][]{Chisholm2020}, which is particularly relevant for the epoch of reionisation where Ly$\alpha$ suffers from IGM attenuation. The line-ratio of the Mg{\sc ii} lines that we measure in the stack is $1.95^{+1.82}_{-0.89}$. While this ratio is uncertain due to the low S/N of the MgII lines, it can still constrain the escape fraction if we cap this ratio to the intrinsic ratio of 2. Following the methodology from \cite{Chisholm2020}, we infer a LyC escape fraction of $f_{\rm esc, LyC} = 6^{+7}_{-3}$ \% for our stack of representative LAEs at $z\approx2$ consistent with our estimate based on Ly$\alpha$.

\section{Discussion} \label{sec:discussion}
We discuss how LAEs compare to the general galaxy population at $z\approx2$ and what this tells us about similar galaxies in the epoch of reionisation.

\subsection{The locations of LAEs on scaling relations}
\subsubsection{The SFR, Mass and gas-phase metallicity of LAEs} \label{sec:MS}
\begin{figure}
\includegraphics[width=8.6cm]{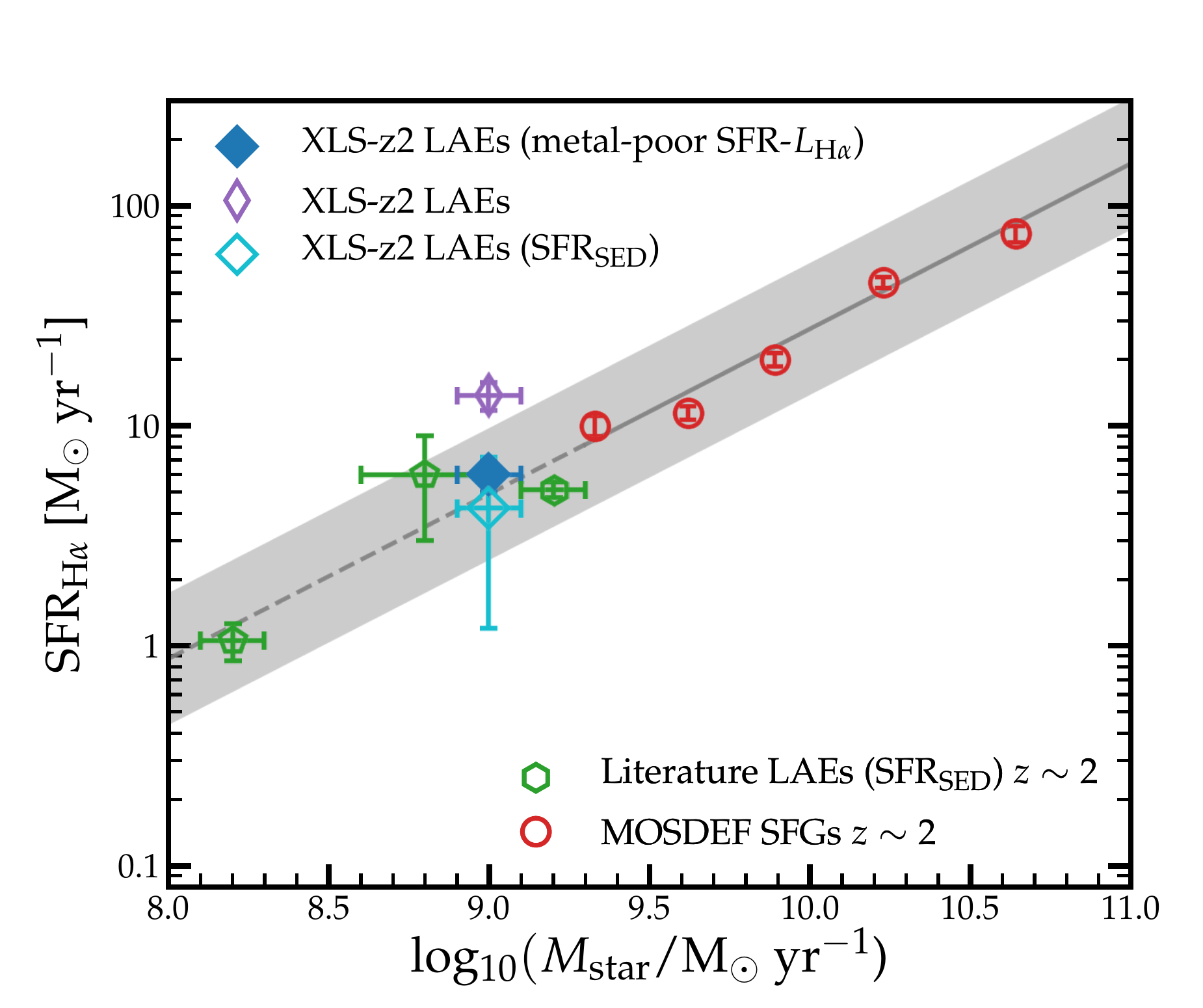}
\caption{The relation between the SFR and stellar mass at $z\approx2.3$. We show the relation (grey line) for general samples of star-forming galaxies measured from the H$\alpha$-based red datapoints from \citet{Sanders2020b}. We assume the typical 0.3 dex scatter on this relation (\citealt{MattheeSchaye2019}) shown as a grey band. The blue diamond shows the XLS stack assuming the metal-poor conversion factor between SFR and H$\alpha$ luminosity, while the open purple diamond shows the standard conversion (that is also used by \citealt{Sanders2020b}) and the open cyan diamond shows the SED-based SFR. Green pentagons show literature measurements of SFRs of stacks of LAEs using SED fitting (\citealt{Guaita2011,Hao2018,Kusakabe2018}).}
\label{fig:sfrmstar}
\end{figure}

Here we compare the location of typical LAEs to other galaxies through well-known scaling relations between SFR, mass and metallicity. In Fig. $\ref{fig:sfrmstar}$, we show the location of the XLS stack on the SFR-M$_{\rm star}$ relation. We also show the general relation found at $z\approx2.3$ for somewhat more massive systems \citep{Sanders2020b} and other stacks of LAEs at $z\sim2$ \citep{Guaita2011,Hao2018,Kusakabe2018}. The relation from \cite{Sanders2020b} uses H$\alpha$-based SFRs, while the literature values on stacks of LAEs are based on SED fitting. We show the location of the XLS stack based on the SED fitting, the standard conversion between SFR and H$\alpha$ luminosity (as also used by \citealt{Sanders2020b}) and the conversion assuming a lower stellar metallicity (see \S $\ref{sec:sfr}$). The H$\alpha$-based SFR from the XLS stack is above the extrapolated relation between SFR and M$_{\rm star}$ using the standard conversion, but would be perfectly on the extrapolated relation using the metal-poor conversion that may be more realistic. The SED based SFR places the XLS stack on a similar location as literature studies of LAEs. 

A more empirical comparison of the specific SFR of LAEs with respect to the general galaxy population can be made through the rest-frame optical emission line EWs, which scale with sSFR (in particular the Balmer lines). \cite{Reddy2018b} provide scaling relations between these EWs and properties such as stellar mass at $z\sim2$. For typical galaxies with masses of $10^9$ M$_{\odot}$, i.e. similar to our stack of LAEs, they find H$\alpha$ and H$\beta$ EWs of 270 and 40 {\AA}, respectively, with a scatter of 0.2 dex. The typical [O{\sc iii}] EW for this mass is 340 {\AA}. These values are a factor $\approx1.5-2$ lower than the measured EWs in the XLS stack (Table $\ref{tab:measurements_stack}$), which suggests that LAEs have a slightly enhanced strength of emission lines compared to the continuum compared to galaxies with similar stellar mass, possibly due to an ongoing starburst.

Galaxies at $z\approx0-3$ are observed to follow a tight relation between mass, SFR and gas-phase metallicity \citep[e.g.][]{Sanders2018,MaiolinoMannucci2019}, such that deviations from the relation between mass and SFR correlate with deviations in the mass-metallicity relation. Extrapolating the mass-metallicity relation at $z\approx2.3$ from \cite{Sanders2020b} we find a gas-phase metallicity that is 0.4 dex higher than the one measured for the XLS stack (using the $T_e$ method). This implies that LAEs have relatively low gas-phase metallicities given their stellar mass. Therefore, in case LAEs follow the fundamental metallicity relation, LAEs should have elevated SFRs at fixed mass.

\subsubsection{Ionisation conditions}
As illustrated in Fig. $\ref{fig:BPT}$, LAEs are among the galaxies with the highest ionisation state known at $z\sim2$. This result is in agreement with earlier results \citep[e.g.][]{Finkelstein2011,Nakajima2013,Song2014,Trainor2016} and also along the lines of those from \cite{Erb2016}, who found that the fraction of galaxies with strong Ly$\alpha$ emission is very high among UV-selected galaxies at $z\approx2$ with extreme [O{\sc iii}]/H$\beta$ ratios, compared to galaxies in their sample with lower values of [O{\sc iii}]/H$\beta$. It is interesting that \cite{Sanders2020b} show that similar [O{\sc iii}]/H$\beta$ ratios as those observed in LAEs are typical for low-mass galaxies at this redshift, while our measured [Ne{\sc iii}]/[O{\sc ii}] is a factor two higher.

Such a high ionisation state can plausibly be driven by a harder ionising spectrum. \cite{Topping2020} show that galaxies in the high [O{\sc iii}]/H$\beta$ and low [N{\sc ii}]/H$\alpha$ part of the BPT diagram have young and more metal poor stellar populations compared to galaxies that share the locus with the SDSS sample. A low stellar metallicity ($Z_{\star} \approx 10^{-3}$) is indeed not unexpected for LAEs. For example, \cite{Cullen2020} report an anti-correlation between Ly$\alpha$ EW and stellar metallicity. The strength of high ionisation UV emission lines also indicates a relatively hard ionising spectrum in LAEs. Fig. $\ref{fig:CIVAGN}$ shows that the C{\sc iv} EW of typical LAEs at $z\sim2$ is slightly lower than those seen in the most extreme systems \citep[e.g.][]{Amorin2017,Nakajima2018} and fainter LAEs at $z\sim3$ \citep{Nakajima2018b,Feltre2020}, and comparable with local dwarf galaxies with low metallicities (12+log(O/H)$\approx7.5-8.0$;  \citealt{Senchyna2017,Senchyna2019,Berg2019}). The measured C{\sc iv} EW from typical galaxies in the VUDS survey at $z\sim3$ \citep{Nakajima2018} is much lower than for typical LAEs. Stacks of brighter continuum-selected galaxies at $z\sim2-3$ \citep{Shapley2003,Rigby2018} report no C{\sc iv} in emission. The C{\sc iv} EW of typical LAEs are much lower than the EWs of $\approx20-40$ {\AA} reported in extreme LAEs at $z\sim7$ \citep{Stark2015_CIV,Mainali2017}.

\subsection{A minority of low-mass galaxies at $z\sim2$ are LAEs} 
We have seen that LAEs at $z\approx2$ have relatively low metallicities, but typical sSFRs and rest-frame optical emission-line EWs given their stellar mass. They show particularly high [O{\sc iii}]/H$\beta$ values, but this result is an indirect consequence of the relatively low mass of LAEs as \cite{Sanders2020} show that [O{\sc iii}]/H$\beta$ increases monotonically with deceasing stellar mass. Ly$\alpha$ surveys typically pick up galaxies with a low mass (Fig. $\ref{fig:sfrmstar}$). This partly reflects that the Ly$\alpha$ escape fraction and typical Ly$\alpha$ EW anti-correlate with mass (see also \citealt{Oyarzun2017,Marchi2019,Santos2020}), as more massive galaxies with a higher SFR (and thus higher intrinsic Ly$\alpha$ luminosity) would otherwise be picked up by Ly$\alpha$ surveys as well \citep{Matthee2016}. In order to understand the relation between LAEs and the galaxy population, the question is thus whether all star-forming galaxies at $z\approx2$ with masses around $10^9$ M$_{\odot}$ are LAEs.

As we showed in \S $\ref{sec:rep}$, the number density of our sample of LAEs is about 15 times lower than general UV-selected systems with similar UV luminosities. In addition, we find that the number densities of H$\beta$+[O{\sc iii}] line-emitters with similar line luminosities as those measured in the XLS stack is $\approx10^{-3}$ cMpc$^{-3}$ \citep{Khostovan2015,Matthee2017Bootes}. This is about three times higher than the number density corresponding to the Ly$\alpha$ luminosity of this stack of $\approx3\times10^{-4}$ cMpc$^{-3}$ \citep{Konno2016,Sobral2016}. This suggests only about one in every 3 to 10 star-forming galaxies selected either through [O{\sc iii}] line-emission or UV emission with comparable luminosity at $z\approx2$ is a LAE with EW of $>25$ {\AA} (in agreement with spectroscopic follow-up of UV selected galaxies at $z\approx2-3$; \citealt{Cassata2015}). This result is also in agreement with the estimate of \cite{Kusakabe2018}, who find that 10 \% of galaxies with mass $\sim10^9$ M$_{\odot}$ are LAEs based on a clustering analysis at $z=2.2$ (see also \citealt{Ouchi2010}).

\begin{figure*}
\includegraphics[width=16.3cm]{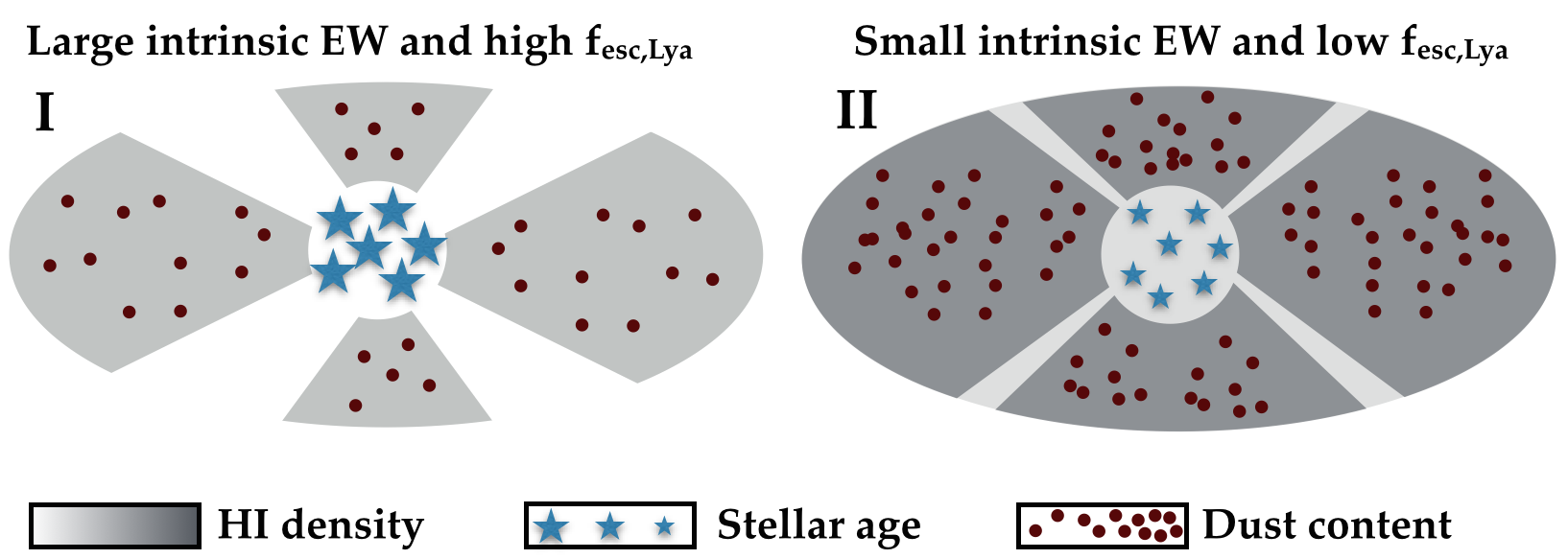}
\caption{Sketches of two extreme star-forming galaxies with high (I) and low (II) {\it likelihood} of being observed as a LAE. The orientations are random with respect to an observer in order to illustrate the relative importances of stochasticity of the viewing angle. Various physical processes that influence the observed Ly$\alpha$ luminosity are illustrated and we stress that the relative importance of these processes remains to be fully understood and that reality is likely between these two examples. Illustrated processes are: 1) the amount of very hot stars (i.e. mostly the age and metallicity of stellar populations) indicated by the size of the blue stars that influence the intrinsic Ly$\alpha$ EW and luminosity, 2) the amount of dust in the ISM indicated by brown dots that influences the amount of Ly$\alpha$ photons that are destroyed, 3) the average H{\sc i} column density and the effective opening angle of low column density channels that determine the optical depth to scattering and hence impact the likelihood of dust destruction and 4) depending on the viewing angle the column density and dust content can vary depending on the specific sightline. Not shown in these sketches are velocity fields of the H{\sc i} that also impact the Ly$\alpha$ line profile or the cause of different column density distributions, such as possible turbulence or outflows. Note that (relatively) low H{\sc i} column density channels should not necessarily have less dust, but we do not show this for illustrative purposes.} 
\label{fig:sketch}
\end{figure*}

\subsection{What causes the differences in observed Ly$\alpha$ output among samples?} \label{sec:discuss}
As discussed in the previous section, only about one in every 3 to 10 low mass galaxies at $z\approx2$ is observed as LAE. In this section we will discuss three possibly complementary hypotheses to explain this \citep[note that these discussions go as far back to at least][]{CharlotFall1993}. Fig. $\ref{fig:sketch}$ sketches the various processes that are invoked in these hypotheses. Compared to other low-mass star-forming galaxies:
\begin{enumerate}
\item[A)] do LAEs have a higher intrinsic EW due to a young and/or low metallicity starburst?
\item[B)] do LAEs have a higher escape fraction due to {\it systematic} variations (for example a lower dust content or the presence of strong outflows that increase the {\it angle-averaged} escape fraction)?
\item[C)] do LAEs have a higher escape fraction due to a fortunate viewing angle?
\end{enumerate}
We focus specifically on low-mass galaxies with M$_{\rm star}\sim10^9$ M$_{\odot}$, as it is very plausible that the low escape fraction of Ly$\alpha$ photons in higher mass galaxies is driving their low Ly$\alpha$ EWs and explains why massive galaxies are typically not present among samples of LAEs. Systematic variations in their dust content may partly drive this trend \citep{Atek2008,GarnBest2010}, but other effects such as a higher H{\sc i} column density (visible through higher Ly$\alpha$ velocity offsets in more massive systems; e.g. \citealt{Steidel2010}), amplifying the effect of variations in the dust content, may also be important.

\subsubsection{Intrinsic EW?}
Compared to other galaxy samples at fixed mass, one possible scenario is that LAEs have a higher intrinsic Ly$\alpha$ luminosity compared to the UV continuum \citep[i.e. a higher intrinsic EW, see e.g.][]{Trainor2019}. In Fig. $\ref{fig:sketch}$ this refers to a higher density of very hot stars (i.e. a younger or more metal-poor stellar population). The intrinsic Ly$\alpha$ EW is related to $\xi_{\rm ion}$ \citep[e.g.][]{SM2019}, but also to the EW of rest-frame optical lines. Indeed, in \S $\ref{sec:MS}$ we found that the rest-frame optical EWs in the XLS stack are a factor $\approx1.5-2$ higher than typical for star-forming galaxies with this mass.

Of particular use is however the comparison to the recent studies by \cite{Du2020} and \cite{Tang2021}. These studies selected reionisation-era analogues at $z\approx2$ based on rest-frame optical emission line properties and measured their Ly$\alpha$ EWs. \cite{Du2020} selected galaxies with  H$\alpha$ EW $>300$ {\AA} and [O{\sc iii}] EW $>300$ {\AA}, completed by a sample of blue, UV-selected systems. \cite{Tang2021} selected extreme [O{\sc iii}] emitters with EW $>300$ {\AA}. The samples in these studies have a similar median stellar mass as our stack of LAEs. For comparable H$\beta$+[O{\sc iii}] EWs as our stack of representative LAEs, these studies report typical Ly$\alpha$ EWs of $\approx10$ {\AA}. These studies also find galaxies with Ly$\alpha$ EWs as high as 70 {\AA} (i.e. comparable to our stack). However, these Ly$\alpha$ EWs are mostly seen for galaxies with much stronger rest-frame optical EWs ($\gtrsim1000$ {\AA}). This suggests that there is significant scatter on the relation between observed Ly$\alpha$ and H$\beta$+[O{\sc iii}] EW and that LAEs are the relatively rare outliers with high Ly$\alpha$ flux given an H$\beta$+[O{\sc iii}] flux. Moreover, UV-selected galaxies with similar luminosity at $z\sim2$ have very comparable $\xi_{\rm ion}$ as the XLS stack \citep{Emami2020}. These comparisons suggest that the variations in observed Ly$\alpha$ EWs among LAEs and other low-mass galaxy samples are not primarily due to significant variations in intrinsic Ly$\alpha$ EWs and thus related to differences in the Ly$\alpha$ escape fraction \citep[e.g.][]{Trainor2019}.

\subsubsection{Systematic variations in $f_{\rm esc, Ly\alpha}$}
The dust attenuation is known to be correlated to the Ly$\alpha$ escape fraction \citep[e.g.][]{Atek2008,Blanc2011,Matthee2016,Runnholm2020} and variations in dust attenuation could therefore potentially explain why only a fraction of low-mass star-forming galaxies is observed as a LAE. Variations in the dust content do not appear to dominate variations in $f_{\rm esc, Ly\alpha}$ at a fixed mass. \cite{Du2020,Tang2021} estimate $E(B-V)\approx0.1$ based on SED fitting, while we measure $E(B-V)=0.2$ for LAEs, based on the Balmer decrement. This implies that dust attenuation is not the only physical driver of $f_{\rm esc, Ly\alpha}$. It is possible that the (stellar) attenuation estimated from their SED fitting under-estimates the nebular attenuation, but it takes a significant underestimation in order to impact this result.

Additionally, studies of low-redshift analogues show that while $f_{\rm esc, Ly\alpha}$ is related to the dust attenuation, the Ly$\alpha$ peak separation is (anti-)correlating equally strongly with the escape fraction  \citep{Henry2015,Yang2017}. The attenuation and the peak separation are not correlated themselves in these systems. The peak separation traces the effective H{\sc i} column density of the path that the Ly$\alpha$ photons encountered while escaping the galaxy \citep[e.g.][]{Neufeld1990,Hashimoto2015}. The amount of low column density channels (and their specific column density) is therefore an additional important parameter in controlling $f_{\rm esc, Ly\alpha}$. This is illustrated as the variations in the greyscale in Fig. $\ref{fig:sketch}$. 

We speculate that the presence of low column density channels could be related to the presence of turbulent kinematics or outflows \citep[e.g.][]{JaskotOey2014,Herenz2016}. Several simulations show that galaxies can be observed as LAE preferentially slightly after star-burst events, such that the stellar birth clouds could have cleared by galactic winds \citep[e.g.][]{Kimm2019,Smith2019}. Even in the very dusty environments of extreme starbursts in the local Universe, strong galactic winds facilitate the escape of Ly$\alpha$ photons \citep{Martin2015}. Moreover, as discussed in \S $\ref{sec:outflows}$, several observations of LAEs show that turbulent gas and outflows are present in these LAEs. As shown in the simulations by \cite{Kimm2019} the blue-to-red Ly$\alpha$ flux ratio of typical LAEs is indeed in agreement with outflowing motions around young stars that allow such low column density pathways to be cleared. Future observations of the presence of such turbulent motions \citep[e.g.][]{Herenz2016,Puschnig2020} and outflows in low-mass galaxies that are not strong LAEs could test this scenario.

\subsubsection{Stochastic variations in $f_{\rm esc, Ly\alpha}$}
Besides systematic variations in the dust attenuation or a physical driver of the presence of low column density channels, the fact that we observe a galaxy as a LAE may also have a stochastic component. This could be the case when the escape fraction depends partly on the viewing angle as found in several simulations \citep[e.g.][]{BehrensBraun2014,ZhengWallace2014,Smith2019}. Galaxies may appear as LAEs when we are observing them along a fortunate sight-line. Due to resonant scattering, the effective low column density channels that Ly$\alpha$ photons traversed are not necessarily single direction-`chimneys' along the line of sight \citep[e.g.][]{Gronke2014,Eide2018,Kimm2019}. Furthermore, the presence of such chimneys can boost the escape fraction significantly more than their fractional opening angle since Ly$\alpha$ photons preferentially escape through them. Regardless, it could be that the {\it effective} angular covering factor of such low column density channels introduces stochastic scatter in the observed values of $f_{\rm esc, Ly\alpha}$ and significantly contributes to the observed Ly$\alpha$ luminosity of a galaxy. This idea is supported somewhat by observations from \cite{Shibuya2014}, \cite{PaulinoAfonso2018} and \cite{Tang2021} who all report higher Ly$\alpha$ EWs in galaxies at $z\sim2$ with smaller ellipticity (i.e. face-on systems). The fraction of 10-30 \% of low-mass star-forming galaxies that is observed as LAE (see \S $\ref{sec:rep}$) in this picture is effectively an upper limit of the typical angular covering factor of low column density channels.

Summarising these various hypotheses with the sketches in Fig. $\ref{fig:sketch}$, we find that it is plausible that several complementary processes determine why only a fraction of low-mass galaxies is observed as LAE at $z\approx2$. While the dust attenuation needs to be low and there need to be sufficient amounts of young stars to yield a high intrinsic EW, these requirements are likely present in the majority of low mass galaxies. Additionally, pathways of low column density through which Ly$\alpha$ photons can escape need to be present. Their presence and properties may be correlated with turbulent motions and outflows powered by feedback, but as such channels do not span the full solid angle, stochasticity of our viewing angle may be induced \citep[e.g.][]{Hagen2016}. Further studies are required to determine the relative importance of each of these effects \citep[e.g.][]{Nakajima2012,Matthee2016}.

\subsection{LAEs are rare galaxies at $z\approx2$, but representative galaxies at $z\approx6$}
Finally, we discuss how our results are relevant for galaxies in the very early Universe, in particular the epoch of reionisation. 
Similar to the analysis in \S $\ref{sec:rep}$ we explore which relative Ly$\alpha$ and UV luminosities are required to match the abundances on the respective luminosity functions. We use UV LFs measured by \cite{Bouwens2015} at $z\approx3, 4, 5, 6, 7$ and the global Ly$\alpha$ LF at $z=3-6$ from \cite{Sobral2018}. The Ly$\alpha$ LF is relatively constant over this redshift \citep[e.g.][]{Ouchi2008,Hayes2011,Sobral2018,Herenz2019}, while the normalisation of the UV luminosity function decreases and the faint-end slope increases at higher lookback times. As a consequence, the population-averaged Ly$\alpha$ output and the fraction of strong Ly$\alpha$ lines among UV-selected galaxies \citep[e.g.][]{Stark2011,Cassata2015} increase. This may be a result of a higher escape fraction, for example due to a lower dust content \citep[e.g.][]{Hayes2011,Konno2016} and/or due to a higher intrinsic Ly$\alpha$ EW \citep{Sobral2018} and/or a higher covering factor of low column density channels.  

\begin{figure}
\includegraphics[width=8.6cm]{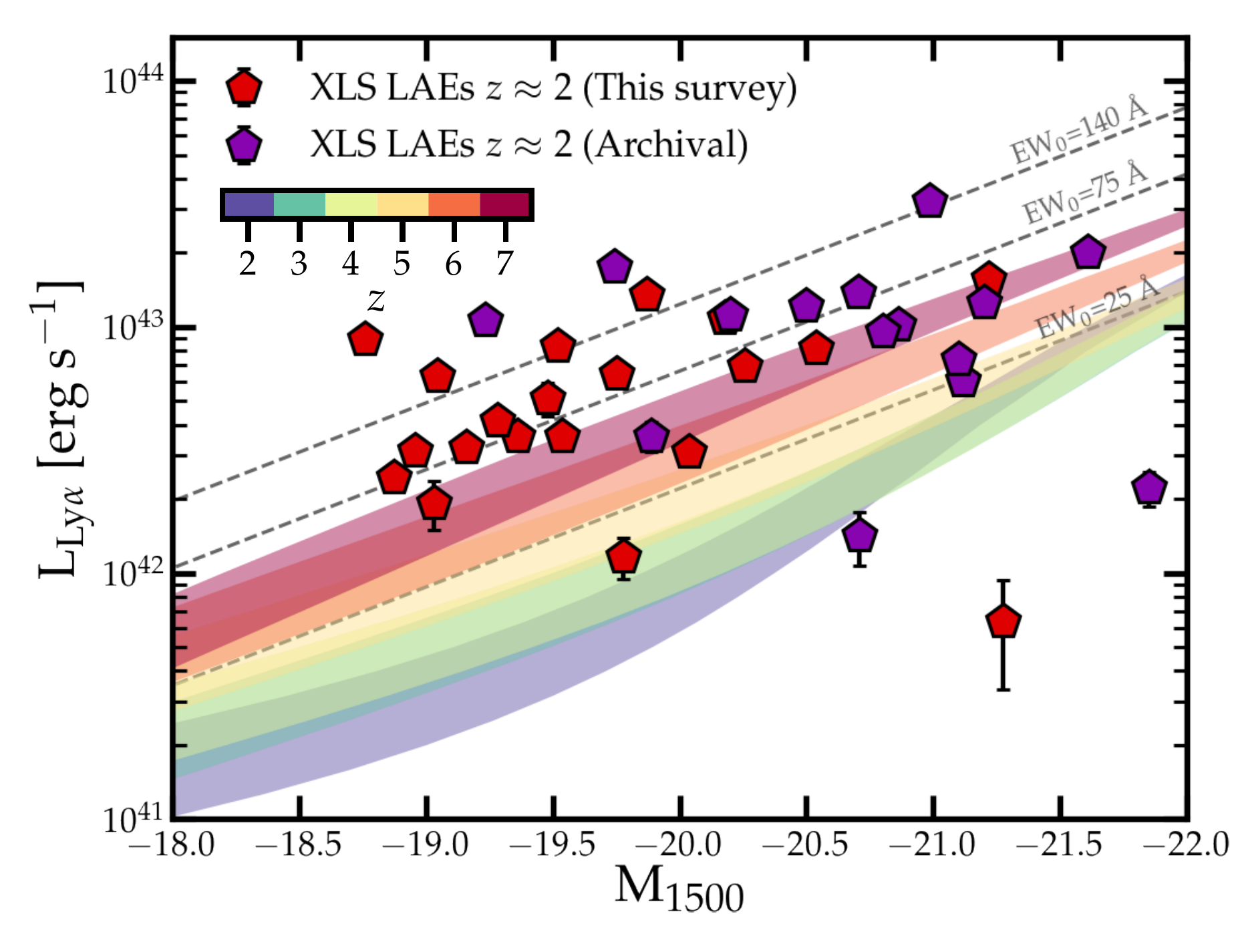}
\caption{The relation between UV and Ly$\alpha$ luminosity for the XLS sample of LAEs at $z\approx2$. The coloured shaded regions are derived as in Fig. $\ref{fig:sample}$. These show the relation between the UV and Ly$\alpha$ luminosity at which the number densities of the respective LFs at $z=2-7$ are within a factor two of each other. Most of the evolution is seen for fainter galaxies. The steepening faint end of the UV LF with increasing redshift yields a higher typical Ly$\alpha$ EW for faint galaxies. Dashed lines indicate lines of fixed EW for a given UV slope $\beta=-2.0$.}
\label{fig:representative_evolution}
\end{figure}

The resulting UV-to-Ly$\alpha$ luminosity relation that is required to match the abundances is shown as coloured shaded regions in Fig. $\ref{fig:representative_evolution}$. The evolution of the LFs implies that particularly UV-faint galaxies on average have a higher Ly$\alpha$ luminosity with increasing redshift. This means that LAEs become more representative of the general galaxy population with increasing redshift \citep[see also][]{Santos2020}. If we assume the constant Ly$\alpha$ LF could be extrapolated to $z>6$ (ignoring any impact of the rapidly increasing opacity from the IGM; e.g. \citealt{Laursen2011}) we would find that a fully representative sample of the UV galaxy population at $z\approx7.5$ has similar Ly$\alpha$ EWs as the XLS sample. In addition, the measured H$\beta$+[O{\sc iii}] EWs, O32 and [O{\sc iii}]/H$\beta$ ratios of LAEs at $z\approx2$ are comparable to values expected in typical galaxies at $z\approx7$ \citep{Faisst2016,deBarros2019,Endsley2020}. Moreover, the rest-frame UV sizes of LAEs are relatively constant in the population of LAEs over $z=2-6$ and these sizes are similar to the sizes of the general population of star-forming galaxies at $z\approx7$ \citep{PaulinoAfonso2018}. These comparisons suggest that the properties of the general population of star-forming galaxies in the epoch of reionisation resembles the properties of typical LAEs at $z\approx2$. Indeed, the SED modelling indicates light-weighted ages of $250\pm200$ Myr, consistent with plausible ages for galaxies in the very early Universe. 

Our discussion in \S $\ref{sec:discuss}$ can also be applied to explain the relative redshift evolution of the Ly$\alpha$ and UV LFs. It is plausible that the increasing strength of Ly$\alpha$ compared to the UV continuum can be explained following the build-up of galaxies that constitute the LF over time. Towards higher redshift, the population-averaged dust attenuation will decrease as the fractional contribution of low mass galaxies is higher, leading to a higher volumetric Ly$\alpha$ escape fraction \citep[e.g.][]{Hayes2011}. Moreover, galaxies will on average be younger, leading to a higher ionising photon efficiency and higher intrinsic Ly$\alpha$ EW \citep{Matthee2017GALEX,Sobral2018}. It is however also possible that there is, in addition, evolution in the H{\sc i} column density (i.e. a larger covering factor of low-density channels) or presence of outflows, which has currently not been investigated in a systematic way observationally \citep[c.f.][]{Cassata2020}. Such evolution could have implications on the escape fraction of ionising photons, but also on inferences of reionisation that rely on assuming the Ly$\alpha$ profile escaping the ISM of galaxies \citep[e.g.][]{Mason2018b}. Future extensions of XLS-$z2$ to $z=3-7$, for example by combining blind spectroscopy with VLT/MUSE and {\it JWST}, that can measure $f_{\rm esc, Ly\alpha}$, the Ly$\alpha$ profile and ISM conditions are required to test the relative importance of these various mechanisms.

\section{Summary} \label{sec:summary}
In this paper we presented the first results of the X-SHOOTER Lyman-$\alpha$ Survey at $z=2$ (XLS-$z2$) which is a deep spectroscopic survey of 35 LAEs at $z\approx2$ covering the rest-frame UV to rest-frame optical ($\lambda_0=0.1-0.7 \mu$m). 

The selected LAEs have stellar masses ranging from $10^{8-10}$ M$_{\odot}$, Ly$\alpha$ luminosities $0.2-10\times L^{\star}_{\rm Ly\alpha}$, UV luminosities $0.2-6\times L^{\star}_{\rm UV}$ and are typically small ($r_{1/2, \rm UV}\approx1$ kpc) and blue ($\beta\approx-2.0$). The rest-frame Ly$\alpha$ EWs range from $\approx1 - 300$ {\AA}, with the majority $>25$ {\AA} and $70$ {\AA} on average. These galaxies constitute a rare, $\approx1/15$, fraction of the UV-selected galaxy population at $z\approx2$ with UV luminosity M$_{1500}\approx-20\pm1$. Targets were observed with VLT/X-SHOOTER for typically $\approx3$ hours. We presented the observing strategy, the data reduction and the methodology of extracting 1D spectra. 

We measured systemic redshifts in 33 of the 35 targets. Most of these come from the [O{\sc iii}]$_{4960,5008}$ doublet with FWHM$\approx130$ km s$^{-1}$ and consistent redshifts for H$\alpha$ are found. In the objects with best S/N, we identify complex features in the [O{\sc iii}] profile, such as multiple narrow or a broad  (FWHM$\approx280$ km s$^{-1}$) component. Systemic redshifts are on average $205\pm5$ km s$^{-1}$ bluewards of the red peak of the Ly$\alpha$ line.

In this paper we focused on exploring the stack of 20 {\it representative} LAEs to establish the various typical properties of LAEs at $z\approx2$ and address the question: what makes galaxies LAEs?. The stack is shown in Fig. $\ref{fig:stack}$ and the measurements and their derived properties are listed in Tables $\ref{tab:measurements_stack}$ and $\ref{tab:measurements_stack_UV}$. The results can be summarised as:

\begin{itemize}
\item The stacked spectrum of LAEs is emission-line dominated with high Ly$\alpha$, H$\beta$, [O{\sc iii}] and H$\alpha$ EWs. The UV continuum is blue ($\beta=-2.1$) and shows several high-ionisation emission lines (C{\sc iv}, He{\sc ii}, O{\sc iii}]) with EWs $\approx1$ {\AA} and C{\sc iii}] and Mg{\sc ii} with EWs $\approx5$ {\AA}. 

\item The rest-frame optical line ratios show that LAEs are powered by star formation, a picture corroborated by rest-frame UV line-ratios. Based on the EWs of H$\alpha$ we show that the typical star-burst is young (6 Myr for a single burst and 40 Myr for a constant star formation history) and the emission-line ratios indicate that the stellar metallicity plausibly is as low as $Z_{\star}\approx10^{-3}$ ($0.07\,Z_{\odot}$), while the direct-method gas-phase metallicity is $0.13\,Z_{\odot}$. 

\item The ISM is characterised by an electron temperature of $14,800^{+700}_{-800}$ K, a moderately low attenuation (E$(B-V)_{\rm gas} = 0.22\pm0.06$) and a high ionisation state. The gas-phase metallicity is low (12+log(O/H)$_{\rm T_e} = 7.83^{+0.06}_{-0.05}$) and the C/O and N/O abundances (log$_{10}$(C/O) = $-0.8^{+0.2}_{-0.2}$, log$_{10}$(N/O) $<-0.8$) are consistent with little chemical enrichment from evolved stars.

\item LAEs have a clumpy morphological structure in the rest-frame UV and show turbulent and outflowing kinematics. The broad [O{\sc iii}] components suggest outflows, but we also detect them more unambiguously through blue-shifted Si{\sc ii} absorption. The average Ly$\alpha$ spectrum has a clear double-peaked profile with a blue peak that is 0.3 times the flux of the dominant red peak. This shape suggests Ly$\alpha$ photons generally scatter through an outflowing medium while escaping galactic environments.

\item We measure a Ly$\alpha$ escape fraction of $27\pm4$ \% based on dust-corrected H$\alpha$ luminosity, consistent with earlier measurements in LAEs. A non-negligible amount of Ly$\alpha$ photons escape directly at the systemic redshift which suggests the presence of low column density channels that also allow the escape of LyC photons. Based on empirical comparisons of the Ly$\alpha$ and Mg{\sc ii} line profiles with those in low-redshift analogues we estimate $f_{\rm esc, LyC}\approx10$ \%.
\end{itemize}

We use these results to discuss how LAEs compare to the general galaxy population at $z\approx2$. As illustrated in Fig. $\ref{fig:sketch}$, we identify several mechanisms that are likely important in determining whether galaxies can be observed as LAEs. These include a low dust attenuation, a high intrinsic Ly$\alpha$ EW due to the presence of young hot stars, but also a low H{\sc i} column density that may be related to gas turbulence and outflows and possibly the viewing angle. 

We argue that the low mass of typical LAEs suggests that dust attenuation prevents the majority of massive galaxies to be bright LAEs. We further show, based on number density analyses, that only a fraction of $\approx10$ \% of low-mass galaxies is observed as LAE at $z\approx2$, consistent with duty cycles inferred from clustering analyses. LAEs appear to have a similar SFR and attenuation as typical galaxies with their mass, suggesting that effective H{\sc i} column density may mostly determine whether we observe low-mass galaxies as LAEs. It is plausible that low H{\sc i} column densities may be related to systematic changes in gas turbulence or outflows, but stochastic differences such as the specific viewing angle may be equally or more important. In future work we will investigate the relative importance of the various processes by exploring the variation within the XLS sample.

Finally, we discuss that while LAEs constitute a relatively rare sample of low-mass galaxies at $z\approx2$, galaxies that resemble LAEs become increasingly more common at higher redshifts. Indeed, galaxies with these Ly$\alpha$ to UV luminosity ratios will be the norm in the epoch of reionisation even if the neutral IGM may prevent us realising that. 

It is of extreme importance in understanding the early phases of galaxy formation and the late phases of the epoch of reionisation to verify whether this evolution is a mere effect of the evolving stellar mass function (and the related evolution of the typical dust attenuation), or whether an increasing Ly$\alpha$ output or higher occurrence of low column density H{\sc i} gas contribute to the observed global evolution. In upcoming XLS-$z2$ papers we will focus on the variation in the Ly$\alpha$ profiles and how these are related to the production and escape of Ly$\alpha$ and LyC photons.

\section*{Acknowledgements}
We thank the referee for constructive comments and suggestions. We thank Dawn Erb, Ruari Mackenzie, Ivan Oteo, Ryan Sanders and Johannes Zabl for useful discussions and suggestions. It is a pleasure to thank the ESO User Support, in particular Giacomo Beccari, Carlo Manara, John Pritchard, Marina Rejkuba and Lowell Tacconi-Garman for assistance in the preparation and execution of the observations.
Based on observations obtained with the Very Large Telescope, programs 084.A-0303, 088.A-0672, 091.A-0413, 091.A-0546, 092.A-0774, 097.A-0153, 098.A-0819, 099.A-0758, 099.A-0254, 101.B-0779 and 102.A-0652.
Based on data products from observations made with ESO Telescopes at the La Silla Paranal Observatory under ESO programme ID 179.A-2005 and on data products produced by CALET and the Cambridge Astronomy Survey Unit on behalf of the UltraVISTA consortium.
Based on observations made with the NASA/ESA Hubble Space Telescope through programs 9133, 9367, 11694 and 12471, and obtained from the Hubble Legacy Archive, which is a collaboration between the Space Telescope Science Institute (STScI/NASA), the Space Telescope European Coordinating Facility (ST-ECF/ESA) and the Canadian Astronomy Data Centre (CADC/NRC/CSA).
This work is based on observations taken by the CANDELS Multi-Cycle Treasury Program with the NASA/ESA HST, which is operated by the Association of Universities for Research in Astronomy, Inc., under NASA contract NAS5-26555.
MG was supported by NASA through the NASA Hubble Fellowship grant HST-HF2-51409 and acknowledges support from HST grants HST-GO-15643.017-A, HST-AR-15039.003-A, and XSEDE grant TG-AST180036.
GP acknowledges support from the Netherlands Research School for Astronomy (NOVA). RA acknowledges the support of ANID FONDECYT Regular Grant 1202007.
We gratefully acknowledge the {\sc Python} programming language, its {\sc numpy, matplotlib, scipy, lmfit} \citep{Scipy,Hunter2007,Numpy}, {\sc pandas} \citep{pandas} and {\sc astropy} \citep{Astropy} packages and the {\sc Topcat} analysis tool \citep{Topcat}. Dedicated to the memory of A.C.J. Matthee (1953-2020).


\section*{Data availability}
The data underlying this article were accessed from the ESO archive. The raw ESO data can be accessed through http://archive.eso.org/cms.html. The derived data generated in this research will be shared on reasonable request to the corresponding author.



\bibliographystyle{mnras}
\bibliography{bib_XLS} 

\begin{thebibliography}{}
\makeatletter
\relax
\def\mn@urlcharsother{\let\do\@makeother \do\$\do\&\do\#\do\^\do\_\do\%\do\~}
\def\mn@doi{\begingroup\mn@urlcharsother \@ifnextchar [ {\mn@doi@}
  {\mn@doi@[]}}
\def\mn@doi@[#1]#2{\def\@tempa{#1}\ifx\@tempa\@empty \href
  {http://dx.doi.org/#2} {doi:#2}\else \href {http://dx.doi.org/#2} {#1}\fi
  \endgroup}
\def\mn@eprint#1#2{\mn@eprint@#1:#2::\@nil}
\def\mn@eprint@arXiv#1{\href {http://arxiv.org/abs/#1} {{\tt arXiv:#1}}}
\def\mn@eprint@dblp#1{\href {http://dblp.uni-trier.de/rec/bibtex/#1.xml}
  {dblp:#1}}
\def\mn@eprint@#1:#2:#3:#4\@nil{\def\@tempa {#1}\def\@tempb {#2}\def\@tempc
  {#3}\ifx \@tempc \@empty \let \@tempc \@tempb \let \@tempb \@tempa \fi \ifx
  \@tempb \@empty \def\@tempb {arXiv}\fi \@ifundefined
  {mn@eprint@\@tempb}{\@tempb:\@tempc}{\expandafter \expandafter \csname
  mn@eprint@\@tempb\endcsname \expandafter{\@tempc}}}

\bibitem[\protect\citeauthoryear{{Aihara} et~al.}{{Aihara}
  et~al.}{2019}]{HSC_SURVEYPAPER}
{Aihara} H.,  et~al., 2019, \mn@doi [PASJ] {10.1093/pasj/psz103}, \href
  {https://ui.adsabs.harvard.edu/abs/2019PASJ..tmp..106A} {p.~106}

\bibitem[\protect\citeauthoryear{{Alam} et~al.}{{Alam} et~al.}{2015}]{Alam2015}
{Alam} S.,  et~al., 2015, \mn@doi [ApJS] {10.1088/0067-0049/219/1/12}, \href
  {http://adsabs.harvard.edu/abs/2015ApJS..219...12A} {219, 12}

\bibitem[\protect\citeauthoryear{{Amor{\'\i}n}, {P{\'e}rez-Montero},
  {V{\'\i}lchez}  \& {Papaderos}}{{Amor{\'\i}n} et~al.}{2012a}]{Amorin2012a}
{Amor{\'\i}n} R.,  {P{\'e}rez-Montero} E.,  {V{\'\i}lchez} J.~M.,   {Papaderos}
  P.,  2012a, \mn@doi [ApJ] {10.1088/0004-637X/749/2/185}, \href
  {https://ui.adsabs.harvard.edu/abs/2012ApJ...749..185A} {749, 185}

\bibitem[\protect\citeauthoryear{{Amor{\'\i}n}, {V{\'\i}lchez}, {H{\"a}gele},
  {Firpo}, {P{\'e}rez-Montero}  \& {Papaderos}}{{Amor{\'\i}n}
  et~al.}{2012b}]{Amorin2012}
{Amor{\'\i}n} R.,  {V{\'\i}lchez} J.~M.,  {H{\"a}gele} G.~F.,  {Firpo} V.,
  {P{\'e}rez-Montero} E.,   {Papaderos} P.,  2012b, \mn@doi [ApJL]
  {10.1088/2041-8205/754/2/L22}, \href
  {https://ui.adsabs.harvard.edu/abs/2012ApJ...754L..22A} {754, L22}

\bibitem[\protect\citeauthoryear{{Amor{\'{\i}}n} et~al.}{{Amor{\'{\i}}n}
  et~al.}{2017}]{Amorin2017}
{Amor{\'{\i}}n} R.,  et~al., 2017, \mn@doi [Nature Astronomy]
  {10.1038/s41550-017-0052}, \href
  {http://adsabs.harvard.edu/abs/2017NatAs...1E..52A} {1, 0052}

\bibitem[\protect\citeauthoryear{{Astropy Collaboration} et~al.}{{Astropy
  Collaboration} et~al.}{2013}]{Astropy}
{Astropy Collaboration} et~al., 2013, \mn@doi [AAP]
  {10.1051/0004-6361/201322068}, \href
  {https://ui.adsabs.harvard.edu/abs/2013A%26A...558A..33A} {558, A33}

\bibitem[\protect\citeauthoryear{{Atek}, {Kunth}, {Hayes}, {{\"O}stlin}  \&
  {Mas-Hesse}}{{Atek} et~al.}{2008}]{Atek2008}
{Atek} H.,  {Kunth} D.,  {Hayes} M.,  {{\"O}stlin} G.,   {Mas-Hesse} J.~M.,
  2008, \mn@doi [AAP] {10.1051/0004-6361:200809527}, \href
  {http://adsabs.harvard.edu/abs/2008A%26A...488..491A} {488, 491}

\bibitem[\protect\citeauthoryear{{Baldwin}, {Phillips}  \&
  {Terlevich}}{{Baldwin} et~al.}{1981}]{BPT}
{Baldwin} J.~A.,  {Phillips} M.~M.,   {Terlevich} R.,  1981, \mn@doi [PASP]
  {10.1086/130766}, \href
  {https://ui.adsabs.harvard.edu/abs/1981PASP...93....5B} {93, 5}

\bibitem[\protect\citeauthoryear{{Barnes}, {Haehnelt}, {Tescari}  \&
  {Viel}}{{Barnes} et~al.}{2011}]{Barnes2011}
{Barnes} L.~A.,  {Haehnelt} M.~G.,  {Tescari} E.,   {Viel} M.,  2011, \mn@doi
  [MNRAS] {10.1111/j.1365-2966.2011.18789.x}, \href
  {http://adsabs.harvard.edu/abs/2011MNRAS.416.1723B} {416, 1723}

\bibitem[\protect\citeauthoryear{{Behrens} \& {Braun}}{{Behrens} \&
  {Braun}}{2014}]{BehrensBraun2014}
{Behrens} C.,  {Braun} H.,  2014, \mn@doi [AAP] {10.1051/0004-6361/201424755},
  \href {https://ui.adsabs.harvard.edu/abs/2014A&A...572A..74B} {572, A74}

\bibitem[\protect\citeauthoryear{{Berg}, {Erb}, {Auger}, {Pettini}  \&
  {Brammer}}{{Berg} et~al.}{2018}]{Berg2018}
{Berg} D.~A.,  {Erb} D.~K.,  {Auger} M.~W.,  {Pettini} M.,   {Brammer} G.~B.,
  2018, \mn@doi [ApJ] {10.3847/1538-4357/aab7fa}, \href
  {https://ui.adsabs.harvard.edu/abs/2018ApJ...859..164B} {859, 164}

\bibitem[\protect\citeauthoryear{{Berg}, {Erb}, {Henry}, {Skillman}  \&
  {McQuinn}}{{Berg} et~al.}{2019a}]{Berg2019b}
{Berg} D.~A.,  {Erb} D.~K.,  {Henry} R. B.~C.,  {Skillman} E.~D.,   {McQuinn}
  K. B.~W.,  2019a, \mn@doi [ApJ] {10.3847/1538-4357/ab020a}, \href
  {https://ui.adsabs.harvard.edu/abs/2019ApJ...874...93B} {874, 93}

\bibitem[\protect\citeauthoryear{{Berg}, {Chisholm}, {Erb}, {Pogge}, {Henry}
  \& {Olivier}}{{Berg} et~al.}{2019b}]{Berg2019}
{Berg} D.~A.,  {Chisholm} J.,  {Erb} D.~K.,  {Pogge} R.,  {Henry} A.,
  {Olivier} G.~M.,  2019b, \mn@doi [ApJL] {10.3847/2041-8213/ab21dc}, \href
  {https://ui.adsabs.harvard.edu/abs/2019ApJ...878L...3B} {878, L3}

\bibitem[\protect\citeauthoryear{{Bian}, {Kewley}  \& {Dopita}}{{Bian}
  et~al.}{2018}]{Bian2018}
{Bian} F.,  {Kewley} L.~J.,   {Dopita} M.~A.,  2018, \mn@doi [ApJ]
  {10.3847/1538-4357/aabd74}, \href
  {https://ui.adsabs.harvard.edu/abs/2018ApJ...859..175B} {859, 175}

\bibitem[\protect\citeauthoryear{{Blanc} et~al.}{{Blanc}
  et~al.}{2011}]{Blanc2011}
{Blanc} G.~A.,  et~al., 2011, \mn@doi [ApJ] {10.1088/0004-637X/736/1/31}, \href
  {https://ui.adsabs.harvard.edu/abs/2011ApJ...736...31B} {736, 31}

\bibitem[\protect\citeauthoryear{{Bosch} et~al.}{{Bosch}
  et~al.}{2019}]{Bosch2019}
{Bosch} G.,  et~al., 2019, \mn@doi [MNRAS] {10.1093/mnras/stz2230}, \href
  {https://ui.adsabs.harvard.edu/abs/2019MNRAS.489.1787B} {489, 1787}

\bibitem[\protect\citeauthoryear{{Bouwens} et~al.,}{{Bouwens}
  et~al.}{2015}]{Bouwens2015}
{Bouwens} R.~J.,  et~al., 2015, \mn@doi [ApJ] {10.1088/0004-637X/803/1/34},
  \href {http://adsabs.harvard.edu/abs/2015ApJ...803...34B} {803, 34}

\bibitem[\protect\citeauthoryear{{Bouwens}, {Smit}, {Labb{\'e}}, {Franx},
  {Caruana}, {Oesch}, {Stefanon}  \& {Rasappu}}{{Bouwens}
  et~al.}{2016}]{BouwensXION}
{Bouwens} R.~J.,  {Smit} R.,  {Labb{\'e}} I.,  {Franx} M.,  {Caruana} J.,
  {Oesch} P.,  {Stefanon} M.,   {Rasappu} N.,  2016, \mn@doi [ApJ]
  {10.3847/0004-637X/831/2/176}, \href
  {http://adsabs.harvard.edu/abs/2016ApJ...831..176B} {831, 176}

\bibitem[\protect\citeauthoryear{{Brinchmann}, {Charlot}, {White}, {Tremonti},
  {Kauffmann}, {Heckman}  \& {Brinkmann}}{{Brinchmann}
  et~al.}{2004}]{Brinchmann2004}
{Brinchmann} J.,  {Charlot} S.,  {White} S.~D.~M.,  {Tremonti} C.,  {Kauffmann}
  G.,  {Heckman} T.,   {Brinkmann} J.,  2004, \mn@doi [MNRAS]
  {10.1111/j.1365-2966.2004.07881.x}, \href
  {http://adsabs.harvard.edu/abs/2004MNRAS.351.1151B} {351, 1151}

\bibitem[\protect\citeauthoryear{{Brinchmann}, {Pettini}  \&
  {Charlot}}{{Brinchmann} et~al.}{2008}]{Brinchmann2008}
{Brinchmann} J.,  {Pettini} M.,   {Charlot} S.,  2008, \mn@doi [MNRAS]
  {10.1111/j.1365-2966.2008.12914.x}, \href
  {https://ui.adsabs.harvard.edu/abs/2008MNRAS.385..769B} {385, 769}

\bibitem[\protect\citeauthoryear{{Bruzual} \& {Charlot}}{{Bruzual} \&
  {Charlot}}{2003}]{BruzualCharlot2003}
{Bruzual} G.,  {Charlot} S.,  2003, \mn@doi [MNRAS]
  {10.1046/j.1365-8711.2003.06897.x}, \href
  {https://ui.adsabs.harvard.edu/abs/2003MNRAS.344.1000B} {344, 1000}

\bibitem[\protect\citeauthoryear{{Byrohl} \& {Gronke}}{{Byrohl} \&
  {Gronke}}{2020}]{Byrohl2020}
{Byrohl} C.,  {Gronke} M.,  2020, \mn@doi [AAP] {10.1051/0004-6361/202038685},
  \href {https://ui.adsabs.harvard.edu/abs/2020A&A...642L..16B} {642, L16}

\bibitem[\protect\citeauthoryear{{Calhau} et~al.}{{Calhau}
  et~al.}{2020}]{Calhau2020}
{Calhau} J.,  et~al., 2020, \mn@doi [MNRAS] {10.1093/mnras/staa476}, \href
  {https://ui.adsabs.harvard.edu/abs/2020MNRAS.493.3341C} {493, 3341}

\bibitem[\protect\citeauthoryear{{Calzetti}, {Armus}, {Bohlin}, {Kinney},
  {Koornneef}  \& {Storchi-Bergmann}}{{Calzetti} et~al.}{2000}]{Calzetti2000}
{Calzetti} D.,  {Armus} L.,  {Bohlin} R.~C.,  {Kinney} A.~L.,  {Koornneef} J.,
   {Storchi-Bergmann} T.,  2000, \mn@doi [ApJ] {10.1086/308692}, \href
  {https://ui.adsabs.harvard.edu/abs/2000ApJ...533..682C} {533, 682}

\bibitem[\protect\citeauthoryear{{Cantalupo}, {Lilly}  \&
  {Haehnelt}}{{Cantalupo} et~al.}{2012}]{Cantalupo2012}
{Cantalupo} S.,  {Lilly} S.~J.,   {Haehnelt} M.~G.,  2012, \mn@doi [MNRAS]
  {10.1111/j.1365-2966.2012.21529.x}, \href
  {https://ui.adsabs.harvard.edu/abs/2012MNRAS.425.1992C} {425, 1992}

\bibitem[\protect\citeauthoryear{{Cardamone} et~al.}{{Cardamone}
  et~al.}{2010}]{Cardamone2010}
{Cardamone} C.~N.,  et~al., 2010, \mn@doi [ApJS] {10.1088/0067-0049/189/2/270},
  \href {https://ui.adsabs.harvard.edu/abs/2010ApJS..189..270C} {189, 270}

\bibitem[\protect\citeauthoryear{{Cardelli}, {Clayton}  \& {Mathis}}{{Cardelli}
  et~al.}{1989}]{Cardelli1989}
{Cardelli} J.~A.,  {Clayton} G.~C.,   {Mathis} J.~S.,  1989, \mn@doi [ApJ]
  {10.1086/167900}, \href
  {https://ui.adsabs.harvard.edu/abs/1989ApJ...345..245C} {345, 245}

\bibitem[\protect\citeauthoryear{{Cassata} et~al.}{{Cassata}
  et~al.}{2015}]{Cassata2015}
{Cassata} P.,  et~al., 2015, \mn@doi [AAP] {10.1051/0004-6361/201423824}, \href
  {http://adsabs.harvard.edu/abs/2015A%26A...573A..24C} {573, A24}

\bibitem[\protect\citeauthoryear{{Cassata} et~al.}{{Cassata}
  et~al.}{2020}]{Cassata2020}
{Cassata} P.,  et~al., 2020, \mn@doi [AAP] {10.1051/0004-6361/202037517}, \href
  {https://ui.adsabs.harvard.edu/abs/2020A&A...643A...6C} {643, A6}

\bibitem[\protect\citeauthoryear{{Chabrier}}{{Chabrier}}{2003}]{Chabrier2003}
{Chabrier} G.,  2003, \mn@doi [PASP] {10.1086/376392}, \href
  {https://ui.adsabs.harvard.edu/abs/2003PASP..115..763C} {115, 763}

\bibitem[\protect\citeauthoryear{{Charlot} \& {Fall}}{{Charlot} \&
  {Fall}}{1993}]{CharlotFall1993}
{Charlot} S.,  {Fall} S.~M.,  1993, \mn@doi [ApJ] {10.1086/173187}, \href
  {http://adsabs.harvard.edu/abs/1993ApJ...415..580C} {415, 580}

\bibitem[\protect\citeauthoryear{{Charlot} \& {Fall}}{{Charlot} \&
  {Fall}}{2000}]{CharlotFall2000}
{Charlot} S.,  {Fall} S.~M.,  2000, \mn@doi [ApJ] {10.1086/309250}, \href
  {https://ui.adsabs.harvard.edu/abs/2000ApJ...539..718C} {539, 718}

\bibitem[\protect\citeauthoryear{{Chisholm}, {Rigby}, {Bayliss}, {Berg},
  {Dahle}, {Gladders}  \& {Sharon}}{{Chisholm} et~al.}{2019}]{Chisholm2019}
{Chisholm} J.,  {Rigby} J.~R.,  {Bayliss} M.,  {Berg} D.~A.,  {Dahle} H.,
  {Gladders} M.,   {Sharon} K.,  2019, \mn@doi [ApJ]
  {10.3847/1538-4357/ab3104}, \href
  {https://ui.adsabs.harvard.edu/abs/2019ApJ...882..182C} {882, 182}

\bibitem[\protect\citeauthoryear{{Chisholm}, {Prochaska}, {Schaerer},
  {Gazagnes}  \& {Henry}}{{Chisholm} et~al.}{2020}]{Chisholm2020}
{Chisholm} J.,  {Prochaska} J.~X.,  {Schaerer} D.,  {Gazagnes} S.,   {Henry}
  A.,  2020, \mn@doi [MNRAS] {10.1093/mnras/staa2470}, \href
  {https://ui.adsabs.harvard.edu/abs/2020MNRAS.498.2554C} {498, 2554}

\bibitem[\protect\citeauthoryear{{Cornachione} et~al.,}{{Cornachione}
  et~al.}{2018}]{Cornachione2018}
{Cornachione} M.~A.,  et~al., 2018, \mn@doi [ApJ] {10.3847/1538-4357/aaa412},
  \href {https://ui.adsabs.harvard.edu/abs/2018ApJ...853..148C} {853, 148}

\bibitem[\protect\citeauthoryear{{Cullen} et~al.}{{Cullen}
  et~al.}{2020}]{Cullen2020}
{Cullen} F.,  et~al., 2020, \mn@doi [MNRAS] {10.1093/mnras/staa1260}, \href
  {https://ui.adsabs.harvard.edu/abs/2020MNRAS.495.1501C} {495, 1501}

\bibitem[\protect\citeauthoryear{{Davies} et~al.}{{Davies}
  et~al.}{2019}]{Davies2019}
{Davies} R.~L.,  et~al., 2019, \mn@doi [ApJ] {10.3847/1538-4357/ab06f1}, \href
  {https://ui.adsabs.harvard.edu/abs/2019ApJ...873..122D} {873, 122}

\bibitem[\protect\citeauthoryear{{De Barros}, {Oesch}, {Labb{\'e}}, {Stefanon},
  {Gonz{\'a}lez}, {Smit}, {Bouwens}  \& {Illingworth}}{{De Barros}
  et~al.}{2019}]{deBarros2019}
{De Barros} S.,  {Oesch} P.~A.,  {Labb{\'e}} I.,  {Stefanon} M.,
  {Gonz{\'a}lez} V.,  {Smit} R.,  {Bouwens} R.~J.,   {Illingworth} G.~D.,
  2019, \mn@doi [MNRAS] {10.1093/mnras/stz940}, \href
  {https://ui.adsabs.harvard.edu/abs/2019MNRAS.489.2355D} {489, 2355}

\bibitem[\protect\citeauthoryear{{Dijkstra}}{{Dijkstra}}{2014}]{DijkstraReview}
{Dijkstra} M.,  2014, \mn@doi [PASA] {10.1017/pasa.2014.33}, \href
  {http://adsabs.harvard.edu/abs/2014PASA...31...40D} {31, 40}

\bibitem[\protect\citeauthoryear{{Dijkstra}, {Haiman}  \& {Spaans}}{{Dijkstra}
  et~al.}{2006}]{Dijkstra2006}
{Dijkstra} M.,  {Haiman} Z.,   {Spaans} M.,  2006, \mn@doi [ApJ]
  {10.1086/506243}, \href
  {https://ui.adsabs.harvard.edu/abs/2006ApJ...649...14D} {649, 14}

\bibitem[\protect\citeauthoryear{{Dijkstra}, {Wyithe}, {Haiman}, {Mesinger}  \&
  {Pentericci}}{{Dijkstra} et~al.}{2014}]{Dijkstra2014}
{Dijkstra} M.,  {Wyithe} S.,  {Haiman} Z.,  {Mesinger} A.,   {Pentericci} L.,
  2014, \mn@doi [MNRAS] {10.1093/mnras/stu531}, \href
  {http://adsabs.harvard.edu/abs/2014MNRAS.440.3309D} {440, 3309}

\bibitem[\protect\citeauthoryear{{Dijkstra}, {Gronke}  \&
  {Venkatesan}}{{Dijkstra} et~al.}{2016}]{Dijkstra2016}
{Dijkstra} M.,  {Gronke} M.,   {Venkatesan} A.,  2016, \mn@doi [ApJ]
  {10.3847/0004-637X/828/2/71}, \href
  {http://adsabs.harvard.edu/abs/2016ApJ...828...71D} {828, 71}

\bibitem[\protect\citeauthoryear{{Dopita} et~al.}{{Dopita}
  et~al.}{2006}]{Dopita2006}
{Dopita} M.~A.,  et~al., 2006, \mn@doi [ApJ] {10.1086/505418}, \href
  {https://ui.adsabs.harvard.edu/abs/2006ApJ...647..244D} {647, 244}

\bibitem[\protect\citeauthoryear{{Drake} et~al.}{{Drake}
  et~al.}{2017}]{Drake2017}
{Drake} A.~B.,  et~al., 2017, \mn@doi [AAP] {10.1051/0004-6361/201731431},
  \href {http://adsabs.harvard.edu/abs/2017A%26A...608A...6D} {608, A6}

\bibitem[\protect\citeauthoryear{{Du}, {Shapley}, {Tang}, {Stark}, {Martin},
  {Mobasher}, {Topping}  \& {Chevallard}}{{Du} et~al.}{2020}]{Du2020}
{Du} X.,  {Shapley} A.~E.,  {Tang} M.,  {Stark} D.~P.,  {Martin} C.~L.,
  {Mobasher} B.,  {Topping} M.~W.,   {Chevallard} J.,  2020, \mn@doi [ApJ]
  {10.3847/1538-4357/ab67b8}, \href
  {https://ui.adsabs.harvard.edu/abs/2020ApJ...890...65D} {890, 65}

\bibitem[\protect\citeauthoryear{{Eide}, {Gronke}, {Dijkstra}  \&
  {Hayes}}{{Eide} et~al.}{2018}]{Eide2018}
{Eide} M.~B.,  {Gronke} M.,  {Dijkstra} M.,   {Hayes} M.,  2018, \mn@doi [ApJ]
  {10.3847/1538-4357/aab5b7}, \href
  {https://ui.adsabs.harvard.edu/abs/2018ApJ...856..156E} {856, 156}

\bibitem[\protect\citeauthoryear{{Emami}, {Siana}, {Alavi}, {Gburek},
  {Freeman}, {Richard}, {Weisz}  \& {Stark}}{{Emami} et~al.}{2020}]{Emami2020}
{Emami} N.,  {Siana} B.,  {Alavi} A.,  {Gburek} T.,  {Freeman} W.~R.,
  {Richard} J.,  {Weisz} D.~R.,   {Stark} D.~P.,  2020, \mn@doi [ApJ]
  {10.3847/1538-4357/ab8f97}, \href
  {https://ui.adsabs.harvard.edu/abs/2020ApJ...895..116E} {895, 116}

\bibitem[\protect\citeauthoryear{{Endsley}, {Stark}, {Chevallard}  \&
  {Charlot}}{{Endsley} et~al.}{2021}]{Endsley2020}
{Endsley} R.,  {Stark} D.~P.,  {Chevallard} J.,   {Charlot} S.,  2021, \mn@doi
  [MNRAS] {10.1093/mnras/staa3370}, \href
  {https://ui.adsabs.harvard.edu/abs/2021MNRAS.500.5229E} {500, 5229}

\bibitem[\protect\citeauthoryear{{Erb}}{{Erb}}{2015}]{Erb2015}
{Erb} D.~K.,  2015, \mn@doi [Nature] {10.1038/nature14454}, \href
  {https://ui.adsabs.harvard.edu/abs/2015Natur.523..169E} {523, 169}

\bibitem[\protect\citeauthoryear{{Erb}, {Steidel}, {Shapley}, {Pettini},
  {Reddy}  \& {Adelberger}}{{Erb} et~al.}{2006a}]{Erb2006_Mass}
{Erb} D.~K.,  {Steidel} C.~C.,  {Shapley} A.~E.,  {Pettini} M.,  {Reddy} N.~A.,
    {Adelberger} K.~L.,  2006a, \mn@doi [ApJ] {10.1086/504891}, \href
  {https://ui.adsabs.harvard.edu/abs/2006ApJ...646..107E} {646, 107}

\bibitem[\protect\citeauthoryear{{Erb}, {Steidel}, {Shapley}, {Pettini},
  {Reddy}  \& {Adelberger}}{{Erb} et~al.}{2006b}]{Erb2006}
{Erb} D.~K.,  {Steidel} C.~C.,  {Shapley} A.~E.,  {Pettini} M.,  {Reddy} N.~A.,
    {Adelberger} K.~L.,  2006b, \mn@doi [ApJ] {10.1086/505341}, \href
  {https://ui.adsabs.harvard.edu/abs/2006ApJ...647..128E} {647, 128}

\bibitem[\protect\citeauthoryear{{Erb}, {Pettini}, {Shapley}, {Steidel}, {Law}
  \& {Reddy}}{{Erb} et~al.}{2010}]{Erb2010}
{Erb} D.~K.,  {Pettini} M.,  {Shapley} A.~E.,  {Steidel} C.~C.,  {Law} D.~R.,
  {Reddy} N.~A.,  2010, \mn@doi [ApJ] {10.1088/0004-637X/719/2/1168}, \href
  {http://adsabs.harvard.edu/abs/2010ApJ...719.1168E} {719, 1168}

\bibitem[\protect\citeauthoryear{{Erb}, {Pettini}, {Steidel}, {Strom}, {Rudie},
  {Trainor}, {Shapley}  \& {Reddy}}{{Erb} et~al.}{2016}]{Erb2016}
{Erb} D.~K.,  {Pettini} M.,  {Steidel} C.~C.,  {Strom} A.~L.,  {Rudie} G.~C.,
  {Trainor} R.~F.,  {Shapley} A.~E.,   {Reddy} N.~A.,  2016, \mn@doi [ApJ]
  {10.3847/0004-637X/830/1/52}, \href
  {http://adsabs.harvard.edu/abs/2016ApJ...830...52E} {830, 52}

\bibitem[\protect\citeauthoryear{{Faisst}}{{Faisst}}{2016}]{Faisst2016}
{Faisst} A.~L.,  2016, \mn@doi [ApJ] {10.3847/0004-637X/829/2/99}, \href
  {http://adsabs.harvard.edu/abs/2016ApJ...829...99F} {829, 99}

\bibitem[\protect\citeauthoryear{{Faisst}, {Capak}, {Emami}, {Tacchella}  \&
  {Larson}}{{Faisst} et~al.}{2019}]{Faisst2019}
{Faisst} A.~L.,  {Capak} P.~L.,  {Emami} N.,  {Tacchella} S.,   {Larson} K.~L.,
   2019, \mn@doi [ApJ] {10.3847/1538-4357/ab425b}, \href
  {https://ui.adsabs.harvard.edu/abs/2019ApJ...884..133F} {884, 133}

\bibitem[\protect\citeauthoryear{{Feltre}, {Charlot}  \& {Gutkin}}{{Feltre}
  et~al.}{2016}]{Feltre2016}
{Feltre} A.,  {Charlot} S.,   {Gutkin} J.,  2016, \mn@doi [MNRAS]
  {10.1093/mnras/stv2794}, \href
  {http://adsabs.harvard.edu/abs/2016MNRAS.456.3354F} {456, 3354}

\bibitem[\protect\citeauthoryear{{Feltre} et~al.}{{Feltre}
  et~al.}{2018}]{Feltre2018}
{Feltre} A.,  et~al., 2018, \mn@doi [AAP] {10.1051/0004-6361/201833281}, \href
  {https://ui.adsabs.harvard.edu/abs/2018A&A...617A..62F} {617, A62}

\bibitem[\protect\citeauthoryear{{Feltre} et~al.}{{Feltre}
  et~al.}{2020}]{Feltre2020}
{Feltre} A.,  et~al., 2020, \mn@doi [AAP] {10.1051/0004-6361/202038133}, \href
  {https://ui.adsabs.harvard.edu/abs/2020A&A...641A.118F} {641, A118}

\bibitem[\protect\citeauthoryear{{Finkelstein} et~al.}{{Finkelstein}
  et~al.}{2011}]{Finkelstein2011}
{Finkelstein} S.~L.,  et~al., 2011, \mn@doi [ApJ]
  {10.1088/0004-637X/729/2/140}, \href
  {https://ui.adsabs.harvard.edu/abs/2011ApJ...729..140F} {729, 140}

\bibitem[\protect\citeauthoryear{{Finkelstein} et~al.}{{Finkelstein}
  et~al.}{2013}]{Finkelstein2013}
{Finkelstein} S.~L.,  et~al., 2013, \mn@doi [Nature] {10.1038/nature12657},
  \href {http://adsabs.harvard.edu/abs/2013Natur.502..524F} {502, 524}

\bibitem[\protect\citeauthoryear{{Freeman} et~al.}{{Freeman}
  et~al.}{2019}]{Freeman2019}
{Freeman} W.~R.,  et~al., 2019, \mn@doi [\apj] {10.3847/1538-4357/ab0655},
  \href {https://ui.adsabs.harvard.edu/abs/2019ApJ...873..102F} {873, 102}

\bibitem[\protect\citeauthoryear{{Furusawa} et~al.}{{Furusawa}
  et~al.}{2008}]{Furusawa2008}
{Furusawa} H.,  et~al., 2008, \mn@doi [ApJs] {10.1086/527321}, \href
  {http://adsabs.harvard.edu/abs/2008ApJS..176....1F} {176, 1}

\bibitem[\protect\citeauthoryear{{Gaia Collaboration} et~al.}{{Gaia
  Collaboration} et~al.}{2018}]{GAIADR2}
{Gaia Collaboration} et~al., 2018, \mn@doi [AAP] {10.1051/0004-6361/201833051},
  \href {https://ui.adsabs.harvard.edu/abs/2018A&A...616A...1G} {616, A1}

\bibitem[\protect\citeauthoryear{{Garn} \& {Best}}{{Garn} \&
  {Best}}{2010}]{GarnBest2010}
{Garn} T.,  {Best} P.~N.,  2010, \mn@doi [MNRAS]
  {10.1111/j.1365-2966.2010.17321.x}, \href
  {https://ui.adsabs.harvard.edu/abs/2010MNRAS.409..421G} {409, 421}

\bibitem[\protect\citeauthoryear{{Gawiser} et~al.}{{Gawiser}
  et~al.}{2007}]{Gawiser2007}
{Gawiser} E.,  et~al., 2007, \mn@doi [ApJ] {10.1086/522955}, \href
  {https://ui.adsabs.harvard.edu/abs/2007ApJ...671..278G} {671, 278}

\bibitem[\protect\citeauthoryear{{Gazagnes}, {Chisholm}, {Schaerer}, {Verhamme}
   \& {Izotov}}{{Gazagnes} et~al.}{2020}]{Gazagnes2020}
{Gazagnes} S.,  {Chisholm} J.,  {Schaerer} D.,  {Verhamme} A.,   {Izotov} Y.,
  2020, \mn@doi [AAP] {10.1051/0004-6361/202038096}, \href
  {https://ui.adsabs.harvard.edu/abs/2020A&A...639A..85G} {639, A85}

\bibitem[\protect\citeauthoryear{{G{\"o}tberg}, {de Mink}, {Groh}, {Leitherer}
  \& {Norman}}{{G{\"o}tberg} et~al.}{2019}]{Gotberg2019}
{G{\"o}tberg} Y.,  {de Mink} S.~E.,  {Groh} J.~H.,  {Leitherer} C.,   {Norman}
  C.,  2019, \mn@doi [AAP] {10.1051/0004-6361/201834525}, \href
  {https://ui.adsabs.harvard.edu/abs/2019A&A...629A.134G} {629, A134}

\bibitem[\protect\citeauthoryear{{Gr{\"a}fener} \& {Vink}}{{Gr{\"a}fener} \&
  {Vink}}{2015}]{GrafenerVink2015}
{Gr{\"a}fener} G.,  {Vink} J.~S.,  2015, \mn@doi [AAP]
  {10.1051/0004-6361/201425287}, \href
  {https://ui.adsabs.harvard.edu/abs/2015A&A...578L...2G} {578, L2}

\bibitem[\protect\citeauthoryear{{Grogin} et~al.}{{Grogin}
  et~al.}{2011}]{Grogin2011}
{Grogin} N.~A.,  et~al., 2011, \mn@doi [ApJS] {10.1088/0067-0049/197/2/35},
  \href {https://ui.adsabs.harvard.edu/abs/2011ApJS..197...35G} {197, 35}

\bibitem[\protect\citeauthoryear{{Gronke} \& {Dijkstra}}{{Gronke} \&
  {Dijkstra}}{2014}]{Gronke2014}
{Gronke} M.,  {Dijkstra} M.,  2014, \mn@doi [MNRAS] {10.1093/mnras/stu1513},
  \href {http://adsabs.harvard.edu/abs/2014MNRAS.444.1095G} {444, 1095}

\bibitem[\protect\citeauthoryear{{Gronke} \& {Dijkstra}}{{Gronke} \&
  {Dijkstra}}{2016}]{GronkeDijkstra2016}
{Gronke} M.,  {Dijkstra} M.,  2016, \mn@doi [ApJ] {10.3847/0004-637X/826/1/14},
  \href {https://ui.adsabs.harvard.edu/abs/2016ApJ...826...14G} {826, 14}

\bibitem[\protect\citeauthoryear{{Gronke}, {Dijkstra}, {McCourt}  \&
  {Oh}}{{Gronke} et~al.}{2017}]{Gronke2017b}
{Gronke} M.,  {Dijkstra} M.,  {McCourt} M.,   {Oh} S.~P.,  2017, \mn@doi [AAP]
  {10.1051/0004-6361/201731013}, \href
  {https://ui.adsabs.harvard.edu/abs/2017A&A...607A..71G} {607, A71}

\bibitem[\protect\citeauthoryear{{Gronwall} et~al.}{{Gronwall}
  et~al.}{2007}]{Gronwall2007}
{Gronwall} C.,  et~al., 2007, \mn@doi [ApJ] {10.1086/520324}, \href
  {http://adsabs.harvard.edu/abs/2007ApJ...667...79G} {667, 79}

\bibitem[\protect\citeauthoryear{{Guaita} et~al.}{{Guaita}
  et~al.}{2011}]{Guaita2011}
{Guaita} L.,  et~al., 2011, \mn@doi [ApJ] {10.1088/0004-637X/733/2/114}, \href
  {https://ui.adsabs.harvard.edu/abs/2011ApJ...733..114G} {733, 114}

\bibitem[\protect\citeauthoryear{{Guo} et~al.}{{Guo} et~al.}{2013}]{Guo2013}
{Guo} Y.,  et~al., 2013, \mn@doi [ApJS] {10.1088/0067-0049/207/2/24}, \href
  {https://ui.adsabs.harvard.edu/abs/2013ApJS..207...24G} {207, 24}

\bibitem[\protect\citeauthoryear{{Gurung-L{\'o}pez}, {Orsi}  \&
  {Bonoli}}{{Gurung-L{\'o}pez} et~al.}{2019}]{Gurung2019}
{Gurung-L{\'o}pez} S.,  {Orsi} {\'A}.~A.,   {Bonoli} S.,  2019, \mn@doi [MNRAS]
  {10.1093/mnras/stz2591}, \href
  {https://ui.adsabs.harvard.edu/abs/2019MNRAS.490..733G} {490, 733}

\bibitem[\protect\citeauthoryear{{Hagen} et~al.}{{Hagen}
  et~al.}{2016}]{Hagen2016}
{Hagen} A.,  et~al., 2016, \mn@doi [ApJ] {10.3847/0004-637X/817/1/79}, \href
  {https://ui.adsabs.harvard.edu/abs/2016ApJ...817...79H} {817, 79}

\bibitem[\protect\citeauthoryear{{Hainline}, {Shapley}, {Greene}  \&
  {Steidel}}{{Hainline} et~al.}{2011}]{Hainline2011}
{Hainline} K.~N.,  {Shapley} A.~E.,  {Greene} J.~E.,   {Steidel} C.~C.,  2011,
  \mn@doi [ApJ] {10.1088/0004-637X/733/1/31}, \href
  {https://ui.adsabs.harvard.edu/abs/2011ApJ...733...31H} {733, 31}

\bibitem[\protect\citeauthoryear{{Hao}, {Huang}, {Xia}, {Zheng}, {Jiang}  \&
  {Li}}{{Hao} et~al.}{2018}]{Hao2018}
{Hao} C.-N.,  {Huang} J.-S.,  {Xia} X.,  {Zheng} X.,  {Jiang} C.,   {Li} C.,
  2018, \mn@doi [ApJ] {10.3847/1538-4357/aad80b}, \href
  {https://ui.adsabs.harvard.edu/abs/2018ApJ...864..145H} {864, 145}

\bibitem[\protect\citeauthoryear{{Hashimoto} et~al.}{{Hashimoto}
  et~al.}{2015}]{Hashimoto2015}
{Hashimoto} T.,  et~al., 2015, \mn@doi [ApJ] {10.1088/0004-637X/812/2/157},
  \href {https://ui.adsabs.harvard.edu/abs/2015ApJ...812..157H} {812, 157}

\bibitem[\protect\citeauthoryear{{H{\"a}ussler} et~al.}{{H{\"a}ussler}
  et~al.}{2007}]{Haussler2007}
{H{\"a}ussler} B.,  et~al., 2007, \mn@doi [ApJS] {10.1086/518836}, \href
  {https://ui.adsabs.harvard.edu/abs/2007ApJS..172..615H} {172, 615}

\bibitem[\protect\citeauthoryear{{Hayes}}{{Hayes}}{2015}]{Hayes2015}
{Hayes} M.,  2015, \mn@doi [PASA] {10.1017/pasa.2015.25}, \href
  {http://adsabs.harvard.edu/abs/2015PASA...32...27H} {32, e027}

\bibitem[\protect\citeauthoryear{{Hayes} et~al.}{{Hayes}
  et~al.}{2010}]{Hayes2010}
{Hayes} M.,  et~al., 2010, \mn@doi [Nature] {10.1038/nature08881}, \href
  {http://adsabs.harvard.edu/abs/2010Natur.464..562H} {464, 562}

\bibitem[\protect\citeauthoryear{{Hayes}, {Schaerer}, {{\"O}stlin},
  {Mas-Hesse}, {Atek}  \& {Kunth}}{{Hayes} et~al.}{2011}]{Hayes2011}
{Hayes} M.,  {Schaerer} D.,  {{\"O}stlin} G.,  {Mas-Hesse} J.~M.,  {Atek} H.,
  {Kunth} D.,  2011, \mn@doi [ApJ] {10.1088/0004-637X/730/1/8}, \href
  {http://adsabs.harvard.edu/abs/2011ApJ...730....8H} {730, 8}

\bibitem[\protect\citeauthoryear{{Hayes}, {Runnholm}, {Gronke}  \&
  {Scarlata}}{{Hayes} et~al.}{2021}]{Hayes2020}
{Hayes} M.~J.,  {Runnholm} A.,  {Gronke} M.,   {Scarlata} C.,  2021, \mn@doi
  [ApJ] {10.3847/1538-4357/abd246}, \href
  {https://ui.adsabs.harvard.edu/abs/2021ApJ...908...36H} {908, 36}

\bibitem[\protect\citeauthoryear{{Heckman}, {Armus}  \& {Miley}}{{Heckman}
  et~al.}{1990}]{Heckman1990}
{Heckman} T.~M.,  {Armus} L.,   {Miley} G.~K.,  1990, \mn@doi [ApJS]
  {10.1086/191522}, \href
  {https://ui.adsabs.harvard.edu/abs/1990ApJS...74..833H} {74, 833}

\bibitem[\protect\citeauthoryear{{Heckman}, {Sembach}, {Meurer}, {Leitherer},
  {Calzetti}  \& {Martin}}{{Heckman} et~al.}{2001}]{Heckman2001}
{Heckman} T.~M.,  {Sembach} K.~R.,  {Meurer} G.~R.,  {Leitherer} C.,
  {Calzetti} D.,   {Martin} C.~L.,  2001, \mn@doi [ApJ] {10.1086/322475}, \href
  {https://ui.adsabs.harvard.edu/abs/2001ApJ...558...56H} {558, 56}

\bibitem[\protect\citeauthoryear{{Henry}, {Scarlata}, {Martin}  \&
  {Erb}}{{Henry} et~al.}{2015}]{Henry2015}
{Henry} A.,  {Scarlata} C.,  {Martin} C.~L.,   {Erb} D.,  2015, \mn@doi [ApJ]
  {10.1088/0004-637X/809/1/19}, \href
  {http://adsabs.harvard.edu/abs/2015ApJ...809...19H} {809, 19}

\bibitem[\protect\citeauthoryear{{Herenz} et~al.}{{Herenz}
  et~al.}{2016}]{Herenz2016}
{Herenz} E.~C.,  et~al., 2016, \mn@doi [AAP] {10.1051/0004-6361/201527373},
  \href {https://ui.adsabs.harvard.edu/abs/2016A&A...587A..78H} {587, A78}

\bibitem[\protect\citeauthoryear{{Herenz} et~al.}{{Herenz}
  et~al.}{2019}]{Herenz2019}
{Herenz} E.~C.,  et~al., 2019, \mn@doi [AAP] {10.1051/0004-6361/201834164},
  \href {https://ui.adsabs.harvard.edu/abs/2019A&A...621A.107H} {621, A107}

\bibitem[\protect\citeauthoryear{{Hoag} et~al.}{{Hoag} et~al.}{2019}]{Hoag2019}
{Hoag} A.,  et~al., 2019, \mn@doi [MNRAS] {10.1093/mnras/stz1768}, \href
  {https://ui.adsabs.harvard.edu/abs/2019MNRAS.488..706H} {488, 706}

\bibitem[\protect\citeauthoryear{{Hogarth} et~al.}{{Hogarth}
  et~al.}{2020}]{Hogarth2020}
{Hogarth} L.,  et~al., 2020, \mn@doi [MNRAS] {10.1093/mnras/staa851}, \href
  {https://ui.adsabs.harvard.edu/abs/2020MNRAS.494.3541H} {494, 3541}

\bibitem[\protect\citeauthoryear{{Horne}}{{Horne}}{1986}]{Horne1986}
{Horne} K.,  1986, \mn@doi [\pasp] {10.1086/131801}, \href
  {https://ui.adsabs.harvard.edu/abs/1986PASP...98..609H} {98, 609}

\bibitem[\protect\citeauthoryear{{Hunter}}{{Hunter}}{2007}]{Hunter2007}
{Hunter} J.~D.,  2007, \mn@doi [Computing in Science and Engineering]
  {10.1109/MCSE.2007.55}, \href
  {http://adsabs.harvard.edu/abs/2007CSE.....9...90H} {9, 90}

\bibitem[\protect\citeauthoryear{{Ilbert} et~al.}{{Ilbert}
  et~al.}{2009}]{Ilbert2009}
{Ilbert} O.,  et~al., 2009, \mn@doi [ApJ] {10.1088/0004-637X/690/2/1236}, \href
  {http://adsabs.harvard.edu/abs/2009ApJ...690.1236I} {690, 1236}

\bibitem[\protect\citeauthoryear{{Inoue}}{{Inoue}}{2001}]{Inoue2001}
{Inoue} A.~K.,  2001, \mn@doi [AJ] {10.1086/323095}, \href
  {https://ui.adsabs.harvard.edu/abs/2001AJ....122.1788I} {122, 1788}

\bibitem[\protect\citeauthoryear{{Inoue}, {Shimizu}, {Iwata}  \&
  {Tanaka}}{{Inoue} et~al.}{2014}]{Inoue2014}
{Inoue} A.~K.,  {Shimizu} I.,  {Iwata} I.,   {Tanaka} M.,  2014, \mn@doi
  [MNRAS] {10.1093/mnras/stu936}, \href
  {https://ui.adsabs.harvard.edu/abs/2014MNRAS.442.1805I} {442, 1805}

\bibitem[\protect\citeauthoryear{{Iwata}, {Inoue}, {Micheva}, {Matsuda}  \&
  {Yamada}}{{Iwata} et~al.}{2019}]{Iwata2019}
{Iwata} I.,  {Inoue} A.~K.,  {Micheva} G.,  {Matsuda} Y.,   {Yamada} T.,  2019,
  \mn@doi [MNRAS] {10.1093/mnras/stz2081}, \href
  {https://ui.adsabs.harvard.edu/abs/2019MNRAS.488.5671I} {488, 5671}

\bibitem[\protect\citeauthoryear{{Izotov}, {Worseck}, {Schaerer}, {Guseva},
  {Thuan}, {Fricke}  \& {Orlitov{\'a}}}{{Izotov} et~al.}{2018}]{Izotov2018}
{Izotov} Y.~I.,  {Worseck} G.,  {Schaerer} D.,  {Guseva} N.~G.,  {Thuan} T.~X.,
   {Fricke} Verhamme A.,   {Orlitov{\'a}} I.,  2018, \mn@doi [MNRAS]
  {10.1093/mnras/sty1378}, \href
  {https://ui.adsabs.harvard.edu/abs/2018MNRAS.478.4851I} {478, 4851}

\bibitem[\protect\citeauthoryear{{Jarvis} et~al.}{{Jarvis}
  et~al.}{2013}]{Jarvis2013}
{Jarvis} M.~J.,  et~al., 2013, \mn@doi [MNRAS] {10.1093/mnras/sts118}, \href
  {https://ui.adsabs.harvard.edu/abs/2013MNRAS.428.1281J} {428, 1281}

\bibitem[\protect\citeauthoryear{{Jaskot} \& {Oey}}{{Jaskot} \&
  {Oey}}{2013}]{Jaskot2013}
{Jaskot} A.~E.,  {Oey} M.~S.,  2013, \mn@doi [ApJ]
  {10.1088/0004-637X/766/2/91}, \href
  {https://ui.adsabs.harvard.edu/abs/2013ApJ...766...91J} {766, 91}

\bibitem[\protect\citeauthoryear{{Jaskot} \& {Oey}}{{Jaskot} \&
  {Oey}}{2014}]{JaskotOey2014}
{Jaskot} A.~E.,  {Oey} M.~S.,  2014, \mn@doi [ApJL]
  {10.1088/2041-8205/791/2/L19}, \href
  {https://ui.adsabs.harvard.edu/abs/2014ApJ...791L..19J} {791, L19}

\bibitem[\protect\citeauthoryear{{Jaskot}, {Dowd}, {Oey}, {Scarlata}  \&
  {McKinney}}{{Jaskot} et~al.}{2019}]{Jaskot2019}
{Jaskot} A.~E.,  {Dowd} T.,  {Oey} M.~S.,  {Scarlata} C.,   {McKinney} J.,
  2019, \mn@doi [ApJ] {10.3847/1538-4357/ab3d3b}, \href
  {https://ui.adsabs.harvard.edu/abs/2019ApJ...885...96J} {885, 96}

\bibitem[\protect\citeauthoryear{Jones, Oliphant, Peterson  et~al.}{Jones
  et~al.}{2001}]{Scipy}
Jones E.,  Oliphant T.,  Peterson P.,   et~al., 2001, {SciPy}: Open source
  scientific tools for {Python}, \url {http://www.scipy.org/}

\bibitem[\protect\citeauthoryear{{Jung} et~al.,}{{Jung}
  et~al.}{2018}]{Jung2018}
{Jung} I.,  et~al., 2018, \mn@doi [ApJ] {10.3847/1538-4357/aad686}, \href
  {https://ui.adsabs.harvard.edu/abs/2018ApJ...864..103J} {864, 103}

\bibitem[\protect\citeauthoryear{{Kakiichi} \& {Gronke}}{{Kakiichi} \&
  {Gronke}}{2019}]{KakiichiGronke2019}
{Kakiichi} K.,  {Gronke} M.,  2019, arXiv e-prints, \href
  {https://ui.adsabs.harvard.edu/abs/2019arXiv190502480K} {p. arXiv:1905.02480}

\bibitem[\protect\citeauthoryear{{Kashikawa} et~al.}{{Kashikawa}
  et~al.}{2011}]{Kashikawa2011}
{Kashikawa} N.,  et~al., 2011, \mn@doi [ApJ] {10.1088/0004-637X/734/2/119},
  \href {http://adsabs.harvard.edu/abs/2011ApJ...734..119K} {734, 119}

\bibitem[\protect\citeauthoryear{{Kashino} et~al.}{{Kashino}
  et~al.}{2013}]{Kashino2013}
{Kashino} D.,  et~al., 2013, \mn@doi [ApJL] {10.1088/2041-8205/777/1/L8}, \href
  {https://ui.adsabs.harvard.edu/abs/2013ApJ...777L...8K} {777, L8}

\bibitem[\protect\citeauthoryear{{Kashino} et~al.}{{Kashino}
  et~al.}{2017}]{Kashino2017}
{Kashino} D.,  et~al., 2017, \mn@doi [ApJ] {10.3847/1538-4357/835/1/88}, \href
  {https://ui.adsabs.harvard.edu/abs/2017ApJ...835...88K} {835, 88}

\bibitem[\protect\citeauthoryear{{Kauffmann} et~al.}{{Kauffmann}
  et~al.}{2003}]{Kauffmann2003}
{Kauffmann} G.,  et~al., 2003, \mn@doi [MNRAS]
  {10.1111/j.1365-2966.2003.07154.x}, \href
  {https://ui.adsabs.harvard.edu/abs/2003MNRAS.346.1055K} {346, 1055}

\bibitem[\protect\citeauthoryear{{Kausch} et~al.}{{Kausch}
  et~al.}{2015}]{molecfit2}
{Kausch} W.,  et~al., 2015, \mn@doi [AAP] {10.1051/0004-6361/201423909}, \href
  {https://ui.adsabs.harvard.edu/abs/2015A&A...576A..78K} {576, A78}

\bibitem[\protect\citeauthoryear{{Keenan}, {Feibelman}  \&
  {Berrington}}{{Keenan} et~al.}{1992}]{Keenan1992}
{Keenan} F.~P.,  {Feibelman} W.~A.,   {Berrington} K.~A.,  1992, \mn@doi [ApJ]
  {10.1086/171220}, \href
  {https://ui.adsabs.harvard.edu/abs/1992ApJ...389..443K} {389, 443}

\bibitem[\protect\citeauthoryear{{Kehrig}, {V{\'\i}lchez}, {P{\'e}rez-Montero},
  {Iglesias-P{\'a}ramo}, {Brinchmann}, {Kunth}, {Durret}  \& {Bayo}}{{Kehrig}
  et~al.}{2015}]{Kehrig2015}
{Kehrig} C.,  {V{\'\i}lchez} J.~M.,  {P{\'e}rez-Montero} E.,
  {Iglesias-P{\'a}ramo} J.,  {Brinchmann} J.,  {Kunth} D.,  {Durret} F.,
  {Bayo} F.~M.,  2015, \mn@doi [ApJL] {10.1088/2041-8205/801/2/L28}, \href
  {https://ui.adsabs.harvard.edu/abs/2015ApJ...801L..28K} {801, L28}

\bibitem[\protect\citeauthoryear{{Kennicutt} \& {Evans}}{{Kennicutt} \&
  {Evans}}{2012}]{KennicuttEvans2012}
{Kennicutt} R.~C.,  {Evans} N.~J.,  2012, \mn@doi [ARAA]
  {10.1146/annurev-astro-081811-125610}, \href
  {http://adsabs.harvard.edu/abs/2012ARA%26A..50..531K} {50, 531}

\bibitem[\protect\citeauthoryear{{Kewley} \& {Dopita}}{{Kewley} \&
  {Dopita}}{2002}]{KewleyDopita2002}
{Kewley} L.~J.,  {Dopita} M.~A.,  2002, \mn@doi [ApJS] {10.1086/341326}, \href
  {https://ui.adsabs.harvard.edu/abs/2002ApJS..142...35K} {142, 35}

\bibitem[\protect\citeauthoryear{{Khostovan}, {Sobral}, {Mobasher}, {Best},
  {Smail}, {Stott}, {Hemmati}  \& {Nayyeri}}{{Khostovan}
  et~al.}{2015}]{Khostovan2015}
{Khostovan} A.~A.,  {Sobral} D.,  {Mobasher} B.,  {Best} P.~N.,  {Smail} I.,
  {Stott} J.~P.,  {Hemmati} S.,   {Nayyeri} H.,  2015, \mn@doi [MNRAS]
  {10.1093/mnras/stv1474}, \href
  {https://ui.adsabs.harvard.edu/abs/2015MNRAS.452.3948K} {452, 3948}

\bibitem[\protect\citeauthoryear{{Kimm}, {Blaizot}, {Garel}, {Michel-Dansac},
  {Katz}, {Rosdahl}, {Verhamme}  \& {Haehnelt}}{{Kimm} et~al.}{2019}]{Kimm2019}
{Kimm} T.,  {Blaizot} J.,  {Garel} T.,  {Michel-Dansac} L.,  {Katz} H.,
  {Rosdahl} J.,  {Verhamme} A.,   {Haehnelt} M.,  2019, \mn@doi [MNRAS]
  {10.1093/mnras/stz989}, \href
  {https://ui.adsabs.harvard.edu/abs/2019MNRAS.486.2215K} {486, 2215}

\bibitem[\protect\citeauthoryear{{Koekemoer} et~al.}{{Koekemoer}
  et~al.}{2007}]{Koekemoer2007}
{Koekemoer} A.~M.,  et~al., 2007, \mn@doi [ApJS] {10.1086/520086}, \href
  {http://adsabs.harvard.edu/abs/2007ApJS..172..196K} {172, 196}

\bibitem[\protect\citeauthoryear{{Konno}, {Ouchi}, {Nakajima}, {Duval},
  {Kusakabe}, {Ono}  \& {Shimasaku}}{{Konno} et~al.}{2016}]{Konno2016}
{Konno} A.,  {Ouchi} M.,  {Nakajima} K.,  {Duval} F.,  {Kusakabe} H.,  {Ono}
  Y.,   {Shimasaku} K.,  2016, \mn@doi [ApJ] {10.3847/0004-637X/823/1/20},
  \href {https://ui.adsabs.harvard.edu/abs/2016ApJ...823...20K} {823, 20}

\bibitem[\protect\citeauthoryear{{Konno} et~al.}{{Konno}
  et~al.}{2018}]{Konno2018}
{Konno} A.,  et~al., 2018, \mn@doi [PASJ] {10.1093/pasj/psx131}, \href
  {http://adsabs.harvard.edu/abs/2018PASJ...70S..16K} {70, S16}

\bibitem[\protect\citeauthoryear{{Kulas}, {Shapley}, {Kollmeier}, {Zheng},
  {Steidel}  \& {Hainline}}{{Kulas} et~al.}{2012}]{Kulas2012}
{Kulas} K.~R.,  {Shapley} A.~E.,  {Kollmeier} J.~A.,  {Zheng} Z.,  {Steidel}
  C.~C.,   {Hainline} K.~N.,  2012, \mn@doi [ApJ] {10.1088/0004-637X/745/1/33},
  \href {https://ui.adsabs.harvard.edu/abs/2012ApJ...745...33K} {745, 33}

\bibitem[\protect\citeauthoryear{{Kusakabe} et~al.}{{Kusakabe}
  et~al.}{2018}]{Kusakabe2018}
{Kusakabe} H.,  et~al., 2018, \mn@doi [PASJ] {10.1093/pasj/psx148}, \href
  {https://ui.adsabs.harvard.edu/abs/2018PASJ...70....4K} {70, 4}

\bibitem[\protect\citeauthoryear{{Laigle} et~al.}{{Laigle}
  et~al.}{2016}]{Laigle2016}
{Laigle} C.,  et~al., 2016, \mn@doi [ApJS] {10.3847/0067-0049/224/2/24}, \href
  {http://adsabs.harvard.edu/abs/2016ApJS..224...24L} {224, 24}

\bibitem[\protect\citeauthoryear{{Laursen}, {Sommer-Larsen}  \&
  {Razoumov}}{{Laursen} et~al.}{2011}]{Laursen2011}
{Laursen} P.,  {Sommer-Larsen} J.,   {Razoumov} A.~O.,  2011, \mn@doi [ApJ]
  {10.1088/0004-637X/728/1/52}, \href
  {http://adsabs.harvard.edu/abs/2011ApJ...728...52L} {728, 52}

\bibitem[\protect\citeauthoryear{{Lawrence} et~al.}{{Lawrence}
  et~al.}{2007}]{Lawrence2007}
{Lawrence} A.,  et~al., 2007, \mn@doi [MNRAS]
  {10.1111/j.1365-2966.2007.12040.x}, \href
  {http://adsabs.harvard.edu/abs/2007MNRAS.379.1599L} {379, 1599}

\bibitem[\protect\citeauthoryear{{Leitherer} et~al.}{{Leitherer}
  et~al.}{1999}]{Leitherer1999}
{Leitherer} C.,  et~al., 1999, \mn@doi [ApJS] {10.1086/313233}, \href
  {https://ui.adsabs.harvard.edu/abs/1999ApJS..123....3L} {123, 3}

\bibitem[\protect\citeauthoryear{{Levesque} \& {Richardson}}{{Levesque} \&
  {Richardson}}{2014}]{Levesque2014}
{Levesque} E.~M.,  {Richardson} M. L.~A.,  2014, \mn@doi [ApJ]
  {10.1088/0004-637X/780/1/100}, \href
  {https://ui.adsabs.harvard.edu/abs/2014ApJ...780..100L} {780, 100}

\bibitem[\protect\citeauthoryear{{Ma} et~al.}{{Ma} et~al.}{2016}]{Ma2016}
{Ma} X.,  et~al., 2016, \mn@doi [MNRAS] {10.1093/mnras/stw941}, \href
  {https://ui.adsabs.harvard.edu/abs/2016MNRAS.459.3614M} {459, 3614}

\bibitem[\protect\citeauthoryear{{Madau}}{{Madau}}{1995}]{Madau1995}
{Madau} P.,  1995, \mn@doi [ApJ] {10.1086/175332}, \href
  {https://ui.adsabs.harvard.edu/abs/1995ApJ...441...18M} {441, 18}

\bibitem[\protect\citeauthoryear{{Mainali}, {Kollmeier}, {Stark}, {Simcoe},
  {Walth}, {Newman}  \& {Miller}}{{Mainali} et~al.}{2017}]{Mainali2017}
{Mainali} R.,  {Kollmeier} J.~A.,  {Stark} D.~P.,  {Simcoe} R.~A.,  {Walth} G.,
   {Newman} A.~B.,   {Miller} D.~R.,  2017, \mn@doi [ApJL]
  {10.3847/2041-8213/836/1/L14}, \href
  {http://adsabs.harvard.edu/abs/2017ApJ...836L..14M} {836, L14}

\bibitem[\protect\citeauthoryear{{Maiolino} \& {Mannucci}}{{Maiolino} \&
  {Mannucci}}{2019}]{MaiolinoMannucci2019}
{Maiolino} R.,  {Mannucci} F.,  2019, \mn@doi [AAPR]
  {10.1007/s00159-018-0112-2}, \href
  {https://ui.adsabs.harvard.edu/abs/2019A&ARv..27....3M} {27, 3}

\bibitem[\protect\citeauthoryear{{Marchi} et~al.}{{Marchi}
  et~al.}{2018}]{Marchi2018}
{Marchi} F.,  et~al., 2018, \mn@doi [AAP] {10.1051/0004-6361/201732133}, \href
  {https://ui.adsabs.harvard.edu/abs/2018A&A...614A..11M} {614, A11}

\bibitem[\protect\citeauthoryear{{Marchi} et~al.}{{Marchi}
  et~al.}{2019}]{Marchi2019}
{Marchi} F.,  et~al., 2019, \mn@doi [AAP] {10.1051/0004-6361/201935495}, \href
  {https://ui.adsabs.harvard.edu/abs/2019A&A...631A..19M} {631, A19}

\bibitem[\protect\citeauthoryear{{Marino} et~al.}{{Marino}
  et~al.}{2018}]{Marino2018}
{Marino} R.~A.,  et~al., 2018, \mn@doi [ApJ] {10.3847/1538-4357/aab6aa}, \href
  {https://ui.adsabs.harvard.edu/abs/2018ApJ...859...53M} {859, 53}

\bibitem[\protect\citeauthoryear{{Martin}, {Dijkstra}, {Henry}, {Soto},
  {Danforth}  \& {Wong}}{{Martin} et~al.}{2015}]{Martin2015}
{Martin} C.~L.,  {Dijkstra} M.,  {Henry} A.,  {Soto} K.~T.,  {Danforth} C.~W.,
   {Wong} J.,  2015, \mn@doi [ApJ] {10.1088/0004-637X/803/1/6}, \href
  {https://ui.adsabs.harvard.edu/abs/2015ApJ...803....6M} {803, 6}

\bibitem[\protect\citeauthoryear{{Mas-Ribas}, {Hennawi}, {Dijkstra}, {Davies},
  {Stern}  \& {Rix}}{{Mas-Ribas} et~al.}{2017}]{MasRibas2017}
{Mas-Ribas} L.,  {Hennawi} J.~F.,  {Dijkstra} M.,  {Davies} F.~B.,  {Stern} J.,
    {Rix} H.-W.,  2017, \mn@doi [ApJ] {10.3847/1538-4357/aa8328}, \href
  {http://adsabs.harvard.edu/abs/2017ApJ...846...11M} {846, 11}

\bibitem[\protect\citeauthoryear{{Maseda} et~al.}{{Maseda}
  et~al.}{2018}]{Maseda2018}
{Maseda} M.~V.,  et~al., 2018, \mn@doi [ApJL] {10.3847/2041-8213/aade4b}, \href
  {https://ui.adsabs.harvard.edu/abs/2018ApJ...865L...1M} {865, L1}

\bibitem[\protect\citeauthoryear{{Maseda} et~al.}{{Maseda}
  et~al.}{2020}]{Maseda2020}
{Maseda} M.~V.,  et~al., 2020, \mn@doi [MNRAS] {10.1093/mnras/staa622}, \href
  {https://ui.adsabs.harvard.edu/abs/2020MNRAS.493.5120M} {493, 5120}

\bibitem[\protect\citeauthoryear{{Mason}, {Treu}, {Dijkstra}, {Mesinger},
  {Trenti}, {Pentericci}, {de Barros}  \& {Vanzella}}{{Mason}
  et~al.}{2018}]{Mason2018b}
{Mason} C.~A.,  {Treu} T.,  {Dijkstra} M.,  {Mesinger} A.,  {Trenti} M.,
  {Pentericci} L.,  {de Barros} S.,   {Vanzella} E.,  2018, \mn@doi [ApJ]
  {10.3847/1538-4357/aab0a7}, \href
  {https://ui.adsabs.harvard.edu/abs/2018ApJ...856....2M} {856, 2}

\bibitem[\protect\citeauthoryear{{Matthee} \& {Schaye}}{{Matthee} \&
  {Schaye}}{2018}]{MattheeSchaye2018}
{Matthee} J.,  {Schaye} J.,  2018, \mn@doi [MNRAS] {10.1093/mnrasl/sly093},
  \href {https://ui.adsabs.harvard.edu/abs/2018MNRAS.479L..34M} {479, L34}

\bibitem[\protect\citeauthoryear{{Matthee} \& {Schaye}}{{Matthee} \&
  {Schaye}}{2019}]{MattheeSchaye2019}
{Matthee} J.,  {Schaye} J.,  2019, \mn@doi [MNRAS] {10.1093/mnras/stz030},
  \href {https://ui.adsabs.harvard.edu/abs/2019MNRAS.484..915M} {484, 915}

\bibitem[\protect\citeauthoryear{{Matthee}, {Sobral}, {Santos},
  {R{\"o}ttgering}, {Darvish}  \& {Mobasher}}{{Matthee}
  et~al.}{2015}]{Matthee2015}
{Matthee} J.,  {Sobral} D.,  {Santos} S.,  {R{\"o}ttgering} H.,  {Darvish} B.,
   {Mobasher} B.,  2015, \mn@doi [MNRAS] {10.1093/mnras/stv947}, \href
  {http://adsabs.harvard.edu/abs/2015MNRAS.451..400M} {451, 400}

\bibitem[\protect\citeauthoryear{{Matthee}, {Sobral}, {Oteo}, {Best}, {Smail},
  {R{\"o}ttgering}  \& {Paulino-Afonso}}{{Matthee} et~al.}{2016}]{Matthee2016}
{Matthee} J.,  {Sobral} D.,  {Oteo} I.,  {Best} P.,  {Smail} I.,
  {R{\"o}ttgering} H.,   {Paulino-Afonso} A.,  2016, \mn@doi [MNRAS]
  {10.1093/mnras/stw322}, \href
  {http://adsabs.harvard.edu/abs/2016MNRAS.458..449M} {458, 449}

\bibitem[\protect\citeauthoryear{{Matthee}, {Sobral}, {Best}, {Khostovan},
  {Oteo}, {Bouwens}  \& {R{\"o}ttgering}}{{Matthee}
  et~al.}{2017a}]{Matthee2017GALEX}
{Matthee} J.,  {Sobral} D.,  {Best} P.,  {Khostovan} A.~A.,  {Oteo} I.,
  {Bouwens} R.,   {R{\"o}ttgering} H.,  2017a, \mn@doi [MNRAS]
  {10.1093/mnras/stw2973}, \href
  {http://adsabs.harvard.edu/abs/2017MNRAS.465.3637M} {465, 3637}

\bibitem[\protect\citeauthoryear{{Matthee}, {Sobral}, {Best}, {Smail}, {Bian},
  {Darvish}, {R{\"o}ttgering}  \& {Fan}}{{Matthee}
  et~al.}{2017b}]{Matthee2017Bootes}
{Matthee} J.,  {Sobral} D.,  {Best} P.,  {Smail} I.,  {Bian} F.,  {Darvish} B.,
   {R{\"o}ttgering} H.,   {Fan} X.,  2017b, \mn@doi [MNRAS]
  {10.1093/mnras/stx1569}, \href
  {https://ui.adsabs.harvard.edu/abs/2017MNRAS.471..629M} {471, 629}

\bibitem[\protect\citeauthoryear{{Matthee}, {Sobral}, {Gronke},
  {Paulino-Afonso}, {Stefanon}  \& {R{\"o}ttgering}}{{Matthee}
  et~al.}{2018}]{Matthee2018}
{Matthee} J.,  {Sobral} D.,  {Gronke} M.,  {Paulino-Afonso} A.,  {Stefanon} M.,
    {R{\"o}ttgering} H.,  2018, \mn@doi [AAP] {10.1051/0004-6361/201833528},
  \href {http://adsabs.harvard.edu/abs/2018A%26A...619A.136M} {619, A136}

\bibitem[\protect\citeauthoryear{{McCracken} et~al.}{{McCracken}
  et~al.}{2012}]{McCracken2012}
{McCracken} H.~J.,  et~al., 2012, \mn@doi [AAP] {10.1051/0004-6361/201219507},
  \href {http://adsabs.harvard.edu/abs/2012A%26A...544A.156M} {544, A156}

\bibitem[\protect\citeauthoryear{{M}c{K}inney}{{M}c{K}inney}{2010}]{pandas}
{M}c{K}inney W.,  2010, in {S}t\'efan van~der {W}alt {J}arrod {M}illman eds,
  {P}roceedings of the 9th {P}ython in {S}cience {C}onference. pp 56 -- 61,
  \mn@doi{10.25080/Majora-92bf1922-00a}

\bibitem[\protect\citeauthoryear{{McKinney}, {Jaskot}, {Oey}, {Yun}, {Dowd}  \&
  {Lowenthal}}{{McKinney} et~al.}{2019}]{McKinney2019}
{McKinney} J.~H.,  {Jaskot} A.~E.,  {Oey} M.~S.,  {Yun} M.~S.,  {Dowd} T.,
  {Lowenthal} J.~D.,  2019, \mn@doi [ApJ] {10.3847/1538-4357/ab08eb}, \href
  {https://ui.adsabs.harvard.edu/abs/2019ApJ...874...52M} {874, 52}

\bibitem[\protect\citeauthoryear{{Mehta} et~al.}{{Mehta}
  et~al.}{2018}]{Mehta2018}
{Mehta} V.,  et~al., 2018, \mn@doi [ApJS] {10.3847/1538-4365/aab60c}, \href
  {https://ui.adsabs.harvard.edu/abs/2018ApJS..235...36M} {235, 36}

\bibitem[\protect\citeauthoryear{{Micheva}, {Iwata}, {Inoue}, {Matsuda},
  {Yamada}  \& {Hayashino}}{{Micheva} et~al.}{2017}]{Micheva2017}
{Micheva} G.,  {Iwata} I.,  {Inoue} A.~K.,  {Matsuda} Y.,  {Yamada} T.,
  {Hayashino} T.,  2017, \mn@doi [MNRAS] {10.1093/mnras/stw2700}, \href
  {https://ui.adsabs.harvard.edu/abs/2017MNRAS.465..316M} {465, 316}

\bibitem[\protect\citeauthoryear{{Modigliani} et~al.}{{Modigliani}
  et~al.}{2010}]{Modigliani2010}
{Modigliani} A.,  et~al., 2010, in Observatory Operations: Strategies,
  Processes, and Systems III. p. 773728, \mn@doi{10.1117/12.857211}

\bibitem[\protect\citeauthoryear{{Murphy} et~al.}{{Murphy}
  et~al.}{2011}]{Murphy2011}
{Murphy} E.~J.,  et~al., 2011, \mn@doi [ApJ] {10.1088/0004-637X/737/2/67},
  \href {https://ui.adsabs.harvard.edu/abs/2011ApJ...737...67M} {737, 67}

\bibitem[\protect\citeauthoryear{{Muzahid} et~al.}{{Muzahid}
  et~al.}{2020}]{Muzahid2020}
{Muzahid} S.,  et~al., 2020, \mn@doi [MNRAS] {10.1093/mnras/staa1347}, \href
  {https://ui.adsabs.harvard.edu/abs/2020MNRAS.496.1013M} {496, 1013}

\bibitem[\protect\citeauthoryear{{Naidu} et~al.}{{Naidu}
  et~al.}{2017}]{Naidu2017}
{Naidu} R.~P.,  et~al., 2017, \mn@doi [ApJ] {10.3847/1538-4357/aa8863}, \href
  {https://ui.adsabs.harvard.edu/abs/2017ApJ...847...12N} {847, 12}

\bibitem[\protect\citeauthoryear{{Naidu}, {Tacchella}, {Mason}, {Bose}, {Oesch}
   \& {Conroy}}{{Naidu} et~al.}{2020}]{Naidu2019}
{Naidu} R.~P.,  {Tacchella} S.,  {Mason} C.~A.,  {Bose} S.,  {Oesch} P.~A.,
  {Conroy} C.,  2020, \mn@doi [ApJ] {10.3847/1538-4357/ab7cc9}, \href
  {https://ui.adsabs.harvard.edu/abs/2020ApJ...892..109N} {892, 109}

\bibitem[\protect\citeauthoryear{{Nakajima} \& {Ouchi}}{{Nakajima} \&
  {Ouchi}}{2014}]{NakajimaOuchi2014}
{Nakajima} K.,  {Ouchi} M.,  2014, \mn@doi [MNRAS] {10.1093/mnras/stu902},
  \href {https://ui.adsabs.harvard.edu/abs/2014MNRAS.442..900N} {442, 900}

\bibitem[\protect\citeauthoryear{{Nakajima} et~al.}{{Nakajima}
  et~al.}{2012}]{Nakajima2012}
{Nakajima} K.,  et~al., 2012, \mn@doi [ApJ] {10.1088/0004-637X/745/1/12}, \href
  {http://adsabs.harvard.edu/abs/2012ApJ...745...12N} {745, 12}

\bibitem[\protect\citeauthoryear{{Nakajima}, {Ouchi}, {Shimasaku}, {Hashimoto},
  {Ono}  \& {Lee}}{{Nakajima} et~al.}{2013}]{Nakajima2013}
{Nakajima} K.,  {Ouchi} M.,  {Shimasaku} K.,  {Hashimoto} T.,  {Ono} Y.,
  {Lee} J.~C.,  2013, \mn@doi [ApJ] {10.1088/0004-637X/769/1/3}, \href
  {https://ui.adsabs.harvard.edu/abs/2013ApJ...769....3N} {769, 3}

\bibitem[\protect\citeauthoryear{{Nakajima}, {Fletcher}, {Ellis}, {Robertson}
  \& {Iwata}}{{Nakajima} et~al.}{2018a}]{Nakajima2018b}
{Nakajima} K.,  {Fletcher} T.,  {Ellis} R.~S.,  {Robertson} B.~E.,   {Iwata}
  I.,  2018a, \mn@doi [MNRAS] {10.1093/mnras/sty750}, \href
  {https://ui.adsabs.harvard.edu/abs/2018MNRAS.477.2098N} {477, 2098}

\bibitem[\protect\citeauthoryear{{Nakajima} et~al.}{{Nakajima}
  et~al.}{2018b}]{Nakajima2018}
{Nakajima} K.,  et~al., 2018b, \mn@doi [AAP] {10.1051/0004-6361/201731935},
  \href {https://ui.adsabs.harvard.edu/abs/2018A&A...612A..94N} {612, A94}

\bibitem[\protect\citeauthoryear{{Nanayakkara} et~al.}{{Nanayakkara}
  et~al.}{2019}]{Nanayakkara2019}
{Nanayakkara} T.,  et~al., 2019, \mn@doi [AAP] {10.1051/0004-6361/201834565},
  \href {https://ui.adsabs.harvard.edu/abs/2019A&A...624A..89N} {624, A89}

\bibitem[\protect\citeauthoryear{{Neufeld}}{{Neufeld}}{1990}]{Neufeld1990}
{Neufeld} D.~A.,  1990, \mn@doi [ApJ] {10.1086/168375}, \href
  {https://ui.adsabs.harvard.edu/abs/1990ApJ...350..216N} {350, 216}

\bibitem[\protect\citeauthoryear{{Nilsson}, {Tapken}, {M{\o}ller}, {Freudling},
  {Fynbo}, {Meisenheimer}, {Laursen}  \& {{\"O}stlin}}{{Nilsson}
  et~al.}{2009}]{Nilsson2009}
{Nilsson} K.~K.,  {Tapken} C.,  {M{\o}ller} P.,  {Freudling} W.,  {Fynbo}
  J.~P.~U.,  {Meisenheimer} K.,  {Laursen} P.,   {{\"O}stlin} G.,  2009,
  \mn@doi [AAP] {10.1051/0004-6361/200810881}, \href
  {https://ui.adsabs.harvard.edu/abs/2009A&A...498...13N} {498, 13}

\bibitem[\protect\citeauthoryear{{Nomoto}, {Tominaga}, {Umeda}, {Kobayashi}  \&
  {Maeda}}{{Nomoto} et~al.}{2006}]{Nomoto2006}
{Nomoto} K.,  {Tominaga} N.,  {Umeda} H.,  {Kobayashi} C.,   {Maeda} K.,  2006,
  \mn@doi [Nuclear Physics A] {10.1016/j.nuclphysa.2006.05.008}, \href
  {http://adsabs.harvard.edu/abs/2006NuPhA.777..424N} {777, 424}

\bibitem[\protect\citeauthoryear{{Oesch} et~al.}{{Oesch}
  et~al.}{2015}]{Oesch2015}
{Oesch} P.~A.,  et~al., 2015, \mn@doi [ApJL] {10.1088/2041-8205/804/2/L30},
  \href {http://adsabs.harvard.edu/abs/2015ApJ...804L..30O} {804, L30}

\bibitem[\protect\citeauthoryear{{Oesch} et~al.,}{{Oesch}
  et~al.}{2018}]{Oesch2018}
{Oesch} P.~A.,  et~al., 2018, \mn@doi [ApJS] {10.3847/1538-4365/aacb30}, \href
  {https://ui.adsabs.harvard.edu/abs/2018ApJS..237...12O} {237, 12}

\bibitem[\protect\citeauthoryear{{Osterbrock}}{{Osterbrock}}{1989}]{Osterbrock1989}
{Osterbrock} D.~E.,  1989, {Astrophysics of gaseous nebulae and active galactic
  nuclei}

\bibitem[\protect\citeauthoryear{{{\"O}stlin}, {Marquart}, {Cumming}, {Fathi},
  {Bergvall}, {Adamo}, {Amram}  \& {Hayes}}{{{\"O}stlin}
  et~al.}{2015}]{Ostlin2015}
{{\"O}stlin} G.,  {Marquart} T.,  {Cumming} R.~J.,  {Fathi} K.,  {Bergvall} N.,
   {Adamo} A.,  {Amram} P.,   {Hayes} M.,  2015, \mn@doi [AAP]
  {10.1051/0004-6361/201323233}, \href
  {https://ui.adsabs.harvard.edu/abs/2015A&A...583A..55O} {583, A55}

\bibitem[\protect\citeauthoryear{{Ouchi} et~al.}{{Ouchi}
  et~al.}{2008}]{Ouchi2008}
{Ouchi} M.,  et~al., 2008, \mn@doi [ApJs] {10.1086/527673}, \href
  {http://adsabs.harvard.edu/abs/2008ApJS..176..301O} {176, 301}

\bibitem[\protect\citeauthoryear{{Ouchi} et~al.}{{Ouchi}
  et~al.}{2010}]{Ouchi2010}
{Ouchi} M.,  et~al., 2010, \mn@doi [ApJ] {10.1088/0004-637X/723/1/869}, \href
  {http://adsabs.harvard.edu/abs/2010ApJ...723..869O} {723, 869}

\bibitem[\protect\citeauthoryear{{Oyarz{\'u}n}, {Blanc}, {Gonz{\'a}lez},
  {Mateo}  \& {Bailey}}{{Oyarz{\'u}n} et~al.}{2017}]{Oyarzun2017}
{Oyarz{\'u}n} G.~A.,  {Blanc} G.~A.,  {Gonz{\'a}lez} V.,  {Mateo} M.,
  {Bailey} John~I. I.,  2017, \mn@doi [ApJ] {10.3847/1538-4357/aa7552}, \href
  {https://ui.adsabs.harvard.edu/abs/2017ApJ...843..133O} {843, 133}

\bibitem[\protect\citeauthoryear{{Parsa}, {Dunlop}, {McLure}  \&
  {Mortlock}}{{Parsa} et~al.}{2016}]{Parsa2016}
{Parsa} S.,  {Dunlop} J.~S.,  {McLure} R.~J.,   {Mortlock} A.,  2016, \mn@doi
  [MNRAS] {10.1093/mnras/stv2857}, \href
  {https://ui.adsabs.harvard.edu/abs/2016MNRAS.456.3194P} {456, 3194}

\bibitem[\protect\citeauthoryear{{Partridge} \& {Peebles}}{{Partridge} \&
  {Peebles}}{1967}]{PartridgePeebles1967}
{Partridge} R.~B.,  {Peebles} P.~J.~E.,  1967, \mn@doi [ApJ] {10.1086/149079},
  \href {http://adsabs.harvard.edu/abs/1967ApJ...147..868P} {147, 868}

\bibitem[\protect\citeauthoryear{{Patr{\'\i}cio} et~al.}{{Patr{\'\i}cio}
  et~al.}{2016}]{Patricio2016}
{Patr{\'\i}cio} V.,  et~al., 2016, \mn@doi [MNRAS] {10.1093/mnras/stv2859},
  \href {https://ui.adsabs.harvard.edu/abs/2016MNRAS.456.4191P} {456, 4191}

\bibitem[\protect\citeauthoryear{{Paulino-Afonso} et~al.,}{{Paulino-Afonso}
  et~al.}{2018}]{PaulinoAfonso2018}
{Paulino-Afonso} A.,  et~al., 2018, \mn@doi [MNRAS] {10.1093/mnras/sty281},
  \href {https://ui.adsabs.harvard.edu/abs/2018MNRAS.476.5479P} {476, 5479}

\bibitem[\protect\citeauthoryear{{Pe{\~n}a-Guerrero} \&
  {Leitherer}}{{Pe{\~n}a-Guerrero} \& {Leitherer}}{2013}]{PenaGuerrero2013}
{Pe{\~n}a-Guerrero} M.~A.,  {Leitherer} C.,  2013, \mn@doi [AJ]
  {10.1088/0004-6256/146/6/158}, \href
  {https://ui.adsabs.harvard.edu/abs/2013AJ....146..158P} {146, 158}

\bibitem[\protect\citeauthoryear{{P{\'e}rez-Montero}}{{P{\'e}rez-Montero}}{2014}]{PerezMontero2014}
{P{\'e}rez-Montero} E.,  2014, \mn@doi [MNRAS] {10.1093/mnras/stu753}, \href
  {https://ui.adsabs.harvard.edu/abs/2014MNRAS.441.2663P} {441, 2663}

\bibitem[\protect\citeauthoryear{{P{\'e}rez-Montero} \&
  {Amor{\'\i}n}}{{P{\'e}rez-Montero} \& {Amor{\'\i}n}}{2017}]{PerezMontero2017}
{P{\'e}rez-Montero} E.,  {Amor{\'\i}n} R.,  2017, \mn@doi [MNRAS]
  {10.1093/mnras/stx186}, \href
  {https://ui.adsabs.harvard.edu/abs/2017MNRAS.467.1287P} {467, 1287}

\bibitem[\protect\citeauthoryear{{Plat}, {Charlot}, {Bruzual}, {Feltre},
  {Vidal-Garc{\'\i}a}, {Morisset}, {Chevallard}  \& {Todt}}{{Plat}
  et~al.}{2019}]{Plat2019}
{Plat} A.,  {Charlot} S.,  {Bruzual} G.,  {Feltre} A.,  {Vidal-Garc{\'\i}a} A.,
   {Morisset} C.,  {Chevallard} J.,   {Todt} H.,  2019, \mn@doi [MNRAS]
  {10.1093/mnras/stz2616}, \href
  {https://ui.adsabs.harvard.edu/abs/2019MNRAS.490..978P} {490, 978}

\bibitem[\protect\citeauthoryear{{Puschnig} et~al.,}{{Puschnig}
  et~al.}{2020}]{Puschnig2020}
{Puschnig} J.,  et~al., 2020, \mn@doi [AAP] {10.1051/0004-6361/201936768},
  \href {https://ui.adsabs.harvard.edu/abs/2020A&A...644A..10P} {644, A10}

\bibitem[\protect\citeauthoryear{{Raiter}, {Schaerer}  \& {Fosbury}}{{Raiter}
  et~al.}{2010}]{Raiter2010}
{Raiter} A.,  {Schaerer} D.,   {Fosbury} R.~A.~E.,  2010, \mn@doi [AAP]
  {10.1051/0004-6361/201015236}, \href
  {https://ui.adsabs.harvard.edu/\#abs/2010A&A...523A..64R} {523, A64}

\bibitem[\protect\citeauthoryear{{Reddy} et~al.}{{Reddy}
  et~al.}{2015}]{Reddy2015}
{Reddy} N.~A.,  et~al., 2015, \mn@doi [ApJ] {10.1088/0004-637X/806/2/259},
  \href {https://ui.adsabs.harvard.edu/abs/2015ApJ...806..259R} {806, 259}

\bibitem[\protect\citeauthoryear{{Reddy}, {Steidel}, {Pettini}  \&
  {Bogosavljevi{\'c}}}{{Reddy} et~al.}{2016a}]{Reddy2016}
{Reddy} N.~A.,  {Steidel} C.~C.,  {Pettini} M.,   {Bogosavljevi{\'c}} M.,
  2016a, \mn@doi [ApJ] {10.3847/0004-637X/828/2/107}, \href
  {https://ui.adsabs.harvard.edu/abs/2016ApJ...828..107R} {828, 107}

\bibitem[\protect\citeauthoryear{{Reddy}, {Steidel}, {Pettini},
  {Bogosavljevi{\'c}}  \& {Shapley}}{{Reddy} et~al.}{2016b}]{Reddy2016b}
{Reddy} N.~A.,  {Steidel} C.~C.,  {Pettini} M.,  {Bogosavljevi{\'c}} M.,
  {Shapley} A.~E.,  2016b, \mn@doi [ApJ] {10.3847/0004-637X/828/2/108}, \href
  {https://ui.adsabs.harvard.edu/abs/2016ApJ...828..108R} {828, 108}

\bibitem[\protect\citeauthoryear{{Reddy} et~al.}{{Reddy}
  et~al.}{2018}]{Reddy2018b}
{Reddy} N.~A.,  et~al., 2018, \mn@doi [ApJ] {10.3847/1538-4357/aaed1e}, \href
  {https://ui.adsabs.harvard.edu/abs/2018ApJ...869...92R} {869, 92}

\bibitem[\protect\citeauthoryear{{Reddy} et~al.}{{Reddy}
  et~al.}{2020}]{Reddy2020}
{Reddy} N.~A.,  et~al., 2020, \mn@doi [ApJ] {10.3847/1538-4357/abb674}, \href
  {https://ui.adsabs.harvard.edu/abs/2020ApJ...902..123R} {902, 123}

\bibitem[\protect\citeauthoryear{{Rhoads}, {Malhotra}, {Dey}, {Stern},
  {Spinrad}  \& {Jannuzi}}{{Rhoads} et~al.}{2000}]{Rhoads2000}
{Rhoads} J.~E.,  {Malhotra} S.,  {Dey} A.,  {Stern} D.,  {Spinrad} H.,
  {Jannuzi} B.~T.,  2000, \mn@doi [ApJL] {10.1086/317874}, \href
  {http://adsabs.harvard.edu/abs/2000ApJ...545L..85R} {545, L85}

\bibitem[\protect\citeauthoryear{{Rhoads}, {Malhotra}, {Richardson},
  {Finkelstein}, {Fynbo}, {McLinden}  \& {Tilvi}}{{Rhoads}
  et~al.}{2014}]{Rhoads2014}
{Rhoads} J.~E.,  {Malhotra} S.,  {Richardson} M. L.~A.,  {Finkelstein} S.~L.,
  {Fynbo} J. P.~U.,  {McLinden} E.~M.,   {Tilvi} V.~S.,  2014, \mn@doi [ApJ]
  {10.1088/0004-637X/780/1/20}, \href
  {https://ui.adsabs.harvard.edu/abs/2014ApJ...780...20R} {780, 20}

\bibitem[\protect\citeauthoryear{{Ribeiro} et~al.}{{Ribeiro}
  et~al.}{2020}]{Ribeiro2020}
{Ribeiro} B.,  et~al., 2020, arXiv e-prints, \href
  {https://ui.adsabs.harvard.edu/abs/2020arXiv200701322R} {p. arXiv:2007.01322}

\bibitem[\protect\citeauthoryear{{Rigby} et~al.,}{{Rigby}
  et~al.}{2018}]{Rigby2018}
{Rigby} J.~R.,  et~al., 2018, \mn@doi [ApJ] {10.3847/1538-4357/aaa2fc}, \href
  {https://ui.adsabs.harvard.edu/abs/2018ApJ...853...87R} {853, 87}

\bibitem[\protect\citeauthoryear{{Rigby} et~al.}{{Rigby}
  et~al.}{2021}]{Rigby2020}
{Rigby} J.~R.,  et~al., 2021, \mn@doi [ApJ] {10.3847/1538-4357/abcfc9}, \href
  {https://ui.adsabs.harvard.edu/abs/2021ApJ...908..154R} {908, 154}

\bibitem[\protect\citeauthoryear{{Rivera-Thorsen} et~al.}{{Rivera-Thorsen}
  et~al.}{2015}]{RiveraThorsen2015}
{Rivera-Thorsen} T.~E.,  et~al., 2015, \mn@doi [ApJ]
  {10.1088/0004-637X/805/1/14}, \href
  {http://adsabs.harvard.edu/abs/2015ApJ...805...14R} {805, 14}

\bibitem[\protect\citeauthoryear{{Rivera-Thorsen} et~al.,}{{Rivera-Thorsen}
  et~al.}{2017}]{RiveraThorsen2017}
{Rivera-Thorsen} T.~E.,  et~al., 2017, \mn@doi [AAP]
  {10.1051/0004-6361/201732173}, \href
  {https://ui.adsabs.harvard.edu/abs/2017A&A...608L...4R} {608, L4}

\bibitem[\protect\citeauthoryear{{Rix} et~al.}{{Rix} et~al.}{2004}]{Rix2004}
{Rix} H.-W.,  et~al., 2004, \mn@doi [ApJS] {10.1086/420885}, \href
  {https://ui.adsabs.harvard.edu/abs/2004ApJS..152..163R} {152, 163}

\bibitem[\protect\citeauthoryear{{Robertson} et~al.}{{Robertson}
  et~al.}{2013}]{Robertson2013}
{Robertson} B.~E.,  et~al., 2013, \mn@doi [ApJ] {10.1088/0004-637X/768/1/71},
  \href {http://adsabs.harvard.edu/abs/2013ApJ...768...71R} {768, 71}

\bibitem[\protect\citeauthoryear{{Runnholm}, {Hayes}, {Melinder},
  {Rivera-Thorsen}, {{\"O}stlin}, {Cannon}  \& {Kunth}}{{Runnholm}
  et~al.}{2020}]{Runnholm2020}
{Runnholm} A.,  {Hayes} M.,  {Melinder} J.,  {Rivera-Thorsen} E.,  {{\"O}stlin}
  G.,  {Cannon} J.,   {Kunth} D.,  2020, \mn@doi [ApJ]
  {10.3847/1538-4357/ab7a91}, \href
  {https://ui.adsabs.harvard.edu/abs/2020ApJ...892...48R} {892, 48}

\bibitem[\protect\citeauthoryear{{Sanders} et~al.}{{Sanders}
  et~al.}{2016}]{Sanders2016}
{Sanders} R.~L.,  et~al., 2016, \mn@doi [ApJ] {10.3847/0004-637X/816/1/23},
  \href {https://ui.adsabs.harvard.edu/abs/2016ApJ...816...23S} {816, 23}

\bibitem[\protect\citeauthoryear{{Sanders} et~al.}{{Sanders}
  et~al.}{2018}]{Sanders2018}
{Sanders} R.~L.,  et~al., 2018, \mn@doi [ApJ] {10.3847/1538-4357/aabcbd}, \href
  {https://ui.adsabs.harvard.edu/abs/2018ApJ...858...99S} {858, 99}

\bibitem[\protect\citeauthoryear{{Sanders} et~al.}{{Sanders}
  et~al.}{2020a}]{Sanders2020b}
{Sanders} R.~L.,  et~al., 2020a, arXiv e-prints, \href
  {https://ui.adsabs.harvard.edu/abs/2020arXiv200907292S} {p. arXiv:2009.07292}

\bibitem[\protect\citeauthoryear{{Sanders} et~al.}{{Sanders}
  et~al.}{2020b}]{Sanders2020}
{Sanders} R.~L.,  et~al., 2020b, \mn@doi [MNRAS] {10.1093/mnras/stz3032}, \href
  {https://ui.adsabs.harvard.edu/abs/2020MNRAS.491.1427S} {491, 1427}

\bibitem[\protect\citeauthoryear{{Santos} et~al.,}{{Santos}
  et~al.}{2020}]{Santos2020}
{Santos} S.,  et~al., 2020, \mn@doi [MNRAS] {10.1093/mnras/staa093}, \href
  {https://ui.adsabs.harvard.edu/abs/2020MNRAS.493..141S} {493, 141}

\bibitem[\protect\citeauthoryear{{Saxena} et~al.}{{Saxena}
  et~al.}{2020}]{Saxena2020}
{Saxena} A.,  et~al., 2020, \mn@doi [AAP] {10.1051/0004-6361/201937170}, \href
  {https://ui.adsabs.harvard.edu/abs/2020A&A...636A..47S} {636, A47}

\bibitem[\protect\citeauthoryear{{Scarlata} et~al.}{{Scarlata}
  et~al.}{2009}]{Scarlata2009}
{Scarlata} C.,  et~al., 2009, \mn@doi [ApJL] {10.1088/0004-637X/704/2/L98},
  \href {https://ui.adsabs.harvard.edu/abs/2009ApJ...704L..98S} {704, L98}

\bibitem[\protect\citeauthoryear{{Schaerer}}{{Schaerer}}{2003}]{Schaerer2003}
{Schaerer} D.,  2003, \mn@doi [AAP] {10.1051/0004-6361:20021525}, \href
  {http://adsabs.harvard.edu/abs/2003A%26A...397..527S} {397, 527}

\bibitem[\protect\citeauthoryear{{Schaerer} \& {de Barros}}{{Schaerer} \& {de
  Barros}}{2009}]{SchaererBarros2009}
{Schaerer} D.,  {de Barros} S.,  2009, \mn@doi [AAP]
  {10.1051/0004-6361/200911781}, \href
  {http://adsabs.harvard.edu/abs/2009A%26A...502..423S} {502, 423}

\bibitem[\protect\citeauthoryear{{Schaerer}, {Fragos}  \& {Izotov}}{{Schaerer}
  et~al.}{2019}]{Schaerer2019}
{Schaerer} D.,  {Fragos} T.,   {Izotov} Y.~I.,  2019, \mn@doi [AAP]
  {10.1051/0004-6361/201935005}, \href
  {https://ui.adsabs.harvard.edu/abs/2019A&A...622L..10S} {622, L10}

\bibitem[\protect\citeauthoryear{{Senchyna} et~al.,}{{Senchyna}
  et~al.}{2017}]{Senchyna2017}
{Senchyna} P.,  et~al., 2017, \mn@doi [MNRAS] {10.1093/mnras/stx2059}, \href
  {https://ui.adsabs.harvard.edu/abs/2017MNRAS.472.2608S} {472, 2608}

\bibitem[\protect\citeauthoryear{{Senchyna}, {Stark}, {Chevallard}, {Charlot},
  {Jones}  \& {Vidal-Garc{\'\i}a}}{{Senchyna} et~al.}{2019}]{Senchyna2019}
{Senchyna} P.,  {Stark} D.~P.,  {Chevallard} J.,  {Charlot} S.,  {Jones} T.,
  {Vidal-Garc{\'\i}a} A.,  2019, \mn@doi [MNRAS] {10.1093/mnras/stz1907}, \href
  {https://ui.adsabs.harvard.edu/abs/2019MNRAS.488.3492S} {488, 3492}

\bibitem[\protect\citeauthoryear{{Shapley}, {Steidel}, {Pettini}  \&
  {Adelberger}}{{Shapley} et~al.}{2003}]{Shapley2003}
{Shapley} A.~E.,  {Steidel} C.~C.,  {Pettini} M.,   {Adelberger} K.~L.,  2003,
  \mn@doi [ApJ] {10.1086/373922}, \href
  {http://adsabs.harvard.edu/abs/2003ApJ...588...65S} {588, 65}

\bibitem[\protect\citeauthoryear{{Shapley} et~al.}{{Shapley}
  et~al.}{2015}]{Shapley2015}
{Shapley} A.~E.,  et~al., 2015, \mn@doi [ApJ] {10.1088/0004-637X/801/2/88},
  \href {https://ui.adsabs.harvard.edu/abs/2015ApJ...801...88S} {801, 88}

\bibitem[\protect\citeauthoryear{{Shibuya} et~al.}{{Shibuya}
  et~al.}{2014}]{Shibuya2014}
{Shibuya} T.,  et~al., 2014, \mn@doi [ApJ] {10.1088/0004-637X/788/1/74}, \href
  {https://ui.adsabs.harvard.edu/abs/2014ApJ...788...74S} {788, 74}

\bibitem[\protect\citeauthoryear{{Shibuya}, {Ouchi}, {Harikane}  \&
  {Nakajima}}{{Shibuya} et~al.}{2019}]{Shibuya2019}
{Shibuya} T.,  {Ouchi} M.,  {Harikane} Y.,   {Nakajima} K.,  2019, \mn@doi
  [ApJ] {10.3847/1538-4357/aaf64b}, \href
  {https://ui.adsabs.harvard.edu/abs/2019ApJ...871..164S} {871, 164}

\bibitem[\protect\citeauthoryear{{Shirazi}, {Brinchmann}  \&
  {Rahmati}}{{Shirazi} et~al.}{2014}]{Shirazi2014}
{Shirazi} M.,  {Brinchmann} J.,   {Rahmati} A.,  2014, \mn@doi [ApJ]
  {10.1088/0004-637X/787/2/120}, \href
  {https://ui.adsabs.harvard.edu/abs/2014ApJ...787..120S} {787, 120}

\bibitem[\protect\citeauthoryear{{Shivaei} et~al.}{{Shivaei}
  et~al.}{2020}]{Shivaei2020}
{Shivaei} I.,  et~al., 2020, \mn@doi [ApJ] {10.3847/1538-4357/aba35e}, \href
  {https://ui.adsabs.harvard.edu/abs/2020ApJ...899..117S} {899, 117}

\bibitem[\protect\citeauthoryear{{Smette} et~al.}{{Smette}
  et~al.}{2015}]{molecfit1}
{Smette} A.,  et~al., 2015, \mn@doi [AAP] {10.1051/0004-6361/201423932}, \href
  {https://ui.adsabs.harvard.edu/abs/2015A&A...576A..77S} {576, A77}

\bibitem[\protect\citeauthoryear{{Smith}, {Ma}, {Bromm}, {Finkelstein},
  {Hopkins}, {Faucher-Gigu{\`e}re}  \& {Kere{\v{s}}}}{{Smith}
  et~al.}{2019}]{Smith2019}
{Smith} A.,  {Ma} X.,  {Bromm} V.,  {Finkelstein} S.~L.,  {Hopkins} P.~F.,
  {Faucher-Gigu{\`e}re} C.-A.,   {Kere{\v{s}}} D.,  2019, \mn@doi [MNRAS]
  {10.1093/mnras/sty3483}, \href
  {https://ui.adsabs.harvard.edu/abs/2019MNRAS.484...39S} {484, 39}

\bibitem[\protect\citeauthoryear{{Sobral} \& {Matthee}}{{Sobral} \&
  {Matthee}}{2019}]{SM2019}
{Sobral} D.,  {Matthee} J.,  2019, \mn@doi [AAP] {10.1051/0004-6361/201833075},
  \href {https://ui.adsabs.harvard.edu/abs/2019A&A...623A.157S} {623, A157}

\bibitem[\protect\citeauthoryear{{Sobral}, {Smail}, {Best}, {Geach}, {Matsuda},
  {Stott}, {Cirasuolo}  \& {Kurk}}{{Sobral} et~al.}{2013}]{Sobral2013}
{Sobral} D.,  {Smail} I.,  {Best} P.~N.,  {Geach} J.~E.,  {Matsuda} Y.,
  {Stott} J.~P.,  {Cirasuolo} M.,   {Kurk} J.,  2013, \mn@doi [MNRAS]
  {10.1093/mnras/sts096}, \href
  {http://adsabs.harvard.edu/abs/2013MNRAS.428.1128S} {428, 1128}

\bibitem[\protect\citeauthoryear{{Sobral}, {Matthee}, {Darvish}, {Schaerer},
  {Mobasher}, {R{\"o}ttgering}, {Santos}  \& {Hemmati}}{{Sobral}
  et~al.}{2015}]{Sobral2015}
{Sobral} D.,  {Matthee} J.,  {Darvish} B.,  {Schaerer} D.,  {Mobasher} B.,
  {R{\"o}ttgering} H.~J.~A.,  {Santos} S.,   {Hemmati} S.,  2015, \mn@doi [ApJ]
  {10.1088/0004-637X/808/2/139}, \href
  {http://adsabs.harvard.edu/abs/2015ApJ...808..139S} {808, 139}

\bibitem[\protect\citeauthoryear{{Sobral} et~al.,}{{Sobral}
  et~al.}{2017}]{Sobral2016}
{Sobral} D.,  et~al., 2017, \mn@doi [MNRAS] {10.1093/mnras/stw3090}, \href
  {http://adsabs.harvard.edu/abs/2017MNRAS.466.1242S} {466, 1242}

\bibitem[\protect\citeauthoryear{{Sobral}, {Santos}, {Matthee},
  {Paulino-Afonso}, {Ribeiro}, {Calhau}  \& {Khostovan}}{{Sobral}
  et~al.}{2018a}]{SC4K}
{Sobral} D.,  {Santos} S.,  {Matthee} J.,  {Paulino-Afonso} A.,  {Ribeiro} B.,
  {Calhau} J.,   {Khostovan} A.~A.,  2018a, \mn@doi [MNRAS]
  {10.1093/mnras/sty378}, \href
  {https://ui.adsabs.harvard.edu/abs/2018MNRAS.476.4725S} {476, 4725}

\bibitem[\protect\citeauthoryear{{Sobral} et~al.}{{Sobral}
  et~al.}{2018b}]{Sobral2018}
{Sobral} D.,  et~al., 2018b, \mn@doi [MNRAS] {10.1093/mnras/sty782}, \href
  {http://adsabs.harvard.edu/abs/2018MNRAS.477.2817S} {477, 2817}

\bibitem[\protect\citeauthoryear{{Song} et~al.}{{Song} et~al.}{2014}]{Song2014}
{Song} M.,  et~al., 2014, \mn@doi [ApJ] {10.1088/0004-637X/791/1/3}, \href
  {http://adsabs.harvard.edu/abs/2014ApJ...791....3S} {791, 3}

\bibitem[\protect\citeauthoryear{{Stanway} \& {Eldridge}}{{Stanway} \&
  {Eldridge}}{2018}]{BPASS2018}
{Stanway} E.~R.,  {Eldridge} J.~J.,  2018, \mn@doi [\mnras]
  {10.1093/mnras/sty1353}, \href
  {https://ui.adsabs.harvard.edu/abs/2018MNRAS.479...75S} {479, 75}

\bibitem[\protect\citeauthoryear{{Stanway}, {Chrimes}, {Eldridge}  \&
  {Stevance}}{{Stanway} et~al.}{2020}]{Stanway2020}
{Stanway} E.~R.,  {Chrimes} A.~A.,  {Eldridge} J.~J.,   {Stevance} H.~F.,
  2020, \mn@doi [MNRAS] {10.1093/mnras/staa1166}, \href
  {https://ui.adsabs.harvard.edu/abs/2020MNRAS.495.4605S} {495, 4605}

\bibitem[\protect\citeauthoryear{{Stark}, {Ellis}, {Chiu}, {Ouchi}  \&
  {Bunker}}{{Stark} et~al.}{2010}]{Stark2010}
{Stark} D.~P.,  {Ellis} R.~S.,  {Chiu} K.,  {Ouchi} M.,   {Bunker} A.,  2010,
  \mn@doi [MNRAS] {10.1111/j.1365-2966.2010.17227.x}, \href
  {http://adsabs.harvard.edu/abs/2010MNRAS.408.1628S} {408, 1628}

\bibitem[\protect\citeauthoryear{{Stark}, {Ellis}  \& {Ouchi}}{{Stark}
  et~al.}{2011}]{Stark2011}
{Stark} D.~P.,  {Ellis} R.~S.,   {Ouchi} M.,  2011, \mn@doi [ApJL]
  {10.1088/2041-8205/728/1/L2}, \href
  {http://adsabs.harvard.edu/abs/2011ApJ...728L...2S} {728, L2}

\bibitem[\protect\citeauthoryear{{Stark} et~al.}{{Stark}
  et~al.}{2015}]{Stark2015_CIV}
{Stark} D.~P.,  et~al., 2015, \mn@doi [MNRAS] {10.1093/mnras/stv1907}, \href
  {http://adsabs.harvard.edu/abs/2015MNRAS.454.1393S} {454, 1393}

\bibitem[\protect\citeauthoryear{{Stasi{\'n}ska}, {Izotov}, {Morisset}  \&
  {Guseva}}{{Stasi{\'n}ska} et~al.}{2015}]{Stasinska2015}
{Stasi{\'n}ska} G.,  {Izotov} Y.,  {Morisset} C.,   {Guseva} N.,  2015, \mn@doi
  [AAP] {10.1051/0004-6361/201425389}, \href
  {https://ui.adsabs.harvard.edu/abs/2015A&A...576A..83S} {576, A83}

\bibitem[\protect\citeauthoryear{{Steidel}, {Shapley}, {Pettini}, {Adelberger},
  {Erb}, {Reddy}  \& {Hunt}}{{Steidel} et~al.}{2004}]{Steidel2004}
{Steidel} C.~C.,  {Shapley} A.~E.,  {Pettini} M.,  {Adelberger} K.~L.,  {Erb}
  D.~K.,  {Reddy} N.~A.,   {Hunt} M.~P.,  2004, \mn@doi [ApJ] {10.1086/381960},
  \href {https://ui.adsabs.harvard.edu/abs/2004ApJ...604..534S} {604, 534}

\bibitem[\protect\citeauthoryear{{Steidel}, {Erb}, {Shapley}, {Pettini},
  {Reddy}, {Bogosavljevi{\'c}}, {Rudie}  \& {Rakic}}{{Steidel}
  et~al.}{2010}]{Steidel2010}
{Steidel} C.~C.,  {Erb} D.~K.,  {Shapley} A.~E.,  {Pettini} M.,  {Reddy} N.,
  {Bogosavljevi{\'c}} M.,  {Rudie} G.~C.,   {Rakic} O.,  2010, \mn@doi [ApJ]
  {10.1088/0004-637X/717/1/289}, \href
  {http://adsabs.harvard.edu/abs/2010ApJ...717..289S} {717, 289}

\bibitem[\protect\citeauthoryear{{Steidel} et~al.}{{Steidel}
  et~al.}{2014}]{Steidel2014}
{Steidel} C.~C.,  et~al., 2014, \mn@doi [ApJ] {10.1088/0004-637X/795/2/165},
  \href {https://ui.adsabs.harvard.edu/abs/2014ApJ...795..165S} {795, 165}

\bibitem[\protect\citeauthoryear{{Steidel}, {Strom}, {Pettini}, {Rudie},
  {Reddy}  \& {Trainor}}{{Steidel} et~al.}{2016}]{Steidel2016}
{Steidel} C.~C.,  {Strom} A.~L.,  {Pettini} M.,  {Rudie} G.~C.,  {Reddy} N.~A.,
    {Trainor} R.~F.,  2016, \mn@doi [ApJ] {10.3847/0004-637X/826/2/159}, \href
  {https://ui.adsabs.harvard.edu/abs/2016ApJ...826..159S} {826, 159}

\bibitem[\protect\citeauthoryear{{Steidel}, {Bogosavljevi{\'c}}, {Shapley},
  {Reddy}, {Rudie}, {Pettini}, {Trainor}  \& {Strom}}{{Steidel}
  et~al.}{2018}]{Steidel2018}
{Steidel} C.~C.,  {Bogosavljevi{\'c}} M.,  {Shapley} A.~E.,  {Reddy} N.~A.,
  {Rudie} G.~C.,  {Pettini} M.,  {Trainor} R.~F.,   {Strom} A.~L.,  2018,
  \mn@doi [ApJ] {10.3847/1538-4357/aaed28}, \href
  {https://ui.adsabs.harvard.edu/abs/2018ApJ...869..123S} {869, 123}

\bibitem[\protect\citeauthoryear{{Strom}, {Steidel}, {Rudie}, {Trainor},
  {Pettini}  \& {Reddy}}{{Strom} et~al.}{2017}]{Strom2017}
{Strom} A.~L.,  {Steidel} C.~C.,  {Rudie} G.~C.,  {Trainor} R.~F.,  {Pettini}
  M.,   {Reddy} N.~A.,  2017, \mn@doi [ApJ] {10.3847/1538-4357/836/2/164},
  \href {https://ui.adsabs.harvard.edu/abs/2017ApJ...836..164S} {836, 164}

\bibitem[\protect\citeauthoryear{{Strom}, {Steidel}, {Rudie}, {Trainor}  \&
  {Pettini}}{{Strom} et~al.}{2018}]{Strom2018}
{Strom} A.~L.,  {Steidel} C.~C.,  {Rudie} G.~C.,  {Trainor} R.~F.,   {Pettini}
  M.,  2018, \mn@doi [ApJ] {10.3847/1538-4357/aae1a5}, \href
  {https://ui.adsabs.harvard.edu/abs/2018ApJ...868..117S} {868, 117}

\bibitem[\protect\citeauthoryear{{Sz{\'e}csi}, {Langer}, {Yoon}, {Sanyal}, {de
  Mink}, {Evans}  \& {Dermine}}{{Sz{\'e}csi} et~al.}{2015}]{Szecsi2015}
{Sz{\'e}csi} D.,  {Langer} N.,  {Yoon} S.-C.,  {Sanyal} D.,  {de Mink} S.,
  {Evans} C.~J.,   {Dermine} T.,  2015, \mn@doi [AAP]
  {10.1051/0004-6361/201526617}, \href
  {https://ui.adsabs.harvard.edu/abs/2015A&A...581A..15S} {581, A15}

\bibitem[\protect\citeauthoryear{{Tang}, {Stark}, {Chevallard}, {Charlot},
  {Endsley}  \& {Congiu}}{{Tang} et~al.}{2020}]{Tang2021}
{Tang} M.,  {Stark} D.,  {Chevallard} J.,  {Charlot} S.,  {Endsley} R.,
  {Congiu} E.,  2020, arXiv e-prints, \href
  {https://ui.adsabs.harvard.edu/abs/2020arXiv201204697T} {p. arXiv:2012.04697}

\bibitem[\protect\citeauthoryear{{Taylor}}{{Taylor}}{2013}]{Topcat}
{Taylor} M.,  2013, Starlink User Note, \href
  {http://adsabs.harvard.edu/abs/2013StaUN.253.....T} {253}

\bibitem[\protect\citeauthoryear{{Taylor}, {Barger}, {Cowie}, {Hu}  \&
  {Songaila}}{{Taylor} et~al.}{2020}]{Taylor2020}
{Taylor} A.~J.,  {Barger} A.~J.,  {Cowie} L.~L.,  {Hu} E.~M.,   {Songaila} A.,
  2020, \mn@doi [ApJ] {10.3847/1538-4357/ab8ada}, \href
  {https://ui.adsabs.harvard.edu/abs/2020ApJ...895..132T} {895, 132}

\bibitem[\protect\citeauthoryear{{Terlevich}, {Terlevich}, {Melnick},
  {Ch{\'a}vez}, {Plionis}, {Bresolin}  \& {Basilakos}}{{Terlevich}
  et~al.}{2015}]{Terlevich2015}
{Terlevich} R.,  {Terlevich} E.,  {Melnick} J.,  {Ch{\'a}vez} R.,  {Plionis}
  M.,  {Bresolin} F.,   {Basilakos} S.,  2015, \mn@doi [MNRAS]
  {10.1093/mnras/stv1128}, \href
  {https://ui.adsabs.harvard.edu/abs/2015MNRAS.451.3001T} {451, 3001}

\bibitem[\protect\citeauthoryear{{Theios}, {Steidel}, {Strom}, {Rudie},
  {Trainor}  \& {Reddy}}{{Theios} et~al.}{2019}]{Theios2019}
{Theios} R.~L.,  {Steidel} C.~C.,  {Strom} A.~L.,  {Rudie} G.~C.,  {Trainor}
  R.~F.,   {Reddy} N.~A.,  2019, \mn@doi [ApJ] {10.3847/1538-4357/aaf386},
  \href {https://ui.adsabs.harvard.edu/abs/2019ApJ...871..128T} {871, 128}

\bibitem[\protect\citeauthoryear{{Topping}, {Shapley}, {Reddy}, {Sanders},
  {Coil}, {Kriek}, {Mobasher}  \& {Siana}}{{Topping}
  et~al.}{2020}]{Topping2020}
{Topping} M.~W.,  {Shapley} A.~E.,  {Reddy} N.~A.,  {Sanders} R.~L.,  {Coil}
  A.~L.,  {Kriek} M.,  {Mobasher} B.,   {Siana} B.,  2020, \mn@doi [MNRAS]
  {10.1093/mnras/staa1410}, \href
  {https://ui.adsabs.harvard.edu/abs/2020MNRAS.495.4430T} {495, 4430}

\bibitem[\protect\citeauthoryear{{Trainor}, {Steidel}, {Strom}  \&
  {Rudie}}{{Trainor} et~al.}{2015}]{Trainor2015}
{Trainor} R.~F.,  {Steidel} C.~C.,  {Strom} A.~L.,   {Rudie} G.~C.,  2015,
  \mn@doi [ApJ] {10.1088/0004-637X/809/1/89}, \href
  {http://adsabs.harvard.edu/abs/2015ApJ...809...89T} {809, 89}

\bibitem[\protect\citeauthoryear{{Trainor}, {Strom}, {Steidel}  \&
  {Rudie}}{{Trainor} et~al.}{2016}]{Trainor2016}
{Trainor} R.~F.,  {Strom} A.~L.,  {Steidel} C.~C.,   {Rudie} G.~C.,  2016,
  \mn@doi [ApJ] {10.3847/0004-637X/832/2/171}, \href
  {http://adsabs.harvard.edu/abs/2016ApJ...832..171T} {832, 171}

\bibitem[\protect\citeauthoryear{{Trainor}, {Strom}, {Steidel}, {Rudie}, {Chen}
   \& {Theios}}{{Trainor} et~al.}{2019}]{Trainor2019}
{Trainor} R.~F.,  {Strom} A.~L.,  {Steidel} C.~C.,  {Rudie} G.~C.,  {Chen} Y.,
   {Theios} R.~L.,  2019, \mn@doi [ApJ] {10.3847/1538-4357/ab4993}, \href
  {https://ui.adsabs.harvard.edu/abs/2019ApJ...887...85T} {887, 85}

\bibitem[\protect\citeauthoryear{{Vanzella} et~al.}{{Vanzella}
  et~al.}{2016}]{Vanzella2016}
{Vanzella} E.,  et~al., 2016, \mn@doi [ApJL] {10.3847/2041-8205/821/2/L27},
  \href {http://adsabs.harvard.edu/abs/2016ApJ...821L..27V} {821, L27}

\bibitem[\protect\citeauthoryear{{Vanzella} et~al.}{{Vanzella}
  et~al.}{2018}]{Vanzella2018}
{Vanzella} E.,  et~al., 2018, \mn@doi [MNRAS] {10.1093/mnrasl/sly023}, \href
  {https://ui.adsabs.harvard.edu/abs/2018MNRAS.476L..15V} {476, L15}

\bibitem[\protect\citeauthoryear{{Veilleux}, {Cecil}  \&
  {Bland-Hawthorn}}{{Veilleux} et~al.}{2005}]{Veilleux2005}
{Veilleux} S.,  {Cecil} G.,   {Bland-Hawthorn} J.,  2005, \mn@doi [ARAA]
  {10.1146/annurev.astro.43.072103.150610}, \href
  {https://ui.adsabs.harvard.edu/abs/2005ARA&A..43..769V} {43, 769}

\bibitem[\protect\citeauthoryear{{Verhamme}, {Schaerer}  \&
  {Maselli}}{{Verhamme} et~al.}{2006}]{Verhamme2006}
{Verhamme} A.,  {Schaerer} D.,   {Maselli} A.,  2006, \mn@doi [AAP]
  {10.1051/0004-6361:20065554}, \href
  {https://ui.adsabs.harvard.edu/abs/2006A%26A...460..397V} {460, 397}

\bibitem[\protect\citeauthoryear{{Verhamme}, {Orlitov{\'a}}, {Schaerer}  \&
  {Hayes}}{{Verhamme} et~al.}{2015}]{Verhamme2015}
{Verhamme} A.,  {Orlitov{\'a}} I.,  {Schaerer} D.,   {Hayes} M.,  2015, \mn@doi
  [AAP] {10.1051/0004-6361/201423978}, \href
  {http://adsabs.harvard.edu/abs/2015A%26A...578A...7V} {578, A7}

\bibitem[\protect\citeauthoryear{{Verhamme} et~al.}{{Verhamme}
  et~al.}{2018}]{Verhamme2018}
{Verhamme} A.,  et~al., 2018, \mn@doi [MNRAS] {10.1093/mnrasl/sly058}, \href
  {http://adsabs.harvard.edu/abs/2018MNRAS.478L..60V} {478, L60}

\bibitem[\protect\citeauthoryear{{Vernet} et~al.}{{Vernet}
  et~al.}{2011}]{Vernet2011}
{Vernet} J.,  et~al., 2011, \mn@doi [AAP] {10.1051/0004-6361/201117752}, \href
  {http://adsabs.harvard.edu/abs/2011A%26A...536A.105V} {536, A105}

\bibitem[\protect\citeauthoryear{{Vidal-Garc{\'\i}a}, {Charlot}, {Bruzual}  \&
  {Hubeny}}{{Vidal-Garc{\'\i}a} et~al.}{2017}]{VidalGarcia2017}
{Vidal-Garc{\'\i}a} A.,  {Charlot} S.,  {Bruzual} G.,   {Hubeny} I.,  2017,
  \mn@doi [MNRAS] {10.1093/mnras/stx1324}, \href
  {https://ui.adsabs.harvard.edu/abs/2017MNRAS.470.3532V} {470, 3532}

\bibitem[\protect\citeauthoryear{{Westmoquette}, {Exter}, {Smith}  \&
  {Gallagher}}{{Westmoquette} et~al.}{2007}]{Westmoquette2007}
{Westmoquette} M.~S.,  {Exter} K.~M.,  {Smith} L.~J.,   {Gallagher} J.~S.,
  2007, \mn@doi [MNRAS] {10.1111/j.1365-2966.2007.12346.x}, \href
  {https://ui.adsabs.harvard.edu/abs/2007MNRAS.381..894W} {381, 894}

\bibitem[\protect\citeauthoryear{{Wofford}, {Vidal-Garc{\'\i}a}, {Feltre},
  {Chevallard}, {Charlot}, {Stark}, {Herenz}  \& {Hayes}}{{Wofford}
  et~al.}{2020}]{Wofford2020}
{Wofford} A.,  {Vidal-Garc{\'\i}a} A.,  {Feltre} A.,  {Chevallard} J.,
  {Charlot} S.,  {Stark} D.~P.,  {Herenz} E.~C.,   {Hayes} M.,  2020, \mn@doi
  [MNRAS] {10.1093/mnras/staa3365}, \href
  {https://ui.adsabs.harvard.edu/abs/2020MNRAS.500.2908W} {500, 2908}

\bibitem[\protect\citeauthoryear{{Xiao}, {Stanway}  \& {Eldridge}}{{Xiao}
  et~al.}{2018}]{Xiao2018}
{Xiao} L.,  {Stanway} E.~R.,   {Eldridge} J.~J.,  2018, \mn@doi [MNRAS]
  {10.1093/mnras/sty646}, \href
  {https://ui.adsabs.harvard.edu/abs/2018MNRAS.477..904X} {477, 904}

\bibitem[\protect\citeauthoryear{{Yang} et~al.}{{Yang} et~al.}{2017}]{Yang2017}
{Yang} H.,  et~al., 2017, \mn@doi [ApJ] {10.3847/1538-4357/aa7d4d}, \href
  {https://ui.adsabs.harvard.edu/abs/2017ApJ...844..171Y} {844, 171}

\bibitem[\protect\citeauthoryear{{Zheng} \& {Wallace}}{{Zheng} \&
  {Wallace}}{2014}]{ZhengWallace2014}
{Zheng} Z.,  {Wallace} J.,  2014, \mn@doi [ApJ] {10.1088/0004-637X/794/2/116},
  \href {https://ui.adsabs.harvard.edu/abs/2014ApJ...794..116Z} {794, 116}

\bibitem[\protect\citeauthoryear{{Zheng} et~al.}{{Zheng}
  et~al.}{2017}]{Zheng2017}
{Zheng} Z.-Y.,  et~al., 2017, \mn@doi [ApJL] {10.3847/2041-8213/aa794f}, \href
  {https://ui.adsabs.harvard.edu/abs/2017ApJ...842L..22Z} {842, L22}

\bibitem[\protect\citeauthoryear{{Zitrin} et~al.}{{Zitrin}
  et~al.}{2015}]{Zitrin2015}
{Zitrin} A.,  et~al., 2015, \mn@doi [ApJL] {10.1088/2041-8205/810/1/L12}, \href
  {http://adsabs.harvard.edu/abs/2015ApJ...810L..12Z} {810, L12}

\bibitem[\protect\citeauthoryear{{da Cunha}, {Charlot}  \& {Elbaz}}{{da Cunha}
  et~al.}{2008}]{daCunha2008}
{da Cunha} E.,  {Charlot} S.,   {Elbaz} D.,  2008, \mn@doi [MNRAS]
  {10.1111/j.1365-2966.2008.13535.x}, \href
  {https://ui.adsabs.harvard.edu/abs/2008MNRAS.388.1595D} {388, 1595}

\bibitem[\protect\citeauthoryear{{van der Walt}, {Colbert}  \&
  {Varoquaux}}{{van der Walt} et~al.}{2011}]{Numpy}
{van der Walt} S.,  {Colbert} S.~C.,   {Varoquaux} G.,  2011, \mn@doi
  [Computing in Science and Engineering] {10.1109/MCSE.2011.37}, \href
  {https://ui.adsabs.harvard.edu/abs/2011CSE....13b..22V} {13, 22}

\bibitem[\protect\citeauthoryear{{van der Wel} et~al.}{{van der Wel}
  et~al.}{2012}]{vdWel2012}
{van der Wel} A.,  et~al., 2012, \mn@doi [ApJS] {10.1088/0067-0049/203/2/24},
  \href {https://ui.adsabs.harvard.edu/abs/2012ApJS..203...24V} {203, 24}

\makeatother
\end{thebibliography}



\appendix
\section*{Appendix}
\renewcommand\thetable{A.\arabic{table}}    
\renewcommand\thefigure{A.\arabic{figure}}

\begin{figure*}
\includegraphics[width=17.8cm]{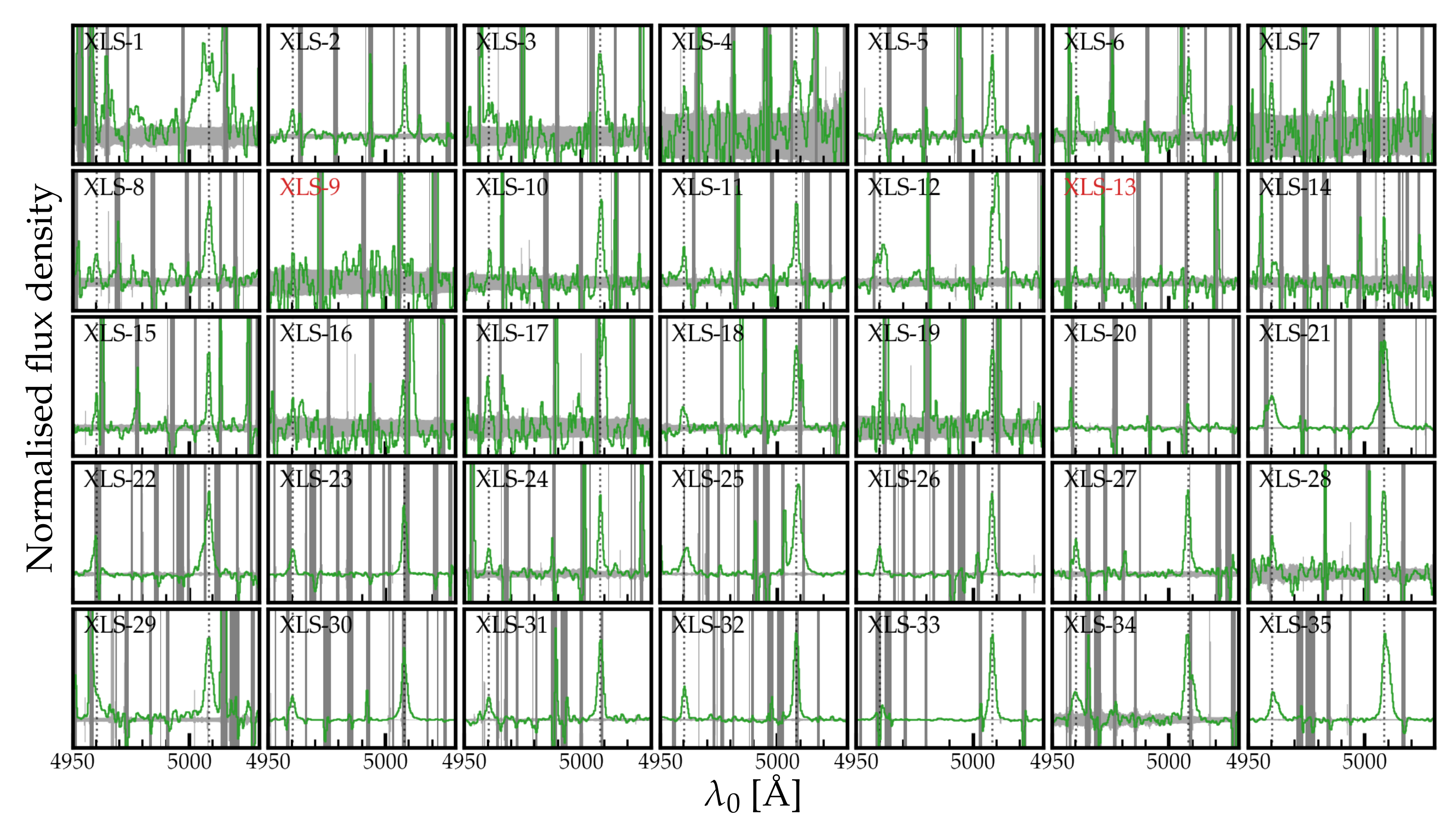} \vspace{-0.8cm}
\caption{[O{\sc iii}]$_{4960,5008}$ spectra of the XLS sample, binned by a factor two and smoothed for visualisation purposes. Spectra are normalised to the peak [O{\sc iii}]$_{5008}$ flux density. The locations of the [O{\sc iii}] lines are highlighted by dotted vertical lines. No [O{\sc iii}] emission is detected in XLS-9 and XLS-13 and we normalise them arbitrarily. The grey shaded region along the horizontal axis shows the noise level, while the vertical stripes mark the locations of skylines. }
\label{fig:O3spec}
\end{figure*}

\begin{table}
\centering
\caption{Galaxy IDs that have previously been used for the galaxies in the XLS sample.}
\begin{tabular}{lr|lr}
XLS-ID & Previous ID & XLS-ID & Previous ID \\ \hline
  XLS-1 & CALYMHA-S16-278 & XLS-21 & VUDS510583858\\
  XLS-2 & CALYMHA-S16-147 &  XLS-22 & VUDS5100998761\\
  XLS-3 & CALYMHA-S16-138  & XLS-23 & VUDS5101421970\\
  XLS-4 & CALYMHA-S16-136 &  XLS-24 & LAE47\\
  XLS-5 & CALYMHA-S16-134 &  XLS-25 & CDFS\_03865\\
  XLS-6 & CALYMHA-S16-129 &   XLS-26 & DBL1349, GS30668\\
  XLS-7 & CALYMHA-S16-126 &   XLS-27 & GS14633\\
  XLS-8 & CALYMHA-S16-122 &   XLS-28 & CDFS\_06482\\
  XLS-9 & CALYMHA-S16-115 &   XLS-29 & Q2343-BX436\\
  XLS-10 & CALYMHA-S16-113 &   XLS-30 & Q2343-BX418\\
  XLS-11 & CALYMHA-S16-95 &   XLS-31 & Q2343-BX660\\
  XLS-12 & CALYMHA-S16-67  & XLS-32 & Q0207-BX144\\
  XLS-13 & CALYMHA-S16-28 &   XLS-33 & Q0207-BX87\\
  XLS-14 & COSMOS{\_}LYA &   XLS-34 & Q0207-BX74\\
  XLS-15 & CALYMHA-S16-389 &  XLS-35 & Q0142-BX165\\
  XLS-16 & CALYMHA-S16-386 & &\\
  XLS-17 & CALYMHA-S16-378 & &\\
  XLS-18 & CALYMHA-S16-373 & &\\
  XLS-19 & CALYMHA-S16-369 & &\\
  XLS-20 & LYRS-S7 & &\\
\end{tabular}
\label{tab:IDs}
\end{table}

\begin{figure*}
\begin{tabular}{cccc}
\includegraphics[width=4.8cm]{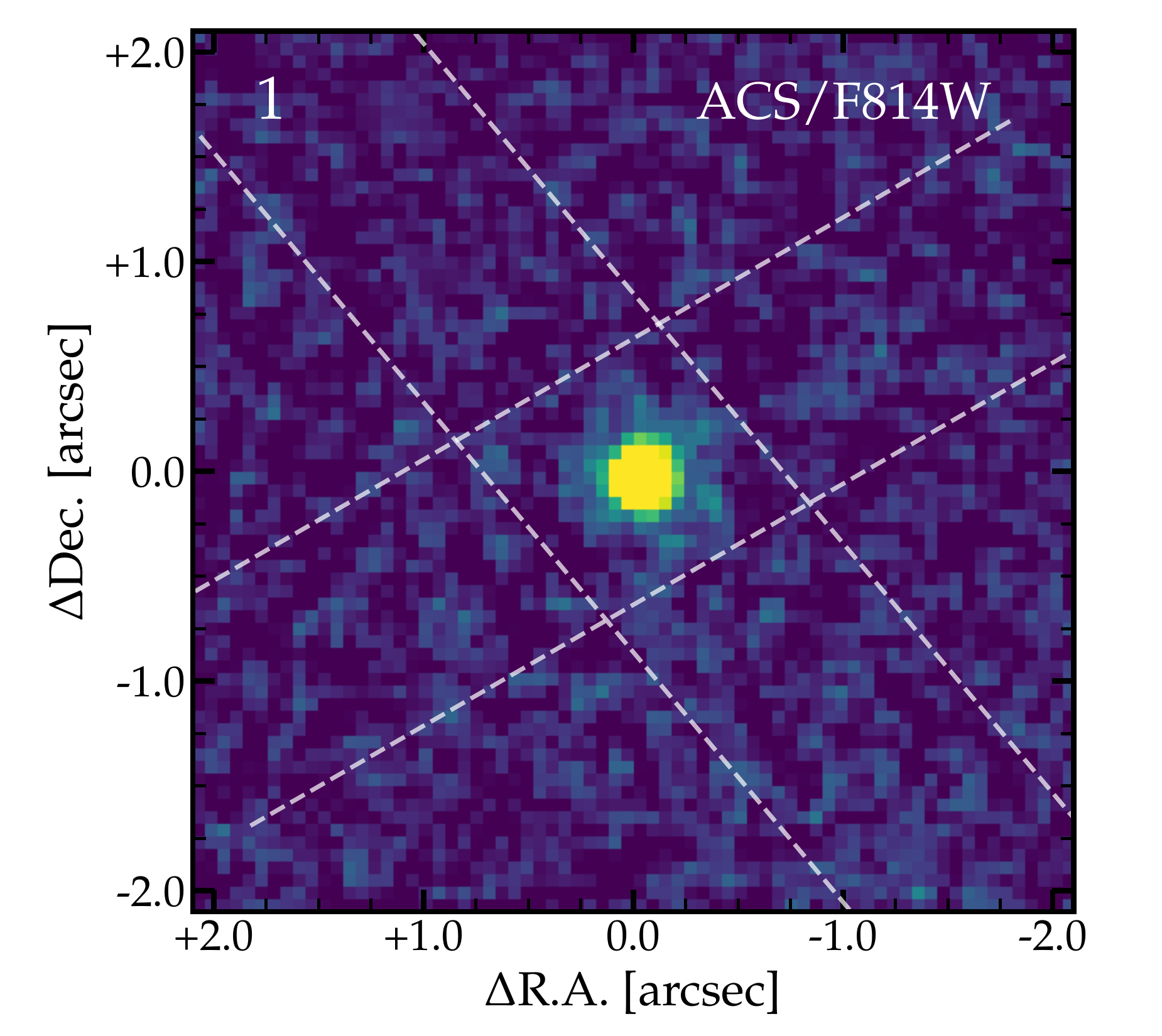} &
\hspace{-1cm}\includegraphics[width=4.8cm]{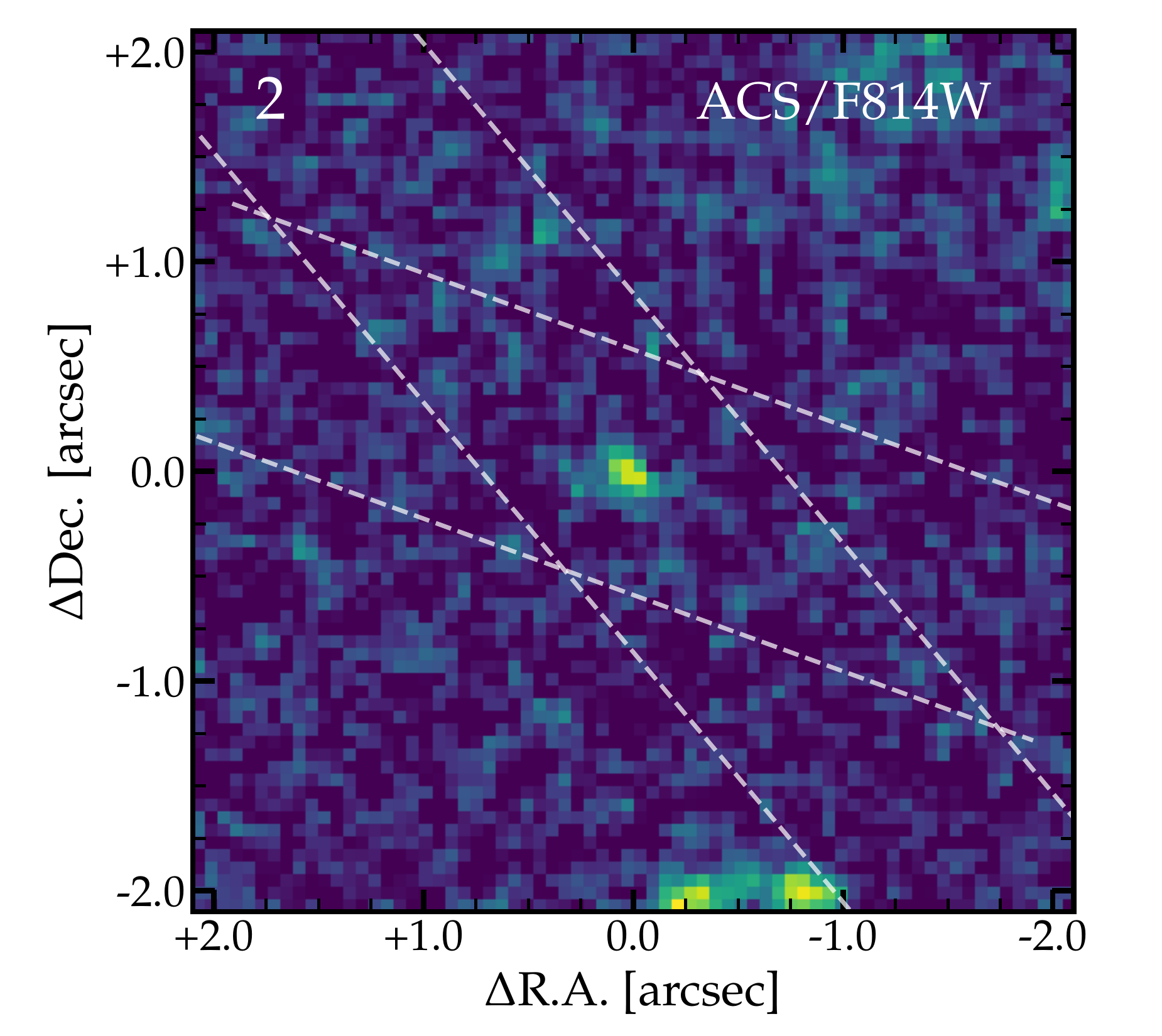} &
\hspace{-1cm}\includegraphics[width=4.8cm]{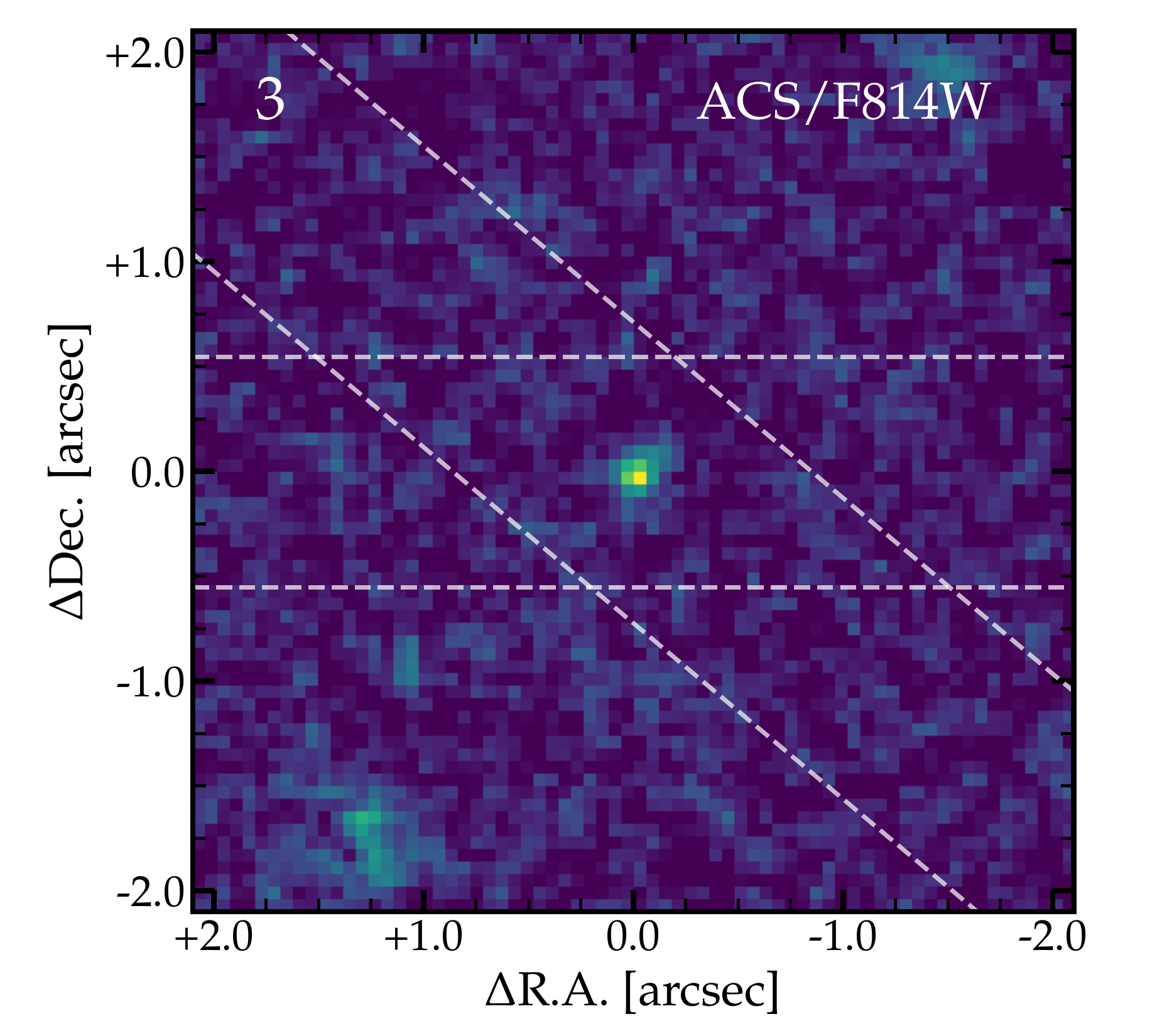} &
\hspace{-1cm}\includegraphics[width=4.8cm]{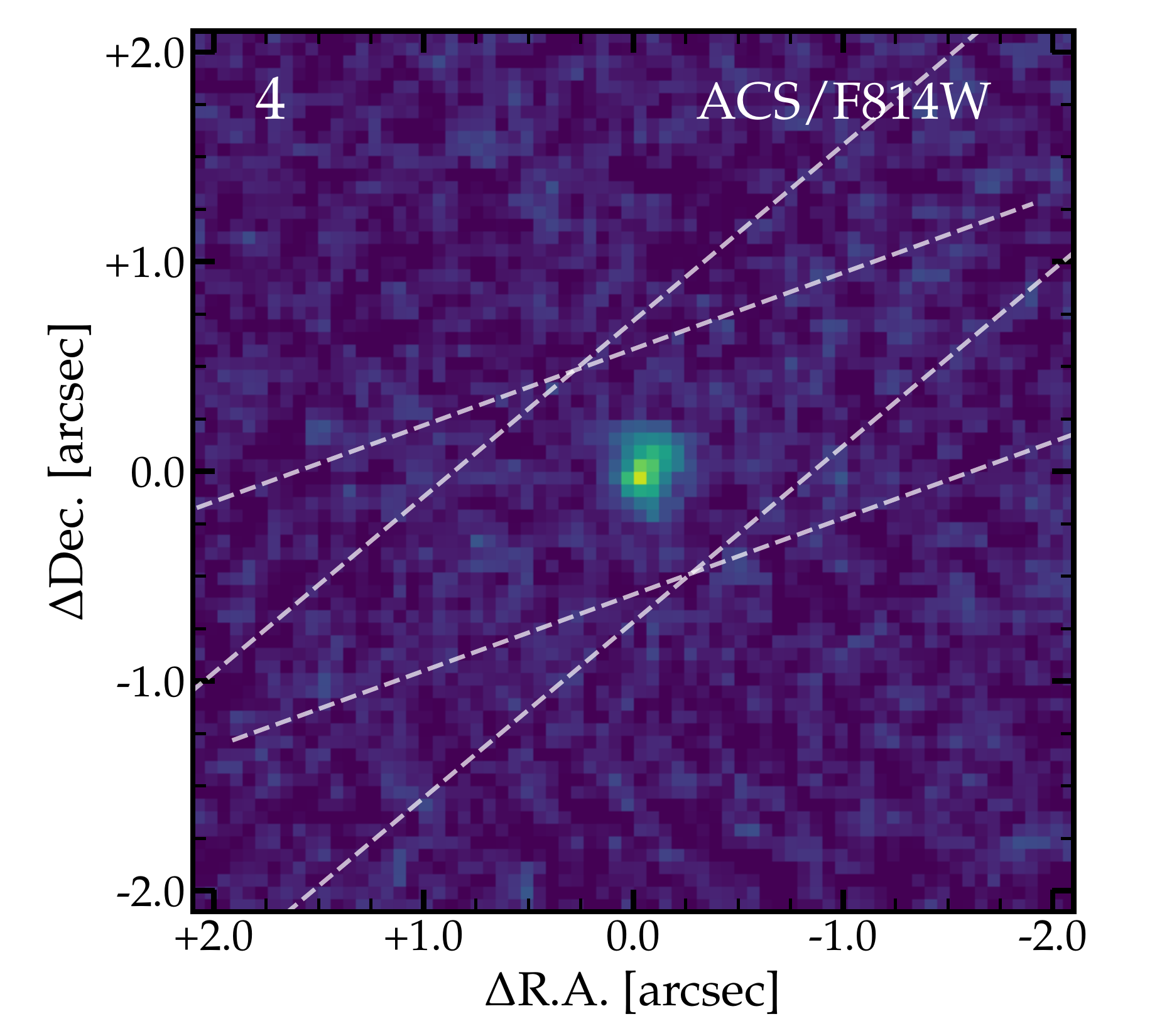} \\
\includegraphics[width=4.8cm]{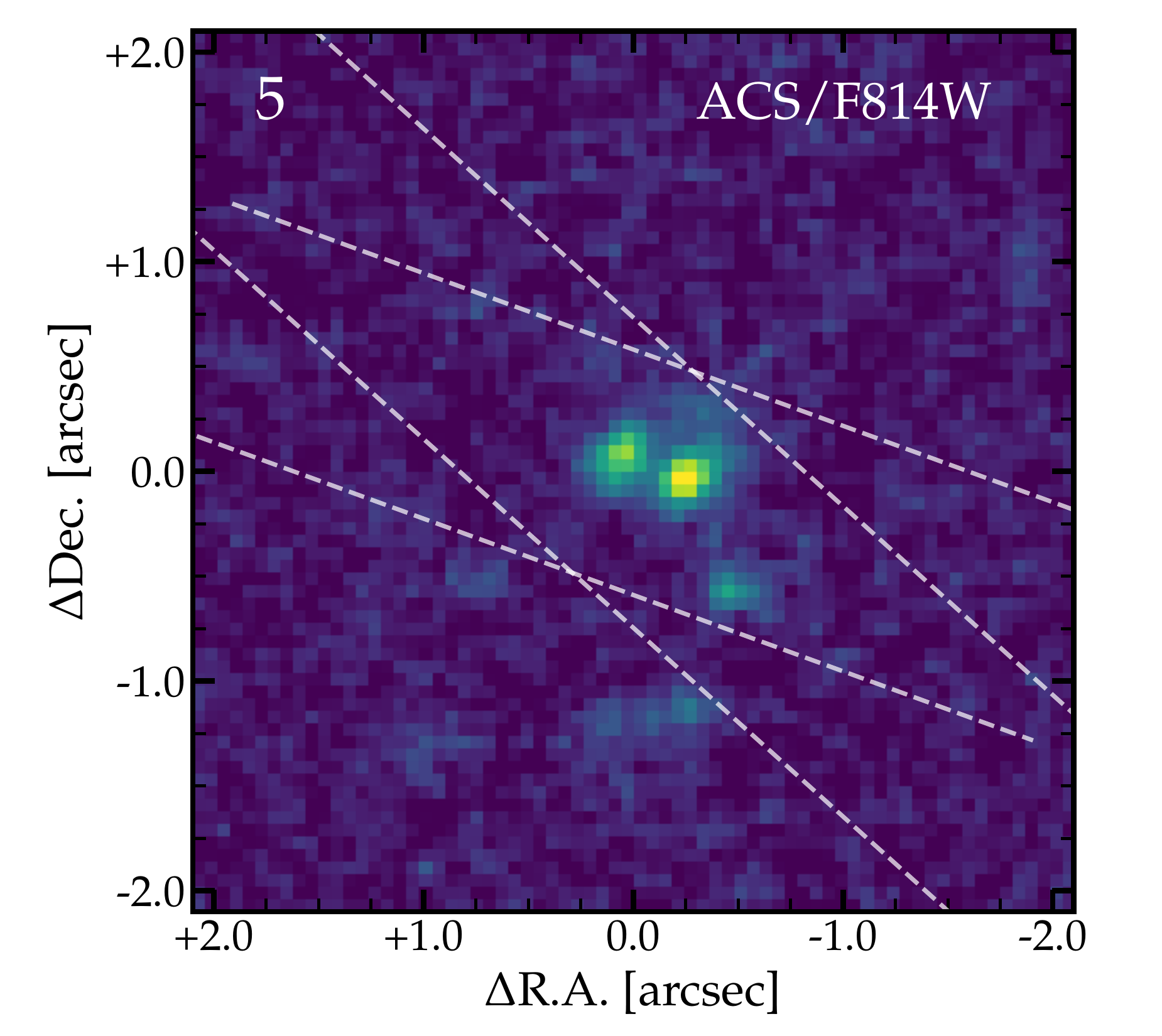} &
\hspace{-1cm}\includegraphics[width=4.8cm]{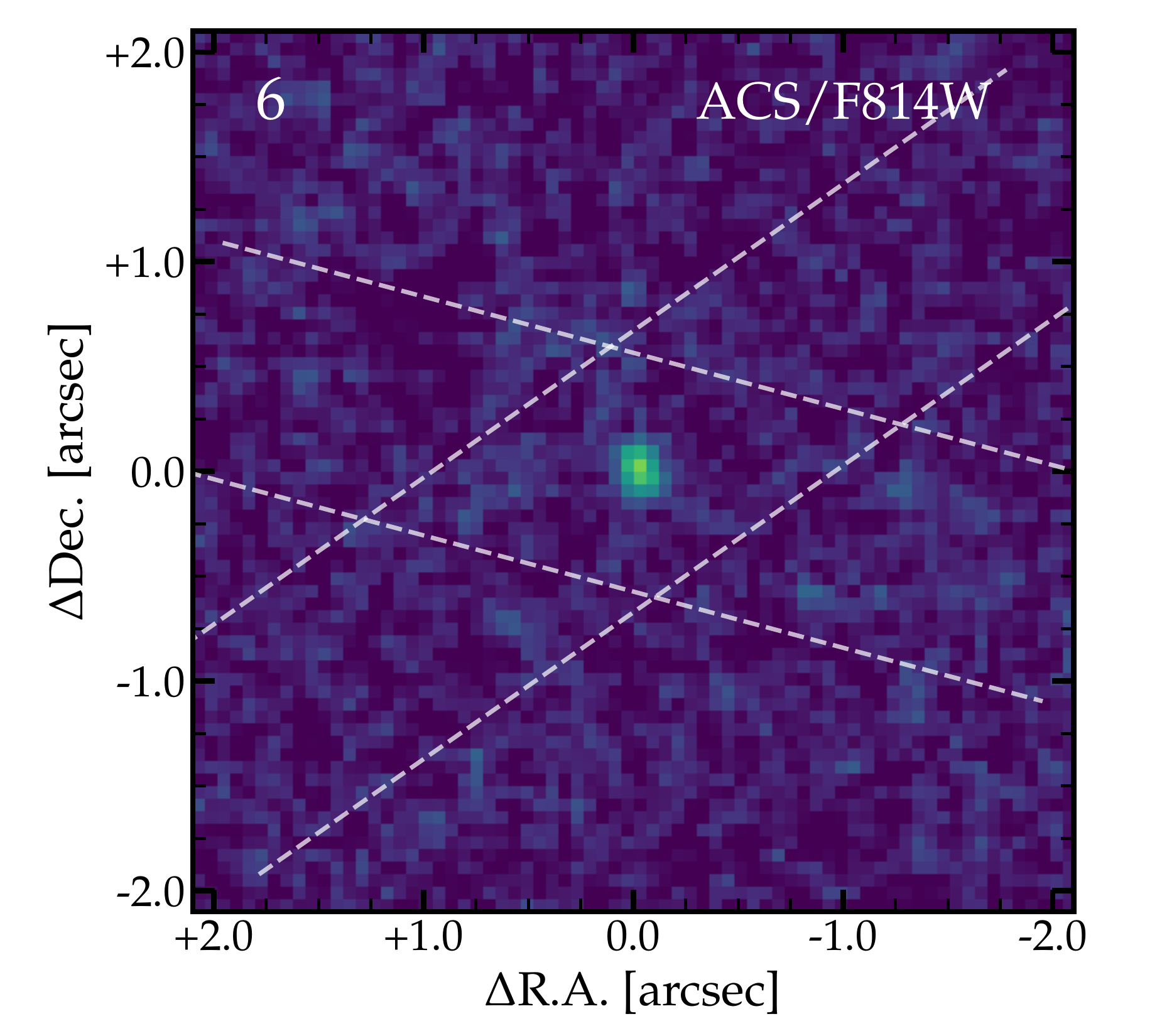} &
\hspace{-1cm}\includegraphics[width=4.8cm]{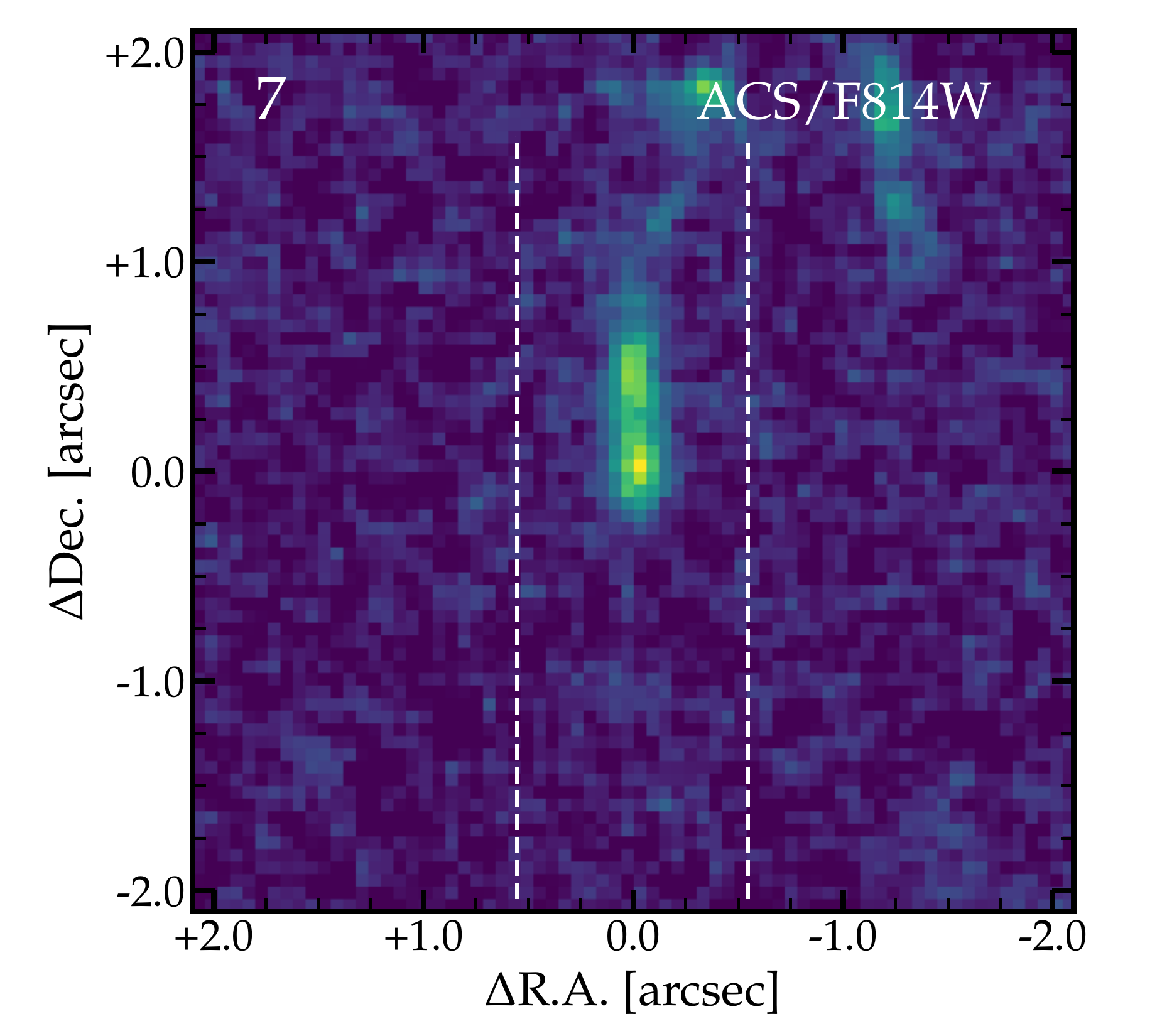} &
\hspace{-1cm}\includegraphics[width=4.8cm]{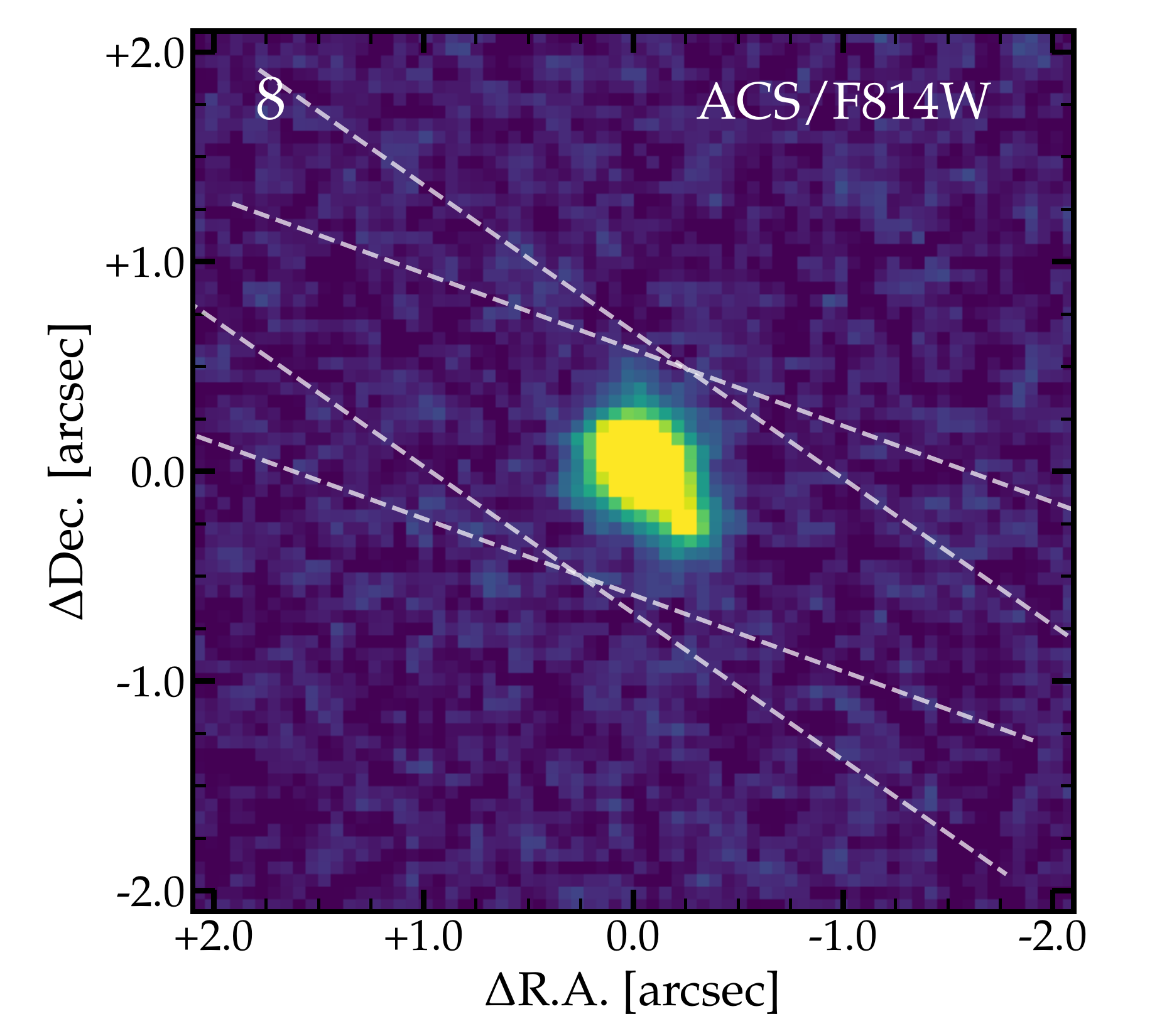} \\
\includegraphics[width=4.8cm]{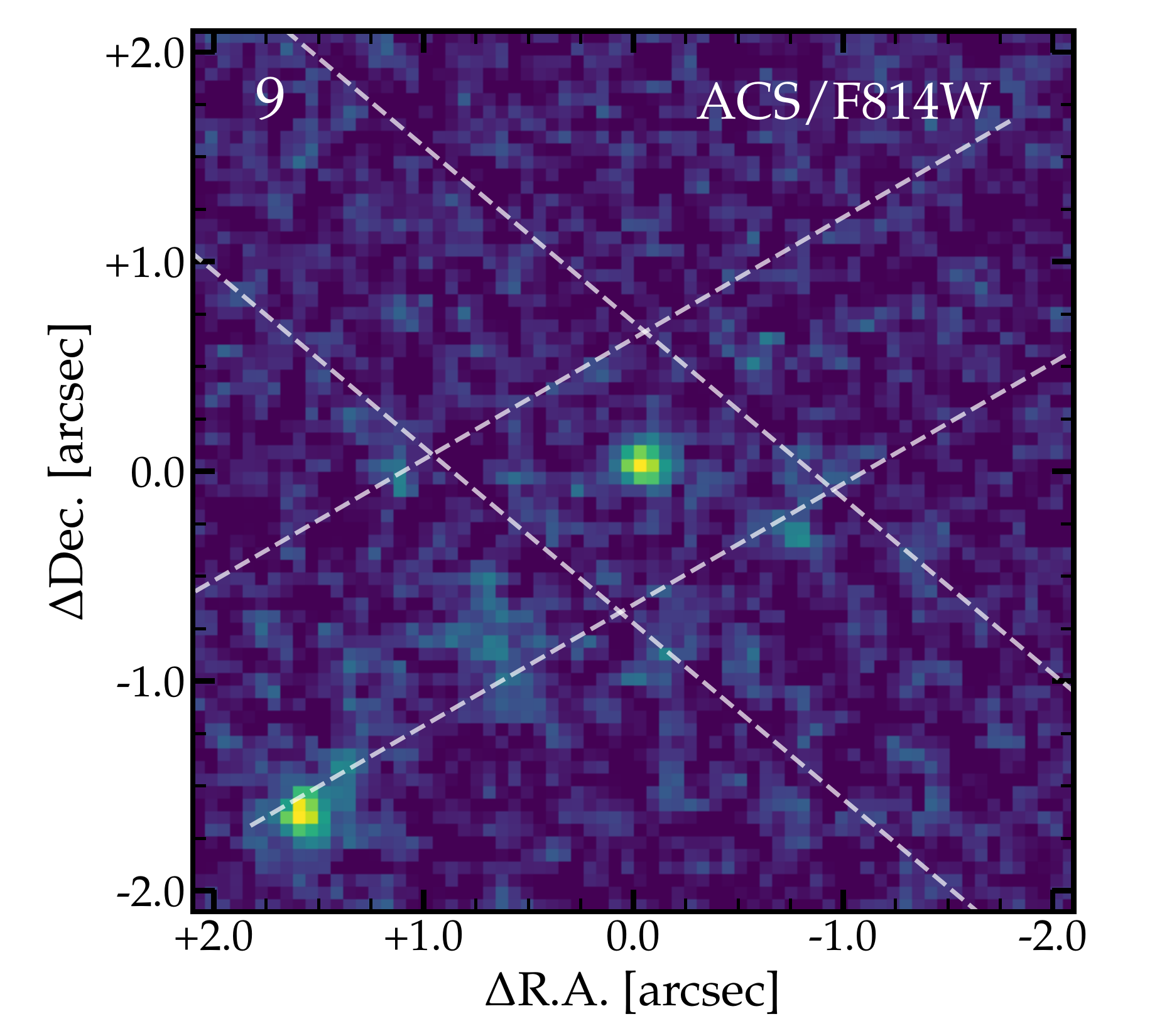} &
\hspace{-1cm}\includegraphics[width=4.8cm]{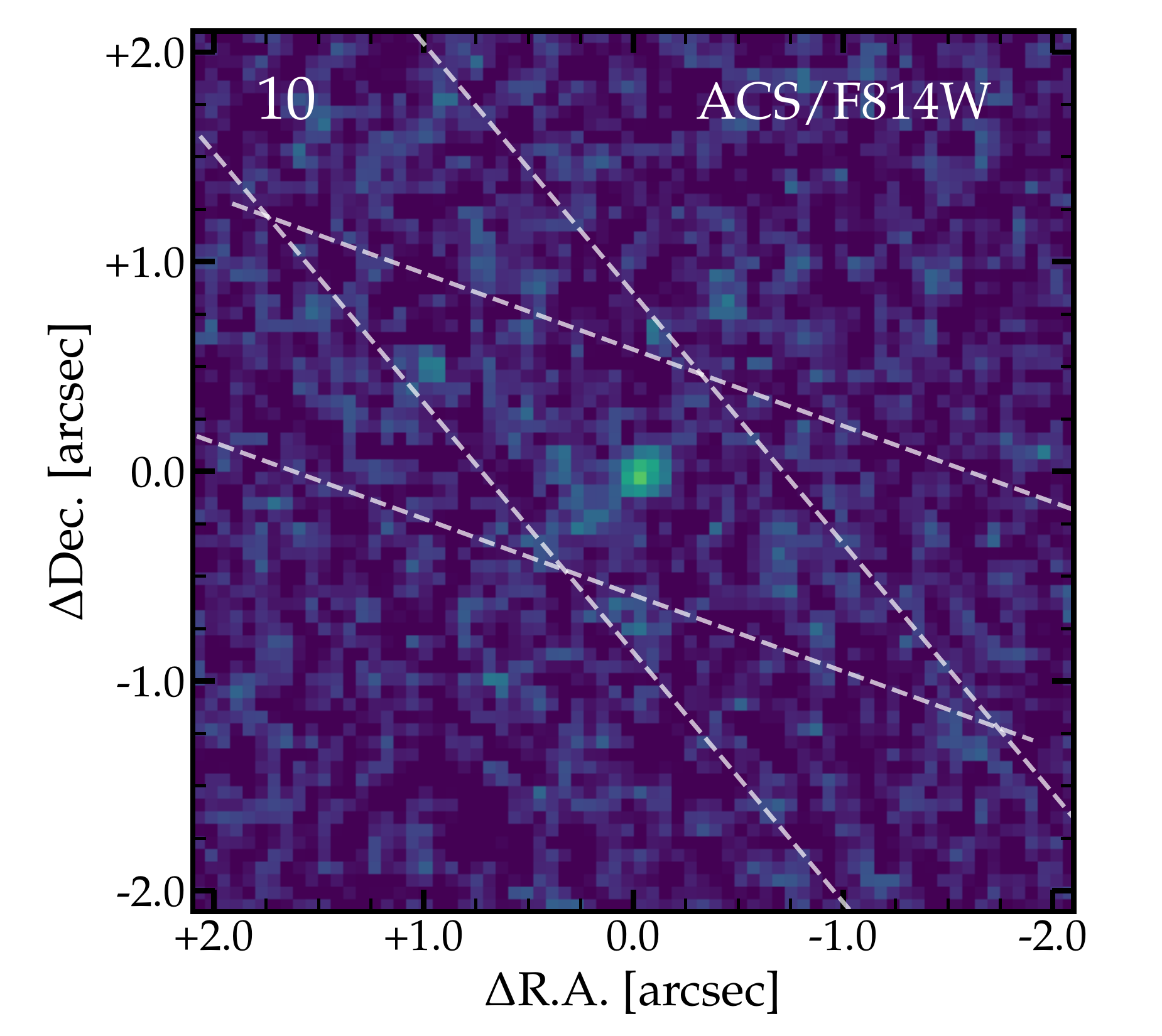} &
\hspace{-1cm}\includegraphics[width=4.8cm]{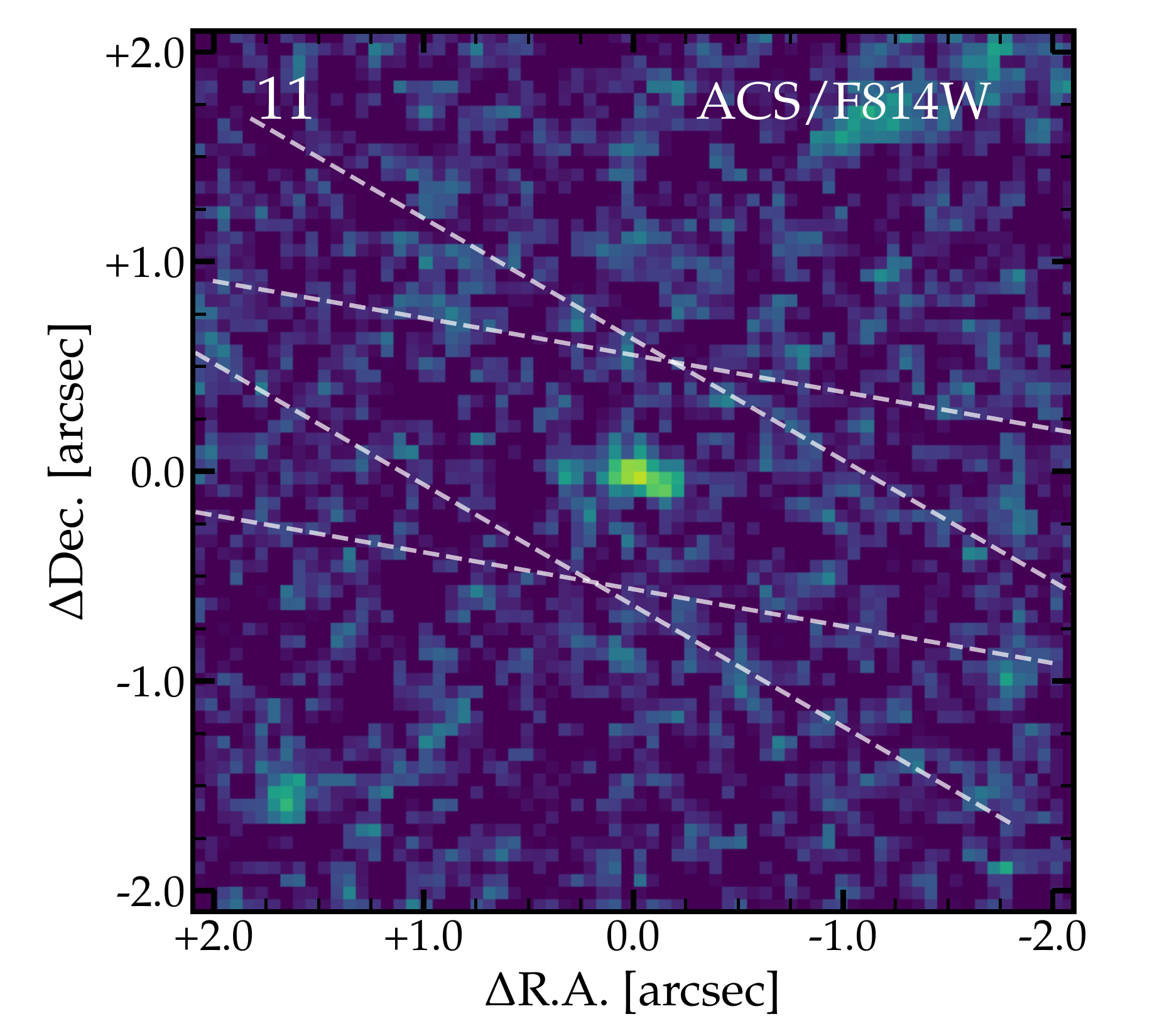} &
\hspace{-1cm}\includegraphics[width=4.8cm]{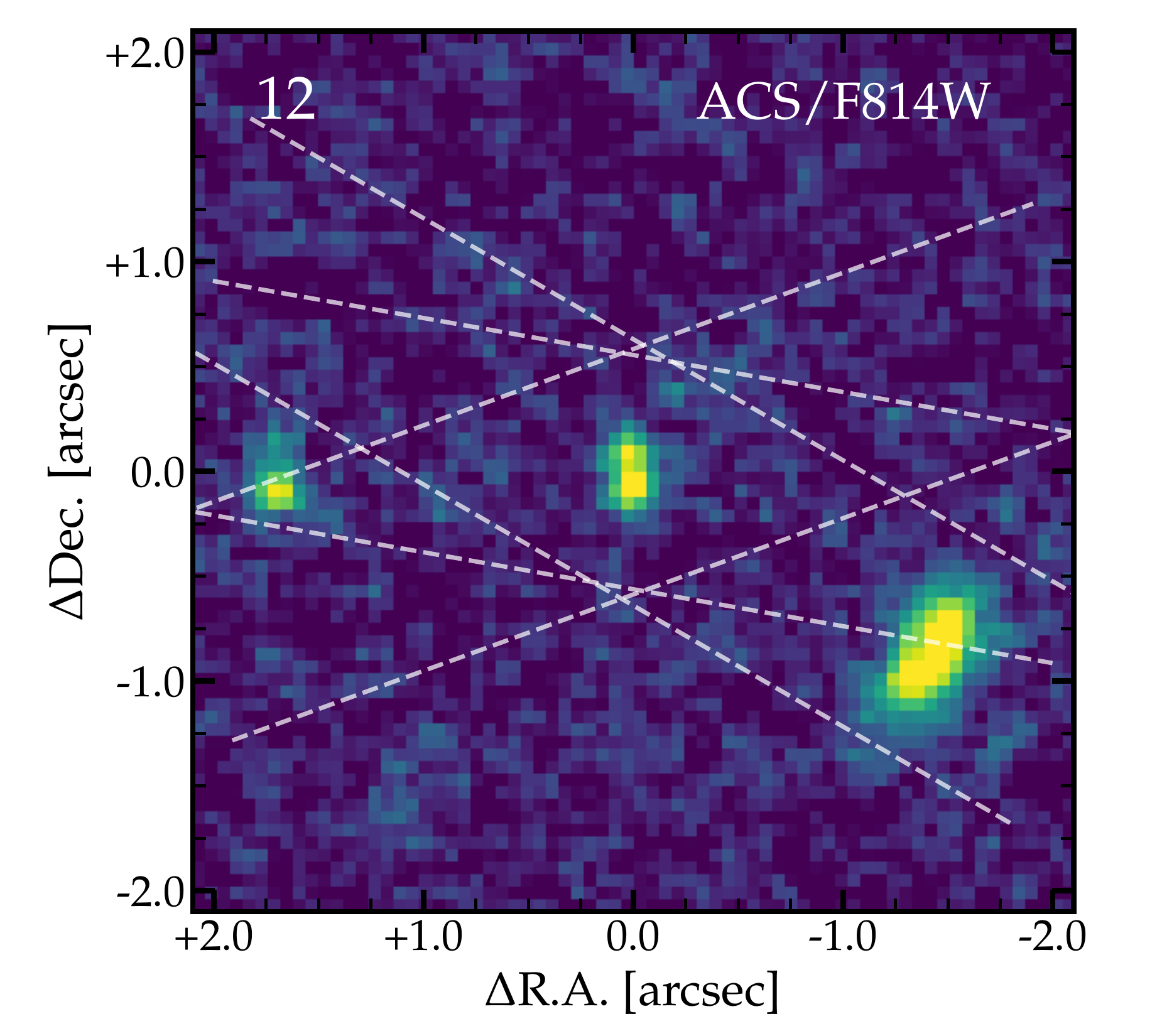} \\
\includegraphics[width=4.8cm]{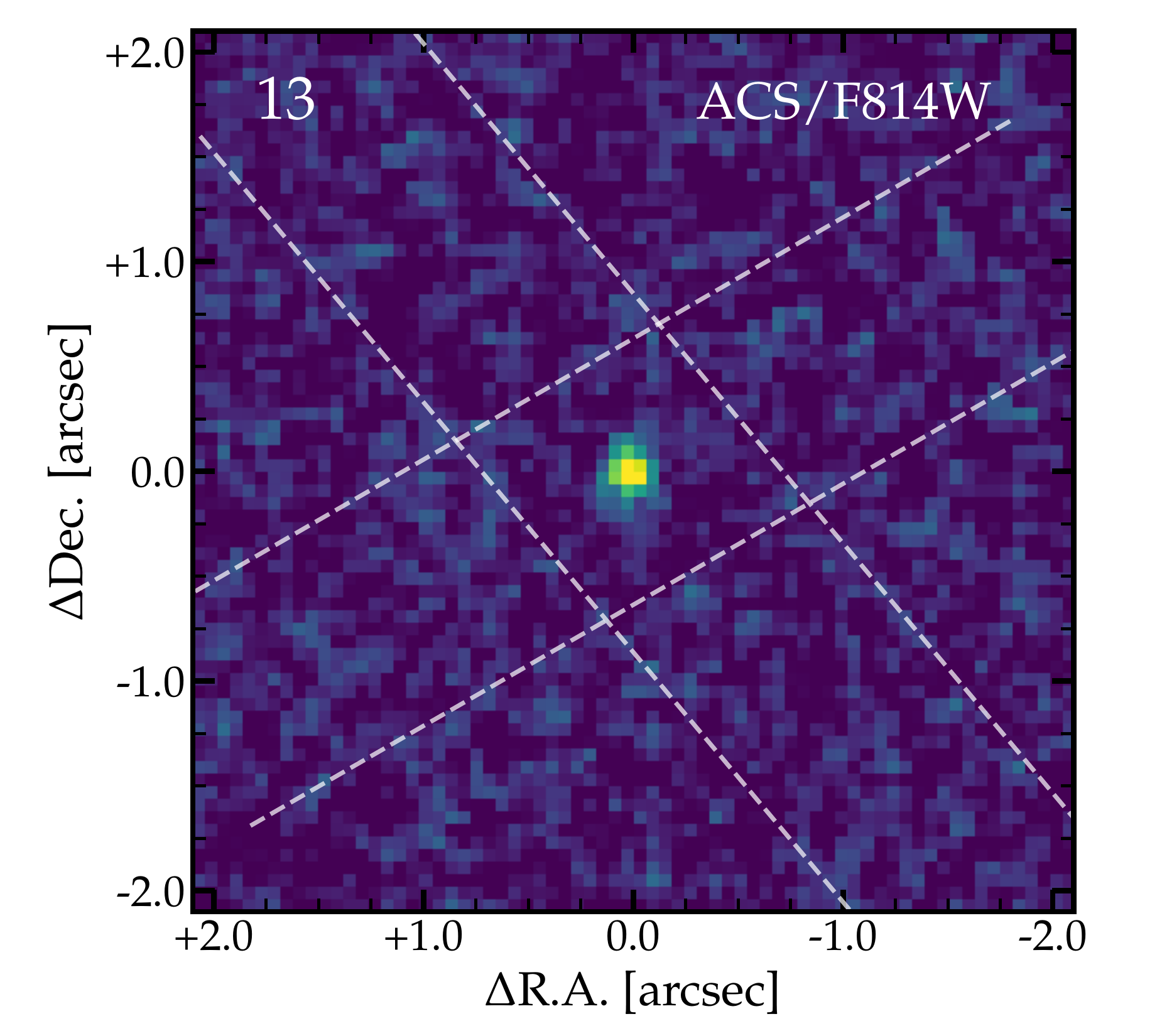} &
\hspace{-1cm}\includegraphics[width=4.8cm]{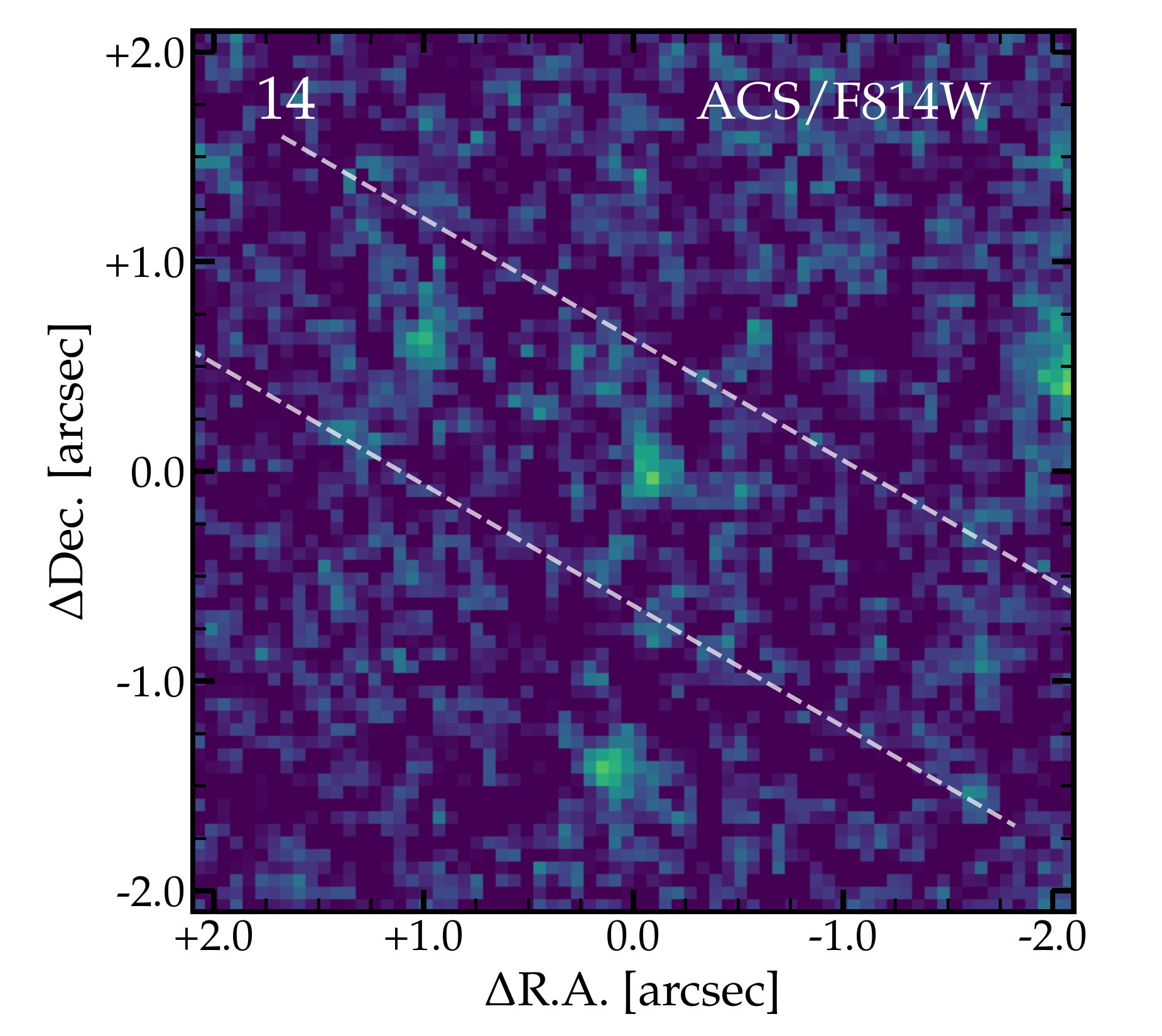} &
\hspace{-1cm}\includegraphics[width=4.8cm]{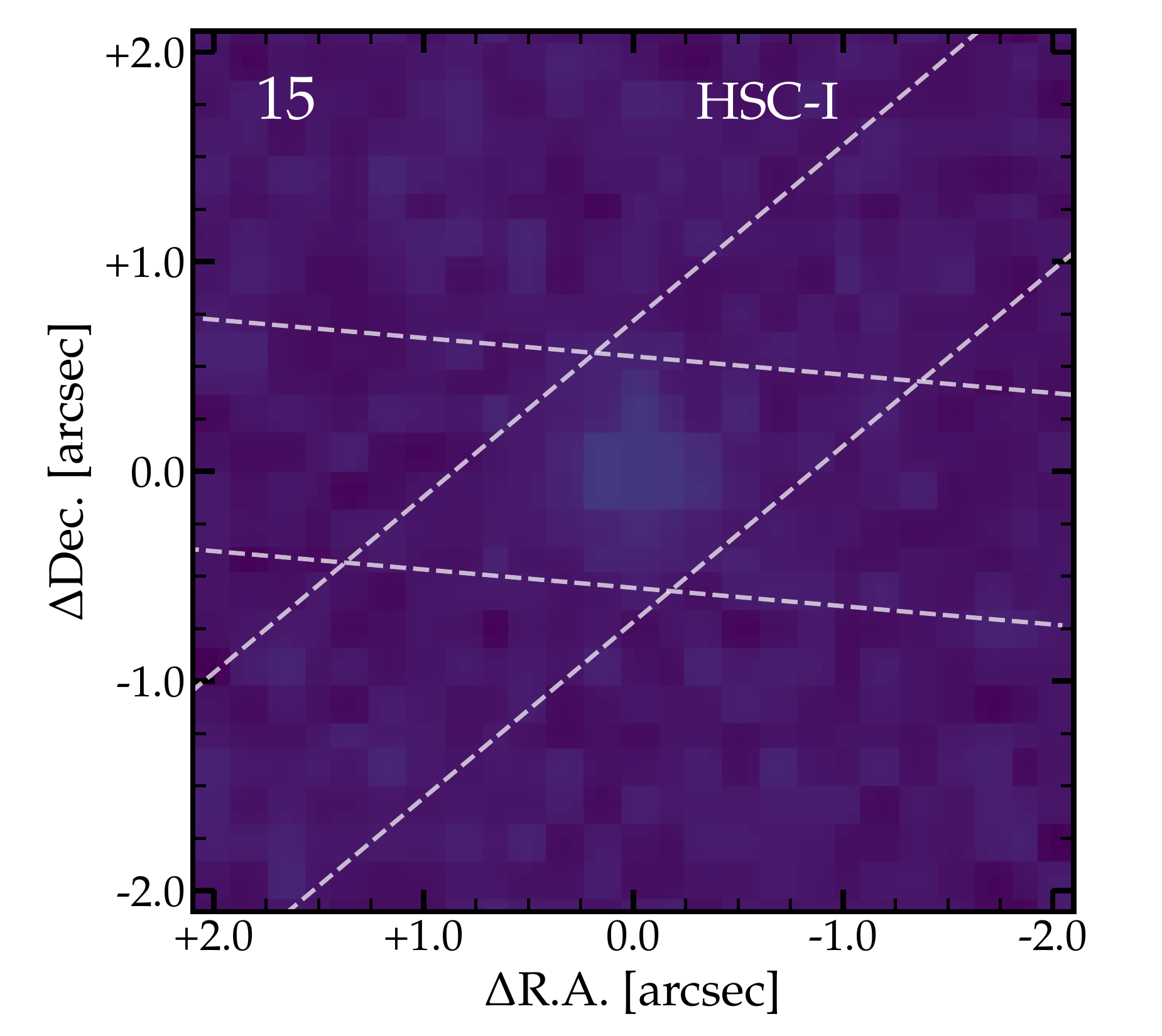} &
\hspace{-1cm}\includegraphics[width=4.8cm]{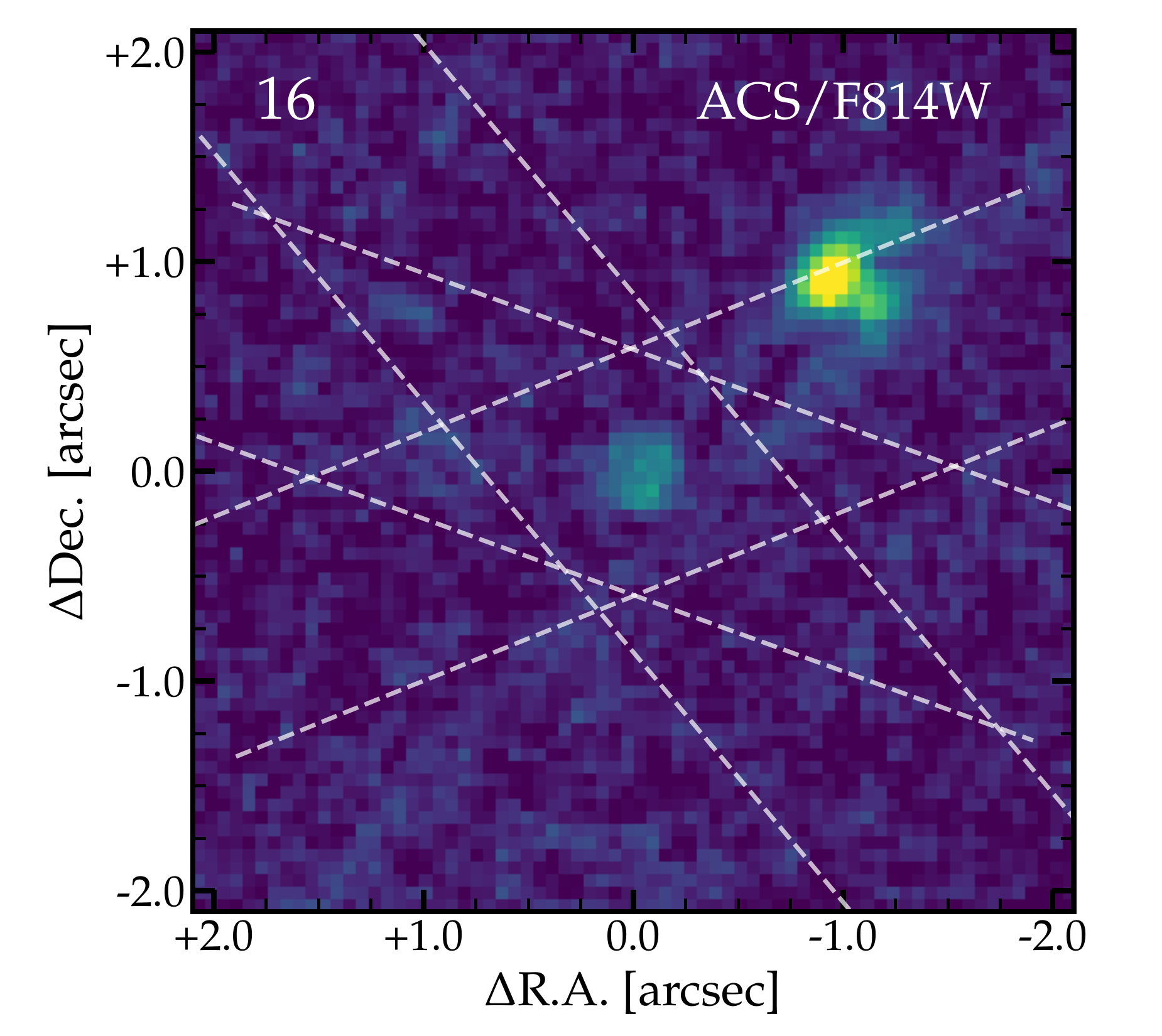} \\
\includegraphics[width=4.8cm]{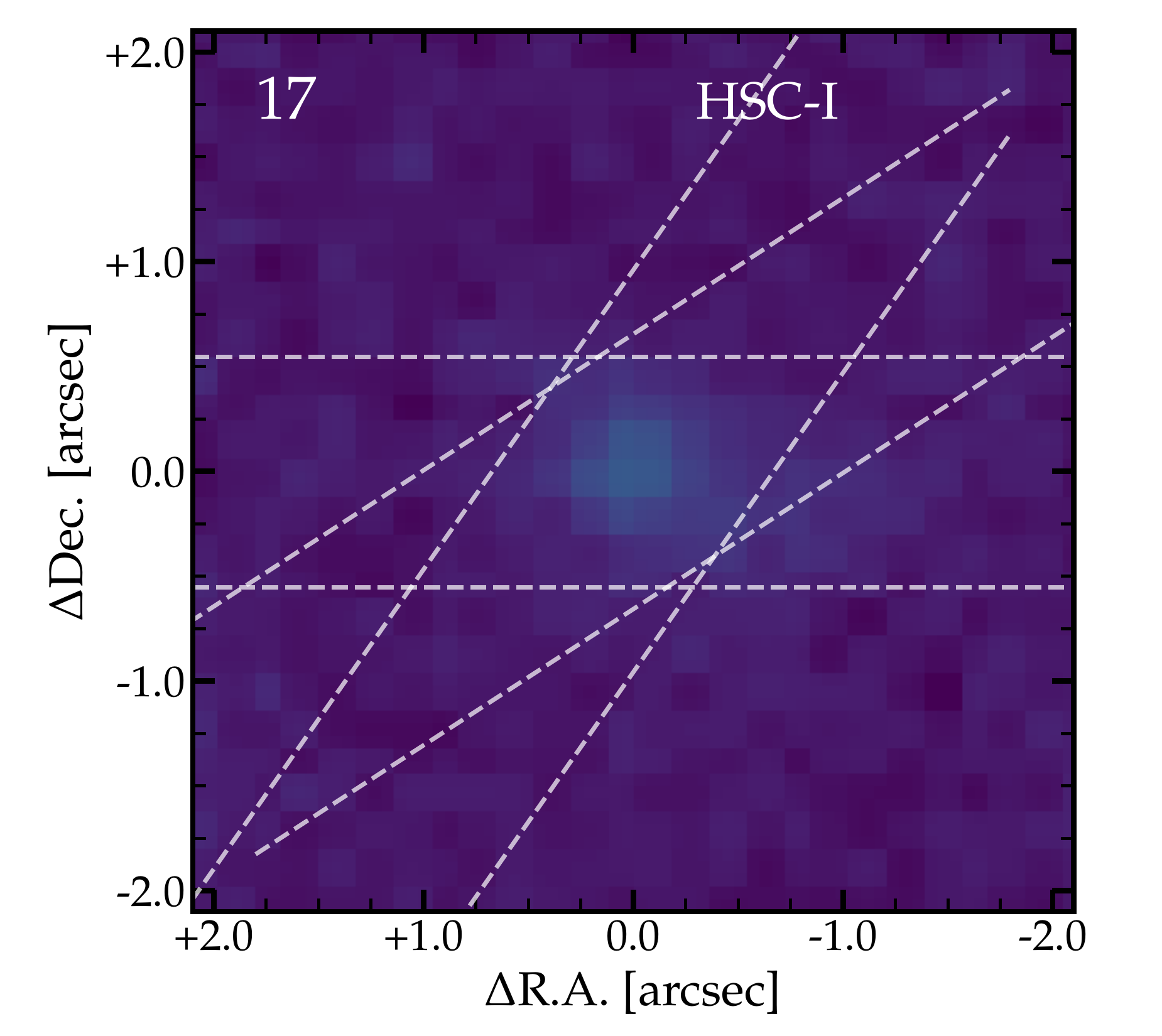} &
\hspace{-1cm}\includegraphics[width=4.8cm]{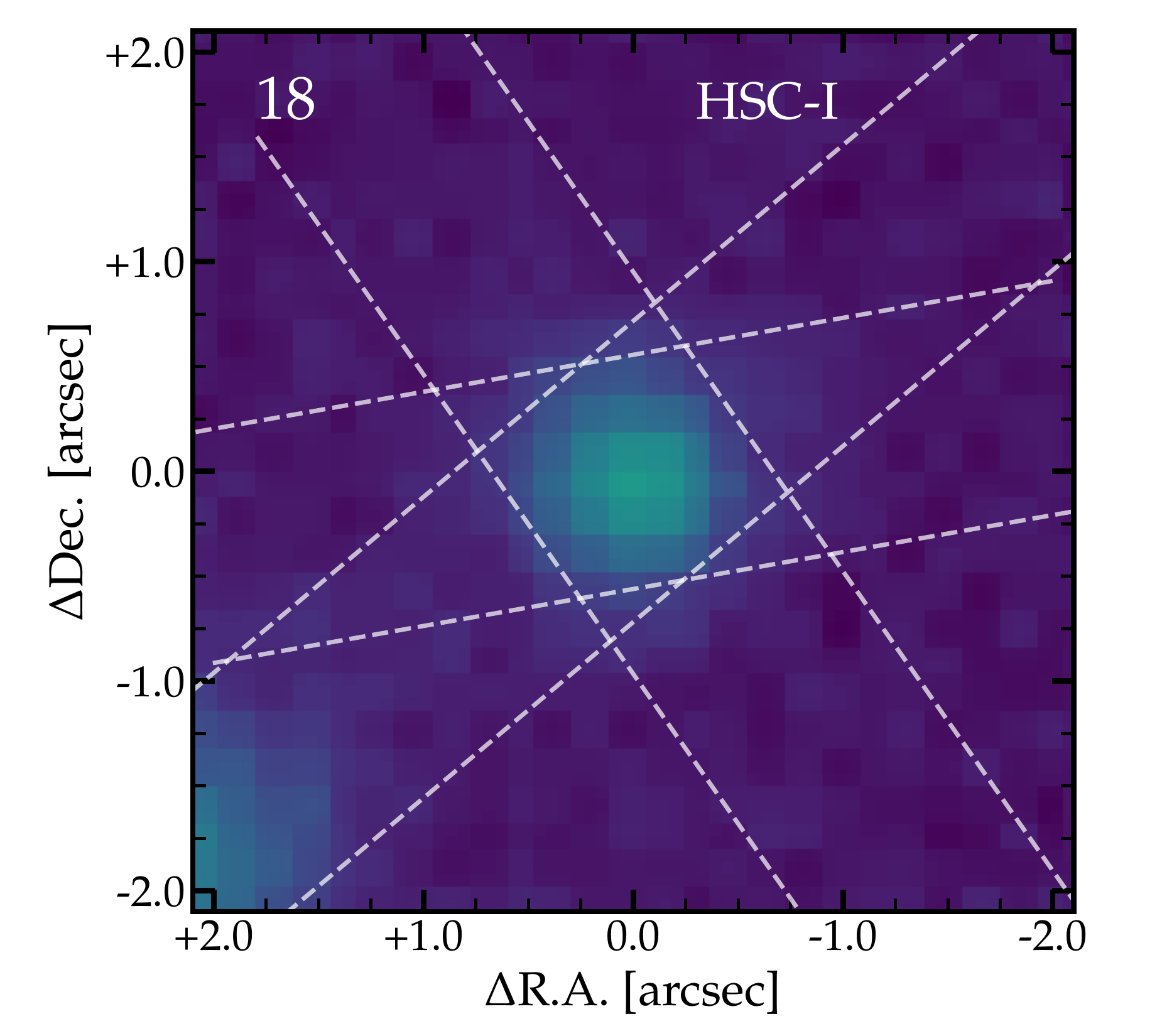} &
\hspace{-1cm}\includegraphics[width=4.8cm]{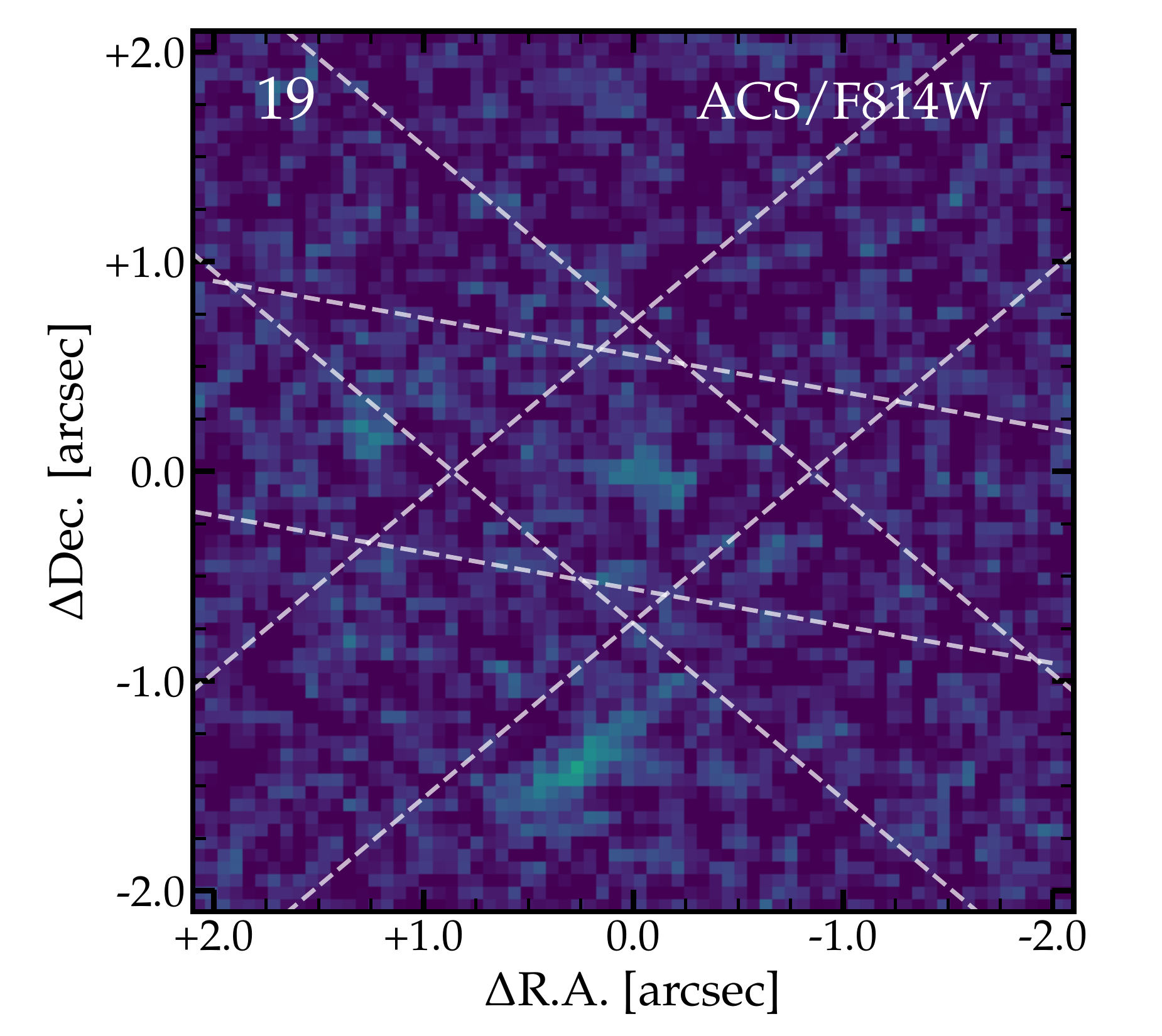} &
\hspace{-1cm}\includegraphics[width=4.8cm]{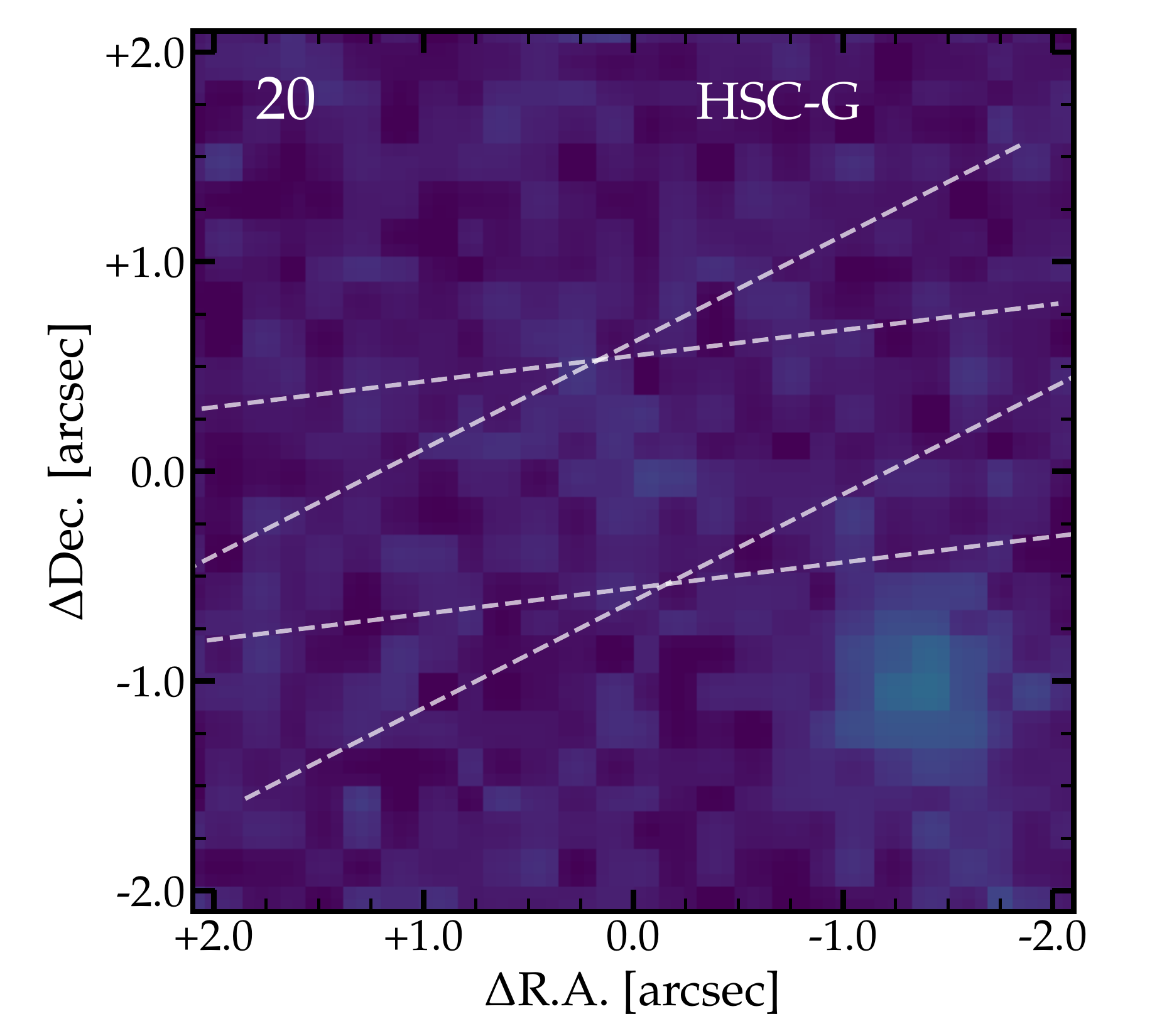} \\
\end{tabular}
\caption{Cut-out images of the {\it HST} data in the ACS/F814W filter that traces the rest-frame UV centred on the XLS objects. White dashed lines show the orientations of the slits in the various OBs. For XLS-15, 17, 18, 20 there is no {\it HST} data available such that we show ground-based data in the $g$ band from the HSC survey. These data are shallower and have a lower resolution compared to the {\it HST} data. }
\label{fig:thumbnails}
\end{figure*}

\begin{figure*}
\begin{tabular}{cccc}
\includegraphics[width=4.8cm]{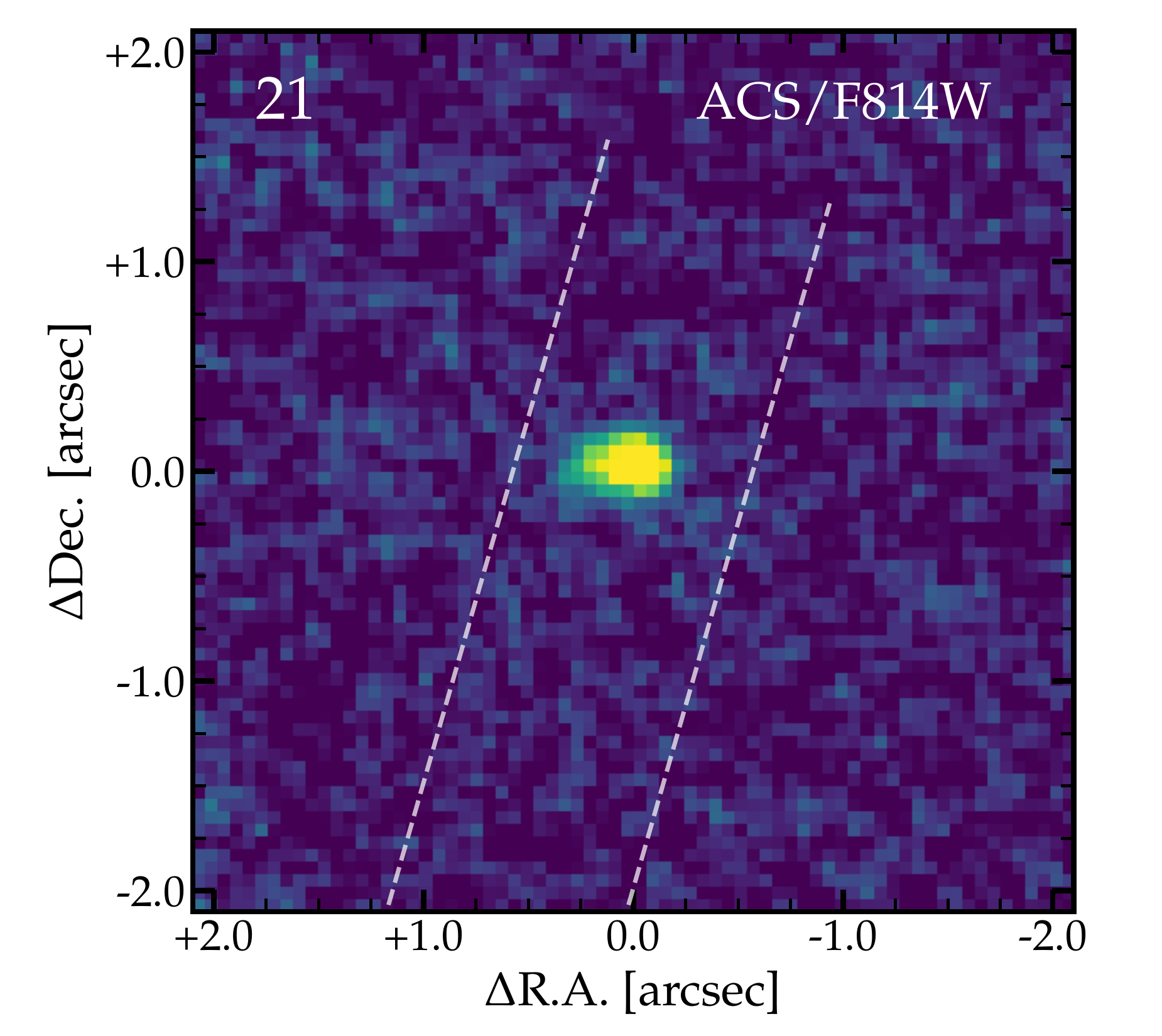} &
\hspace{-1cm}\includegraphics[width=4.8cm]{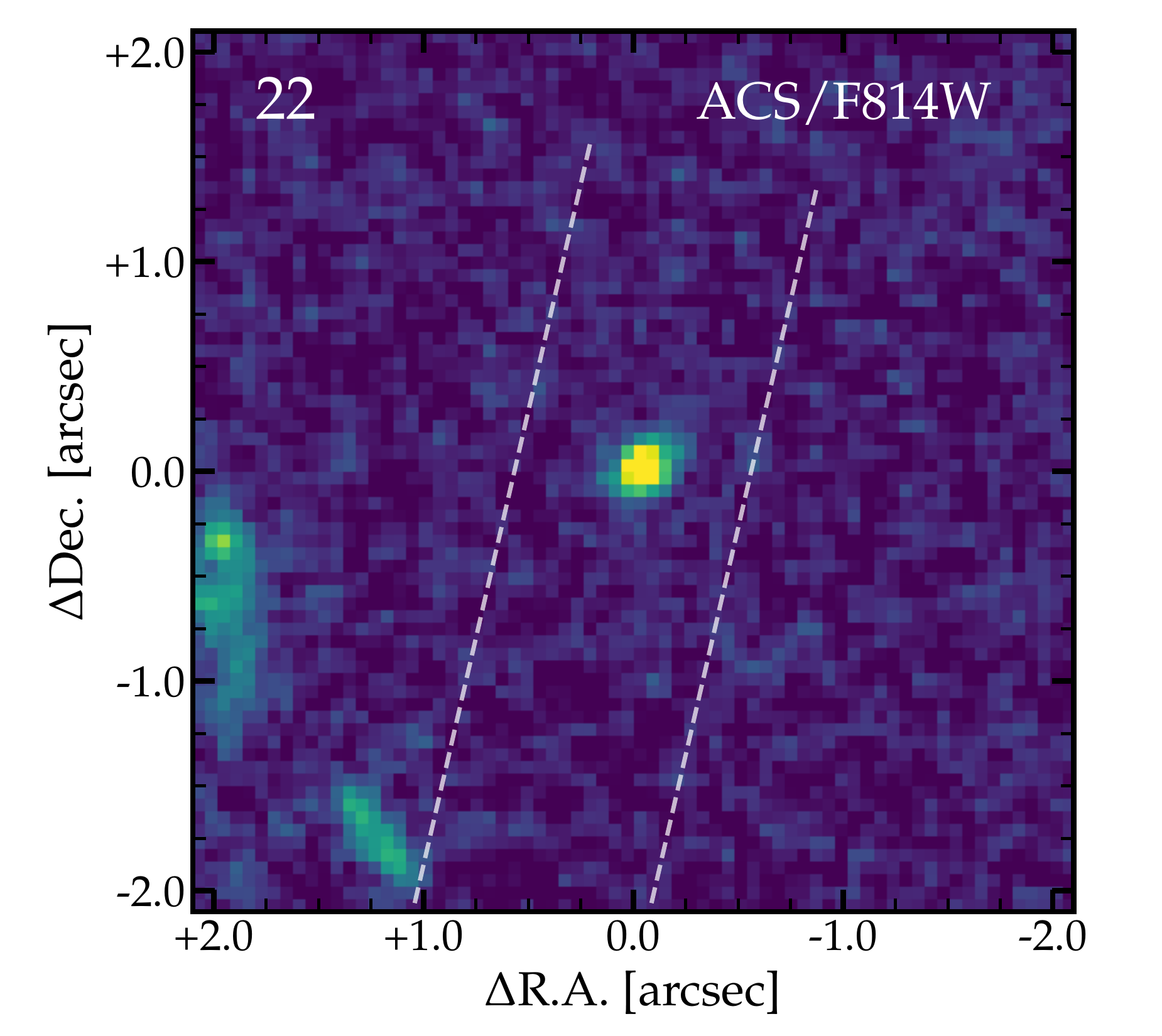} &
\hspace{-1cm}\includegraphics[width=4.8cm]{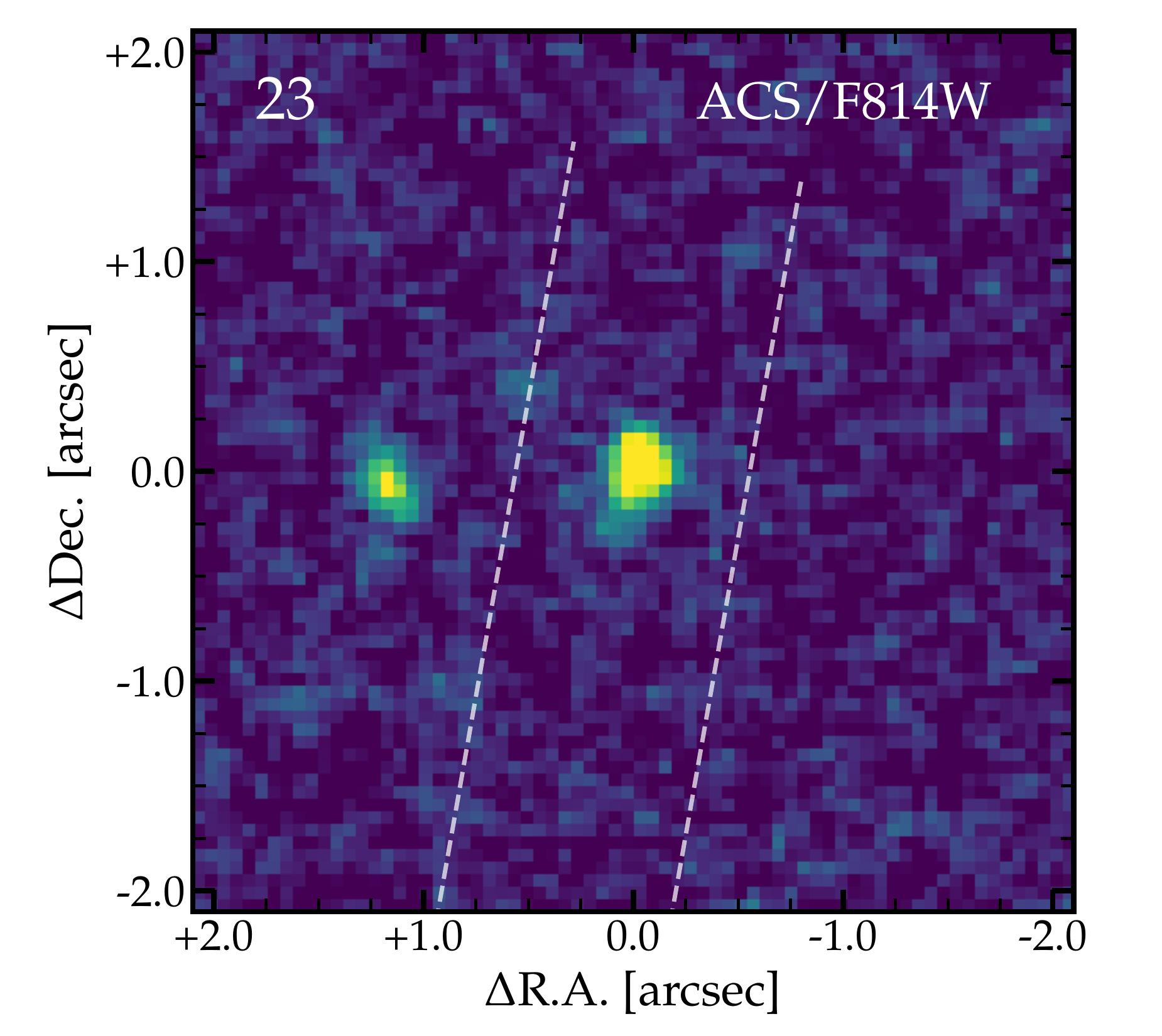} &
\hspace{-1cm}\includegraphics[width=4.8cm]{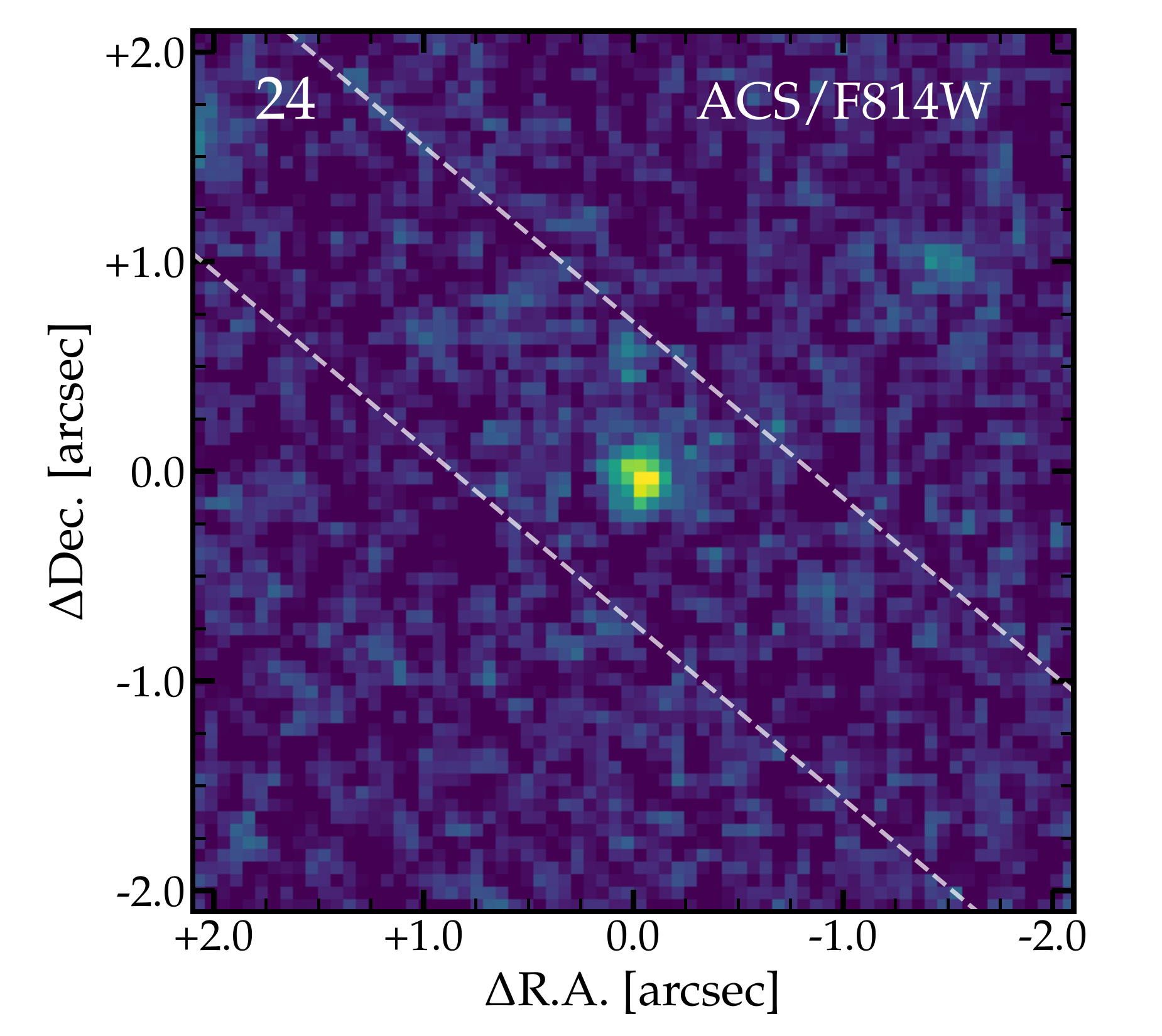} \\
\includegraphics[width=4.8cm]{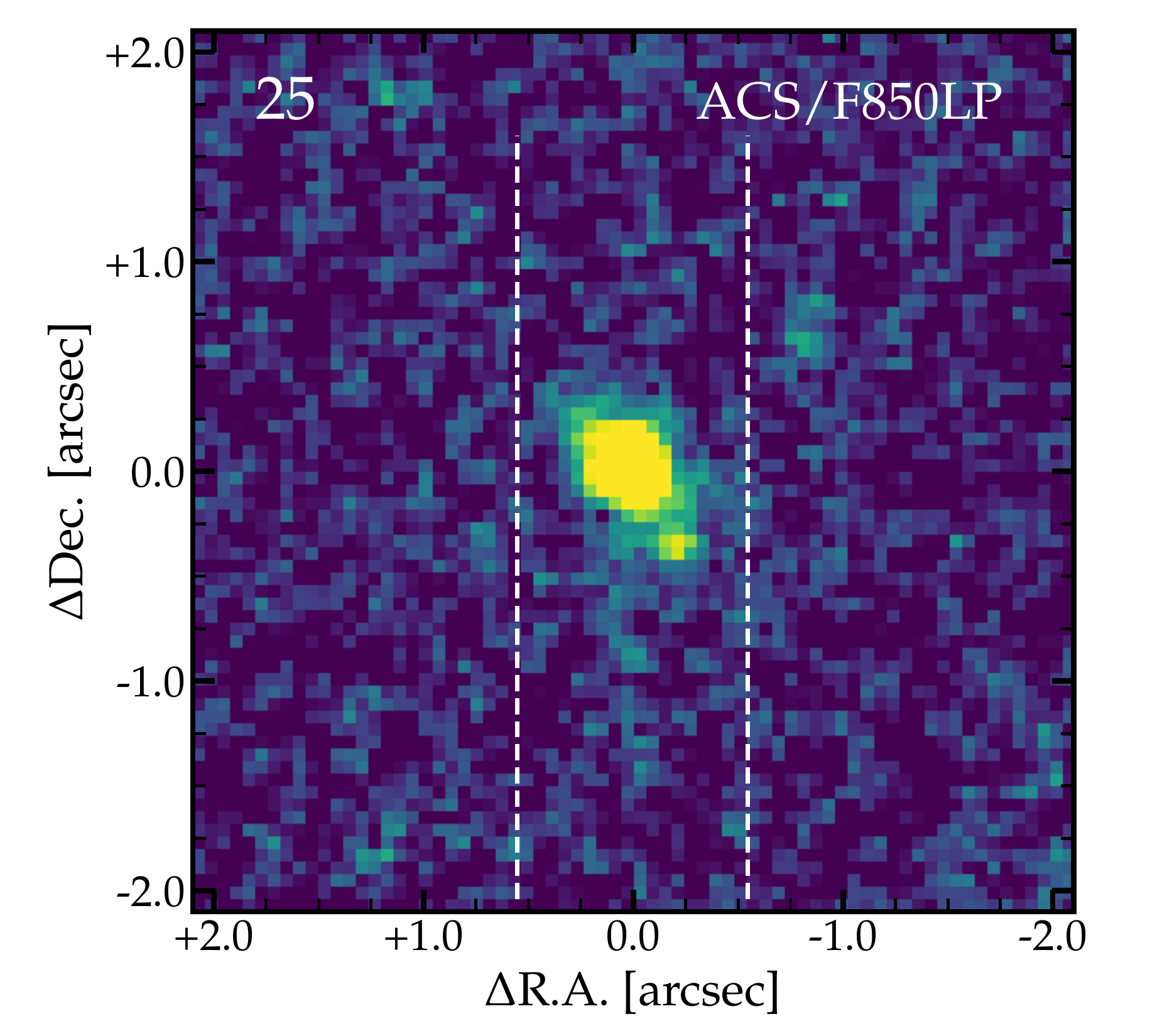} &
\hspace{-1cm}\includegraphics[width=4.8cm]{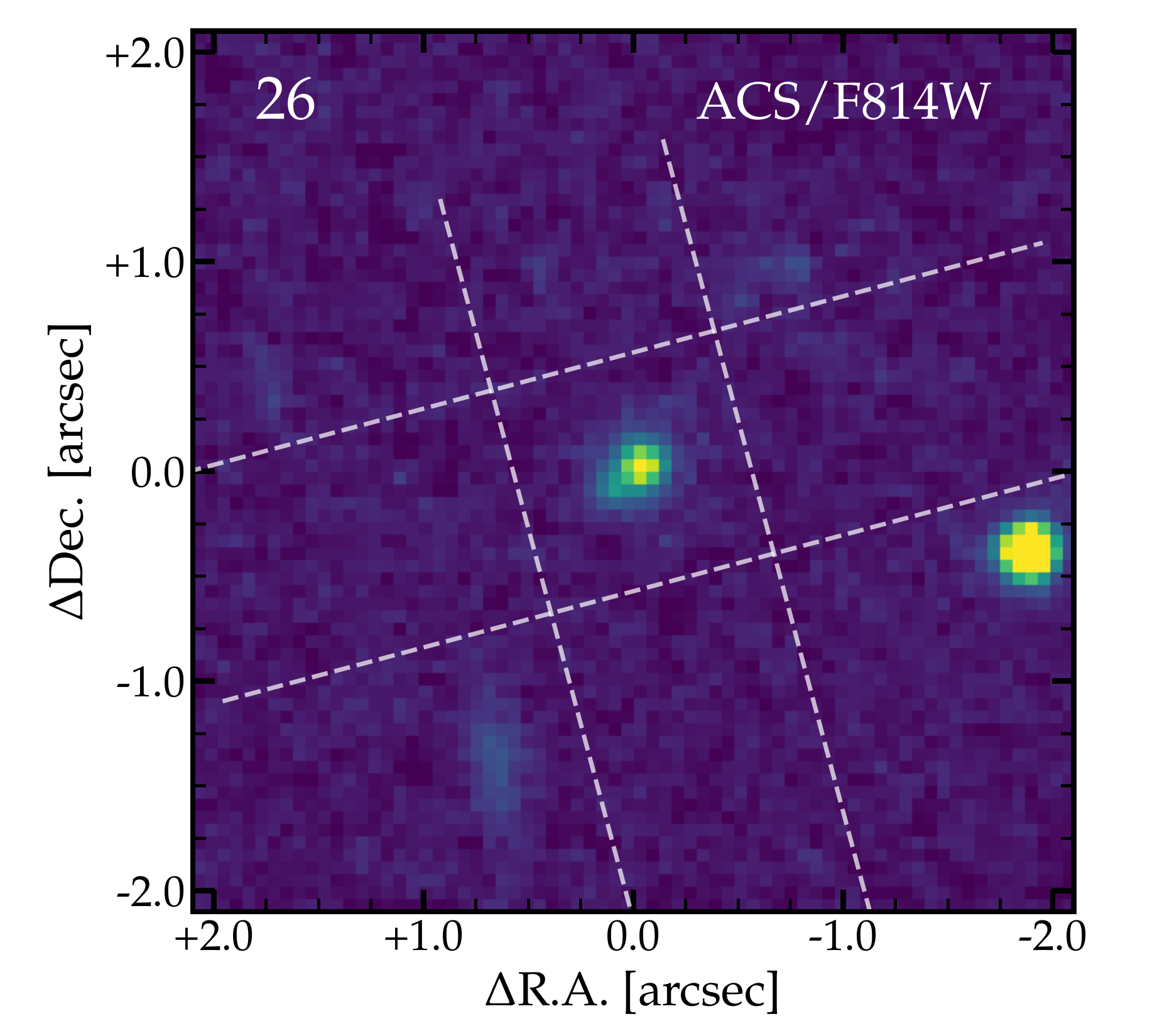} &
\hspace{-1cm}\includegraphics[width=4.8cm]{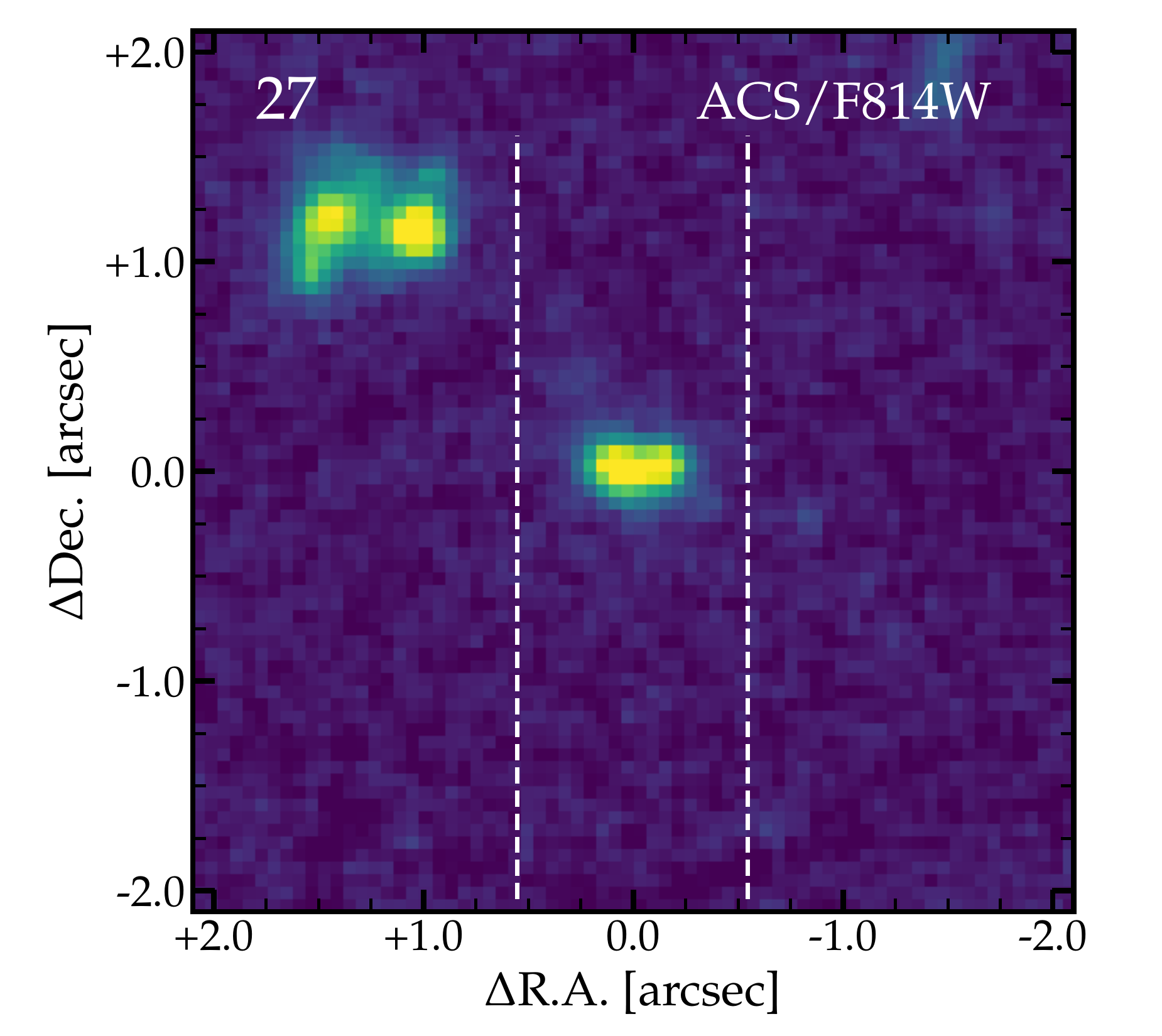} &
\hspace{-1cm}\includegraphics[width=4.8cm]{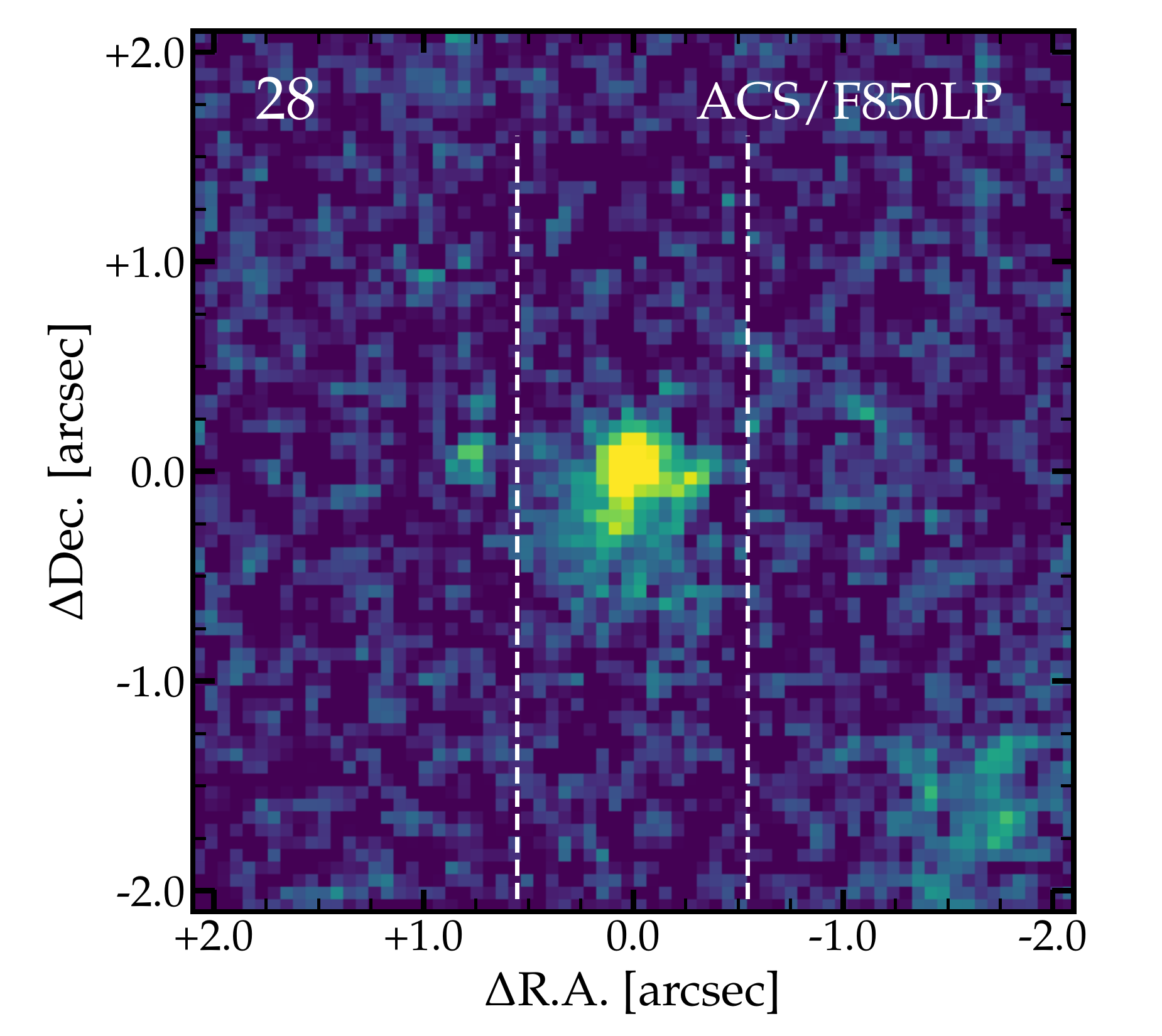} \\
\includegraphics[width=4.8cm]{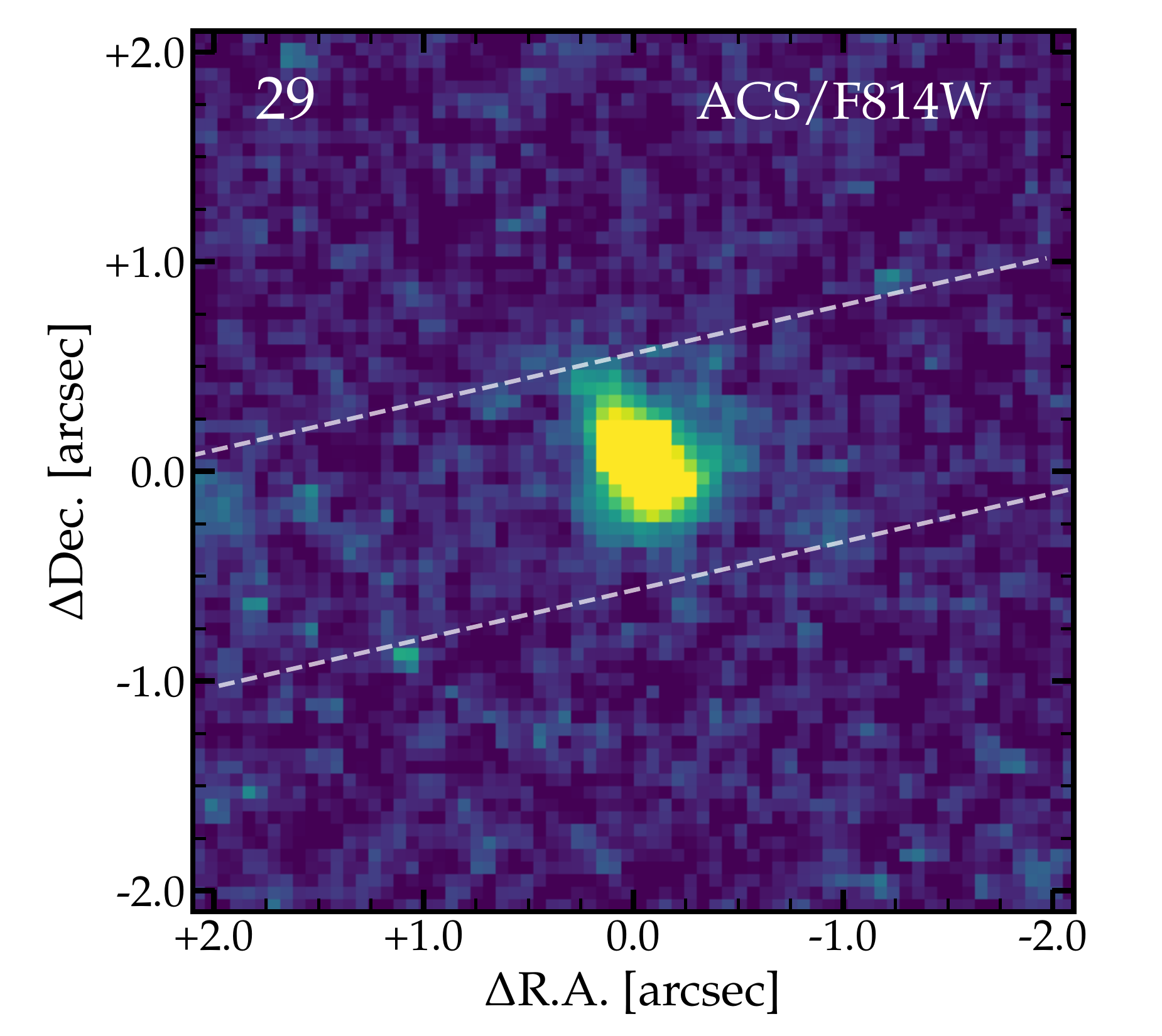} &
\hspace{-1cm}\includegraphics[width=4.8cm]{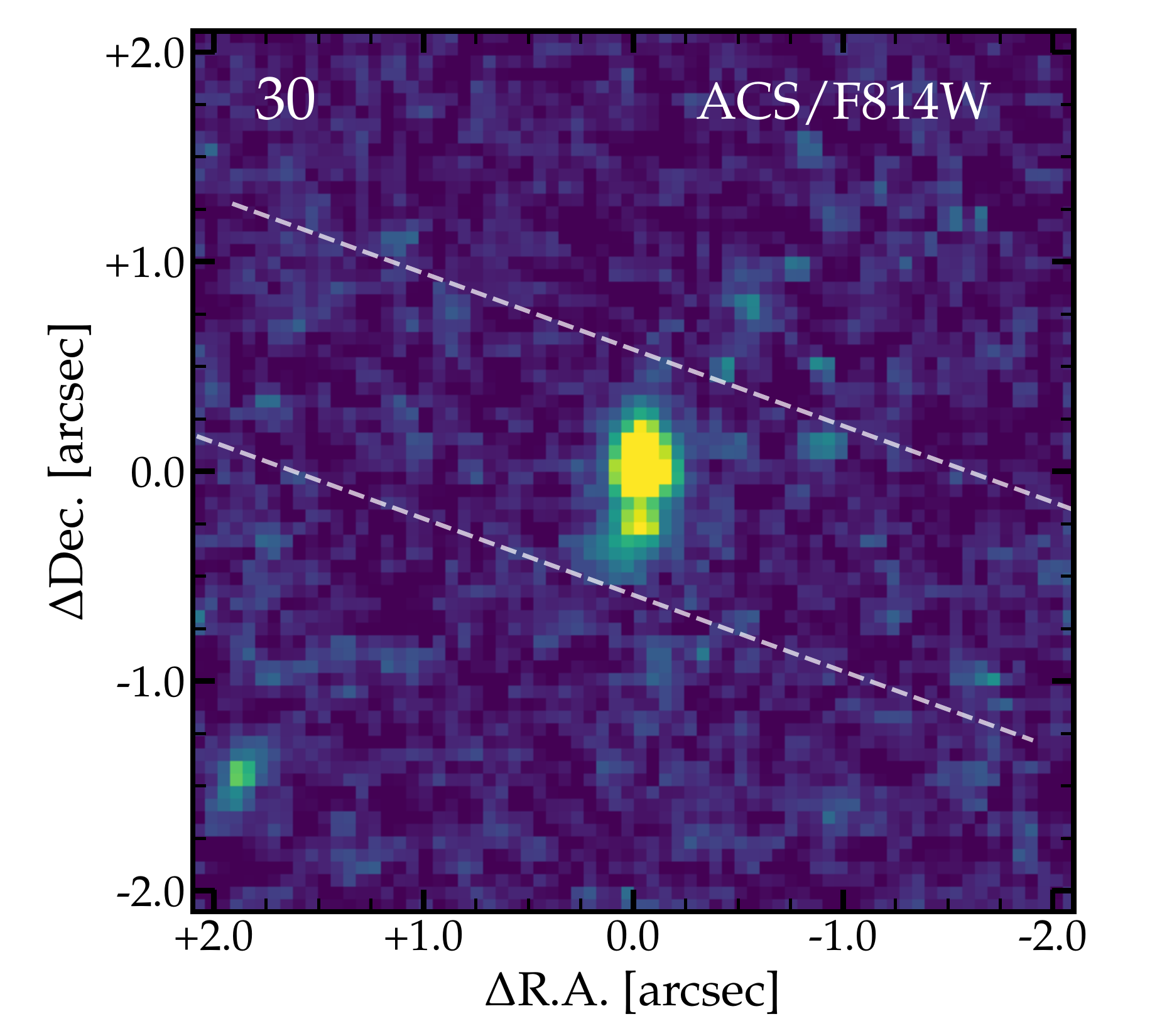} &
\hspace{-1cm}\includegraphics[width=4.8cm]{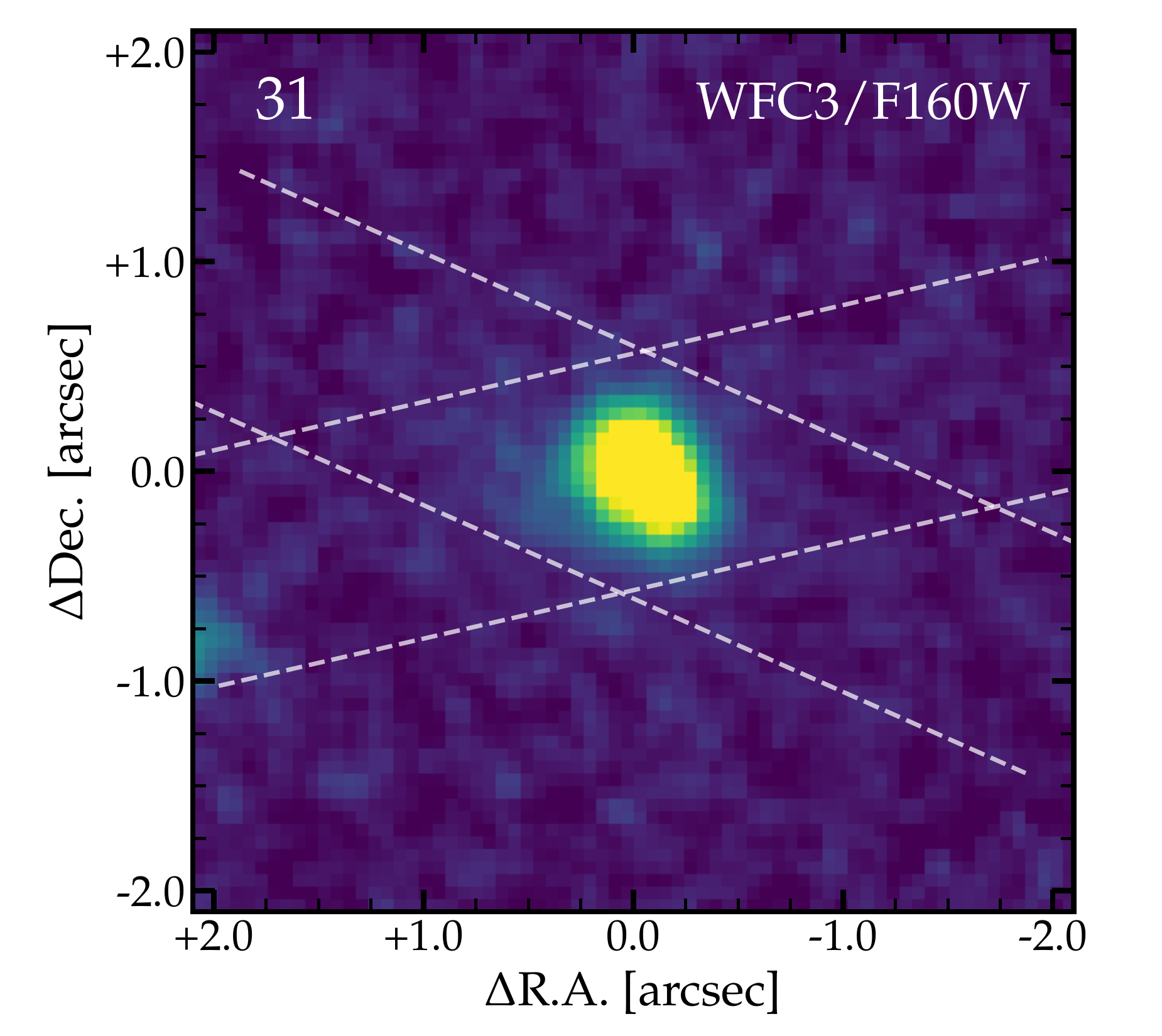} &
\hspace{-1cm}\includegraphics[width=4.8cm]{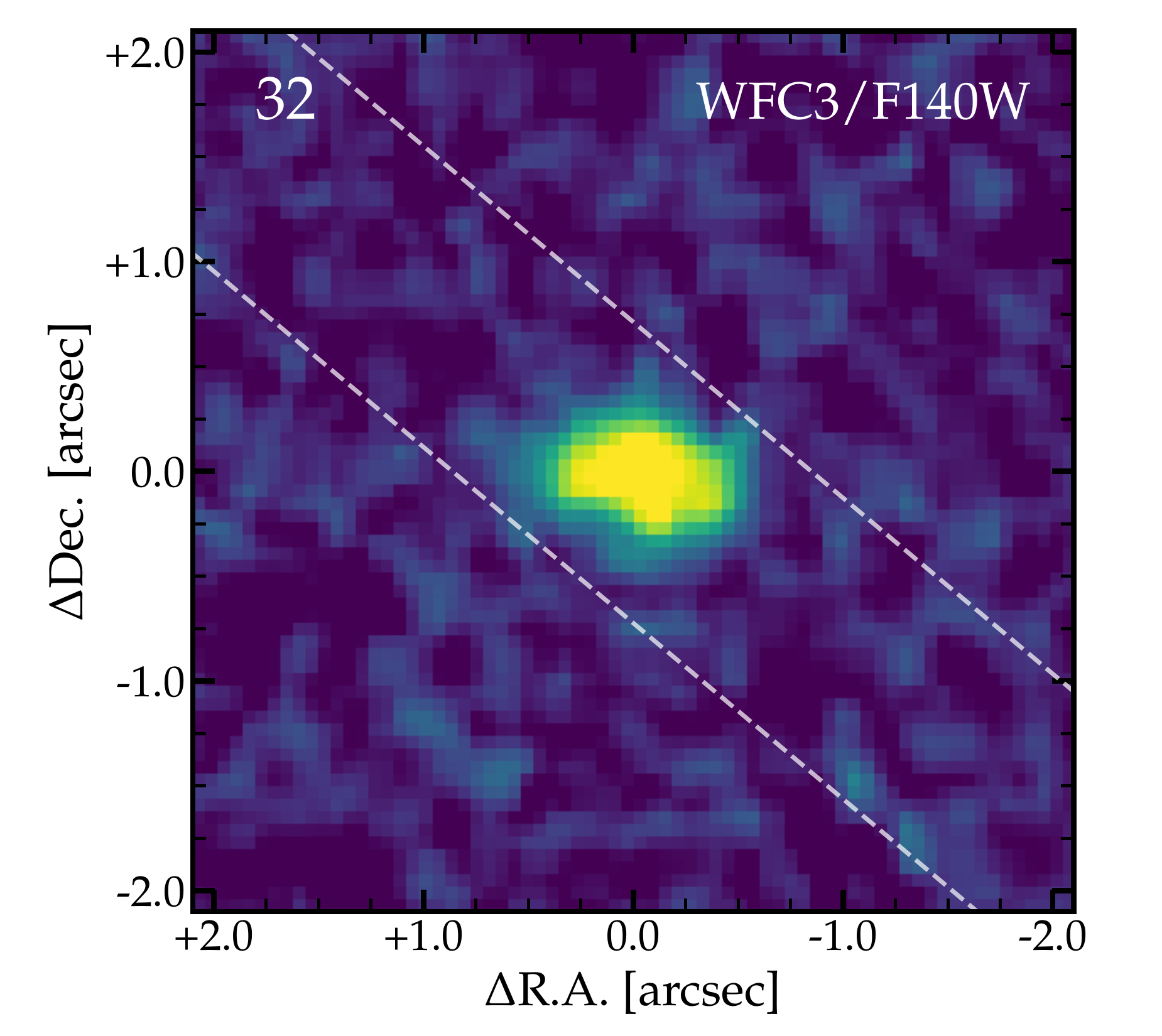} \\
\includegraphics[width=4.8cm]{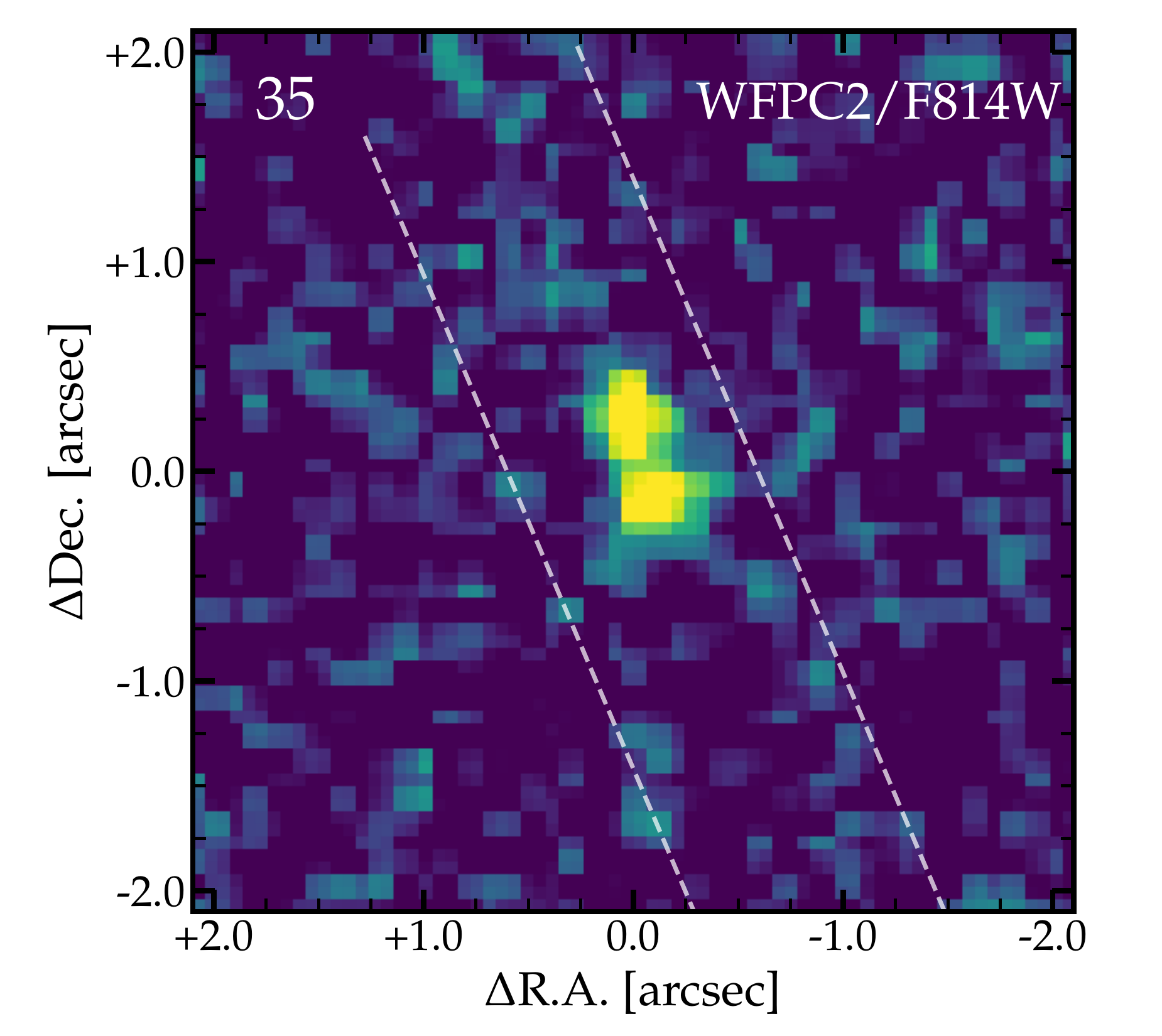} & \hspace{-1cm} \\
\end{tabular}
\caption{As Fig. $\ref{fig:thumbnails}$. For XLS-31 and 32 we can only show rest-frame optical data taken with {\it HST}/WFC3 instead of rest-frame UV data. The images for XLS-25 and 28 have a slightly different filter (F850LP) and the image for XLS-35 has been taken with WFPC2.}
\label{fig:thumbnails2}
\end{figure*}


\bsp	
\label{lastpage}
\end{document}